\documentclass[onecolumn]{aastex631}

\shorttitle{CH$_3$CCH in warm molecular gas
}
\shortauthors{Li et al.}

\begin{document}

\title{CH$_3$CCH as a thermometer in warm molecular gas}

\author[0000-0002-2243-6038]{Yuqiang Li}\thanks{E-mail: lyq@kasi.re.kr}
\affiliation{Shanghai Astronomical Observatory, Chinese Academy of Sciences, No. 80 Nandan Road, Shanghai, 200030, People’s Republic of China}
\affiliation{Korea Astronomy and Space Science Institute, No. 776, Daedeok-daero, Yuseong-gu, Daejeon, Republic of Korea}
\affiliation{School of Astronomy and Space Sciences, University of Chinese Academy of Sciences, No. 19A Yuquan Road, Beijing 100049, People’s Republic of China}
\affiliation{State Key Laboratory of Radio Astronomy and Technology, A20 Datun Road, Chaoyang District, Beijing, 100101, P. R. China}
\author[0000-0001-6106-1171]{Junzhi Wang}\thanks{E-mail: junzhiwang@gxu.edu.cn}
\affiliation{Guangxi Key Laboratory for Relativistic Astrophysics, Department of Physics, Guangxi University, Nanning 530004, People’s Republic of China}
\author[0000-0003-3520-6191]{Juan Li}\thanks{E-mail: lijuan@shao.ac.cn}
\affiliation{Shanghai Astronomical Observatory, Chinese Academy of Sciences, No. 80 Nandan Road, Shanghai, 200030, People’s Republic of China}
\affiliation{State Key Laboratory of Radio Astronomy and Technology, A20 Datun Road, Chaoyang District, Beijing, 100101, P. R. China}
\author[0000-0003-2619-9305]{Xing Lu}
\affiliation{Shanghai Astronomical Observatory, Chinese Academy of Sciences, No. 80 Nandan Road, Shanghai, 200030, People’s Republic of China}
\author[0000-0001-9047-846X]{Siqi Zheng}
\affiliation{Shanghai Astronomical Observatory, Chinese Academy of Sciences, No. 80 Nandan Road, Shanghai, 200030, People’s Republic of China}
\affiliation{School of Astronomy and Space Sciences, University of Chinese Academy of Sciences, No. 19A Yuquan Road, Beijing 100049, People’s Republic of China}
\affiliation{I. Physikalisches Institut, Universit\"{a}t zu Köln, Z\"{u}lpicher Str. 77, 50937 K\"{o}ln, Germany}
\affiliation{State Key Laboratory of Radio Astronomy and Technology, A20 Datun Road, Chaoyang District, Beijing, 100101, P. R. China}
\author[0000-0003-0128-4570]{Chao Ou}
\affiliation{Guangxi Key Laboratory for Relativistic Astrophysics, Department of Physics, Guangxi University, Nanning 530004, People’s Republic of China}
\author{Qian Huang}
\affiliation{Guangxi Key Laboratory for Relativistic Astrophysics, Department of Physics, Guangxi University, Nanning 530004, People’s Republic of China}
\author[0000-0002-7338-0986]{Miguel Santander-Garc\'ia}
\affiliation{Observatorio Astron\'omico Nacional (OAN, IGN), C/ Alfonso XII 3, 28014 Madrid, Spain}  
\affiliation{Observatorio de Yebes (IGN). Cerro de la Palera s/n, 19141 Yebes, Guadalajara, Spain}
\author[0000-0002-6732-8540]{Jos\'e Jairo D\'iaz-Luis}
\affiliation{Observatorio Astron\'omico Nacional (OAN, IGN), C/ Alfonso XII 3, 28014 Madrid, Spain}  
\affiliation{Observatorio de Yebes (IGN). Cerro de la Palera s/n, 19141 Yebes, Guadalajara, Spain}
\author[0000-0002-0226-9295]{Seokho Lee}
\affiliation{Korea Astronomy and Space Science Institute, No. 776, Daedeok-daero, Yuseong-gu, Daejeon, Republic of Korea}
\author[0000-0002-5286-2564]{Tie Liu}
\affiliation{Shanghai Astronomical Observatory, Chinese Academy of Sciences, No. 80 Nandan Road, Shanghai, 200030, People’s Republic of China}
\author[0000-0003-3540-8746]{Zhiqiang Shen}
\affiliation{Shanghai Astronomical Observatory, Chinese Academy of Sciences, No. 80 Nandan Road, Shanghai, 200030, People’s Republic of China}
 


\begin{abstract}
Kinetic temperature is a fundamental parameter in molecular clouds. 
Symmetric top molecules, such as  NH$_3$ and CH$_3$CCH,  are often used as thermometers.
However, at high temperatures, NH$_3$(2,2) can be collisionally excited to NH$_3$(2,1) and rapidly decay to NH$_3$(1,1), which can lead to an underestimation of the kinetic temperature when using rotation temperatures derived from NH$_3$(1,1) and NH$_3$(2,2).   
In contrast, CH$_3$CCH is a symmetric top molecule with lower critical densities of its rotational levels than those of NH$_3$, which can be thermalized close to the kinetic temperature at relatively low densities of about 10$^{4}$ cm$^{-3}$. 
To compare the rotation temperatures derived from NH$_3$(1,1)\&(2,2) and CH$_3$CCH rotational levels in warm molecular gas, we used observational data toward 55 massive star-forming regions obtained with Yebes 40m and TMRT 65m. Our results show that rotation temperatures derived from NH$_3$(1,1)\&(2,2) are systematically lower than those from CH$_3$CCH 5-4. This suggests that CH$_3$CCH rotational lines with the same $J$+1$\rightarrow$$J$ quantum number may be a more reliable thermometer than NH$_3$(1,1)\&(2,2) in warm molecular gas located in the surroundings of massive young stellar objects or, more generally, in massive star-forming regions.

\end{abstract}



\section{Introduction}
\label{Introduction}
Kinetic temperature is one of the key parameters in molecular clouds \citep{1999ARA&A..37..311E}. It significantly influences the chemical evolution of molecular clouds \citep{2012A&ARv..20...56C}, as it regulates gas-phase reaction rates \citep[e.g.,][]{2010SSRv..156...13W,2008ApJ...682..283G} and release of molecules from icy mantles into the gas phase \citep[e.g.,][]{1999MNRAS.305..755V,2008ApJ...682..283G,2009ARA&A..47..427H}, thereby determining which chemical pathways dominate \citep[e.g.,][]{2006A&A...457..927G,2015ARPC...66...43K,2020A&A...635A...4H}. Consequently, variations in kinetic temperature, together with other factors such as gas density, radiation field, metallicity, and cosmic-ray ionization rate, strongly affect the relative abundances of complex organic and simple species in star-forming regions. In addition, the kinetic temperature is a key factor in the star formation process, as it controls the balance between thermal pressure and gravity, thereby influencing the fragmentation of molecular clouds and the formation of dense cores \citep[e.g.,][]{2007ARA&A..45..339B,2007ARA&A..45..565M,2023A&A...674A.160G,2024ARA&A..62..369S}. Accurate measurements of this parameter are essential for understanding the physical and chemical environments of the interstellar medium \citep[ISM; e.g.,][]{1999ARA&A..37..311E,2007ARA&A..45..565M,2020ARA&A..58..727J}.

Carbon monoxide (CO) and its isotopologues, mainly $^{13}$CO, emission are widely used to estimate gas temperature under the assumptions of local thermodynamic equilibrium (LTE) and optically thick conditions \citep{1998AJ....116..336N}. But the rotation transitions of CO often suffer from self-absorption effects, which can lead to inaccurate estimates of the gas temperature \citep[e.g.,][]{1990A&A...234..469C,2021A&A...650A.164M,2023ApJS..264...16P}. 
Slightly asymmetric top molecules, such as formaldehyde (H$_2$CO, \citealt{1993ApJS...89..123M}), are considered a thermometer due to their nearly symmetric structures, which make their rotational level populations sensitive to kinetic temperature \citep[e.g.,][]{1987ApJ...318..392M,1993ApJS...89..123M,2018A&A...609A..16T,2024A&A...687A.207Z}. However, H$_2$CO is difficult to thermalize in most of the molecular clouds, because of its relatively high critical density ($\sim$10$^6$ cm$^{-3}$, \citealt{2011A&A...534A..49G,2019ApJ...883..202F}). 
Accurate temperature determination using H$_2$CO requires non-LTE radiative transfer modeling \citep[e.g.,][]{1993ApJS...89..123M,2017A&A...600A..16T,2021A&A...655A..12T,2023ApJ...953...77M,2024A&A...686A..49G,2025A&A...699A..70H}, which involves many input parameters (such as gas density, background radiation temperature and source geometry) and introduces considerable uncertainties in the derived kinetic temperatures.

\begin{figure}[h]
\centering
\includegraphics[width=0.6\textwidth]{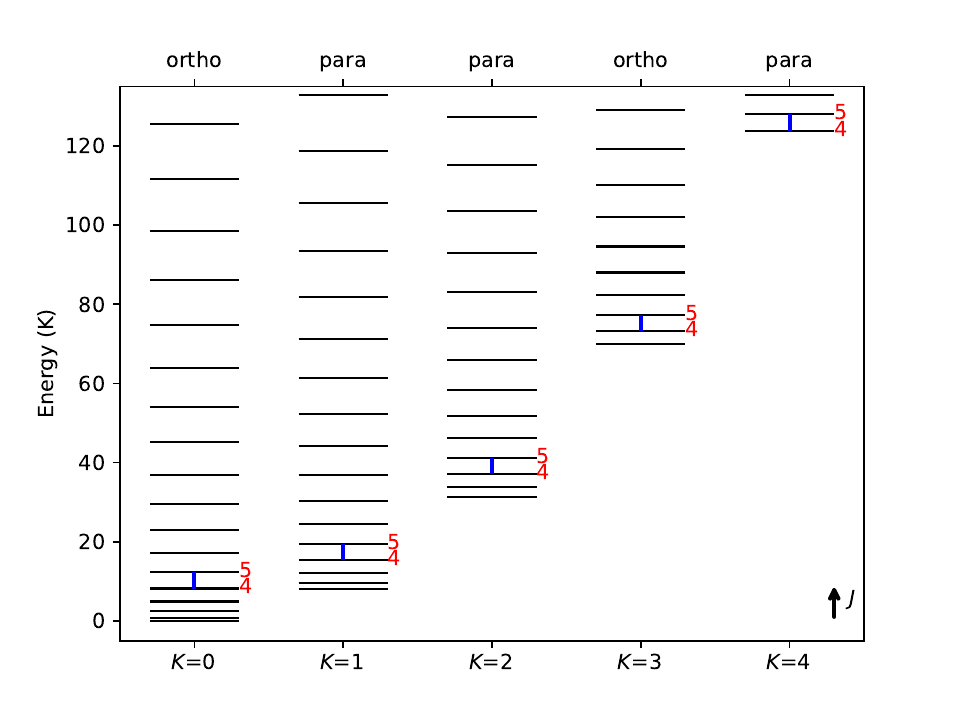}
\caption{Energy levels of CH$_3$CCH. The nuclear spin symmetries and their “para” and “ortho” appellations are indicated. The blue line between two levels indicates the $J=5-4$ transitions, which are used in this work. The level energies were obtained from CDMS.}
\label{Energy levels}
\end{figure}

Symmetric top molecules, such as ammonia (NH$_3$, \citealt{1983ARA&A..21..239H}), acetonitrile (CH$_3$CN, \citealt{1984A&A...130..311A}) and methyl acetylene (CH$_3$CCH, \citealt{1994ApJ...431..674B}), are a class of molecules characterized by having a rotational symmetry axis. Their rotational energy levels are described by two quantum numbers: $J$, representing the total angular momentum, and $K$, the projection of $J$ along the symmetry axis \citep{1955misp.book.....T,2002A&A...389..603F,2017A&A...603A..33G}. Because the dipole moment of symmetric top molecules is parallel to the symmetry axis, the radiative transitions between different $K$ levels are forbidden and the relative populations of different $K$ levels are only related to collisions \citep[e.g.,][]{1955misp.book.....T,1969ApJ...157L..13C,1979JPCRD...8..537B}. These molecules follow specific selection rules for rotational transitions: $\Delta J=\pm 1$ and $\Delta K=0$. 
Because the energy differences between the same $J$+1$\rightarrow$$J$ transitions in different $K$-ladders are similar, while the corresponding upper-state energies ($E_{\rm u}$) differ significantly (such as CH$_3$CCH,  see Figure \ref{Energy levels}), symmetric top molecules can exhibit multiple transitions, with same $J$+1$\rightarrow$$J$ and different $K$, at closely spaced frequencies \citep{1994ApJ...431..674B}. These characteristics make them particularly effective thermometers in the ISM, as the relative intensities of these transitions are highly sensitive to the kinetic temperature.

For NH$_3$, energy levels with $J > K$ are referred to as non-metastable, while those with $J=K$ are called metastable \citep{1983ARA&A..21..239H}. The dipole moment of NH$_3$ is approximately 1.47 D \citep{1987JChPh..87.1557T}. The non-metastable levels require very high critical densities ($\sim$10$^8$-10$^9$ cm$^{-3}$) to be excited, causing them to decay rapidly via $\Delta J = 1$ transitions until they stabilize in a metastable state \citep{1983ARA&A..21..239H}. Metastable levels undergo only inversion transitions ($\Delta J=0$, $\Delta K=0$), which are produced by rapid vibrations of NH$_3$, allowing its nitrogen atom to tunnel quantum-mechanically through the plane of the three hydrogen atoms, resulting in microwave inversion transitions \citep[see][chap.~12]{1955misp.book.....T}. Among these levels, NH$_3$(1,1) and NH$_3$(2,2) are often used to derive the kinetic temperature. The excitation between sub-levels of inversion transitions within metastable levels requires much lower critical densities ($\sim$10$^3$-10$^4$ cm$^{-3}$),  making transitions such as NH$_3$(1,1) and NH$_3$(2,2) widely used to study physical conditions in star-forming regions at different evolutionary stages, from infrared dark clouds \citep[IRDCs; e.g.,][]{1989ApJS...71...89B,2012A&A...544A.146W,2014ApJ...790...84L} to hot molecular cores \citep[e.g.,][]{2012A&A...544A.146W,2016AJ....152...92L} and ultracompact H II regions \citep[e.g.,][]{2011MNRAS.418.1689U,2012A&A...544A.146W,2015MNRAS.452.4029U}.
However, as temperature increases, NH$_3$(2,2) can be collisionally excited to the non-metastable NH$_3$(2,1), which can rapidly decay to NH$_3$(1,1) \citep{1969ApJ...157L..13C}. Consequently, at temperatures above $\sim$25 K, the populations of NH$_3$(1,1) and NH$_3$(2,2) no longer reliably reflect the kinetic temperature, limiting their effectiveness as thermometers \citep{1983A&A...122..164W,2004A&A...416..191T}.

In contrast, CH$_3$CCH has a relatively small electric dipole moment ($\mu$=0.78 D, \cite{1979JPCRD...8..537B}), allowing its rotational transitions to be easily thermalized at densities $n$$\gtrsim 10^{4}$cm$^{-3}$ \citep{1984A&A...130..311A,2002A&A...389..603F,2016ApJ...826L...8M}. 
In addition, the energy level differences for $\Delta J=1$ transitions in CH$_3$CCH are smaller than those of NH$_3$, making all transitions of CH$_3$CCH more easily excited in the ISM. At millimeter and submillimeter wavelengths, the $J$+1$\rightarrow$$J$ transitions of CH$_3$CCH exhibit several $K$-ladder components at nearly the same frequencies, which can be observed simultaneously and reduce the uncertainties from telescope calibration and pointing.
Because of these properties, CH$_3$CCH is an excellent thermometer in molecular clouds \citep[e.g.,][]{2017A&A...603A..33G,2019A&A...631A.137C,2022A&A...658A.128L}, better than NH$_3$(1,1) and NH$_3$(2,2) in warm molecular gas. 

Over the past decade, several observational studies used NH$_3$, CH$_3$CN, and CH$_3$CCH as temperature tracers in various star-forming environments, ranging from IRDCs to hot cores \citep[e.g.,][]{2016ApJ...826L...8M,2017A&A...603A..33G,2019A&A...631A.137C,2022A&A...658A.128L}. These different molecular thermometers probe distinct gas components depending on density and excitation conditions \citep[e.g.,][]{2017A&A...603A..33G,2018A&A...609A..16T,2024A&A...687A.207Z}. However, systematic comparisons between temperatures derived from CH$_3$CCH and NH$_3$ using uniform samples remain limited. 
Since, as pointed out in \citet{2004A&A...416..191T}, NH$_3$(1,1)\&(2,2) are primarily sensitive to relatively low temperatures ($\lesssim$20-25 K), it is useful to compare the rotation temperatures derived from NH$_3$(1,1)\&(2,2) with those obtained from a more reliable thermometer in warm gas. 
Therefore, we compare two approaches for calculating rotation temperatures, one using CH$_3$CCH $J$=5-4 and the other using NH$_3$(1,1)\&(2,2).

Using Yebes 40m telescope and Shanghai Tianma 65m radio telescope (TMRT) observations, this work presents a comparison of rotation temperatures derived from CH$_3$CCH $J$=5-4 and NH$_3$(1,1)\&(2,2) toward a sample of late-stage massive star-forming regions. These targets are associated with 6.7 GHz CH$_3$OH maser emission \citep{2014ApJ...783..130R}, indicating the presence of warm gas with kinetic temperatures of 20-100 K \citep{2007ARA&A..45..339B,2011A&A...529A..32T}. The observations are described in Section \ref{observations}, the methods used to derive the rotation temperatures of CH$_3$CCH are given in Section \ref{fitting}, the main results are presented in Section \ref{results}, the discussion is given in Section \ref{discussion}, and a brief summary is given in Section \ref{summary}.


\section{Observations and data reduction}
\label{observations}
The Yebes 40m radio telescope was used to observe CH$_3$CCH $J$=5-4 at 85457.3003 MHz toward 101 massive star-forming regions with 6.7 GHz CH$_3$OH masers, using the same sample as \citet{2014ApJ...783..130R}. The details of each source are listed in Table 
\ref{source}. The position-switch mode was used in our observations with a frequency coverage of about 18.5 GHz bandwidth, spanning from 72 to 90.5 GHz. The Fast Fourier Transform Spectrometer (XFFTS) was used with 65536 channels, corresponding to a frequency resolution of 38 kHz and a velocity resolution of 0.13 km s$^{-1}$ at 85 GHz. The system temperatures ranged from 110 to 390 K. The beam size of Yebes 40m is about 20.5$\arcsec$ at 86.2 GHz. In the following analysis, the velocity resolution is smoothed to about 0.6 km s$^{-1}$ with a frequency resolution of 153 kHz at 85457.3003 MHz. The typical rms was about 15 mK ($T_{\rm A}^*$ scale) at 0.6 km s$^{-1}$ velocity resolution.

CH$_3$CCH 5-4 was detected in 55 targets, which showed distinct CH$_3$CCH ($J$=5-4, $K$=2) emissions above the 3$\sigma$ level. Among these, 18 sources have previously reported NH$_3$(1,1) and NH$_3$(2,2) observations with the Effelsberg 100m telescope \citep{2016AJ....152...92L}, and the reported data are used in this work. 
The TMRT 65m was used to observe NH$_3$(1,1) and NH$_3$(2,2) toward the remaining 37 targets.   
Observations were conducted in the digital backend
system \citep[DIBAS,][]{2016ApJ...824..136L} mode 6, with 131072 channels and a bandwidth of 187.5 MHz, providing a frequency resolution of 1.4 KHz and a velocity resolution of 0.02 km s$^{-1}$ at about 23.7 GHz. The position-switch mode was also used during TMRT observations. The beam size of TMRT is about 50$\arcsec$ at 23 GHz. The system temperatures were 100-200 K during our observation. In the following analysis, the velocity resolution was also smoothed to about 0.6 km s$^{-1}$, with a frequency resolution of 46 kHz at 23694.4955 MHz. The typical rms was about 20 mK ($T_{\rm A}^*$ scale) at 0.6 km s$^{-1}$ velocity resolution.  The main beam efficiency of the Yebes 40m telescope varies by less than 0.1\% between 85431.1745 MHz (corresponding $K$=4 line) to 85457.3003 MHz (corresponding to $K$=0 line)\footnote{\url{https://rt40m.oan.es/rt40m_en.php}} \citep{2021A&A...645A..37T}. Applying the main beam efficiency correction would only rescale the line intensities, affecting the derived column densities but leaving the relative line ratios unchanged. As a result, the derived rotation temperature is not affected. Because our primary aim is to determine $T_{\rm rot}$ rather than absolute column densities, no main-beam efficiency correction was applied to the spectra before the fitting.

Data reduction was conducted with CLASS/GILDAS software\footnote{\url{http://www.iram.fr/IRAMFR/GILDAS}}. First-order baselines were used for all spectra. The $T_{\rm A}^*$ scale was used for all spectra instead of the $T_{\rm mb}$ scale because only the intensity ratios were needed and the frequencies of the line for calculating ratios are close. In the following analysis, we assume that the beam-filling factor is unity. 

\section{CH$_3$CCH spectral fitting}
\label{fitting}

The rotation temperature of CH$_3$CCH 5-4 ($T_{\rm rot}$(CH$_3$CCH)) was obtained by directly fitting the observed spectra with a multi-Gaussian model under the assumption of LTE.  
The modeled spectra were generated using spectroscopic parameters (line frequencies, Einstein coefficients, upper state energy and upper state degeneracy) by CDMS \footnote{\url{https://cdms.astro.uni-koeln.de/classic/predictions/catalog/}} \citep{2001A&A...370L..49M,2005JMoSt.742..215M}. For each source, the Local Standard of Rest velocity ($v_{\rm lsr}$), rotation temperature ($T_{\rm rot}$),  Full Width at Half Maximum (FWHM), and column density are treated as global free parameters and are shared by all the $K$ components. Each $K$ component is modeled as a Gaussian line profile, assuming that all $K$ levels are optically thin.
Synthetic spectra were compared with the observations using the non-linear least-squares minimization routine $curve\_fit$ from $scipy.optimize$ module \footnote{\url{https://scipy.org}} \citep{2020SciPy-NMeth}. The best-fit $T_{\rm rot}$ values were determined by minimizing the residuals between the observed and modeled spectra, and the parameter uncertainties were estimated from the covariance matrix returned by the fit.

\section{Results}
\label{results}
\subsection{Rotation temperature of CH$_3$CCH}

The CH$_3$CCH ($J$=5-4, $K$=2) transition was detected above the 3$\sigma$ level in 55 out of 101 targets observed with the Yebes 40m telescope. 
As shown in Table \ref{CH3CCH parameter}, the detected $K$=2 component of CH$_3$CCH spans a sufficiently wide range in energy to allow the determination of the rotation temperature \citep{1999ApJ...517..209G,2015PASP..127..266M}. 
Although the $K$=3 transition of CH$_3$CCH 5-4 was detected in 36 sources with 3$\sigma$ level, only $K$=0,1,2 transitions of CH$_3$CCH 5-4 were used in the spectra fitting to maintain a uniform analysis. 
An example of the CH$_3$CCH 5-4 fitting result for G005.88-00.39 is presented in Figure \ref{spectra fitting}a, while the results for other sources are shown in Figure \ref{appendix: CH3CCH_fitting_1}-\ref{appendix: CH3CCH_fitting_5}. The derived $T_{\rm rot}$(CH$_3$CCH) are listed in Table \ref{result}. Some of our sources (e.g., G012.80-00.20, G028.86+00.06, and G035.19-00.74) were also reported by \citet{2017A&A...603A..33G}, who obtained rotation temperature of CH$_3$CCH from a simultaneous fit of the 5-4, 6-5, and 20-19 transitions, yielding results consistent with this work. 

\subsection{Statistical equilibrium calculation of CH$_3$CCH}
\label{Statistical equilibrium calculation}

We solved the statistical equilibrium equations for CH$_3$CCH using level energy ($E_{\rm i}$), level degeneracy ($g_{i}$), Einstein emission coefficient ($A_{ij}$) and collisional rate coefficients ($C_{ij}$) provided by \cite{2024A&A...683A..53B}. \cite{2024A&A...683A..53B} provided the collisional data for CH$_3$CCH--He, and the He$\to$H$_2$ scaling from excitation of molecules and atoms for astrophysics database (EMAA)\footnote{\url{https://emaa.osug.fr}} was applied following \citep{2025A&A...700A.266F} to obtain H$_2$ collision data. The level populations were determined by solving:
\begin{equation}
\label{eq: Statistical equilibrium calculation}
\sum_{j\neq i}n_j(C_{ij}+R_{ij})-n_i\sum_{j\neq i}(C_{ij}+R_{ij})=0,
\end{equation} 
where $C_{ij}$ is the collisional rates, $R_{ij}$ is radiative rates and $n_i$ is population of level $i$. We approximate the radiative rates by retaining only spontaneous emission,
\begin{equation}
R_{ij} \simeq
\left\{
\begin{array}{ll}
A_{ij}, & E_i > E_j\ \text{(downward)}\\
0, & E_i < E_j\ \text{(upward)}
\end{array}
\right.
\end{equation}
i.e., we neglect stimulated emission and radiative absorption. For missing reverse collisional rate coefficients, we used detailed balance:
\begin{equation}
\label{eq: detailed balance}
C_{ji}=C_{ij}\frac{g_i}{g_j}exp[-\frac{E_i-E_j}{kT_{\rm kin}}],
\end{equation} 
where $k$ is the Boltzmann constant and $T_{\rm kin}$ is kinetic temperature. Populations were normalized such that $\sum_i n_i=1$ when solving Eq. \ref{eq: Statistical equilibrium calculation}.

\begin{figure}[h]
\centering
\includegraphics[width=0.6\textwidth]{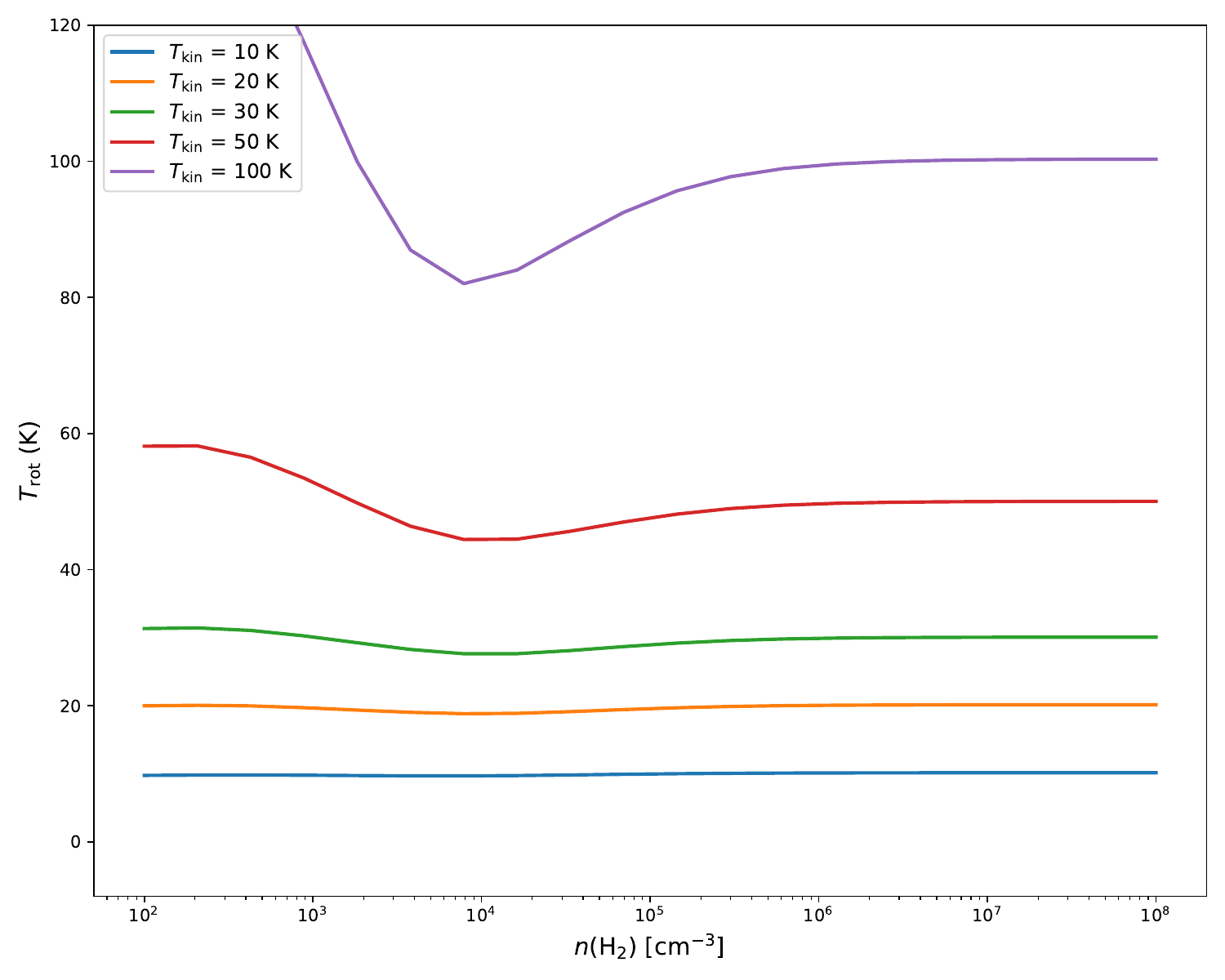}
\caption{CH$_3$CCH $J$=5-4, $K$=0,1,2,3,4 rotation temperature as a function of temperature and density.}
\label{fig: Statistical equilibrium}
\end{figure}

Using these models to predict the population of $J$=5-4, $K$=0,1,2,3,4 transitions, we can obtain the $T_{\rm rot}$(CH$_3$CCH) as a function of both gas density and kinetic temperature. Figure \ref{fig: Statistical equilibrium} shows the results of the statistical equilibrium calculations, which are consistent with those reported by \cite{1994ApJ...431..674B}. When density exceeds 10$^{4}$cm$^{-3}$, the derived $T_{\rm rot}$ approaches the kinetic temperature $T_{\rm kin}$, indicating that the CH$_3$CCH levels are nearly thermalized under such conditions. In our sample of 55 targets with detected CH$_3$CCH ($J$=5-4, $K$=2) emission, the H$^{13}$CN 1-0 was also detected, which has a critical density of about 9$\times$10$^{5}$cm$^{-3}$ at 20 K, calculated using $n_{crit}=A_{ij}/C{ij}$, where the Einstein emission coefficient and collisional rate coefficients are taken from the EMAA database. Therefore, $T_{\rm kin}$ can be reliably traced by $T_{\rm rot}$(CH$_3$CCH) in these sources.

\subsection{Rotation temperature of NH$_3$}
Among these 55 targets, 18 have previously reported rotation temperatures of NH$_3$ ($T_{\rm rot}$(2,2;1,1), derived from the NH$_3$(1,1) and (2,2) inversion transitions, and denoted as $T_{\rm rot}$(NH$_3$)) with Effelsberg 100m observations \citep{2016AJ....152...92L}, which were used in this work. To ensure consistency, three of the sources from \cite{2016AJ....152...92L} were also observed with the TMRT. The rotation temperatures of NH$_3$ derived from TMRT observations are 19.3$\pm$0.5 K, 23.9$\pm$0.5 K, and 24.3$\pm$0.3 K in G081.75+00.59, G109.87+02.11 and G111.54+00.77, respectively. For comparison, the corresponding values reported by \cite{2016AJ....152...92L} are 18.5$\pm$0.3 K, 23.4$\pm$3.0 K and 26.0$\pm$1.7 K, respectively. The results demonstrate good consistency between the TMRT and Effelsberg observations.

For the remaining 37 targets, ammonia inversion transitions NH$_3$(1,1) and NH$_3$(2,2) were observed using TMRT.
In these 37 targets, NH$_3$(1,1) absorption was detected in three sources, G010.62-00.38, G012.80-00.20 and G133.94+01.06, which were also reported by \citet{1988ApJ...324..920K}, \citet{2022A&A...658A..34T} and \citet{1987ApJ...312..830R}, respectively. Additionally, both NH$_3$(1,1) and NH$_3$(2,2) exhibit a low signal-to-noise ratio in G209.00-19.38. For this source, the rms noise is 12 mK at a velocity resolution of 0.6 km s$^{-1}$, and the NH$_3$ hyperfine components are below the 3$\sigma$ level, resulting in unreliable hyperfine fitting and temperature determination. Excluding these four sources, a total of 33 sources with TMRT observations were used in this work. 

\begin{figure}[h]
\gridline{\fig{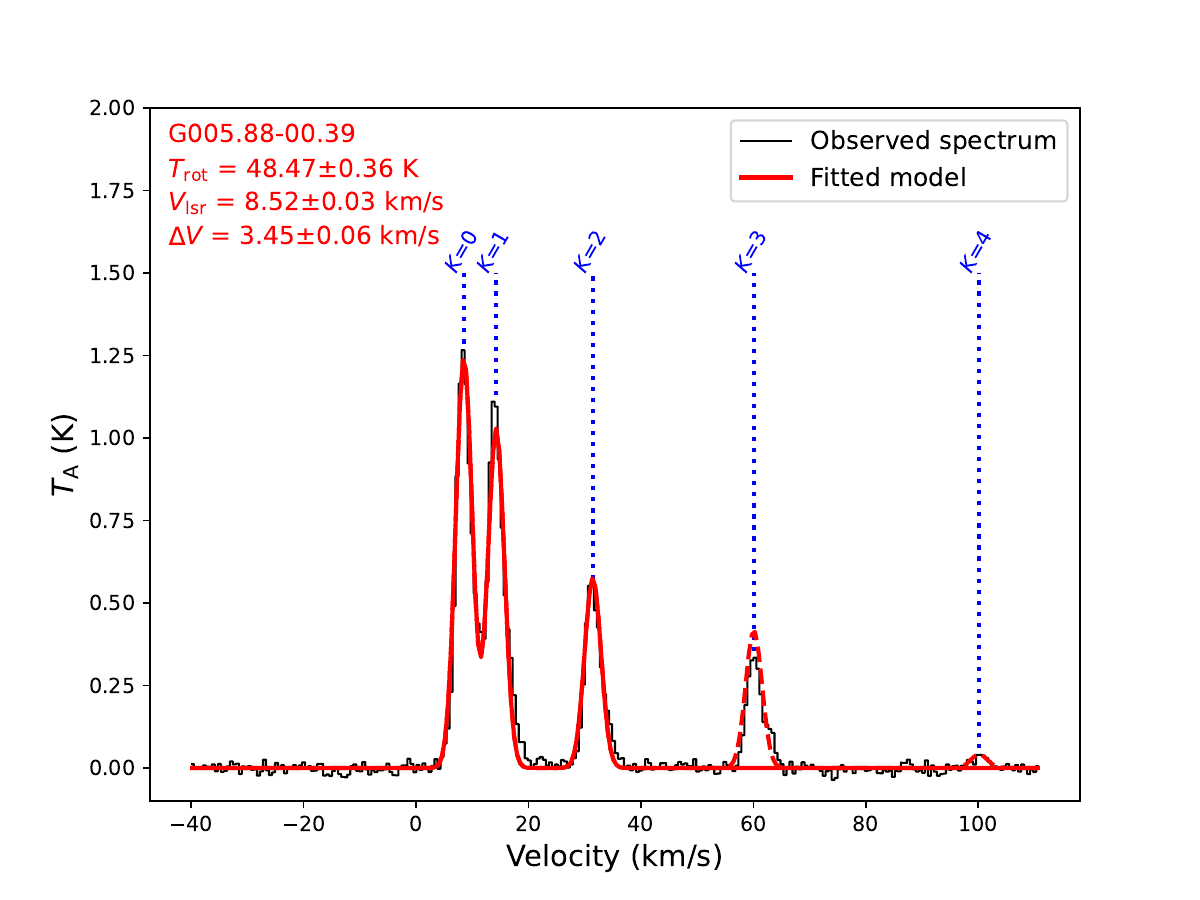}{0.5\textwidth}{(a)}
          \fig{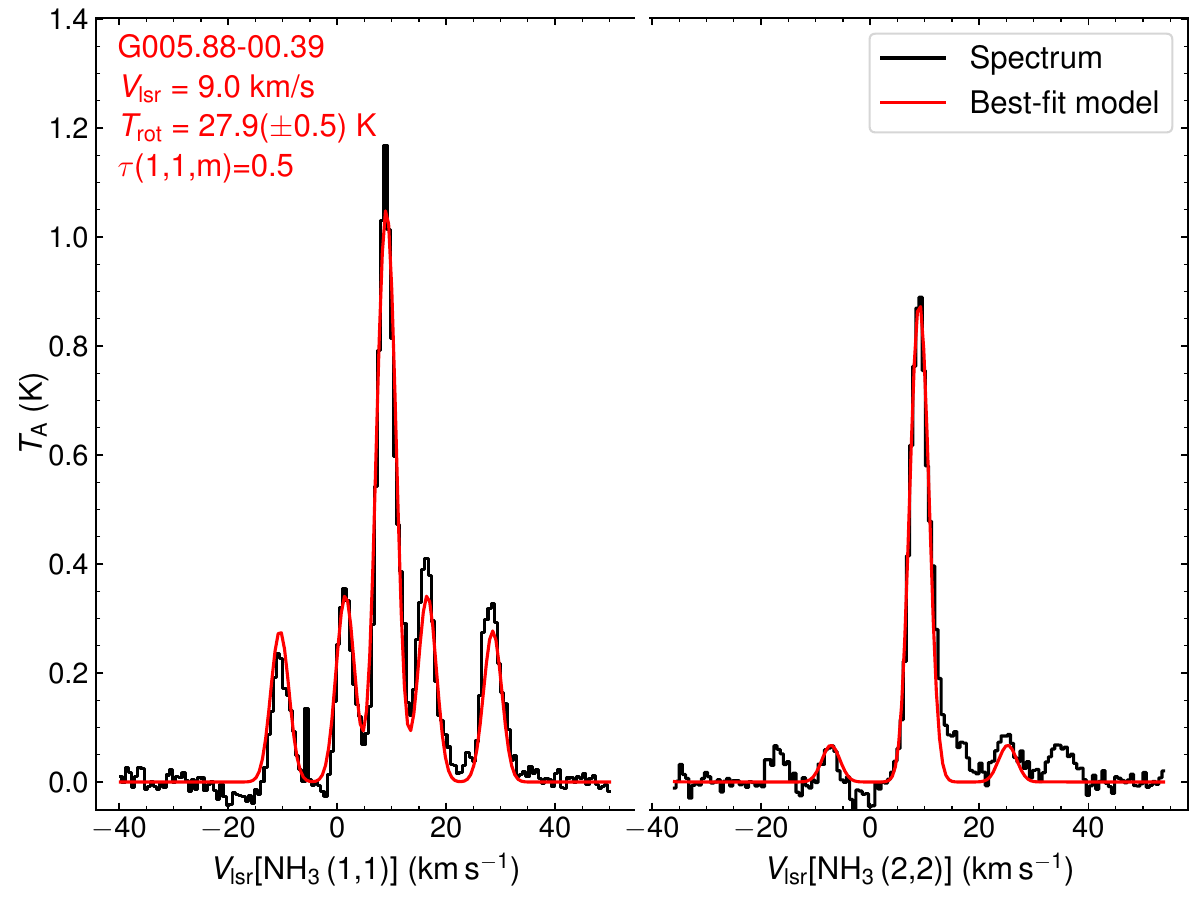}{0.45\textwidth}{(b)}}
\caption{(a) The CH$_3$CCH $J$=5-4 fitting result in G005.88-00.39. The observed spectra are in black, while the red is the best fit from our code. The red dashed line indicates the expected line of the $K=3,4$ transition. (b) The spectra of NH$_3$(1,1) and NH$_3$(2,2) in G005.88-00.39. The observed spectra are in black, while the red are the best fit from the code of \cite{2015ApJ...805..171L}.}
\label{spectra fitting}
\end{figure}

Based on the theoretical frameworks of \cite{1983ARA&A..21..239H} and \cite{1992ApJ...388..467M}, \cite{2015ApJ...805..171L} provided a Python code\footnote{\url{https://github.com/xinglunju/pyAmor}} to fit the NH$_3$(1,1) and NH$_3$(2,2) spectra to derive $T_{\rm rot}$(2,2;1,1). 
Within this framework, the NH$_3$(1,1) and NH$_3$(2,2) lines are fitted simultaneously, assuming Gaussian profiles for all hyperfine structure components.   
For each source, five parameters are treated as free parameters in the fitting, including $T_{\rm rot}$(2,2;1,1), $\tau$(1,1,m), column density ($N_{\rm tot}$(NH$_3$)), $v_{\rm LSR}$ and FWHM. The $T_{\rm rot}$(2,2;1,1), $N_{\rm tot}$(NH$_3$), $v_{\rm LSR}$ and FWHM are assumed to be the same for all hyperfine components, while $\tau$(1,1,m) is determined from the relative intensities of the NH$_3$(1,1) hyperfine structure and is used to correct for opacity effects. 
Using this code, the $T_{\rm rot}$(2,2;1,1) derived from the relative populations of the NH$_3$(1,1) and NH$_3$(2,2) was obtained for 33 sources observed with TMRT. One of the fitting results for G005.88-00.39 is presented in Figure \ref{spectra fitting}b, while other sources are shown in Figure \ref{appendix: NH3_fitting_1}-\ref{appendix: NH3_fitting_3}.

\subsection{Compare the rotation temperature of CH$_3$CCH and NH$_3$}
The range of $T_{\rm rot}$(CH$_3$CCH) spans from 23.0$\pm$0.4 K to 88.8$\pm$3.0 K, while  $T_{\rm rot}$(2,2;1,1) ranges from 15.7$\pm$0.4 K to 29.5$\pm$0.6 K. The highest value of optical depth for the NH$_3$(1,1) main group observed by TMRT is 2.9, which is detected in G012.90-00.24. 
The median value of $T_{\rm rot}$(CH$_3$CCH) is 33.1$\pm$0.6 K, while $T_{\rm rot}$(2,2;1,1) is 20.7$\pm$0.5 K.
A comparison of the rotation temperatures derived from CH$_3$CCH 5-4 and NH$_3$(1,1)\&(2,2) is presented in Figure \ref{Trot comparison}. 
For most sources, $T_{\rm rot}$(CH$_3$CCH) is significantly higher than that obtained from NH$_3$(1,1)\&(2,2) at temperatures higher than 30 K.

\section{Discussion}
\label{discussion}

We compare the rotation temperature derived from CH$_3$CCH 5-4 and NH$_3$(1,1)\&NH$_3$(2,2) in late-stage massive star-forming regions to investigate the physical origin of their differences and to evaluate that CH$_3$CCH is a more reliable thermometer than NH$_3$ in warm gas. The excitation properties and temperature sensitivities of these two tracers have been described in Section \ref{Introduction}. Almost all targets show that $T_{\rm rot}$(CH$_3$CCH) is higher than $T_{\rm rot}$(2,2;1,1), which may help to illustrate the underlying causes of this difference.

\begin{figure}
\centering
\includegraphics[width=0.6\textwidth]{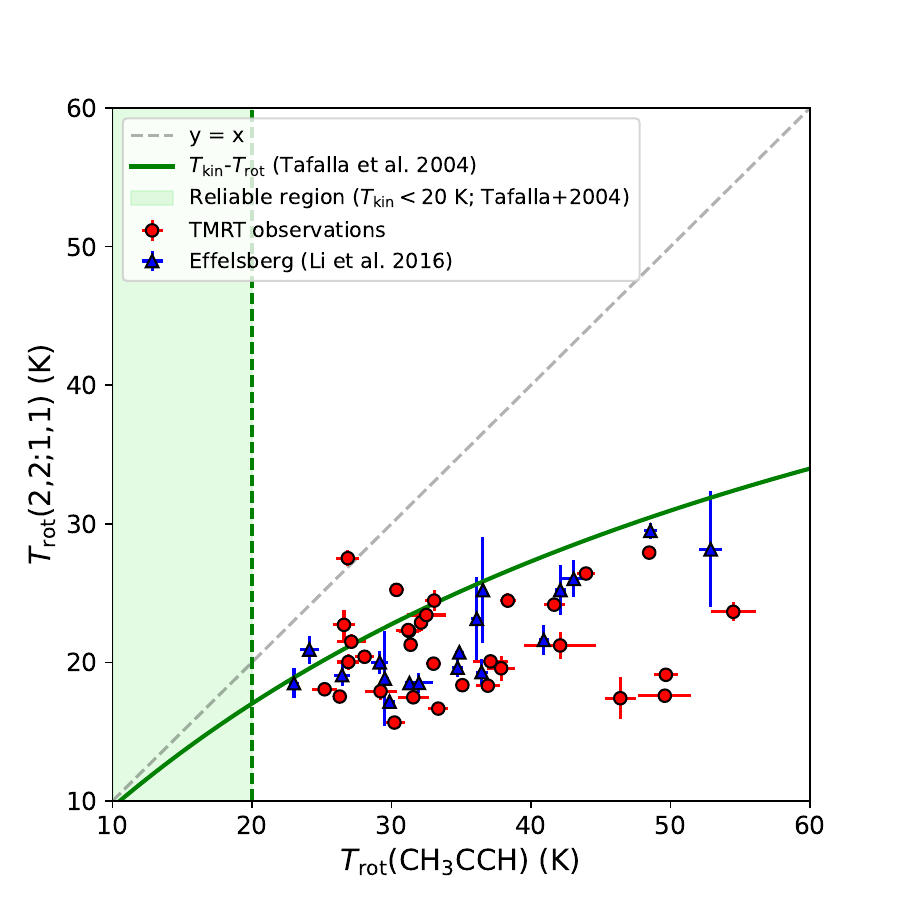}
\caption{Comparison between $T_{\rm rot}$ derived from CH$_3$CCH and $T_{\rm rot}$ derived from NH$_3$. Each point represents one source. The red circles correspond to NH$_3$ data observed with TMRT in this work, while the blue triangles are NH$_3$ data from \cite{2016AJ....152...92L} observed with the Effelsberg. The x-axis shows the rotation temperature derived from CH$_3$CCH, and the y-axis shows that derived from NH$_3$. The green curve represents the relation between $T_{\rm rot}$(2,2;1,1) and $T_{\rm kin}$ derived by \cite{2004A&A...416..191T}, and the light-green shaded area marks the reliable temperature range reported in that work.}
\label{Trot comparison}
\end{figure}

As shown in Figure \ref{Trot comparison}, the range of $T_{\rm rot}$(CH$_3$CCH) is significantly broader than that of $T_{\rm rot}$(2,2;1,1). As $T_{\rm rot}$(CH$_3$CCH) increases, $T_{\rm rot}$(2,2;1,1) shows insignificant increase. When $T_{\rm kin}$ is above 20 K, $T_{\rm rot}$(2,2;1,1) is not sensitive to temperature, indicating that the empirical equation between the $T_{\rm rot}$(2,2;1,1) and the $T_{\rm kin}$ \citep[e.g.,][]{2017PASP..129b5003E,2005ApJ...620..823S,2004A&A...416..191T} may also not provide accurate kinetic temperature of NH$_3$. This suggests that NH$_3$ may not be a reliable thermometer in warm molecular gas.

With the statistical equilibrium calculations, \cite{1983A&A...122..164W} reported that $T_{\rm rot}$(2,2;1,1) deviates from $T_{\rm kin}$ as the temperature increases. \cite{2004A&A...416..191T} improved this method with Monte Carlo models to estimate $T_{\rm kin}$ from $T_{\rm rot}$(2,2;1,1) as a function almost independent of density: 
\begin{equation}
\label{eq: kinetic temperature by Walmsley}
T_{\rm kin}=\frac{T_{\rm rot}{\rm (2,2;1,1)}}{1-\frac{T_{\rm rot}{\rm (2,2;1,1)}}{\Delta E{(2,2;1,1)}}\ln[1+1.1\exp(\frac{-16}{T_{\rm rot}{\rm (2,2;1,1)}})]},
\end{equation} 
where $\Delta E(2,2;1,1)$=[$E(2,2)-E(1,1)$]/$k$ is the temperature associated with energy level difference. 
However, \cite{2004A&A...416..191T} also pointed out that the uncertainty of this function becomes significant when $T_{\rm kin} > 20$ K (see the green line in Figure \ref{Trot comparison}). As shown by the statistical equilibrium calculation in Section \ref{Statistical equilibrium calculation}, our targets can be reasonably assumed to follow $T_{\rm kin}$=$T_{\rm rot}$(CH$_3$CCH). In all targets, the $T_{\rm rot}$(CH$_3$CCH) is higher than 20 K, while $T_{\rm rot}$(2,2;1,1) is lower than the $T_{\rm kin}$ derived from NH$_3$ in almost all targets, particularly when $T_{\rm rot}$(CH$_3$CCH) exceeds 30 K. This indicates that $T_{\rm kin}$ cannot be reliably predicted from $T_{\rm rot}$(2,2;1,1) at higher temperatures, as $T_{\rm rot}$(2,2;1,1) loses sensitivity to temperature variations. An accurate estimation of temperature from NH$_3$ may require higher-$J$ metastable transitions, such as NH$_3$(4,4) and NH$_3$(5,5). But these transitions are hard to detect in most molecular clouds, because they need higher excitation temperatures. Therefore, when using NH$_3$(1,1) and NH$_3$(2,2) as a thermometer in warm molecular gas one should be cautious.

Compared to NH$_3$, CH$_3$CCH is more easily thermalized at gas density $n\gtrsim 10^{4}$cm$^{-3}$ \citep{1984A&A...130..311A}. Statistical equilibrium calculations by \cite{1994ApJ...431..674B} and Section \ref{Statistical equilibrium calculation} further demonstrated that the rotation temperature of CH$_3$CCH closely approximates the kinetic temperature at gas density $n\gtrsim 10^{4}$cm$^{-3}$. Therefore, CH$_3$CCH is a reliable tracer of kinetic temperature in warm gas of star-forming regions. 

A wide range of rotational transitions of CH$_3$CCH have been used to derive kinetic temperatures of molecular clouds. Transitions in the 3 mm and 2 mm bands, such as $J$=5-4, $J$=6-5, $J$=8-7 and $J$=9-8 are often used \citep[e.g.,][]{1983ApJ...272..591C,1994ApJ...431..674B,2017A&A...603A..33G,2023MNRAS.526.3673P}. These mid-$J$ lines cover $E_{\rm u}$ in the range of $\sim$10-100 K across their first four $K$-levels, making them ideal for tracing kinetic temperatures in warm gas ($T$=20-100 K). Higher-$J$ transitions ($J >$11), whose $K$=0 levels have $E_{\rm u}$ exceeding 50 K, are suitable for probing hotter environments ($T > $50 K, e.g., \citealt{2019A&A...631A.137C,2022A&A...658A.128L,2022ApJ...925....3S}). Although $J$=3-2, $J$=4-3, $J$=7-6 and $J$=11-10 transitions are in principle accessible, they are seldom used. Because $J$=3–2, $J$=7–6, and $J$=11–10 transitions lie in frequency ranges with very poor atmospheric transmission, the $J$=4–3 transition lies close to an atmospheric window and can only be observed under exceptionally good conditions. The $J$=1-0 transition is useless for temperature estimation because it includes only the $K$=0 line, while the $J$=2-1 transitions are also rarely used in warm molecular gas, as the energy difference between $K=0$ and $K=1$ levels is 7.2 K, making it insensitive to excitation conditions. However, the $J$=2-1 transitions might be used to derive kinetic temperatures in cold environments ($T\lesssim$20 K), with a similar usable range to that of   $T_{\rm rot}$(2,2;1,1) for NH$_3$.

CH$_3$CN,  as another symmetric top molecule,  had been more widely used as a thermometer than CH$_3$CCH, because of relatively strong emission \citep{1984A&A...130..311A} with a larger dipole moment ($\mu$=3.922 D; \citealt{Gadhi1995}). The properties of CH$_3$CN and CH$_3$CCH are very similar. The two species share many physical similarities: both are symmetric top molecules with nearly identical rotational energy structures consisting of multiple $K$-components within each $J$-transition, and both can be used to estimate gas temperatures. In addition, the separations between $K$-levels in CH$_3$CCH and CH$_3$CN are similar, which allows temperature diagnostics over a broad range of physical conditions. A comparison of the two molecules was reported by \cite{2017A&A...603A..33G}. They derived rotation temperatures of CH$_3$CCH and CH$_3$CN toward a large sample of ATLASGAL sources. They found that $T_{\rm rot}$(CH$_3$CN) is systematically higher than $T_{\rm rot}$(CH$_3$CCH), suggesting that CH$_3$CCH traces the cooler, more extended envelope gas, while CH$_3$CN probes the denser and warmer cores. This result highlights their complementary diagnostic power in probing different temperature conditions of star-forming regions.
However, the relatively strong emission of CH$_3$CN means that the low $K$ levels are often optically thick \citep{1998ApJ...494..636Z}. Moreover, CH$_3$CN is typically concentrated in the immediate vicinity of protostellar objects, tracing more compact and warmer regions than CH$_3$CCH \citep{2008A&A...488..959B}. In this case, CH$_3$CN cannot provide an accurate kinetic temperature in molecular clouds, while CH$_3$CCH is still a reliable thermometer because of its relatively low optical depth for low $K$ level lines. Another disadvantage of CH$_3$CN lines is the higher critical densities of $J>K$ levels as non-metastable ones than those of CH$_3$CCH, because the dipole moment of CH$_3$CN is about 5 times greater than that of CH$_3$CCH. 

Considering the presence of temperature gradients in massive star-forming regions \citep[e.g.,][]{2015ApJ...811...79D,2003ApJ...587..262L,2010A&A...518L..83S}, CH$_3$CCH and NH$_3$ may trace different regions within the envelope \citep{2017A&A...603A..33G}. It could be one of the reasons for the significant discrepancies observed between the rotation temperatures derived from CH$_3$CCH and NH$_3$ in our sources. This requires spatially resolved information of CH$_3$CCH and NH$_3$ for further analysis.
Additionally, the beam sizes for these two molecules differ in the present work. Specifically, the beam size of Yebes 40m is about 20.5$\arcsec$ at 86.2 GHz, while the beam sizes of TMRT and Effelsberg are 50$\arcsec$ and 43$\arcsec$ at 23 GHz, respectively. The larger beam sizes of  NH$_3$ observations may have included more lower-temperature gas from the residual envelope, resulting in $T_{\rm rot}$(CH$_3$CCH) being higher than $T_{\rm rot}$(2,2;1,1). In addition, the heliocentric distances of our targets range from 0.7 to 10.2 kpc, corresponding to physical sizes from 0.07 to 1.02 pc assuming a 20.5$\arcsec$ beam size. This indicates that the derived rotational temperatures may represent averages over different spatial scales. Therefore, using observations with similar beam sizes of CH$_3$CCH and NH$_3$ toward sample sources with similar distances may provide a more reliable comparison of these two temperature tracers.
Overall, our results reinforce previous theoretical predictions that CH$_3$CCH traces the warm gas more reliably than NH$_3$(1,1)\&(2,2), and the rotation temperature difference can be attributed to the collisional excitation of NH$_3$ into non-metastable levels at higher kinetic temperatures.

\section{Conclusions}
\label{summary}
Using Yebes 40m and TMRT 65m observations, we investigate CH$_3$CCH and NH$_3$ as temperature tracers in warm molecular gas. We find that the $T_{\rm rot}$(CH$_3$CCH) is significantly higher than $T_{\rm rot}$(2,2;1,1). This discrepancy may arise because NH$_3$ accumulates in the (1,1) level at high temperatures, leading to an underestimation of the kinetic temperature. Our results suggest that CH$_3$CCH provides a more reliable and sensitive probe of gas temperature than NH$_3$(1,1)\&(2,2), particularly in warm molecular regions with $T_{\rm kin} > 20$ K. Caution should be taken when using $T_{\rm rot}$(2,2;1,1) in such environments.

\begin{acknowledgements}
This work is supported by the National Key R$\&$D Program of China (No. 2022YFA1603101) and the National Natural Science Foundation of China (NSFC, Grant No. 12173067). This work is also supported by 100101 Key Laboratory of Radio Astronomy and Technology (Chinese Academy of Sciences). X.Lu\ acknowledges support from the Strategic Priority Research Program of the Chinese Academy of Sciences (CAS) Grant No.\ XDB0800300, State Key Laboratory of Radio Astronomy and Technology (CAS), the National Natural Science Foundation of China (NSFC) through grant Nos.\ 12273090 and 12322305, the Natural Science Foundation of Shanghai (No.\ 23ZR1482100), and the CAS ``Light of West China'' Program No.\ xbzg-zdsys-202212. Based on observations carried out with the Yebes 40 m telescope (project code 23A013). The 40 m radio telescope at Yebes Observatory is operated by the Spanish Geographic Institute (IGN; Ministerio de Transportes y Movilidad Sostenible). We thank the support from operators and staff at the TMRT during our observations. S. Zheng would like to thank the China Scholarship Council (CSC) for support. This research has made use of spectroscopic and collisional data from the EMAA database (https://emaa.osug.fr and https://dx.doi.org/10.17178/EMAA).
\end{acknowledgements}

%

\vspace{5mm}



\bibliography{sample631}{}

@ARTICLE{2014ApJ...783..130R,
       author = {{Reid}, M.~J. and {Menten}, K.~M. and {Brunthaler}, A. and {Zheng}, X.~W. and {Dame}, T.~M. and {Xu}, Y. and {Wu}, Y. and {Zhang}, B. and {Sanna}, A. and {Sato}, M. and {Hachisuka}, K. and {Choi}, Y.~K. and {Immer}, K. and {Moscadelli}, L. and {Rygl}, K.~L.~J. and {Bartkiewicz}, A.},
        title = "{Trigonometric Parallaxes of High Mass Star Forming Regions: The Structure and Kinematics of the Milky Way}",
      journal = {\apj},
         year = 2014,
        month = mar,
       volume = {783},
       number = {2},
          eid = {130},
        pages = {130},
          doi = {10.1088/0004-637X/783/2/130},
archivePrefix = {arXiv},
       eprint = {1401.5377},
 primaryClass = {astro-ph.GA},
       adsurl = {https://ui.adsabs.harvard.edu/abs/2014ApJ...783..130R}
}

@ARTICLE{2015ApJ...805..171L,
       author = {{Lu}, Xing and {Zhang}, Qizhou and {Wang}, Ke and {Gu}, Qiusheng},
        title = "{Initial Fragmentation in the Infrared Dark Cloud G28.53-0.25}",
      journal = {\apj},
         year = 2015,
        month = jun,
       volume = {805},
       number = {2},
          eid = {171},
        pages = {171},
          doi = {10.1088/0004-637X/805/2/171},
archivePrefix = {arXiv},
       eprint = {1503.08797},
 primaryClass = {astro-ph.GA},
       adsurl = {https://ui.adsabs.harvard.edu/abs/2015ApJ...805..171L}
}

@ARTICLE{2016AJ....152...92L,
       author = {{Li}, F.~C. and {Xu}, Y. and {Wu}, Y.~W. and {Yang}, J. and {Lu}, D.~R. and {Menten}, K.~M. and {Henkel}, C.},
        title = "{Ammonia and CO Outflow around 6.7 GHz Methanol Masers}",
      journal = {\aj},
         year = 2016,
        month = oct,
       volume = {152},
       number = {4},
          eid = {92},
        pages = {92},
          doi = {10.3847/0004-6256/152/4/92},
archivePrefix = {arXiv},
       eprint = {1608.04251},
 primaryClass = {astro-ph.GA},
       adsurl = {https://ui.adsabs.harvard.edu/abs/2016AJ....152...92L}
}

@ARTICLE{2009ARA&A..47..427H,
       author = {{Herbst}, Eric and {van Dishoeck}, Ewine F.},
        title = "{Complex Organic Interstellar Molecules}",
      journal = {\araa},
         year = 2009,
        month = sep,
       volume = {47},
       number = {1},
        pages = {427-480},
          doi = {10.1146/annurev-astro-082708-101654},
       adsurl = {https://ui.adsabs.harvard.edu/abs/2009ARA&A..47..427H}
}

@ARTICLE{1994ApJ...431..674B,
       author = {{Bergin}, Edwin A. and {Goldsmith}, Paul F. and {Snell}, Ronald L. and {Ungerechts}, Hans},
        title = "{CH 3C 2H as a Temperature Probe in Dense Giant Molecular Cloud Cores}",
      journal = {\apj},
         year = 1994,
        month = aug,
       volume = {431},
        pages = {674},
          doi = {10.1086/174518},
       adsurl = {https://ui.adsabs.harvard.edu/abs/1994ApJ...431..674B}
}

@ARTICLE{1983ARA&A..21..239H,
       author = {{Ho}, P.~T.~P. and {Townes}, C.~H.},
        title = "{Interstellar ammonia.}",
      journal = {\araa},
         year = 1983,
        month = jan,
       volume = {21},
        pages = {239-270},
          doi = {10.1146/annurev.aa.21.090183.001323},
       adsurl = {https://ui.adsabs.harvard.edu/abs/1983ARA&A..21..239H}
}

@ARTICLE{1992ApJ...388..467M,
       author = {{Mangum}, Jeffrey G. and {Wootten}, Alwyn and {Mundy}, Lee G.},
        title = "{Synthesis Imaging of the DR 21 (OH) Cluster. II. Thermal Ammonia and Water Maser Emission}",
      journal = {\apj},
         year = 1992,
        month = apr,
       volume = {388},
        pages = {467},
          doi = {10.1086/171167},
       adsurl = {https://ui.adsabs.harvard.edu/abs/1992ApJ...388..467M}
}

@ARTICLE{1983A&A...122..164W,
       author = {{Walmsley}, C.~M. and {Ungerechts}, H.},
        title = "{Ammonia as a molecular cloud thermometer.}",
      journal = {\aap},
         year = 1983,
        month = jun,
       volume = {122},
        pages = {164-170},
       adsurl = {https://ui.adsabs.harvard.edu/abs/1983A&A...122..164W}
}

@ARTICLE{2005ApJ...620..823S,
       author = {{Swift}, Jonathan J. and {Welch}, William J. and {Di Francesco}, James},
        title = "{A Pre-Protostellar Core in L1551}",
      journal = {\apj},
         year = 2005,
        month = feb,
       volume = {620},
       number = {2},
        pages = {823-834},
          doi = {10.1086/427257},
archivePrefix = {arXiv},
       eprint = {astro-ph/0411157},
 primaryClass = {astro-ph},
       adsurl = {https://ui.adsabs.harvard.edu/abs/2005ApJ...620..823S}
}

@BOOK{1955misp.book.....T,
   author = {{Townes}, C.~H. and {Schawlow}, A.~L.},
    title = "{Microwave Spectroscopy}",
booktitle = {Microwave Spectroscopy, New York: McGraw-Hill, 1955},
publisher = {Mcgraw-Hill},
     year = 1955,
   adsurl = {http://adsabs.harvard.edu/abs/1955misp.book.....T}
}

@ARTICLE{1979JPCRD...8..537B,
       author = {{Bauer}, A. and {Boucher}, D. and {Burie}, J. and {Demaison}, J. and {Dubrulle}, A.},
        title = "{Microwave spectra of molecules of astrophysical interest XV. Propyne}",
      journal = {Journal of Physical and Chemical Reference Data},
         year = 1979,
        month = apr,
       volume = {8},
       number = {2},
        pages = {537-558},
          doi = {10.1063/1.555603},
       adsurl = {https://ui.adsabs.harvard.edu/abs/1979JPCRD...8..537B}
}

@ARTICLE{2001A&A...370L..49M,
       author = {{M{\"u}ller}, H.~S.~P. and {Thorwirth}, S. and {Roth}, D.~A. and {Winnewisser}, G.},
        title = "{The Cologne Database for Molecular Spectroscopy, CDMS}",
      journal = {\aap},
         year = 2001,
        month = apr,
       volume = {370},
        pages = {L49-L52},
          doi = {10.1051/0004-6361:20010367},
       adsurl = {https://ui.adsabs.harvard.edu/abs/2001A&A...370L..49M}
}

@ARTICLE{2005JMoSt.742..215M,
       author = {{M{\"u}ller}, Holger S.~P. and {Schl{\"o}der}, Frank and {Stutzki}, J{\"u}rgen and {Winnewisser}, Gisbert},
        title = "{The Cologne Database for Molecular Spectroscopy, CDMS: a useful tool for astronomers and spectroscopists}",
      journal = {Journal of Molecular Structure},
         year = 2005,
        month = may,
       volume = {742},
       number = {1-3},
        pages = {215-227},
          doi = {10.1016/j.molstruc.2005.01.027},
       adsurl = {https://ui.adsabs.harvard.edu/abs/2005JMoSt.742..215M}
}

@ARTICLE{2017PASP..129b5003E,
       author = {{Estalella}, Robert},
        title = "{HfS, Hyperfine Structure Fitting Tool}",
      journal = {\pasp},
         year = 2017,
        month = feb,
       volume = {129},
       number = {972},
        pages = {025003},
          doi = {10.1088/1538-3873/129/972/025003},
archivePrefix = {arXiv},
       eprint = {1608.04088},
 primaryClass = {astro-ph.IM},
       adsurl = {https://ui.adsabs.harvard.edu/abs/2017PASP..129b5003E}
}

@ARTICLE{2004A&A...416..191T,
       author = {{Tafalla}, M. and {Myers}, P.~C. and {Caselli}, P. and {Walmsley}, C.~M.},
        title = "{On the internal structure of starless cores. I. Physical conditions and the distribution of CO, CS, N$_{2}$H$^{+}$, and NH$_{3}$ in L1498 and L1517B}",
      journal = {\aap},
         year = 2004,
        month = mar,
       volume = {416},
        pages = {191-212},
          doi = {10.1051/0004-6361:20031704},
       adsurl = {https://ui.adsabs.harvard.edu/abs/2004A&A...416..191T}
}

@ARTICLE{1999ARA&A..37..311E,
       author = {{Evans}, Neal J., II},
        title = "{Physical Conditions in Regions of Star Formation}",
      journal = {\araa},
         year = 1999,
        month = jan,
       volume = {37},
        pages = {311-362},
          doi = {10.1146/annurev.astro.37.1.311},
archivePrefix = {arXiv},
       eprint = {astro-ph/9905050},
 primaryClass = {astro-ph},
       adsurl = {https://ui.adsabs.harvard.edu/abs/1999ARA&A..37..311E}
}

@ARTICLE{2020SciPy-NMeth,
  author  = {Virtanen, Pauli and Gommers, Ralf and Oliphant, Travis E. and
            Haberland, Matt and Reddy, Tyler and Cournapeau, David and
            Burovski, Evgeni and Peterson, Pearu and Weckesser, Warren and
            Bright, Jonathan and {van der Walt}, St{\'e}fan J. and
            Brett, Matthew and Wilson, Joshua and Millman, K. Jarrod and
            Mayorov, Nikolay and Nelson, Andrew R. J. and Jones, Eric and
            Kern, Robert and Larson, Eric and Carey, C J and
            Polat, {\.I}lhan and Feng, Yu and Moore, Eric W. and
            {VanderPlas}, Jake and Laxalde, Denis and Perktold, Josef and
            Cimrman, Robert and Henriksen, Ian and Quintero, E. A. and
            Harris, Charles R. and Archibald, Anne M. and
            Ribeiro, Ant{\^o}nio H. and Pedregosa, Fabian and
            {van Mulbregt}, Paul and {SciPy 1.0 Contributors}},
  title   = {{{SciPy} 1.0: Fundamental Algorithms for Scientific
            Computing in Python}},
  journal = {Nature Methods},
  year    = {2020},
  volume  = {17},
  pages   = {261--272},
  adsurl  = {https://rdcu.be/b08Wh},
  doi     = {10.1038/s41592-019-0686-2},
}

@ARTICLE{2007ARA&A..45..339B,
       author = {{Bergin}, Edwin A. and {Tafalla}, Mario},
        title = "{Cold Dark Clouds: The Initial Conditions for Star Formation}",
      journal = {\araa},
         year = 2007,
        month = sep,
       volume = {45},
       number = {1},
        pages = {339-396},
          doi = {10.1146/annurev.astro.45.071206.100404},
archivePrefix = {arXiv},
       eprint = {0705.3765},
 primaryClass = {astro-ph},
       adsurl = {https://ui.adsabs.harvard.edu/abs/2007ARA&A..45..339B}
}

@ARTICLE{2017A&A...603A..33G,
       author = {{Giannetti}, A. and {Leurini}, S. and {Wyrowski}, F. and {Urquhart}, J. and {Csengeri}, T. and {Menten}, K.~M. and {K{\"o}nig}, C. and {G{\"u}sten}, R.},
        title = "{ATLASGAL-selected massive clumps in the inner Galaxy. V. Temperature structure and evolution}",
      journal = {\aap},
         year = 2017,
        month = jul,
       volume = {603},
          eid = {A33},
        pages = {A33},
          doi = {10.1051/0004-6361/201630048},
archivePrefix = {arXiv},
       eprint = {1703.08485},
 primaryClass = {astro-ph.GA},
       adsurl = {https://ui.adsabs.harvard.edu/abs/2017A&A...603A..33G}
}

@ARTICLE{2003ApJ...587..262L,
       author = {{Li}, D. and {Goldsmith}, P.~F. and {Menten}, K.},
        title = "{Massive Quiescent Cores in Orion. I. Temperature Structure}",
      journal = {\apj},
         year = 2003,
        month = apr,
       volume = {587},
       number = {1},
        pages = {262-277},
          doi = {10.1086/368078},
archivePrefix = {arXiv},
       eprint = {astro-ph/0301060},
 primaryClass = {astro-ph},
       adsurl = {https://ui.adsabs.harvard.edu/abs/2003ApJ...587..262L}
}

@ARTICLE{2010A&A...518L..83S,
       author = {{Schneider}, N. and {Motte}, F. and {Bontemps}, S. and {Hennemann}, M. and {di Francesco}, J. and {Andr{\'e}}, Ph. and {Zavagno}, A. and {Csengeri}, T. and {Men'shchikov}, A. and {Abergel}, A. and {Baluteau}, J. -P. and {Bernard}, J. -Ph. and {Cox}, P. and {Didelon}, P. and {di Giorgio}, A. -M. and {Gastaud}, R. and {Griffin}, M. and {Hargrave}, P. and {Hill}, T. and {Huang}, M. and {Kirk}, J. and {K{\"o}nyves}, V. and {Leeks}, S. and {Li}, J.~Z. and {Marston}, A. and {Martin}, P. and {Minier}, V. and {Molinari}, S. and {Olofsson}, G. and {Panuzzo}, P. and {Persi}, P. and {Pezzuto}, S. and {Roussel}, H. and {Russeil}, D. and {Sadavoy}, S. and {Saraceno}, P. and {Sauvage}, M. and {Sibthorpe}, B. and {Spinoglio}, L. and {Testi}, L. and {Teyssier}, D. and {Vavrek}, R. and {Ward-Thompson}, D. and {White}, G. and {Wilson}, C.~D. and {Woodcraft}, A.},
        title = "{The Herschel view of star formation in the Rosette molecular cloud under the influence of NGC 2244}",
      journal = {\aap},
         year = 2010,
        month = jul,
       volume = {518},
          eid = {L83},
        pages = {L83},
          doi = {10.1051/0004-6361/201014627},
archivePrefix = {arXiv},
       eprint = {1005.3924},
 primaryClass = {astro-ph.GA},
       adsurl = {https://ui.adsabs.harvard.edu/abs/2010A&A...518L..83S}
}

@ARTICLE{2015ApJ...811...79D,
       author = {{Dewangan}, L.~K. and {Luna}, A. and {Ojha}, D.~K. and {Anandarao}, B.~G. and {Mallick}, K.~K. and {Mayya}, Y.~D.},
        title = "{The Physical Environment of the Massive Star-forming Region W42}",
      journal = {\apj},
         year = 2015,
        month = oct,
       volume = {811},
       number = {2},
          eid = {79},
        pages = {79},
          doi = {10.1088/0004-637X/811/2/79},
archivePrefix = {arXiv},
       eprint = {1508.04425},
 primaryClass = {astro-ph.GA},
       adsurl = {https://ui.adsabs.harvard.edu/abs/2015ApJ...811...79D}
}

@ARTICLE{1969ApJ...157L..13C,
       author = {{Cheung}, A.~C. and {Rank}, D.~M. and {Townes}, C.~H. and {Knowles}, S.~H. and {Sullivan}, III, W.~T.},
        title = "{Distribution of Ammonia Density, Velocity, and Rotational Excitation in the Region of Sagittarius B2}",
      journal = {\apjl},
         year = 1969,
        month = jul,
       volume = {157},
        pages = {L13},
          doi = {10.1086/180374},
       adsurl = {https://ui.adsabs.harvard.edu/abs/1969ApJ...157L..13C}
}

@ARTICLE{1998AJ....116..336N,
       author = {{Nagahama}, Tomoo and {Mizuno}, Akira and {Ogawa}, Hideo and {Fukui}, Yasuo},
        title = "{A Spatially Complete \^13CO J = 1-0 Survey of the Orion A Cloud}",
      journal = {\aj},
         year = 1998,
        month = jul,
       volume = {116},
       number = {1},
        pages = {336-348},
          doi = {10.1086/300392},
       adsurl = {https://ui.adsabs.harvard.edu/abs/1998AJ....116..336N}
}

@ARTICLE{1984A&A...130..311A,
       author = {{Askne}, J. and {Hoglund}, B. and {Hjalmarson}, A. and {Irvine}, W.~M.},
        title = "{Methyl acetylene as a temperature probe in molecular clouds.}",
      journal = {\aap},
         year = 1984,
        month = jan,
       volume = {130},
        pages = {311-318},
       adsurl = {https://ui.adsabs.harvard.edu/abs/1984A&A...130..311A}
}

@ARTICLE{1993ApJS...89..123M,
       author = {{Mangum}, Jeffrey G. and {Wootten}, Alwyn},
        title = "{Formaldehyde as a Probe of Physical Conditions in Dense Molecular Clouds}",
      journal = {\apjs},
         year = 1993,
        month = nov,
       volume = {89},
        pages = {123},
          doi = {10.1086/191841},
       adsurl = {https://ui.adsabs.harvard.edu/abs/1993ApJS...89..123M}
}

@ARTICLE{1987ApJ...318..392M,
       author = {{Mundy}, Lee G. and {Evans}, II, Neal J. and {Snell}, Ronald L. and {Goldsmith}, Paul F.},
        title = "{Models of Molecular Cloud Cores. III. A Multitransition Study of H 2CO}",
      journal = {\apj},
         year = 1987,
        month = jul,
       volume = {318},
        pages = {392},
          doi = {10.1086/165376},
       adsurl = {https://ui.adsabs.harvard.edu/abs/1987ApJ...318..392M}
}

@ARTICLE{2015MNRAS.452.4029U,
       author = {{Urquhart}, J.~S. and {Figura}, C.~C. and {Moore}, T.~J.~T. and {Csengeri}, T. and {Lumsden}, S.~L. and {Pillai}, T. and {Thompson}, M.~A. and {Eden}, D.~J. and {Morgan}, L.~K.},
        title = "{The RMS survey: ammonia mapping of the environment of massive young stellar objects}",
      journal = {\mnras},
         year = 2015,
        month = oct,
       volume = {452},
       number = {4},
        pages = {4029-4053},
          doi = {10.1093/mnras/stv1514},
archivePrefix = {arXiv},
       eprint = {1507.02187},
 primaryClass = {astro-ph.GA},
       adsurl = {https://ui.adsabs.harvard.edu/abs/2015MNRAS.452.4029U}
}

@ARTICLE{1989ApJS...71...89B,
       author = {{Benson}, P.~J. and {Myers}, P.~C.},
        title = "{A Survey for Dense Cores in Dark Clouds}",
      journal = {\apjs},
         year = 1989,
        month = sep,
       volume = {71},
        pages = {89},
          doi = {10.1086/191365},
       adsurl = {https://ui.adsabs.harvard.edu/abs/1989ApJS...71...89B}
}

@ARTICLE{2014ApJ...790...84L,
       author = {{Lu}, Xing and {Zhang}, Qizhou and {Liu}, Hauyu Baobab and {Wang}, Junzhi and {Gu}, Qiusheng},
        title = "{Very Large Array Observations of Ammonia in High-mass Star Formation Regions}",
      journal = {\apj},
         year = 2014,
        month = aug,
       volume = {790},
       number = {2},
          eid = {84},
        pages = {84},
          doi = {10.1088/0004-637X/790/2/84},
archivePrefix = {arXiv},
       eprint = {1405.7933},
 primaryClass = {astro-ph.GA},
       adsurl = {https://ui.adsabs.harvard.edu/abs/2014ApJ...790...84L}
}

@ARTICLE{1990A&A...234..469C,
       author = {{Castets}, A. and {Duvert}, G. and {Dutrey}, A. and {Bally}, J. and {Langer}, W.~D. and {Wilson}, R.~W.},
        title = "{A multi-transition study of carbon monoxide in the Orion A molecular cloud.}",
      journal = {\aap},
         year = 1990,
        month = aug,
       volume = {234},
        pages = {469},
       adsurl = {https://ui.adsabs.harvard.edu/abs/1990A&A...234..469C}
}

@ARTICLE{1998ApJ...494..636Z,
       author = {{Zhang}, Qizhou and {Ho}, Paul T.~P. and {Ohashi}, Nagayoshi},
        title = "{Dynamical Collapse in W51 Massive Cores: CS (3-2) and CH$_{3}$CN Observations}",
      journal = {\apj},
         year = 1998,
        month = feb,
       volume = {494},
       number = {2},
        pages = {636-656},
          doi = {10.1086/305243},
       adsurl = {https://ui.adsabs.harvard.edu/abs/1998ApJ...494..636Z}
}

@article{Gadhi1995,
	author = {{Gadhi}, J and {Lahrouni, A} and {Legrand, J} and {Demaison, J}},
	title = {Moment dipolaire de CH3CN},
	DOI= "10.1051/jcp/1995921984",
	url= "https://doi.org/10.1051/jcp/1995921984",
	journal = {J. Chim. Phys.},
	year = 1995,
	volume = 92,
	pages = "1984-1992",
}

@ARTICLE{2011A&A...534A..49G,
       author = {{Guzm{\'a}n}, V. and {Pety}, J. and {Goicoechea}, J.~R. and {Gerin}, M. and {Roueff}, E.},
        title = "{H$_{2}$CO in the Horsehead PDR: photo-desorption of dust grain ice mantles}",
      journal = {\aap},
         year = 2011,
        month = oct,
       volume = {534},
          eid = {A49},
        pages = {A49},
          doi = {10.1051/0004-6361/201117257},
archivePrefix = {arXiv},
       eprint = {1108.0604},
 primaryClass = {astro-ph.GA},
       adsurl = {https://ui.adsabs.harvard.edu/abs/2011A&A...534A..49G}
}

@ARTICLE{2019ApJ...883..202F,
       author = {{Feng}, S. and {Caselli}, P. and {Wang}, K. and {Lin}, Y. and {Beuther}, H. and {Sipil{\"a}}, O.},
        title = "{The Chemical Structure of Young High-mass Star-forming Clumps. I. Deuteration}",
      journal = {\apj},
         year = 2019,
        month = oct,
       volume = {883},
       number = {2},
          eid = {202},
        pages = {202},
          doi = {10.3847/1538-4357/ab3a42},
archivePrefix = {arXiv},
       eprint = {1909.00209},
 primaryClass = {astro-ph.GA},
       adsurl = {https://ui.adsabs.harvard.edu/abs/2019ApJ...883..202F}
}

@ARTICLE{1983ApJ...272..591C,
       author = {{Churchwell}, E. and {Hollis}, J.~M.},
        title = "{The kinetic temperature and CH3CCH column density profiles in SGR B2,Orion and DR 21}",
      journal = {\apj},
         year = 1983,
        month = sep,
       volume = {272},
        pages = {591-608},
          doi = {10.1086/161322},
       adsurl = {https://ui.adsabs.harvard.edu/abs/1983ApJ...272..591C}
}

@ARTICLE{2023MNRAS.526.3673P,
       author = {{Pazukhin}, A.~G. and {Zinchenko}, I.~I. and {Trofimova}, E.~A. and {Henkel}, C. and {Semenov}, D.~A.},
        title = "{Variations of the HCO$^{+}$, HCN, HNC, N$_{2}$H$^{+}$, and NH$_{3}$ deuterium fractionation in high-mass star-forming regions}",
      journal = {\mnras},
         year = 2023,
        month = dec,
       volume = {526},
       number = {3},
        pages = {3673-3696},
          doi = {10.1093/mnras/stad2976},
archivePrefix = {arXiv},
       eprint = {2309.16510},
 primaryClass = {astro-ph.GA},
       adsurl = {https://ui.adsabs.harvard.edu/abs/2023MNRAS.526.3673P}
}

@ARTICLE{2022ApJ...925....3S,
       author = {{Santos}, Julia C. and {Bronfman}, Leonardo and {Mendoza}, Edgar and {L{\'e}pine}, Jacques R.~D. and {Duronea}, Nicolas U. and {Merello}, Manuel and {Finger}, Ricardo},
        title = "{A Spectral Survey of CH$_{3}$CCH in the Hot Molecular Core G331.512-0.103}",
      journal = {\apj},
         year = 2022,
        month = jan,
       volume = {925},
       number = {1},
          eid = {3},
        pages = {3},
          doi = {10.3847/1538-4357/ac36cc},
archivePrefix = {arXiv},
       eprint = {2201.06330},
 primaryClass = {astro-ph.GA},
       adsurl = {https://ui.adsabs.harvard.edu/abs/2022ApJ...925....3S}
}

@ARTICLE{2012A&ARv..20...56C,
       author = {{Caselli}, Paola and {Ceccarelli}, Cecilia},
        title = "{Our astrochemical heritage}",
      journal = {\aapr},
         year = 2012,
        month = oct,
       volume = {20},
          eid = {56},
        pages = {56},
          doi = {10.1007/s00159-012-0056-x},
archivePrefix = {arXiv},
       eprint = {1210.6368},
 primaryClass = {astro-ph.GA},
       adsurl = {https://ui.adsabs.harvard.edu/abs/2012A&ARv..20...56C}
}

@ARTICLE{2016ApJ...824..136L,
       author = {{Li}, Juan and {Shen}, Zhi-Qiang and {Wang}, Junzhi and {Chen}, Xi and {Wu}, Ya-Jun and {Zhao}, Rong-Bing and {Wang}, Jin-Qing and {Zuo}, Xiu-Ting and {Fan}, Qing-Yuan and {Hong}, Xiao-Yu and {Jiang}, Dong-Rong and {Li}, Bin and {Liang}, Shi-Guang and {Ling}, Quan-Bao and {Liu}, Qing-Hui and {Qian}, Zhi-Han and {Zhang}, Xiu-Zhong and {Zhong}, Wei-Ye and {Ye}, Shu-Hua},
        title = "{TMRT Observations of Carbon-chain Molecules in Serpens South 1a}",
      journal = {\apj},
         year = 2016,
        month = jun,
       volume = {824},
       number = {2},
          eid = {136},
        pages = {136},
          doi = {10.3847/0004-637X/824/2/136},
archivePrefix = {arXiv},
       eprint = {1604.06795},
 primaryClass = {astro-ph.GA},
       adsurl = {https://ui.adsabs.harvard.edu/abs/2016ApJ...824..136L}
}

@ARTICLE{2024A&A...683A..53B,
       author = {{Ben Khalifa}, M. and {Darna}, B. and {Loreau}, J.},
        title = "{Collisional excitation of propyne (CH$_{3}$CCH) by He atoms}",
      journal = {\aap},
         year = 2024,
        month = mar,
       volume = {683},
          eid = {A53},
        pages = {A53},
          doi = {10.1051/0004-6361/202348717},
archivePrefix = {arXiv},
       eprint = {2402.17491},
 primaryClass = {astro-ph.GA},
       adsurl = {https://ui.adsabs.harvard.edu/abs/2024A&A...683A..53B}
}

@ARTICLE{2025A&A...700A.266F,
       author = {{Faure}, A. and {Bacmann}, A. and {Jacquot}, R.},
        title = "{Excitation of Molecules and Atoms for Astrophysics (EMAA): A spectroscopic and collisional database}",
      journal = {\aap},
         year = 2025,
        month = aug,
       volume = {700},
          eid = {A266},
        pages = {A266},
          doi = {10.1051/0004-6361/202554403},
       adsurl = {https://ui.adsabs.harvard.edu/abs/2025A&A...700A.266F}
}

@ARTICLE{2008ApJ...682..283G,
       author = {{Garrod}, Robin T. and {Widicus Weaver}, Susanna L. and {Herbst}, Eric},
        title = "{Complex Chemistry in Star-forming Regions: An Expanded Gas-Grain Warm-up Chemical Model}",
      journal = {\apj},
         year = 2008,
        month = jul,
       volume = {682},
       number = {1},
        pages = {283-302},
          doi = {10.1086/588035},
archivePrefix = {arXiv},
       eprint = {0803.1214},
 primaryClass = {astro-ph},
       adsurl = {https://ui.adsabs.harvard.edu/abs/2008ApJ...682..283G}
}

@ARTICLE{2010SSRv..156...13W,
       author = {{Wakelam}, V. and {Smith}, I.~W.~M. and {Herbst}, E. and {Troe}, J. and {Geppert}, W. and {Linnartz}, H. and {{\"O}berg}, K. and {Roueff}, E. and {Ag{\'u}ndez}, M. and {Pernot}, P. and {Cuppen}, H.~M. and {Loison}, J.~C. and {Talbi}, D.},
        title = "{Reaction Networks for Interstellar Chemical Modelling: Improvements and Challenges}",
      journal = {\ssr},
         year = 2010,
        month = oct,
       volume = {156},
       number = {1-4},
        pages = {13-72},
          doi = {10.1007/s11214-010-9712-5},
archivePrefix = {arXiv},
       eprint = {1011.1184},
 primaryClass = {astro-ph.GA},
       adsurl = {https://ui.adsabs.harvard.edu/abs/2010SSRv..156...13W}
}

@ARTICLE{1999MNRAS.305..755V,
       author = {{Viti}, Serena and {Williams}, David A.},
        title = "{Time-dependent evaporation of icy mantles in hot cores}",
      journal = {\mnras},
         year = 1999,
        month = may,
       volume = {305},
       number = {4},
        pages = {755-762},
          doi = {10.1046/j.1365-8711.1999.02447.x},
       adsurl = {https://ui.adsabs.harvard.edu/abs/1999MNRAS.305..755V}
}

@ARTICLE{2020A&A...635A...4H,
       author = {{Hacar}, A. and {Bosman}, A.~D. and {van Dishoeck}, E.~F.},
        title = "{HCN-to-HNC intensity ratio: a new chemical thermometer for the molecular ISM}",
      journal = {\aap},
         year = 2020,
        month = mar,
       volume = {635},
          eid = {A4},
        pages = {A4},
          doi = {10.1051/0004-6361/201936516},
archivePrefix = {arXiv},
       eprint = {1910.13754},
 primaryClass = {astro-ph.GA},
       adsurl = {https://ui.adsabs.harvard.edu/abs/2020A&A...635A...4H}
}

@ARTICLE{2006A&A...457..927G,
       author = {{Garrod}, R.~T. and {Herbst}, E.},
        title = "{Formation of methyl formate and other organic species in the warm-up phase of hot molecular cores}",
      journal = {\aap},
         year = 2006,
        month = oct,
       volume = {457},
       number = {3},
        pages = {927-936},
          doi = {10.1051/0004-6361:20065560},
archivePrefix = {arXiv},
       eprint = {astro-ph/0607560},
 primaryClass = {astro-ph},
       adsurl = {https://ui.adsabs.harvard.edu/abs/2006A&A...457..927G}
}

@ARTICLE{2015ARPC...66...43K,
       author = {{Kaiser}, Ralf I. and {Parker}, Dorian S.~N. and {Mebel}, Alexander M.},
        title = "{Reaction Dynamics in Astrochemistry: Low-Temperature Pathways to Polycyclic Aromatic Hydrocarbons in the Interstellar Medium}",
      journal = {Annual Review of Physical Chemistry},
         year = 2015,
        month = apr,
       volume = {66},
        pages = {43-67},
          doi = {10.1146/annurev-physchem-040214-121502},
       adsurl = {https://ui.adsabs.harvard.edu/abs/2015ARPC...66...43K}
}

@ARTICLE{2012A&A...544A.146W,
       author = {{Wienen}, M. and {Wyrowski}, F. and {Schuller}, F. and {Menten}, K.~M. and {Walmsley}, C.~M. and {Bronfman}, L. and {Motte}, F.},
        title = "{Ammonia from cold high-mass clumps discovered in the inner Galactic disk by the ATLASGAL survey}",
      journal = {\aap},
         year = 2012,
        month = aug,
       volume = {544},
          eid = {A146},
        pages = {A146},
          doi = {10.1051/0004-6361/201118107},
archivePrefix = {arXiv},
       eprint = {1208.4848},
 primaryClass = {astro-ph.GA},
       adsurl = {https://ui.adsabs.harvard.edu/abs/2012A&A...544A.146W}
}

@ARTICLE{2011MNRAS.418.1689U,
       author = {{Urquhart}, J.~S. and {Morgan}, L.~K. and {Figura}, C.~C. and {Moore}, T.~J.~T. and {Lumsden}, S.~L. and {Hoare}, M.~G. and {Oudmaijer}, R.~D. and {Mottram}, J.~C. and {Davies}, B. and {Dunham}, M.~K.},
        title = "{The Red MSX Source survey: ammonia and water maser analysis of massive star-forming regions}",
      journal = {\mnras},
         year = 2011,
        month = dec,
       volume = {418},
       number = {3},
        pages = {1689-1706},
          doi = {10.1111/j.1365-2966.2011.19594.x},
archivePrefix = {arXiv},
       eprint = {1107.3913},
 primaryClass = {astro-ph.GA},
       adsurl = {https://ui.adsabs.harvard.edu/abs/2011MNRAS.418.1689U}
}

@ARTICLE{2008A&A...488..959B,
       author = {{Bisschop}, S.~E. and {J{\o}rgensen}, J.~K. and {Bourke}, T.~L. and {Bottinelli}, S. and {van Dishoeck}, E.~F.},
        title = "{An interferometric study of the low-mass protostar IRAS 16293-2422: small scale organic chemistry}",
      journal = {\aap},
         year = 2008,
        month = sep,
       volume = {488},
       number = {3},
        pages = {959-968},
          doi = {10.1051/0004-6361:200809673},
archivePrefix = {arXiv},
       eprint = {0807.1447},
 primaryClass = {astro-ph},
       adsurl = {https://ui.adsabs.harvard.edu/abs/2008A&A...488..959B}
}

@ARTICLE{2024MNRAS.527.5049L,
       author = {{Li}, Yuqiang and {Wang}, Junzhi and {Li}, Juan and {Liu}, Shu and {Yang}, Kai and {Zheng}, Siqi and {Lu}, Zhe},
        title = "{Spatial distribution of NH$_{2}$D in massive star-forming regions}",
      journal = {\mnras},
         year = 2024,
        month = jan,
       volume = {527},
       number = {3},
        pages = {5049-5074},
          doi = {10.1093/mnras/stad3480},
archivePrefix = {arXiv},
       eprint = {2311.11534},
 primaryClass = {astro-ph.GA},
       adsurl = {https://ui.adsabs.harvard.edu/abs/2024MNRAS.527.5049L}
}

@ARTICLE{2021A&A...650A.164M,
       author = {{Mazumdar}, P. and {Wyrowski}, F. and {Colombo}, D. and {Urquhart}, J.~S. and {Thompson}, M.~A. and {Menten}, K.~M.},
        title = "{High-resolution LAsMA $^{12}$CO and $^{13}$CO observation of the G305 giant molecular cloud complex. I. Feedback on the molecular gas}",
      journal = {\aap},
         year = 2021,
        month = jun,
       volume = {650},
          eid = {A164},
        pages = {A164},
          doi = {10.1051/0004-6361/202040205},
archivePrefix = {arXiv},
       eprint = {2105.11703},
 primaryClass = {astro-ph.GA},
       adsurl = {https://ui.adsabs.harvard.edu/abs/2021A&A...650A.164M}
}

@ARTICLE{2023ApJS..264...16P,
       author = {{Park}, Geumsook and {Currie}, Malcolm J. and {Thomas}, Holly S. and {Rosolowsky}, Erik and {Dempsey}, Jessica T. and {Kim}, Kee-Tae and {Rigby}, Andrew J. and {Su}, Yang and {Eden}, David J. and {Colombo}, Dario and {Parsons}, Harriet and {Moore}, Toby J.~T.},
        title = "{$^{12}$CO (3-2) High-Resolution Survey (COHRS) of the Galactic Plane: Complete Data Release}",
      journal = {\apjs},
         year = 2023,
        month = jan,
       volume = {264},
       number = {1},
          eid = {16},
        pages = {16},
          doi = {10.3847/1538-4365/ac9b59},
archivePrefix = {arXiv},
       eprint = {2210.05570},
 primaryClass = {astro-ph.GA},
       adsurl = {https://ui.adsabs.harvard.edu/abs/2023ApJS..264...16P}
}

@ARTICLE{2018A&A...609A..16T,
       author = {{Tang}, X.~D. and {Henkel}, C. and {Menten}, K.~M. and {Wyrowski}, F. and {Brinkmann}, N. and {Zheng}, X.~W. and {Gong}, Y. and {Lin}, Y.~X. and {Esimbek}, J. and {Zhou}, J.~J. and {Yuan}, Y. and {Li}, D.~L. and {He}, Y.~X.},
        title = "{Kinetic temperature of massive star-forming molecular clumps measured with formaldehyde. III. The Orion molecular cloud 1}",
      journal = {\aap},
         year = 2018,
        month = jan,
       volume = {609},
          eid = {A16},
        pages = {A16},
          doi = {10.1051/0004-6361/201731849},
archivePrefix = {arXiv},
       eprint = {1709.07694},
 primaryClass = {astro-ph.GA},
       adsurl = {https://ui.adsabs.harvard.edu/abs/2018A&A...609A..16T}
}

@ARTICLE{2021A&A...655A..12T,
       author = {{Tang}, X.~D. and {Henkel}, C. and {Menten}, K.~M. and {Gong}, Y. and {Chen}, C.-H.~R. and {Li}, D.~L. and {Lee}, M.-Y. and {Mangum}, J.~G. and {Ao}, Y.~P. and {M{\"u}hle}, S. and {Aalto}, S. and {Garc{\'\i}a-Burillo}, S. and {Mart{\'\i}n}, S. and {Viti}, S. and {Muller}, S. and {Costagliola}, F. and {Asiri}, H. and {Levshakov}, S.~A. and {Spaans}, M. and {Ott}, J. and {Impellizzeri}, C.~M.~V. and {Fukui}, Y. and {He}, Y.~X. and {Esimbek}, J. and {Zhou}, J.~J. and {Zheng}, X.~W. and {Zhao}, X. and {Li}, J.~S.},
        title = "{Kinetic temperature of massive star-forming molecular clumps measured with formaldehyde. IV. The ALMA view of N113 and N159W in the LMC}",
      journal = {\aap},
         year = 2021,
        month = nov,
       volume = {655},
          eid = {A12},
        pages = {A12},
          doi = {10.1051/0004-6361/202141804},
archivePrefix = {arXiv},
       eprint = {2108.10519},
 primaryClass = {astro-ph.GA},
       adsurl = {https://ui.adsabs.harvard.edu/abs/2021A&A...655A..12T}
}

@ARTICLE{2017A&A...600A..16T,
       author = {{Tang}, X.~D. and {Henkel}, C. and {Chen}, C.-H.~R. and {Menten}, K.~M. and {Indebetouw}, R. and {Zheng}, X.~W. and {Esimbek}, J. and {Zhou}, J.~J. and {Yuan}, Y. and {Li}, D.~L. and {He}, Y.~X.},
        title = "{Kinetic temperature of massive star-forming molecular clumps measured with formaldehyde. II. The Large Magellanic Cloud}",
      journal = {\aap},
         year = 2017,
        month = apr,
       volume = {600},
          eid = {A16},
        pages = {A16},
          doi = {10.1051/0004-6361/201630183},
archivePrefix = {arXiv},
       eprint = {1701.01604},
 primaryClass = {astro-ph.GA},
       adsurl = {https://ui.adsabs.harvard.edu/abs/2017A&A...600A..16T}
}

@ARTICLE{2002A&A...389..603F,
       author = {{Fontani}, F. and {Cesaroni}, R. and {Caselli}, P. and {Olmi}, L.},
        title = "{The structure of molecular clumps around high-mass young stellar objects}",
      journal = {\aap},
         year = 2002,
        month = jul,
       volume = {389},
        pages = {603-617},
          doi = {10.1051/0004-6361:20020579},
archivePrefix = {arXiv},
       eprint = {astro-ph/0204313},
 primaryClass = {astro-ph},
       adsurl = {https://ui.adsabs.harvard.edu/abs/2002A&A...389..603F}
}

@ARTICLE{2016ApJ...826L...8M,
       author = {{Molinari}, S. and {Merello}, M. and {Elia}, D. and {Cesaroni}, R. and {Testi}, L. and {Robitaille}, T.},
        title = "{Calibration of Evolutionary Diagnostics in High-mass Star Formation}",
      journal = {\apjl},
         year = 2016,
        month = jul,
       volume = {826},
       number = {1},
          eid = {L8},
        pages = {L8},
          doi = {10.3847/2041-8205/826/1/L8},
archivePrefix = {arXiv},
       eprint = {1604.06192},
 primaryClass = {astro-ph.GA},
       adsurl = {https://ui.adsabs.harvard.edu/abs/2016ApJ...826L...8M}
}

@ARTICLE{2022A&A...658A.128L,
       author = {{Lin}, Y. and {Wyrowski}, F. and {Liu}, H.~B. and {Izquierdo}, A.~F. and {Csengeri}, T. and {Leurini}, S. and {Menten}, K.~M.},
        title = "{The evolution of temperature and density structures of OB cluster-forming molecular clumps}",
      journal = {\aap},
         year = 2022,
        month = feb,
       volume = {658},
          eid = {A128},
        pages = {A128},
          doi = {10.1051/0004-6361/202142023},
archivePrefix = {arXiv},
       eprint = {2112.01115},
 primaryClass = {astro-ph.GA},
       adsurl = {https://ui.adsabs.harvard.edu/abs/2022A&A...658A.128L}
}

@ARTICLE{2019A&A...631A.137C,
       author = {{Calcutt}, H. and {Willis}, E.~R. and {J{\o}rgensen}, J.~K. and {Bjerkeli}, P. and {Ligterink}, N.~F.~W. and {Coutens}, A. and {M{\"u}ller}, H.~S.~P. and {Garrod}, R.~T. and {Wampfler}, S.~F. and {Drozdovskaya}, M.~N.},
        title = "{The ALMA-PILS survey: propyne (CH$_{3}$CCH) in IRAS 16293-2422}",
      journal = {\aap},
         year = 2019,
        month = nov,
       volume = {631},
          eid = {A137},
        pages = {A137},
          doi = {10.1051/0004-6361/201936323},
archivePrefix = {arXiv},
       eprint = {1909.13329},
 primaryClass = {astro-ph.SR},
       adsurl = {https://ui.adsabs.harvard.edu/abs/2019A&A...631A.137C}
}

@ARTICLE{2024A&A...687A.207Z,
       author = {{Zhao}, X. and {Tang}, X.~D. and {Henkel}, C. and {Gong}, Y. and {Lin}, Y. and {Li}, D.~L. and {He}, Y.~X. and {Ao}, Y.~P. and {Lu}, X. and {Liu}, T. and {Sun}, Y. and {Wang}, K. and {Chen}, X.~P. and {Esimbek}, J. and {Zhou}, J.~J. and {Wu}, J.~W. and {Qiu}, J.~J. and {Zheng}, X.~W. and {Li}, J.~S. and {Luo}, C.~S. and {Zhao}, Q.},
        title = "{Kinetic temperature of massive star-forming molecular clumps measured with formaldehyde. V. The massive filament DR21}",
      journal = {\aap},
         year = 2024,
        month = jul,
       volume = {687},
          eid = {A207},
        pages = {A207},
          doi = {10.1051/0004-6361/202449352},
archivePrefix = {arXiv},
       eprint = {2405.18767},
 primaryClass = {astro-ph.GA},
       adsurl = {https://ui.adsabs.harvard.edu/abs/2024A&A...687A.207Z}
}

@ARTICLE{2023A&A...674A.160G,
       author = {{Gieser}, C. and {Beuther}, H. and {Semenov}, D. and {Ahmadi}, A. and {Henning}, Th. and {Wells}, M.~R.~A.},
        title = "{Physical and chemical complexity in high-mass star-forming regions with ALMA. I. Overview and evolutionary trends of physical properties}",
      journal = {\aap},
         year = 2023,
        month = jun,
       volume = {674},
          eid = {A160},
        pages = {A160},
          doi = {10.1051/0004-6361/202245249},
archivePrefix = {arXiv},
       eprint = {2304.07237},
 primaryClass = {astro-ph.GA},
       adsurl = {https://ui.adsabs.harvard.edu/abs/2023A&A...674A.160G}
}

@ARTICLE{2007ARA&A..45..565M,
       author = {{McKee}, Christopher F. and {Ostriker}, Eve C.},
        title = "{Theory of Star Formation}",
      journal = {\araa},
         year = 2007,
        month = sep,
       volume = {45},
       number = {1},
        pages = {565-687},
          doi = {10.1146/annurev.astro.45.051806.110602},
archivePrefix = {arXiv},
       eprint = {0707.3514},
 primaryClass = {astro-ph},
       adsurl = {https://ui.adsabs.harvard.edu/abs/2007ARA&A..45..565M}
}

@ARTICLE{2024ARA&A..62..369S,
       author = {{Schinnerer}, E. and {Leroy}, A.~K.},
        title = "{Molecular Gas and the Star-Formation Process on Cloud Scales in Nearby Galaxies}",
      journal = {\araa},
         year = 2024,
        month = sep,
       volume = {62},
       number = {1},
        pages = {369-436},
          doi = {10.1146/annurev-astro-071221-052651},
archivePrefix = {arXiv},
       eprint = {2403.19843},
 primaryClass = {astro-ph.GA},
       adsurl = {https://ui.adsabs.harvard.edu/abs/2024ARA&A..62..369S}
}

@ARTICLE{1987JChPh..87.1557T,
       author = {{Tanaka}, Keiichi and {Ito}, Hajime and {Tanaka}, Takehiko},
        title = "{CO$_{2}$ laser-microwave double resonance spectroscopy of NH$_{3}$: Precise measurement of dipole moment in the ground state}",
      journal = {\jcp},
         year = 1987,
        month = aug,
       volume = {87},
       number = {3},
        pages = {1557-1567},
          doi = {10.1063/1.453725},
       adsurl = {https://ui.adsabs.harvard.edu/abs/1987JChPh..87.1557T}
}

@ARTICLE{2011A&A...529A..32T,
       author = {{Torstensson}, K.~J.~E. and {van der Tak}, F.~F.~S. and {van Langevelde}, H.~J. and {Kristensen}, L.~E. and {Vlemmings}, W.~H.~T.},
        title = "{Distribution and excitation of thermal methanol in 6.7 GHz maser bearing star-forming regions. I. The nearby source <ASTROBJ>Cepheus A</ASTROBJ>}",
      journal = {\aap},
         year = 2011,
        month = may,
       volume = {529},
          eid = {A32},
        pages = {A32},
          doi = {10.1051/0004-6361/200913256},
archivePrefix = {arXiv},
       eprint = {1102.3845},
 primaryClass = {astro-ph.SR},
       adsurl = {https://ui.adsabs.harvard.edu/abs/2011A&A...529A..32T}
}

@ARTICLE{2020ARA&A..58..727J,
       author = {{J{\o}rgensen}, Jes K. and {Belloche}, Arnaud and {Garrod}, Robin T.},
        title = "{Astrochemistry During the Formation of Stars}",
      journal = {\araa},
         year = 2020,
        month = aug,
       volume = {58},
        pages = {727-778},
          doi = {10.1146/annurev-astro-032620-021927},
archivePrefix = {arXiv},
       eprint = {2006.07071},
 primaryClass = {astro-ph.SR},
       adsurl = {https://ui.adsabs.harvard.edu/abs/2020ARA&A..58..727J}
}

@ARTICLE{1999ApJ...517..209G,
       author = {{Goldsmith}, Paul F. and {Langer}, William D.},
        title = "{Population Diagram Analysis of Molecular Line Emission}",
      journal = {\apj},
         year = 1999,
        month = may,
       volume = {517},
       number = {1},
        pages = {209-225},
          doi = {10.1086/307195},
       adsurl = {https://ui.adsabs.harvard.edu/abs/1999ApJ...517..209G}
}

@ARTICLE{2021A&A...645A..37T,
       author = {{Tercero}, F. and {L{\'o}pez-P{\'e}rez}, J.~A. and {Gallego}, J.~D. and {Beltr{\'a}n}, F. and {Garc{\'\i}a}, O. and {Patino-Esteban}, M. and {L{\'o}pez-Fern{\'a}ndez}, I. and {G{\'o}mez-Molina}, G. and {Diez}, M. and {Garc{\'\i}a-Carre{\~n}o}, P. and {Malo}, I. and {Amils}, R. and {Serna}, J.~M. and {Albo}, C. and {Hern{\'a}ndez}, J.~M. and {Vaquero}, B. and {Gonz{\'a}lez-Garc{\'\i}a}, J. and {Barbas}, L. and {L{\'o}pez-Fern{\'a}ndez}, J.~A. and {Bujarrabal}, V. and {G{\'o}mez-Garrido}, M. and {Pardo}, J.~R. and {Santander-Garc{\'\i}a}, M. and {Tercero}, B. and {Cernicharo}, J. and {de Vicente}, P.},
        title = "{Yebes 40 m radio telescope and the broad band Nanocosmos receivers at 7 mm and 3 mm for line surveys}",
      journal = {\aap},
         year = 2021,
        month = jan,
       volume = {645},
          eid = {A37},
        pages = {A37},
          doi = {10.1051/0004-6361/202038701},
archivePrefix = {arXiv},
       eprint = {2010.16224},
 primaryClass = {astro-ph.IM},
       adsurl = {https://ui.adsabs.harvard.edu/abs/2021A&A...645A..37T}
}

@ARTICLE{2015PASP..127..266M,
       author = {{Mangum}, Jeffrey G. and {Shirley}, Yancy L.},
        title = "{How to Calculate Molecular Column Density}",
      journal = {\pasp},
         year = 2015,
        month = mar,
       volume = {127},
       number = {949},
        pages = {266},
          doi = {10.1086/680323},
archivePrefix = {arXiv},
       eprint = {1501.01703},
 primaryClass = {astro-ph.IM},
       adsurl = {https://ui.adsabs.harvard.edu/abs/2015PASP..127..266M}
}

@ARTICLE{2025A&A...699A..70H,
       author = {{Huang}, K.-Y. and {Behrens}, E. and {Bouvier}, M. and {Viti}, S. and {Mangum}, J.~G. and {Eibensteiner}, C.},
        title = "{Investigating the chemical link between H$_{2}$CO and CH$_{3}$OH within the central molecular zone of NGC 253}",
      journal = {\aap},
         year = 2025,
        month = jul,
       volume = {699},
          eid = {A70},
        pages = {A70},
          doi = {10.1051/0004-6361/202554156},
archivePrefix = {arXiv},
       eprint = {2505.16255},
 primaryClass = {astro-ph.GA},
       adsurl = {https://ui.adsabs.harvard.edu/abs/2025A&A...699A..70H}
}

@ARTICLE{2023ApJ...953...77M,
       author = {{Mendoza}, Edgar and {Carvajal}, Miguel and {Merello}, Manuel and {Bronfman}, Leonardo and {Boechat-Roberty}, Heloisa M.},
        title = "{Observations and Chemical Modeling of the Isotopologues of Formaldehyde and the Cations of Formyl and Protonated Formaldehyde in the Hot Molecular Core G331.512-0.103}",
      journal = {\apj},
         year = 2023,
        month = aug,
       volume = {953},
       number = {1},
          eid = {77},
        pages = {77},
          doi = {10.3847/1538-4357/ace048},
archivePrefix = {arXiv},
       eprint = {2306.11146},
 primaryClass = {astro-ph.GA},
       adsurl = {https://ui.adsabs.harvard.edu/abs/2023ApJ...953...77M}
}

@ARTICLE{2024A&A...686A..49G,
       author = {{Gerin}, Maryvonne and {Liszt}, Harvey and {Pety}, J{\'e}r{\^o}me and {Faure}, Alexandre},
        title = "{H$_{2}$CO and CS in diffuse clouds: Excitation and abundance}",
      journal = {\aap},
         year = 2024,
        month = jun,
       volume = {686},
          eid = {A49},
        pages = {A49},
          doi = {10.1051/0004-6361/202449152},
archivePrefix = {arXiv},
       eprint = {2403.07075},
 primaryClass = {astro-ph.GA},
       adsurl = {https://ui.adsabs.harvard.edu/abs/2024A&A...686A..49G}
}

@ARTICLE{1988ApJ...324..920K,
       author = {{Keto}, Eric R. and {Ho}, Paul T.~P. and {Haschick}, Aubrey D.},
        title = "{The Observed Structure of the Accretion Flow around G10.6-0.4}",
      journal = {\apj},
         year = 1988,
        month = jan,
       volume = {324},
        pages = {920},
          doi = {10.1086/165949},
       adsurl = {https://ui.adsabs.harvard.edu/abs/1988ApJ...324..920K}
}

@ARTICLE{1987ApJ...312..830R,
       author = {{Reid}, Mark J. and {Myers}, Philip C. and {Bieging}, John H.},
        title = "{The Circumstellar Envelope of W3(OH): NH 3 Observations}",
      journal = {\apj},
         year = 1987,
        month = jan,
       volume = {312},
        pages = {830},
          doi = {10.1086/164929},
       adsurl = {https://ui.adsabs.harvard.edu/abs/1987ApJ...312..830R}
}

@ARTICLE{2022A&A...658A..34T,
       author = {{Tursun}, K. and {Henkel}, C. and {Esimbek}, J. and {Tang}, X.~D. and {Wilson}, T.~L. and {Malawi}, A. and {Alkhuja}, E. and {Wyrowski}, F. and {Mauersberger}, R. and {Immer}, K. and {Asiri}, H. and {Zhou}, J.~J. and {Wu}, G.},
        title = "{Observations of multiple NH$_{3}$ transitions in W33}",
      journal = {\aap},
         year = 2022,
        month = feb,
       volume = {658},
          eid = {A34},
        pages = {A34},
          doi = {10.1051/0004-6361/202141937},
archivePrefix = {arXiv},
       eprint = {2111.06200},
 primaryClass = {astro-ph.GA},
       adsurl = {https://ui.adsabs.harvard.edu/abs/2022A&A...658A..34T}
}
\bibliographystyle{aasjournal}

\startlongtable
\begin{deluxetable*}{ccccccccc}
\tablecaption{Source list}
\label{source}
\tablewidth{0pt}
\tablehead{
\colhead{Source Name} & \colhead{Alias} & \colhead{R.A. (J2000)} & \colhead{Decl. (J2000)} & \colhead{$v_{\rm LSR}$ \tablenotemark{*}}  & \colhead{$D_{\sun}$ \tablenotemark{**}} & \colhead{$K$=2 line of CH$_3$CCH 5-4\tablenotemark{***} } \\
\colhead{} & \colhead{} & \colhead{(hh:mm:ss)} & \colhead{(dd:mm:ss)}  & \colhead{(km s$^{-1}$)} &  \colhead{(kpc)} &
}
\startdata
G000.67-00.03       	&	Sgr B2	&	17:47:20.00    	&	-28:22:40.00   	&	62	$\pm$	5	&	7.8	&	N	\\
G005.88-00.39       	&		&	18:00:30.31    	&	-24:04:04.50   	&	9	$\pm$	3	&	3.0	&	Y	\\
G009.62+00.19       	&		&	18:06:14.66    	&	-20:31:31.70   	&	2	$\pm$	3	&	5.2	&	Y	\\
G010.47+00.02       	&		&	18:08:38.23    	&	-19:51:50.30   	&	69	$\pm$	5	&	8.5	&	N	\\
G010.62-00.38       	&	W 31	&	18:10:28.55    	&	-19:55:48.60   	&	-3	$\pm$	5	&	5.0	&	Y	\\
G011.49-01.48       	&		&	18:16:22.13    	&	-19:41:27.20   	&	11	$\pm$	3	&	1.3	&	Y	\\
G011.91-00.61       	&		&	18:13:58.12    	&	-18:54:20.30   	&	37	$\pm$	5	&	3.4	&	Y	\\
G012.02-00.03       	&		&	18:12:01.84    	&	-18:31:55.80   	&	108	$\pm$	5	&	9.4	&	N	\\
G012.68-00.18       	&		&	18:13:54.75    	&	-18:01:46.60   	&	58	$\pm$	10	&	2.4	&	N	\\
G012.80-00.20       	&		&	18:14:14.23    	&	-17:55:40.50   	&	34	$\pm$	5	&	2.9	&	Y	\\
G012.88+00.48       	&	IRAS 18089-1732	&	18:11:51.42    	&	-17:31:29.00   	&	31	$\pm$	7	&	2.5	&	Y	\\
G012.90-00.24       	&		&	18:14:34.42    	&	-17:51:51.90   	&	36	$\pm$	10	&	2.5	&	Y	\\
G012.90-00.26       	&		&	18:14:39.57    	&	-17:52:00.40   	&	39	$\pm$	10	&	2.5	&	Y	\\
G013.87+00.28       	&		&	18:14:35.83    	&	-16:45:35.90   	&	48	$\pm$	10	&	3.9	&	Y	\\
G014.33-00.64       	&		&	18:18:54.67    	&	-16:47:50.30   	&	22	$\pm$	5	&	1.1	&	Y	\\
G014.63-00.57       	&		&	18:19:15.54    	&	-16:29:45.80   	&	19	$\pm$	5	&	1.8	&	Y	\\
G015.03-00.67       	&	M 17	&	18:20:24.81    	&	-16:11:35.30   	&	22	$\pm$	3	&	2.0	&	Y	\\
G016.58-00.05       	&		&	18:21:09.08    	&	-14:31:48.80   	&	60	$\pm$	5	&	3.6	&	Y	\\
G023.00-00.41       	&		&	18:34:40.20    	&	-09:00:37.00   	&	80	$\pm$	3	&	4.6	&	Y	\\
G023.43-00.18       	&		&	18:34:39.29    	&	-08:31:25.40   	&	97	$\pm$	3	&	5.9	&	Y	\\
G023.65-00.12       	&		&	18:34:51.59    	&	-08:18:21.40   	&	83	$\pm$	3	&	3.2	&	N	\\
G023.70-00.19       	&		&	18:35:12.36    	&	-08:17:39.50   	&	73	$\pm$	5	&	6.2	&	N	\\
G025.70+00.04       	&		&	18:38:03.14    	&	-06:24:15.50   	&	93	$\pm$	5	&	10.2	&	Y	\\
G027.36-00.16       	&		&	18:41:51.06    	&	-05:01:43.40   	&	92	$\pm$	3	&	8.0	&	Y	\\
G028.86+00.06       	&		&	18:43:46.22    	&	-03:35:29.60   	&	100	$\pm$	10	&	7.4	&	Y	\\
G029.86-00.04       	&		&	18:45:59.57    	&	-02:45:06.70   	&	100	$\pm$	3	&	6.2	&	N	\\
G029.95-00.01       	&	W 43S	&	18:46:03.74    	&	-02:39:22.30   	&	98	$\pm$	3	&	5.3	&	Y	\\
G031.28+00.06       	&		&	18:48:12.39    	&	-01:26:30.70   	&	109	$\pm$	3	&	4.3	&	Y	\\
G031.58+00.07       	&	W 43Main	&	18:48:41.68    	&	-01:09:59.00   	&	96	$\pm$	5	&	4.9	&	Y	\\
G032.04+00.05       	&		&	18:49:36.58    	&	-00:45:46.90   	&	97	$\pm$	5	&	5.2	&	Y	\\
G033.64-00.22       	&		&	18:53:32.56    	&	+00:31:39.10   	&	60	$\pm$	3	&	6.5	&	N	\\
G034.39+00.22       	&		&	18:53:19.00    	&	+01:24:08.80   	&	57	$\pm$	5	&	1.6	&	Y	\\
G035.02+00.34       	&		&	18:54:00.67    	&	+02:01:19.20   	&	52	$\pm$	5	&	2.3	&	Y	\\
G035.19-00.74       	&		&	18:58:13.05    	&	+01:40:35.70   	&	30	$\pm$	7	&	2.2	&	Y	\\
G035.20-01.73       	&		&	19:01:45.54    	&	+01:13:32.50   	&	42	$\pm$	3	&	3.3	&	N	\\
G037.43+01.51       	&		&	18:54:14.35    	&	+04:41:41.70   	&	41	$\pm$	3	&	1.9	&	Y	\\
G043.16+00.01       	&	W 49N	&	19:10:13.41    	&	+09:06:12.80   	&	10	$\pm$	5	&	11.1	&	N	\\
G043.79-00.12       	&	OH 43.8-0.1	&	19:11:53.99    	&	+09:35:50.30   	&	44	$\pm$	10	&	6.0	&	N	\\
G043.89-00.78       	&		&	19:14:26.39    	&	+09:22:36.50   	&	54	$\pm$	5	&	8.3	&	N	\\
G045.07+00.13       	&		&	19:13:22.04    	&	+10:50:53.30   	&	59	$\pm$	5	&	8.0	&	N	\\
G045.45+00.05       	&		&	19:14:21.27    	&	+11:09:15.90   	&	55	$\pm$	7	&	8.4	&	N	\\
G048.60+00.02       	&		&	19:20:31.18    	&	+13:55:25.20   	&	18	$\pm$	5	&	10.8	&	N	\\
G049.19-00.33       	&		&	19:22:57.77    	&	+14:16:10.00   	&	67	$\pm$	5	&	5.3	&	Y	\\
G049.48-00.36       	&	W 51 IRS2	&	19:23:39.82    	&	+14:31:05.00   	&	56	$\pm$	3	&	5.1	&	Y	\\
G049.48-00.38       	&	W 51M	&	19:23:43.87    	&	+14:30:29.50   	&	58	$\pm$	4	&	5.4	&	Y	\\
G052.10+01.04       	&	IRAS 19213+1723	&	19:23:37.32    	&	+17:29:10.50   	&	42	$\pm$	5	&	4.0	&	N	\\
G059.78+00.06       	&		&	19:43:11.25    	&	+23:44:03.30   	&	25	$\pm$	3	&	2.2	&	Y	\\
G069.54-00.97       	&	ON 1	&	20:10:09.07    	&	+31:31:36.00   	&	12	$\pm$	5	&	2.5	&	Y	\\
G074.03-01.71       	&		&	20:25:07.11    	&	+34:49:57.60   	&	5	$\pm$	5	&	1.6	&	Y	\\
G075.29+01.32       	&		&	20:16:16.01    	&	+37:35:45.80   	&	-58	$\pm$	5	&	9.3	&	N	\\
G075.76+00.33       	&		&	20:21:41.09    	&	+37:25:29.30   	&	-9	$\pm$	9	&	3.5	&	Y	\\
G075.78+00.34       	&	ON 2N	&	20:21:44.01    	&	+37:26:37.50   	&	1	$\pm$	5	&	3.8	&	Y	\\
G076.38-00.61       	&		&	20:27:25.48    	&	+37:22:48.50   	&	-2	$\pm$	5	&	1.3	&	Y	\\
G078.12+03.63       	&	IRAS 20126+4104	&	20:14:26.07    	&	+41:13:32.70   	&	-4	$\pm$	5	&	1.6	&	Y	\\
G078.88+00.70       	&	AFGL 2591	&	20:29:24.82    	&	+40:11:19.60   	&	-6	$\pm$	7	&	3.3	&	Y	\\
G079.73+00.99       	&	IRAS 20290+4052	&	20:30:50.67    	&	+41:02:27.50   	&	-3	$\pm$	5	&	1.4	&	Y	\\
G079.87+01.17       	&		&	20:30:29.14    	&	+41:15:53.60   	&	-5	$\pm$	10	&	1.6	&	Y	\\
G080.79-01.92       	&	NML Cyg	&	20:46:25.54    	&	+40:06:59.40   	&	-3	$\pm$	3	&	1.6	&	N	\\
G080.86+00.38       	&	DR 20	&	20:37:00.96    	&	+41:34:55.70   	&	-3	$\pm$	5	&	1.5	&	Y	\\
G081.75+00.59       	&	DR 21	&	20:39:01.99    	&	+42:24:59.30   	&	-3	$\pm$	3	&	1.5	&	Y	\\
G081.87+00.78       	&	W 75N	&	20:38:36.43    	&	+42:37:34.80   	&	7	$\pm$	3	&	1.3	&	Y	\\
G090.21+02.32       	&		&	21:02:22.70    	&	+50:03:08.30   	&	-3	$\pm$	5	&	0.7	&	N	\\
G092.67+03.07       	&		&	21:09:21.73    	&	+52:22:37.10   	&	-5	$\pm$	10	&	1.6	&	Y	\\
G094.60-01.79       	&	AFGL 2789	&	21:39:58.27    	&	+50:14:21.00   	&	-46	$\pm$	5	&	3.6	&	N	\\
G095.29-00.93       	&		&	21:39:40.51    	&	+51:20:32.80   	&	-38	$\pm$	5	&	4.9	&	N	\\
G097.53+03.18       	&		&	21:32:12.43    	&	+55:53:49.70   	&	-73	$\pm$	5	&	7.5	&	N	\\
G100.37-03.57       	&		&	22:16:10.37    	&	+52:21:34.10   	&	-37	$\pm$	10	&	3.4	&	N	\\
G105.41+09.87       	&		&	21:43:06.48    	&	+66:06:55.30   	&	-10	$\pm$	5	&	0.9	&	N	\\
G107.29+05.63       	&	IRAS 22198+6336	&	22:21:26.73    	&	+63:51:37.90   	&	-11	$\pm$	5	&	0.8	&	N	\\
G108.18+05.51       	&	L 1206	&	22:28:51.41    	&	+64:13:41.30   	&	-11	$\pm$	3	&	0.8	&	N	\\
G108.20+00.58       	&		&	22:49:31.48    	&	+59:55:42.00   	&	-49	$\pm$	5	&	4.4	&	N	\\
G108.47-02.81       	&		&	23:02:32.08    	&	+56:57:51.40   	&	-54	$\pm$	5	&	3.2	&	N	\\
G108.59+00.49       	&		&	22:52:38.30    	&	+60:00:52.00   	&	-52	$\pm$	5	&	2.5	&	N	\\
G109.87+02.11       	&	Cep A	&	22:56:18.10    	&	+62:01:49.50   	&	-7	$\pm$	5	&	0.7	&	Y	\\
G111.23-01.23       	&		&	23:17:20.79    	&	+59:28:47.00   	&	-53	$\pm$	10	&	3.5	&	N	\\
G111.25-00.76       	&		&	23:16:10.36    	&	+59:55:28.50   	&	-43	$\pm$	5	&	3.4	&	N	\\
G111.54+00.77       	&	NGC 7538	&	23:13:45.36    	&	+61:28:10.60   	&	-57	$\pm$	5	&	2.6	&	Y	\\
G121.29+00.65       	&	L 1287	&	00:36:47.35    	&	+63:29:02.20   	&	-23	$\pm$	5	&	0.9	&	Y	\\
G122.01-07.08       	&	IRAS 00420+5530	&	00:44:58.40    	&	+55:46:47.60   	&	-50	$\pm$	5	&	2.2	&	N	\\
G123.06-06.30       	&	NGC 281	&	00:52:24.70    	&	+56:33:50.50   	&	-30	$\pm$	5	&	2.8	&	Y	\\
G123.06-06.30       	&	NGC 281W	&	00:52:24.20    	&	+56:33:43.20   	&	-29	$\pm$	3	&	2.4	&	Y	\\
G133.94+01.06       	&	W 3OH	&	02:27:03.82    	&	+61:52:25.20   	&	-47	$\pm$	3	&	2.0	&	Y	\\
G134.62-02.19       	&	S Per	&	02:22:51.71    	&	+58:35:11.40   	&	-39	$\pm$	5	&	2.4	&	N	\\
G135.27+02.79       	&	WB 89-437	&	02:43:28.57    	&	+62:57:08.40   	&	-72	$\pm$	3	&	6.0	&	N	\\
G160.14+03.15       	&		&	05:01:40.24    	&	+47:07:19.00   	&	-18	$\pm$	5	&	4.1	&	N	\\
G168.06+00.82       	&	IRAS 05137+3919	&	05:17:13.74    	&	+39:22:19.90   	&	-27	$\pm$	5	&	7.7	&	N	\\
G176.51+00.20       	&		&	05:37:52.14    	&	+32:00:03.90   	&	-17	$\pm$	5	&	1.0	&	Y	\\
G182.67-03.26       	&		&	05:39:28.42    	&	+24:56:32.10   	&	-7	$\pm$	10	&	6.7	&	N	\\
G183.72-03.66       	&		&	05:40:24.23    	&	+23:50:54.70   	&	3	$\pm$	5	&	1.8	&	Y	\\
G188.79+01.03       	&	IRAS 06061+2151	&	06:09:06.97    	&	+21:50:41.40   	&	-5	$\pm$	5	&	2.0	&	N	\\
G188.94+00.88       	&	S 252	&	06:08:53.35    	&	+21:38:28.70   	&	8	$\pm$	5	&	2.1	&	Y	\\
G192.16-03.81       	&		&	05:58:13.53    	&	+16:31:58.90   	&	5	$\pm$	5	&	1.5	&	Y	\\
G192.60-00.04       	&	S 255	&	06:12:54.02    	&	+17:59:23.30   	&	6	$\pm$	5	&	1.6	&	N	\\
G196.45-01.67       	&	S 269	&	06:14:37.08    	&	+13:49:36.70   	&	19	$\pm$	5	&	5.3	&	N	\\
G209.00-19.38       	&	Orion Nebula	&	05:35:15.80    	&	-05:23:14.10   	&	3	$\pm$	5	&	0.4	&	Y	\\
G211.59+01.05       	&		&	06:52:45.32    	&	+01:40:23.10   	&	45	$\pm$	5	&	4.4	&	N	\\
G229.57+00.15       	&		&	07:23:01.84    	&	-14:41:32.80   	&	47	$\pm$	10	&	4.5	&	N	\\
G232.62+00.99       	&		&	07:32:09.78    	&	-16:58:12.80   	&	21	$\pm$	3	&	1.7	&	N	\\
G236.81+01.98       	&		&	07:44:28.24    	&	-20:08:30.20   	&	43	$\pm$	7	&	3.4	&	N	\\
G239.35-05.06       	&	VY CMa	&	07:22:58.33    	&	-25:46:03.10   	&	20	$\pm$	3	&	1.2	&	N	\\
G240.31+00.07       	&		&	07:44:51.92    	&	-24:07:41.50   	&	67	$\pm$	5	&	4.7	&	N	\\
\enddata
\tablenotetext{*}{The $v_{\rm LSR}$ values are from \citet{2014ApJ...783..130R}.}
\tablenotetext{**}{The $D_{\sun}$ values are derived from the trigonometric parallax reported in \citet{2014ApJ...783..130R}.}
\tablenotetext{***}{"Y" indicates that the CH$_3$CCH $K=2$ line was detected above the 3$\sigma$ level, while "N" denotes non-detection.}
\end{deluxetable*}

\newpage
\begin{deluxetable*}{ccccccccc}[h]
\tablecaption{The parameter of CH$_3$CCH $J$=5-4.}
\label{CH3CCH parameter}
\tablewidth{0pt}
\tablehead{
\colhead{Transition} & \colhead{Frequency} & \colhead{A$_{ul}$ \tablenotemark{*}} & \colhead{$g_{\rm u}$ \tablenotemark{*}} & \colhead{$E_{\rm u}$ \tablenotemark{*}} & \colhead{$n_{\rm crit}$ \tablenotemark{**}} \\
\colhead{$J_u$,$K_u$-$J_l$,$K_l$} & \colhead{(MHz)} &  \colhead{(s$^{-1}$)} & \colhead{} & \colhead{(K)} & \colhead{(cm$^{-3}$)} &
}
\startdata
\hline
CH$_3$CCH 5,4-4,4 & 85431.17 & 7.30$\times$10$^{-7}$ & 22 & 127.9 & 1.9$\times10^4$ \\
CH$_3$CCH 5,3-4,3 & 85442.60 & 1.30$\times$10$^{-6}$ & 44 & 77.3 & 4.2$\times10^4$ \\
CH$_3$CCH 5,2-4,2 & 85450.77 & 1.70$\times$10$^{-6}$ & 22 & 41.2 & 5.1$\times10^4$ \\
CH$_3$CCH 5,1-4,1 & 85455.67 & 1.94$\times$10$^{-6}$ & 22 & 19.5 & 8.0$\times10^4$ \\
CH$_3$CCH 5,0-4,0 & 85457.30 & 2.03$\times$10$^{-6}$ & 22 & 12.3 & 6.6$\times10^4$ \\
\enddata
\tablenotetext{*}{Line spectroscopic parameters are given according to CDMS.}
\tablenotetext{**}{The critical density of each transition is derived from the Einstein emission coefficient and collisional rate coefficients at 30 K given by EMAA.}
\end{deluxetable*}

\startlongtable
\begin{deluxetable*}{ccccccccc}
\tabletypesize{\scriptsize}
\tablecaption{$T_{\rm rot}$(CH$_3$CCH) and $T_{\rm rot}$(2,2;1,1) used in this work}
\label{result}
\tablewidth{0pt}
\tablehead{
\colhead{Source Name} & \colhead{$K$=3 line of CH$_3$CCH 5-4\tablenotemark{*} } & \colhead{$T_{\rm rot}$(CH$_3$CCH)} & \colhead{$T_{\rm rot}$(2,2;1,1)} & \colhead{Telescope \tablenotemark{**}} \\
\colhead{} & \colhead{} &  \colhead{(K)} &  \colhead{(K)} &
}
\startdata
G005.88-00.39       	&	Y	&	48.5 	$\pm$	0.4 	&	27.9 	$\pm$	0.5 	&	TMRT	\\
G009.62+00.19       	&	Y	&	36.6 	$\pm$	0.3 	&	25.2 	$\pm$	3.8 	&	Effelsberg	\\
G010.62-00.38       	&	Y	&	42.2 	$\pm$	0.3 	&		...		&		...\\
G011.49-01.48       	&	Y	&	26.9 	$\pm$	0.8 	&	20.0 	$\pm$	0.6 	&	TMRT	\\
G011.91-00.61       	&	Y	&	26.3 	$\pm$	0.3 	&	17.5 	$\pm$	0.4 	&	TMRT	\\
G012.80-00.20       	&	Y	&	44.4 	$\pm$	0.4 	&		...		&		...\\
G012.88+00.48       	&	Y	&	36.5 	$\pm$	0.4 	&	19.3 	$\pm$	0.9 	&	Effelsberg	\\
G012.90-00.24       	&	N	&	30.2 	$\pm$	0.8 	&	15.7 	$\pm$	0.4 	&	TMRT	\\
G012.90-00.26       	&	Y	&	34.7 	$\pm$	0.4 	&	19.6 	$\pm$	0.6 	&	Effelsberg	\\
G013.87+00.28       	&	N	&	31.3 	$\pm$	0.7 	&	22.3 	$\pm$	0.6 	&	TMRT	\\
G014.33-00.64       	&	Y	&	31.4 	$\pm$	0.3 	&	21.3 	$\pm$	0.3 	&	TMRT	\\
G014.63-00.57       	&	Y	&	35.1 	$\pm$	0.5 	&	18.4 	$\pm$	0.4 	&	TMRT	\\
G015.03-00.67       	&	Y	&	52.9 	$\pm$	0.8 	&	28.2 	$\pm$	4.2 	&	Effelsberg	\\
G016.58-00.05       	&	N	&	26.5 	$\pm$	0.6 	&	19.1 	$\pm$	0.8 	&	Effelsberg	\\
G023.00-00.41       	&	Y	&	29.9 	$\pm$	0.4 	&	17.2 	$\pm$	0.5 	&	Effelsberg	\\
G023.43-00.18       	&	N	&	29.2 	$\pm$	0.6 	&	20.0 	$\pm$	0.8 	&	Effelsberg	\\
G025.70+00.04       	&	N	&	23.0 	$\pm$	0.4 	&	18.5 	$\pm$	1.1 	&	Effelsberg	\\
G027.36-00.16       	&	Y	&	28.1 	$\pm$	0.7 	&	20.4 	$\pm$	0.5 	&	TMRT	\\
G028.86+00.06       	&	Y	&	37.9 	$\pm$	1.0 	&	19.6 	$\pm$	0.9 	&	TMRT	\\
G029.95-00.01       	&	Y	&	24.1 	$\pm$	0.7 	&	20.9 	$\pm$	1.0 	&	Effelsberg	\\
G031.28+00.06       	&	Y	&	40.9 	$\pm$	0.4 	&	21.6 	$\pm$	1.1 	&	Effelsberg	\\
G031.58+00.07       	&	Y	&	32.1 	$\pm$	0.5 	&	22.9 	$\pm$	0.6 	&	TMRT	\\
G032.04+00.05       	&	Y	&	33.0 	$\pm$	0.6 	&	19.9 	$\pm$	0.3 	&	TMRT	\\
G034.39+00.22       	&	Y	&	36.9 	$\pm$	0.8 	&	18.3 	$\pm$	0.3 	&	TMRT	\\
G035.02+00.34       	&	Y	&	33.1 	$\pm$	0.6 	&	24.5 	$\pm$	0.8 	&	TMRT	\\
G035.19-00.74       	&	Y	&	34.9 	$\pm$	0.3 	&	20.7 	$\pm$	0.5 	&	Effelsberg	\\
G037.43+01.51       	&	N	&	88.8 	$\pm$	3.0 	&	21.9 	$\pm$	0.6 	&	TMRT	\\
G049.19-00.33       	&	N	&	42.1 	$\pm$	2.6 	&	21.2 	$\pm$	1.0 	&	TMRT	\\
G049.48-00.36       	&	Y	&	42.1 	$\pm$	0.4 	&	25.2 	$\pm$	1.8 	&	Effelsberg	\\
G049.48-00.38       	&	Y	&	48.6 	$\pm$	0.4 	&	29.5 	$\pm$	0.6 	&	Effelsberg	\\
G059.78+00.06       	&	Y	&	49.7 	$\pm$	0.9 	&	19.1 	$\pm$	0.3 	&	TMRT	\\
G069.54-00.97       	&	Y	&	29.5 	$\pm$	0.3 	&	18.8 	$\pm$	3.4 	&	Effelsberg	\\
G074.03-01.71       	&	N	&	33.4 	$\pm$	0.7 	&	16.7 	$\pm$	0.4 	&	TMRT	\\
G075.76+00.33       	&	Y	&	43.9 	$\pm$	0.6 	&	26.4 	$\pm$	0.5 	&	TMRT	\\
G075.78+00.34       	&	Y	&	41.7 	$\pm$	0.7 	&	24.2 	$\pm$	0.5 	&	TMRT	\\
G076.38-00.61       	&	Y	&	54.5 	$\pm$	1.6 	&	23.7 	$\pm$	0.7 	&	TMRT	\\
G078.12+03.63       	&	N	&	32.0 	$\pm$	1.0 	&	18.5 	$\pm$	0.7 	&	Effelsberg	\\
G078.88+00.70       	&	Y	&	26.9 	$\pm$	0.8 	&	27.5 	$\pm$	0.6 	&	TMRT	\\
G079.73+00.99       	&	N	&	29.2 	$\pm$	1.1 	&	17.9 	$\pm$	0.7 	&	TMRT	\\
G079.87+01.17       	&	N	&	26.6 	$\pm$	0.8 	&	22.7 	$\pm$	1.1 	&	TMRT	\\
G080.86+00.38       	&	N	&	32.5 	$\pm$	1.4 	&	23.4 	$\pm$	0.5 	&	TMRT	\\
G081.75+00.59       	&	Y	&	31.3 	$\pm$	0.3 	&	18.5 	$\pm$	0.3 	&	Effelsberg	\\
G081.87+00.78       	&	Y	&	30.4 	$\pm$	0.5 	&	25.2 	$\pm$	0.4 	&	TMRT	\\
G092.67+03.07       	&	Y	&	38.3 	$\pm$	0.6 	&	24.5 	$\pm$	0.6 	&	TMRT	\\
G109.87+02.11       	&	Y	&	36.1 	$\pm$	0.4 	&	23.2 	$\pm$	3.0 	&	Effelsberg	\\
G111.54+00.77       	&	N	&	43.1 	$\pm$	0.9 	&	26.0 	$\pm$	1.3 	&	Effelsberg	\\
G121.29+00.65       	&	N	&	25.2 	$\pm$	0.9 	&	18.1 	$\pm$	0.2 	&	TMRT	\\
G123.06-06.30       	&	N	&	27.1 	$\pm$	1.0 	&	21.5 	$\pm$	0.5 	&	TMRT	\\
G123.06-06.30       	&	Y	&	37.1 	$\pm$	1.2 	&	20.1 	$\pm$	0.5 	&	TMRT	\\
G133.94+01.06       	&	Y	&	38.1 	$\pm$	0.8 	&		...		&		...\\
G176.51+00.20       	&	N	&	49.6 	$\pm$	1.9 	&	17.6 	$\pm$	0.4 	&	TMRT	\\
G183.72-03.66       	&	N	&	31.6 	$\pm$	1.1 	&	17.5 	$\pm$	0.4 	&	TMRT	\\
G188.94+00.88       	&	N	&	31.2 	$\pm$	0.9 	&	22.3 	$\pm$	0.4 	&	TMRT	\\
G192.60-00.04       	&	N	&	46.4 	$\pm$	1.1 	&	17.4 	$\pm$	1.5 	&	TMRT	\\
G209.00-19.38       	&	Y	&	38.1	$\pm$	0.7	&		...			&	...	\\
\enddata
\tablenotetext{*}{"Y" indicates that the CH$_3$CCH $K=2$ line was detected above the 3$\sigma$ level, while "N" denotes non-detection.}
\tablenotetext{**}{Telescope using TMRT indicates NH$_3$ data obtained in this work. The telescope using Effelsberg indicates NH$_3$ data are taken from \citep{2016AJ....152...92L}. Please note that G010.62-00.38, G012.80-00.20, G133.94+01.06 and G209.00-19.38 have no NH$_3$ data because of the absorption or low signal-to-noise ratio.}
\end{deluxetable*}

\clearpage
\newpage
\appendix
\section{The sources with notable temperature differences between $T_{\rm rot}$(CH$_3$CCH) and $T_{\rm rot}$(2,2;1,1)}
Among 51 sources used in this work, five show departures from the overall consistency between $T_{\rm rot}$(CH$_3$CCH) and $T_{\rm rot}$(2,2;1,1).
Only one source, G078.88+00.70, shows a case where the rotation temperature derived from CH$_3$CCH ($T_{\rm rot}$(CH$_3$CCH)) is slightly lower (0.6 K) than that from NH$_3$(1,1)\&(2,2) ($T_{\rm rot}$(2,2;1,1)). 
This indicates that the two temperatures are consistent within the uncertainties. 
One exceptional case is G037.43+01.51, which shows an unusually high $T_{\rm rot}$(CH$_3$CCH) of 88.8 K. However, since the CH$_3$CCH $K$=2 transition in this source was only marginally detected, the temperature estimate was unreliable. Therefore, we exclude this source in our analysis. There are three other notable sources, G059.78+00.06, G176.51+00.20, and G192.60-00.04, which show significantly higher $T_{\rm rot}$(CH$_3$CCH) compared to $T_{\rm rot}$(2,2;1,1). In G176.51+00.20, the CH$_3$CCH $K$=2 transition was marginally detected, which may affect the reliability of the CH$_3$CCH temperature estimate. In G192.60-00.04, the NH$_3$(2,2) was marginally detected, likely leading to an underestimated $T_{\rm rot}$(2,2;1,1). As for G059.78+00.06, the discrepancy might also arise from general observational uncertainties between TMRT and Yebes affecting either or both calculations. 

\clearpage
\newpage
\section{The fitting result of CH$_3$CCH 5-4}
We present the fitting result of CH$_3$CCH 5-4.

\renewcommand{\thefigure}{B\arabic{figure}}  
\setcounter{figure}{0}
\begin{figure}[ht!]
\centering
\gridline{
\fig{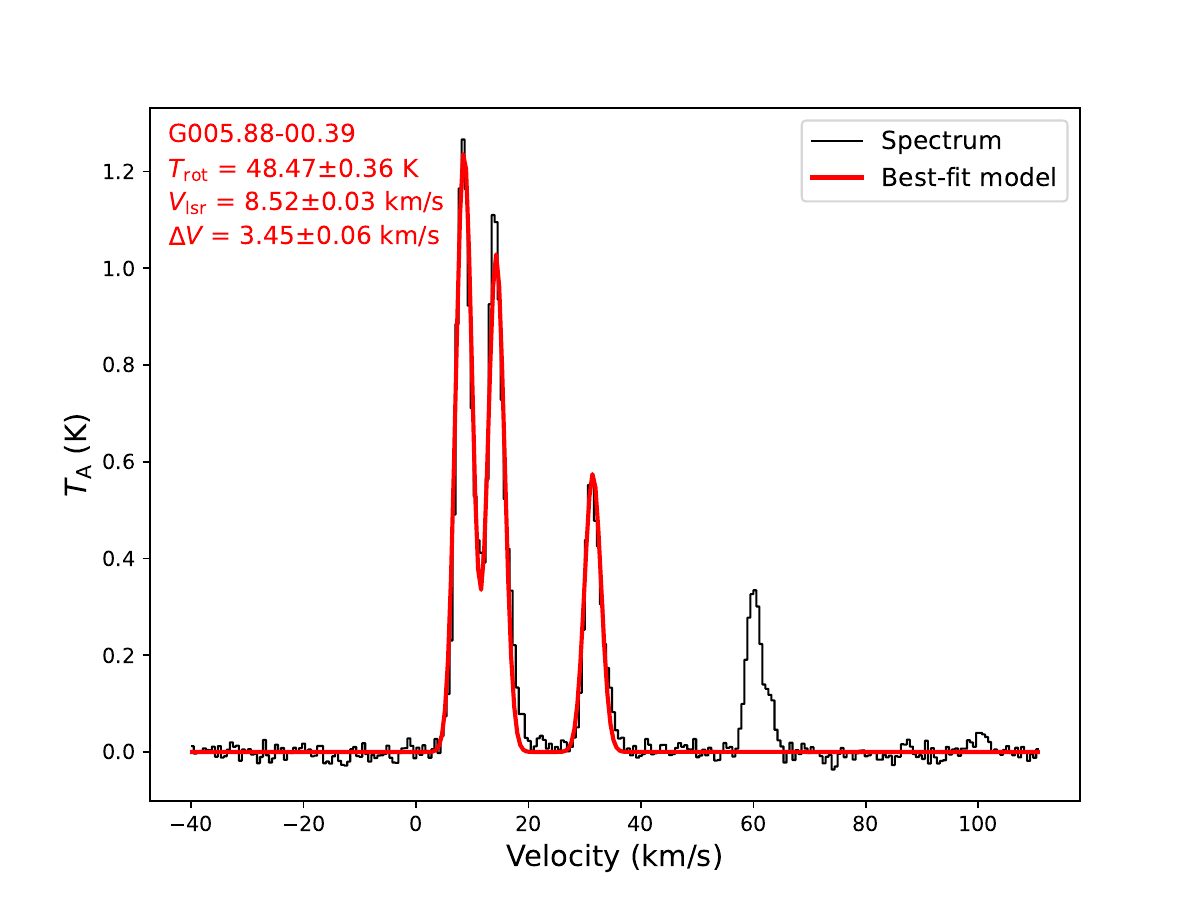}{0.31\textwidth}{}
\fig{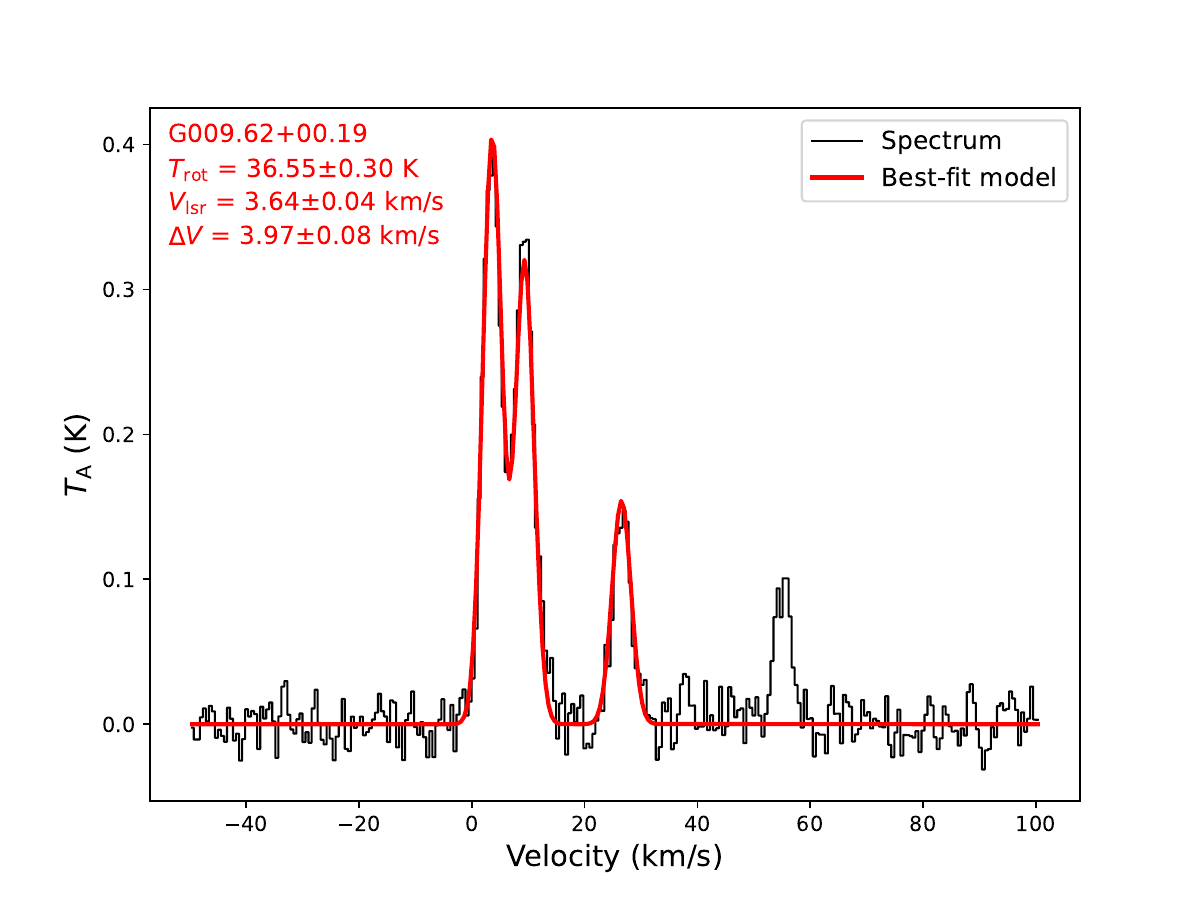}{0.31\textwidth}{}
\fig{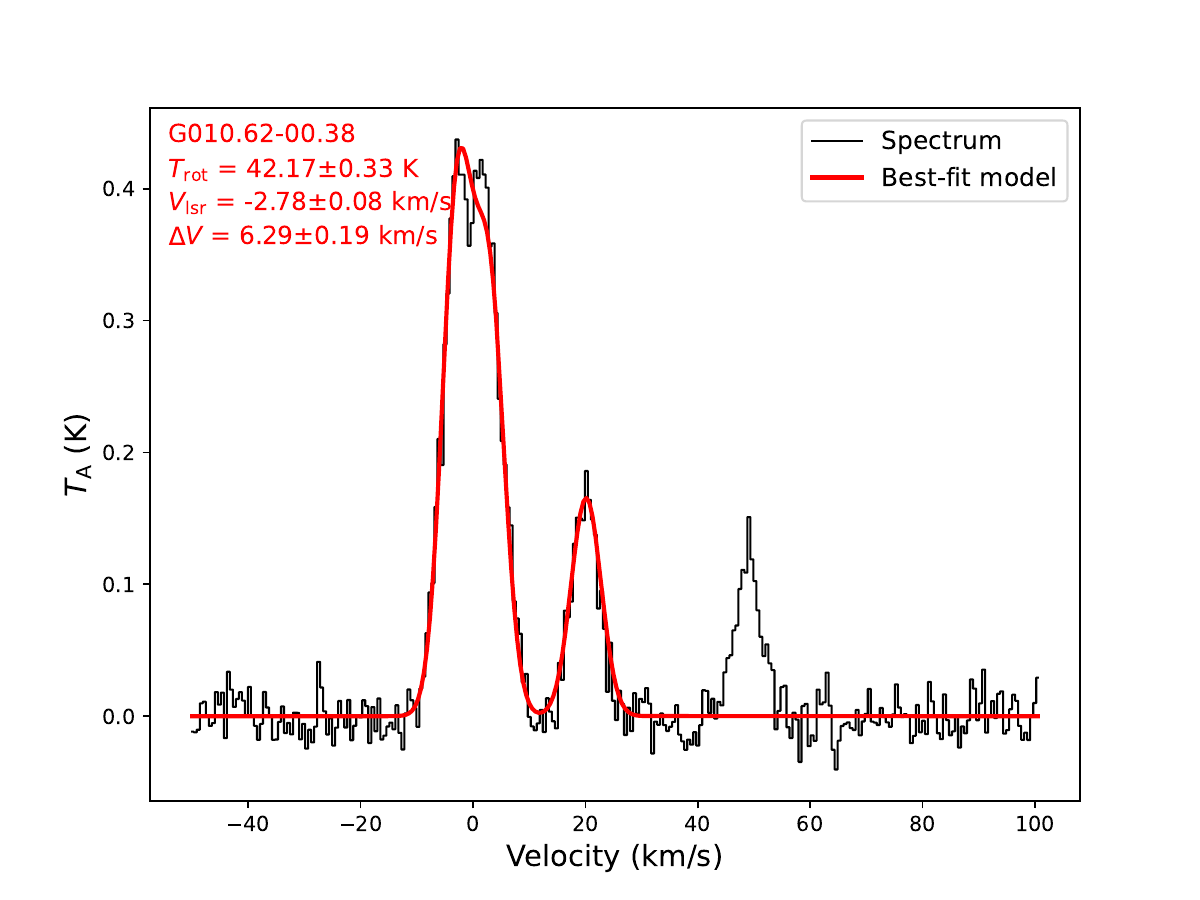}{0.31\textwidth}{}
}
\gridline{
\fig{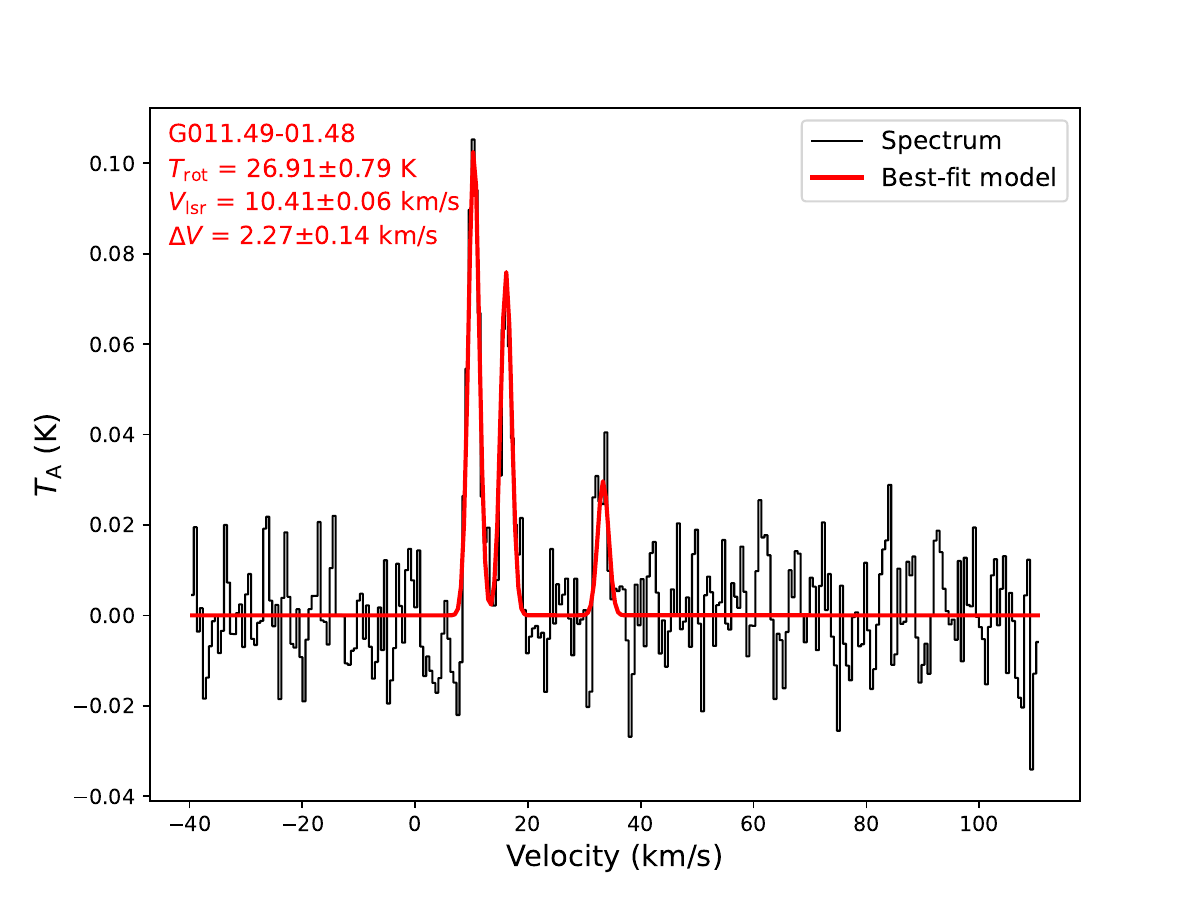}{0.31\textwidth}{}
\fig{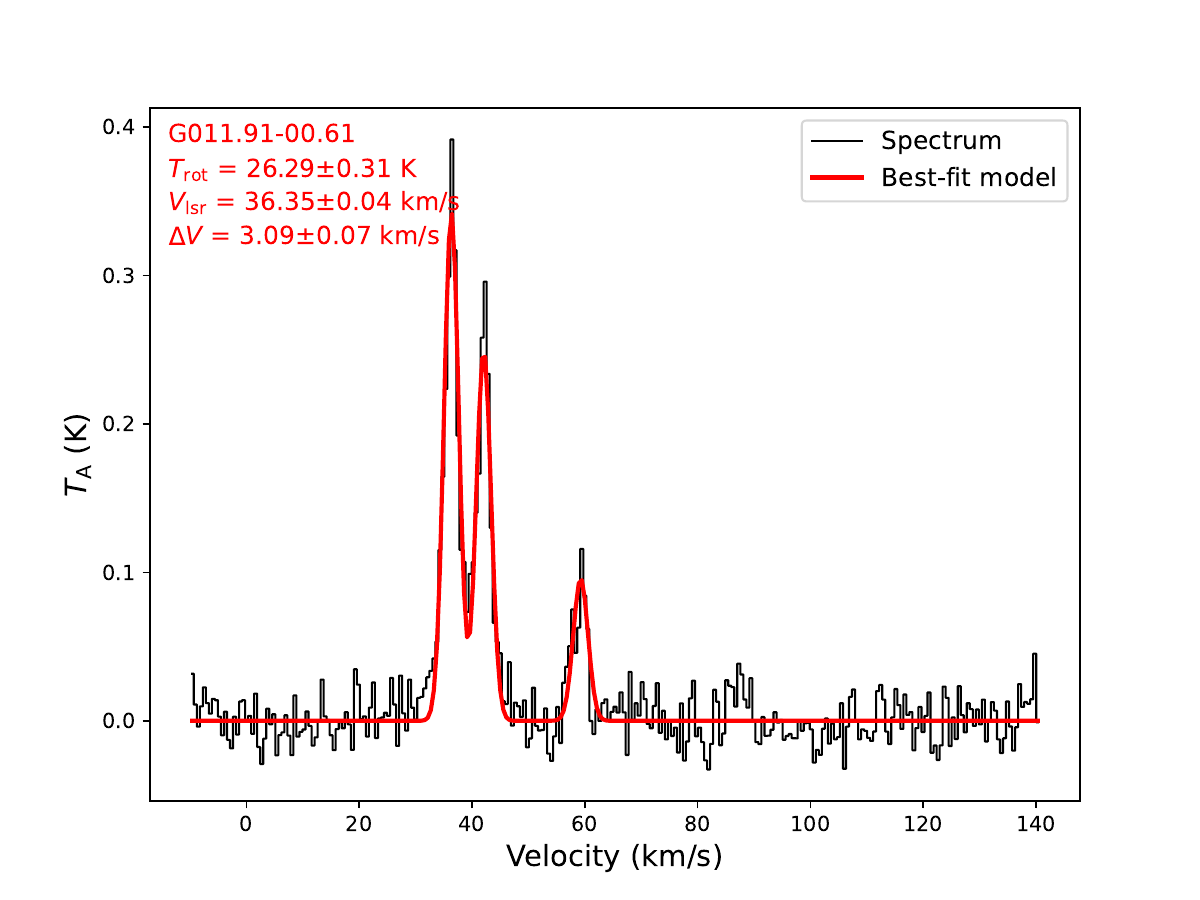}{0.31\textwidth}{}
\fig{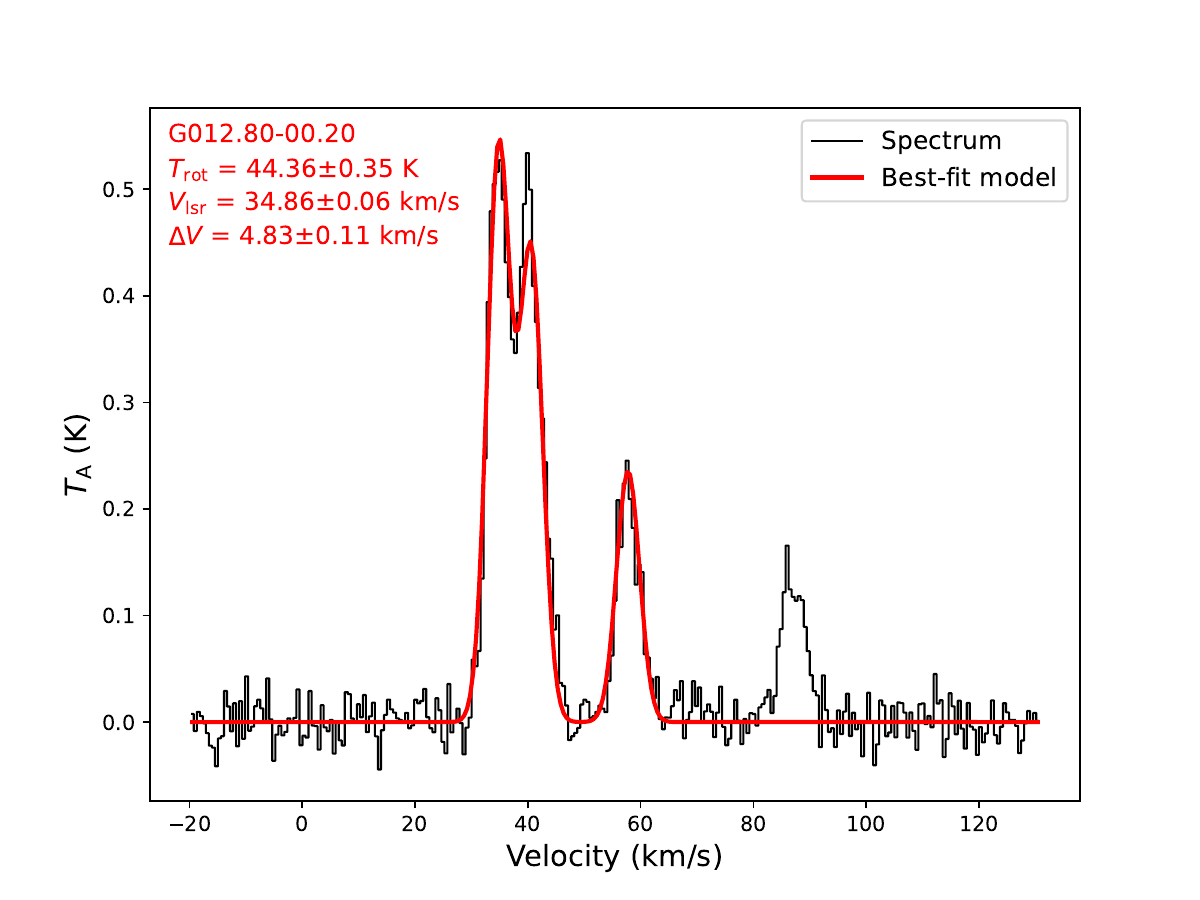}{0.31\textwidth}{}
}
\gridline{
\fig{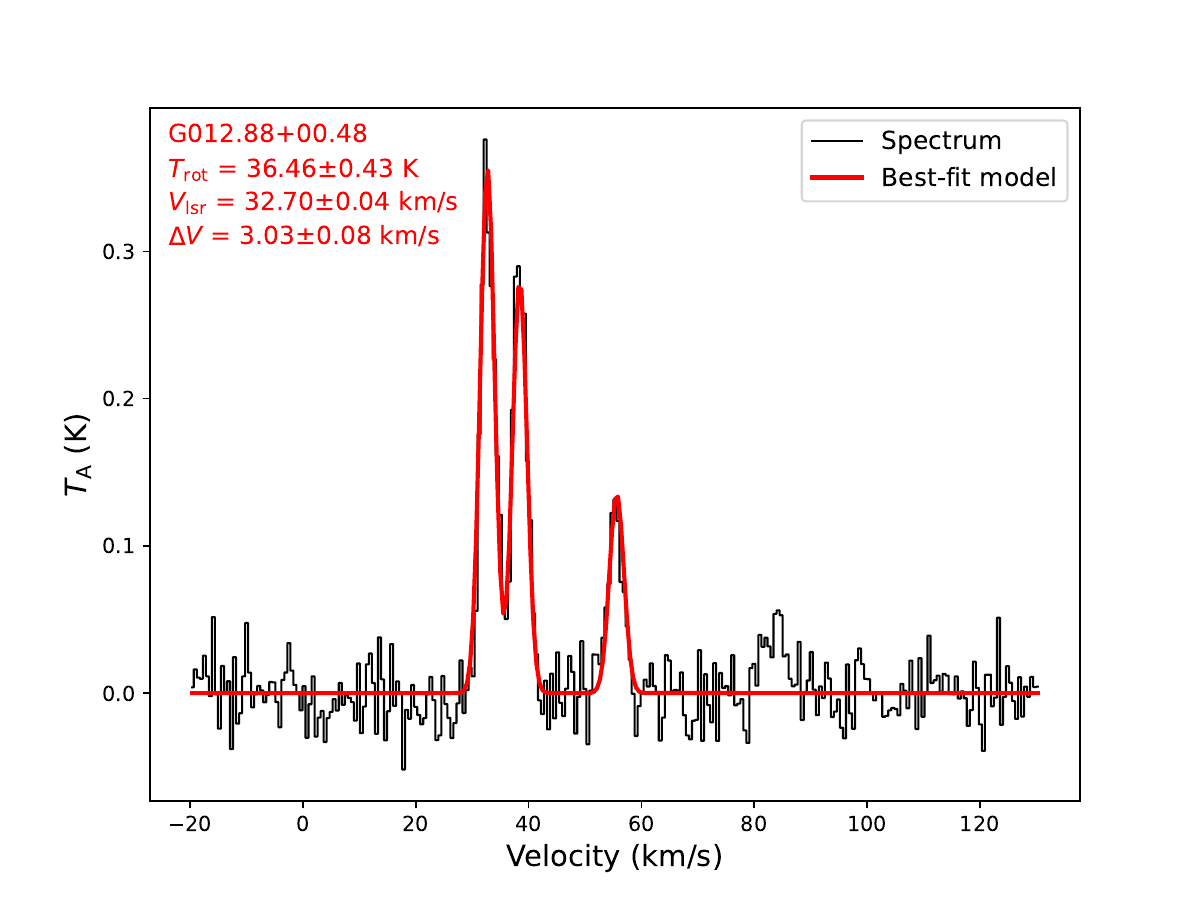}{0.31\textwidth}{}
\fig{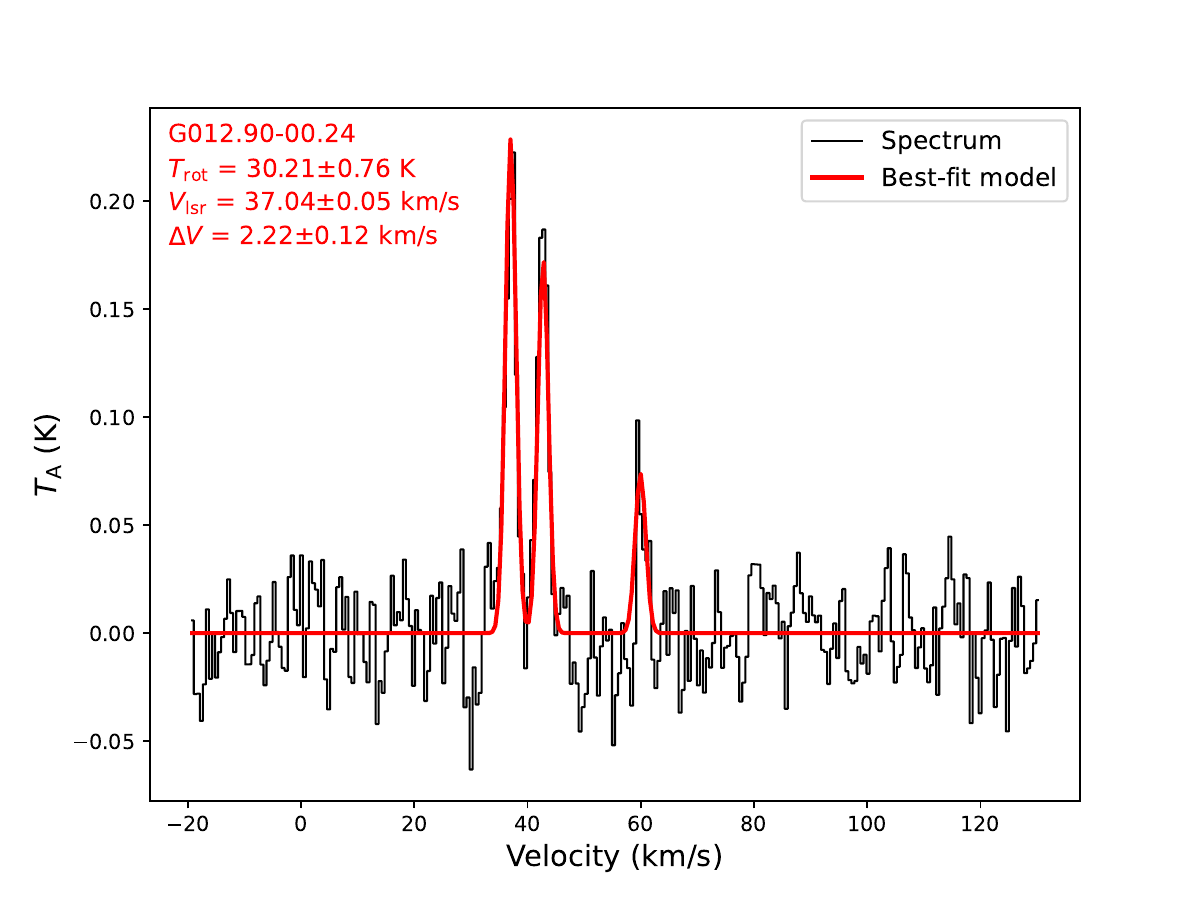}{0.31\textwidth}{}
\fig{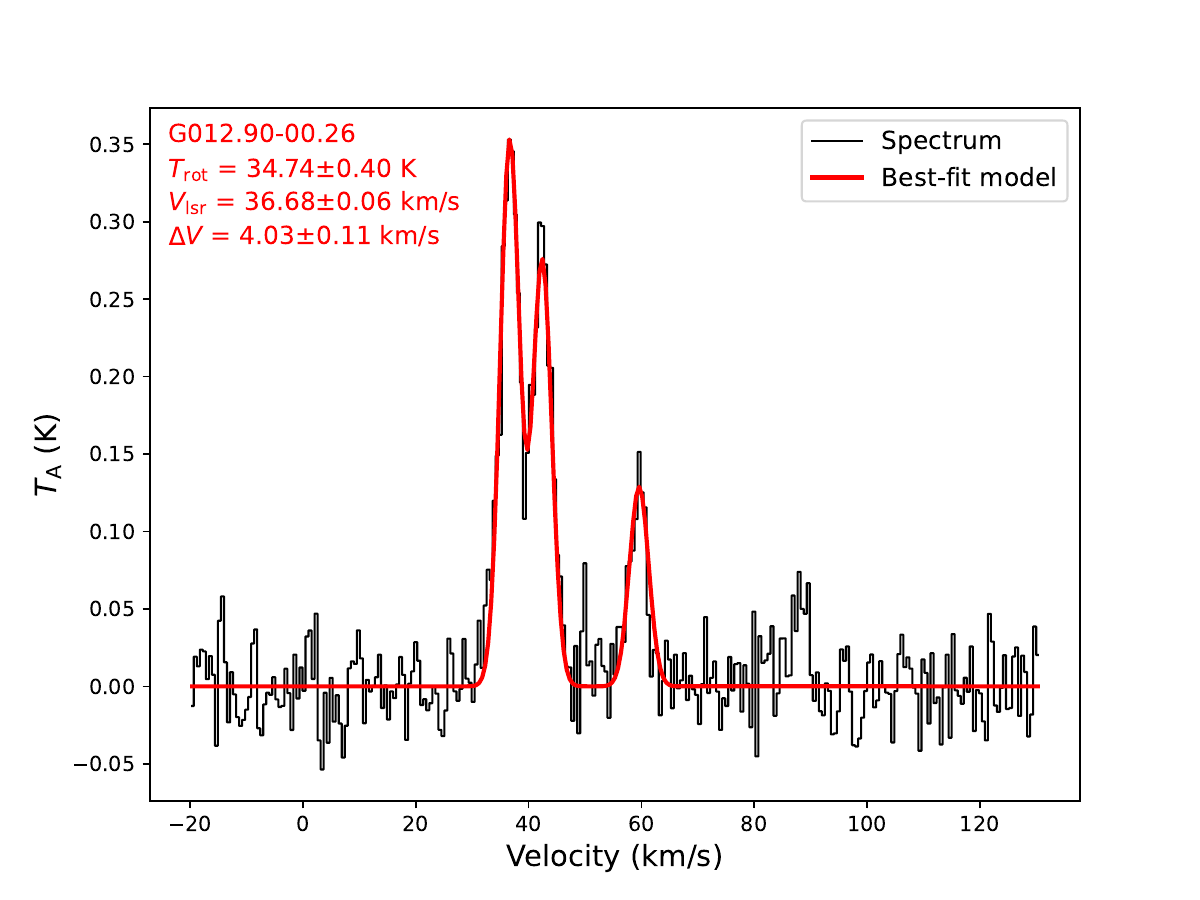}{0.31\textwidth}{}
}
\gridline{
\fig{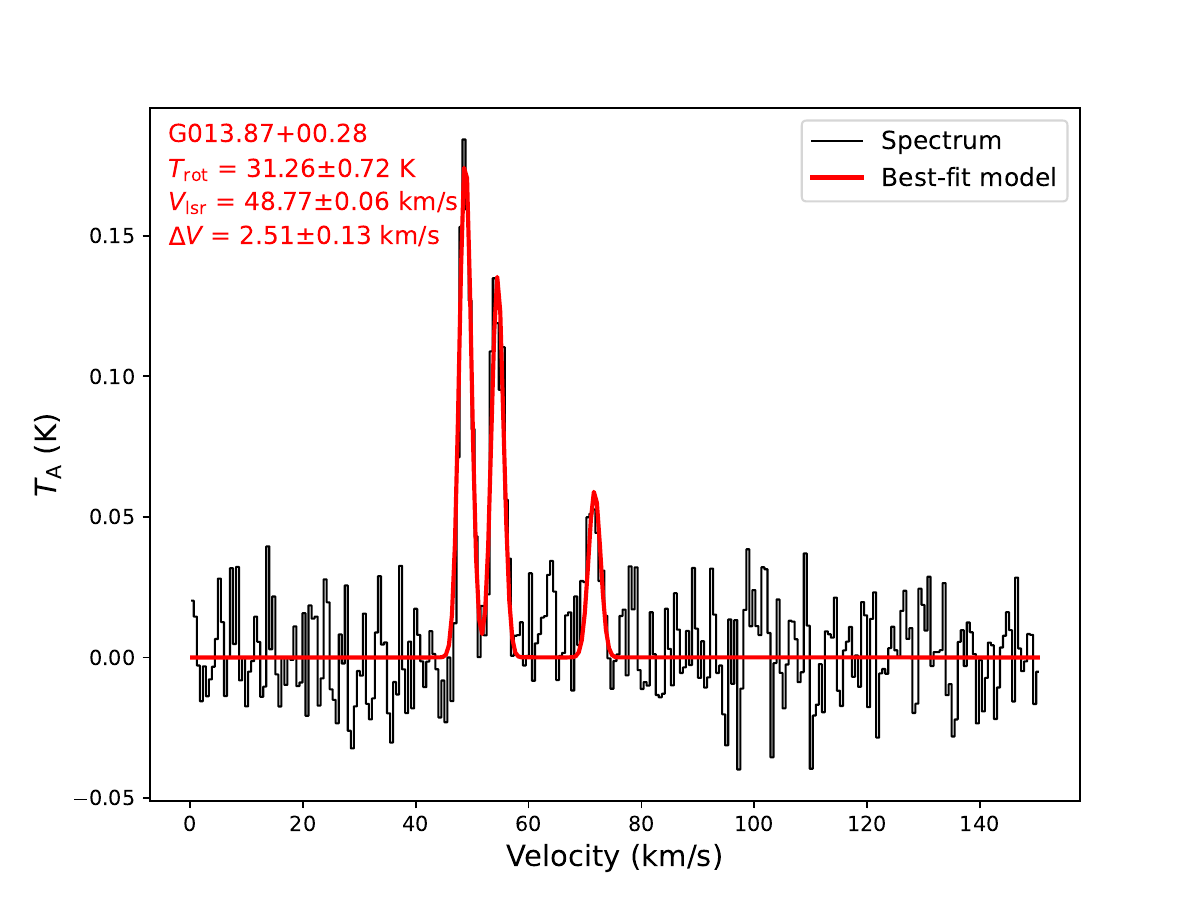}{0.31\textwidth}{}
\fig{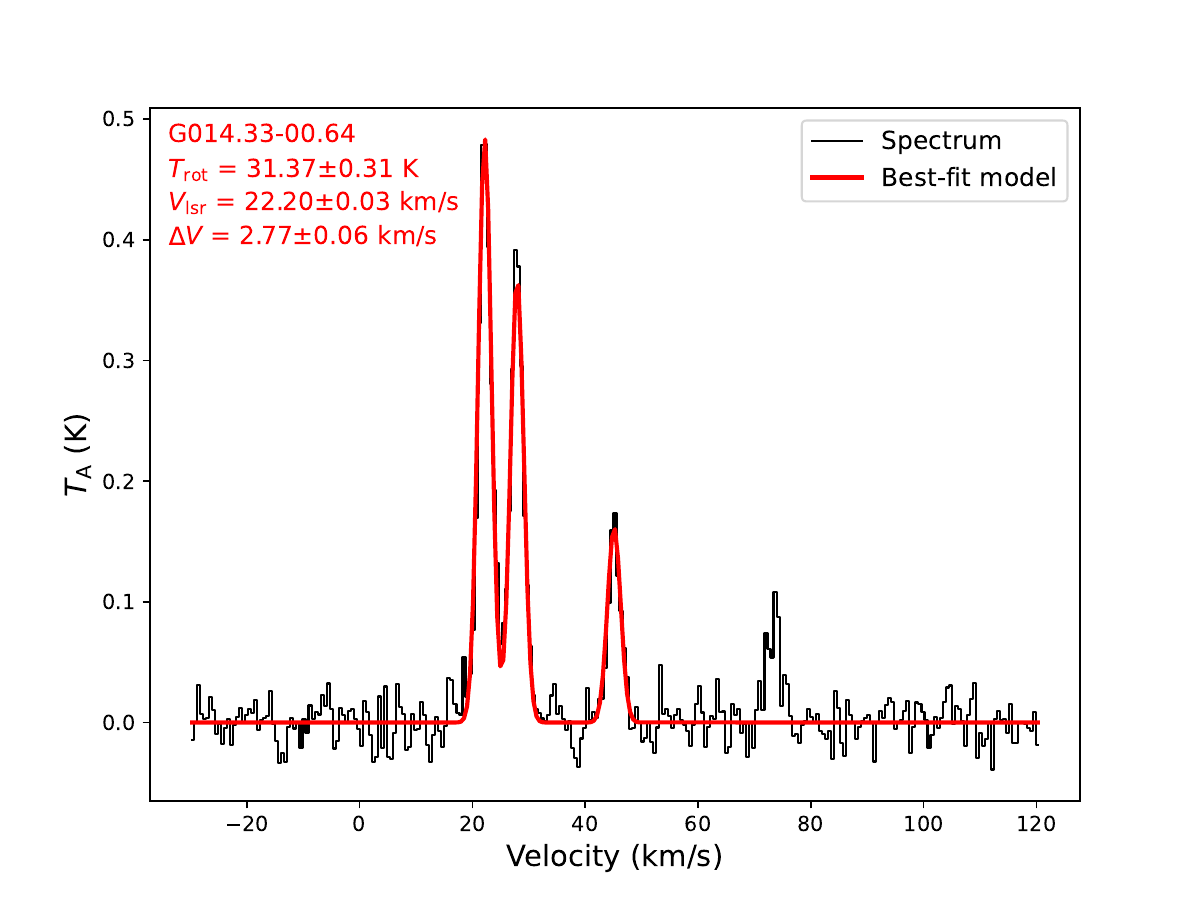}{0.31\textwidth}{}
\fig{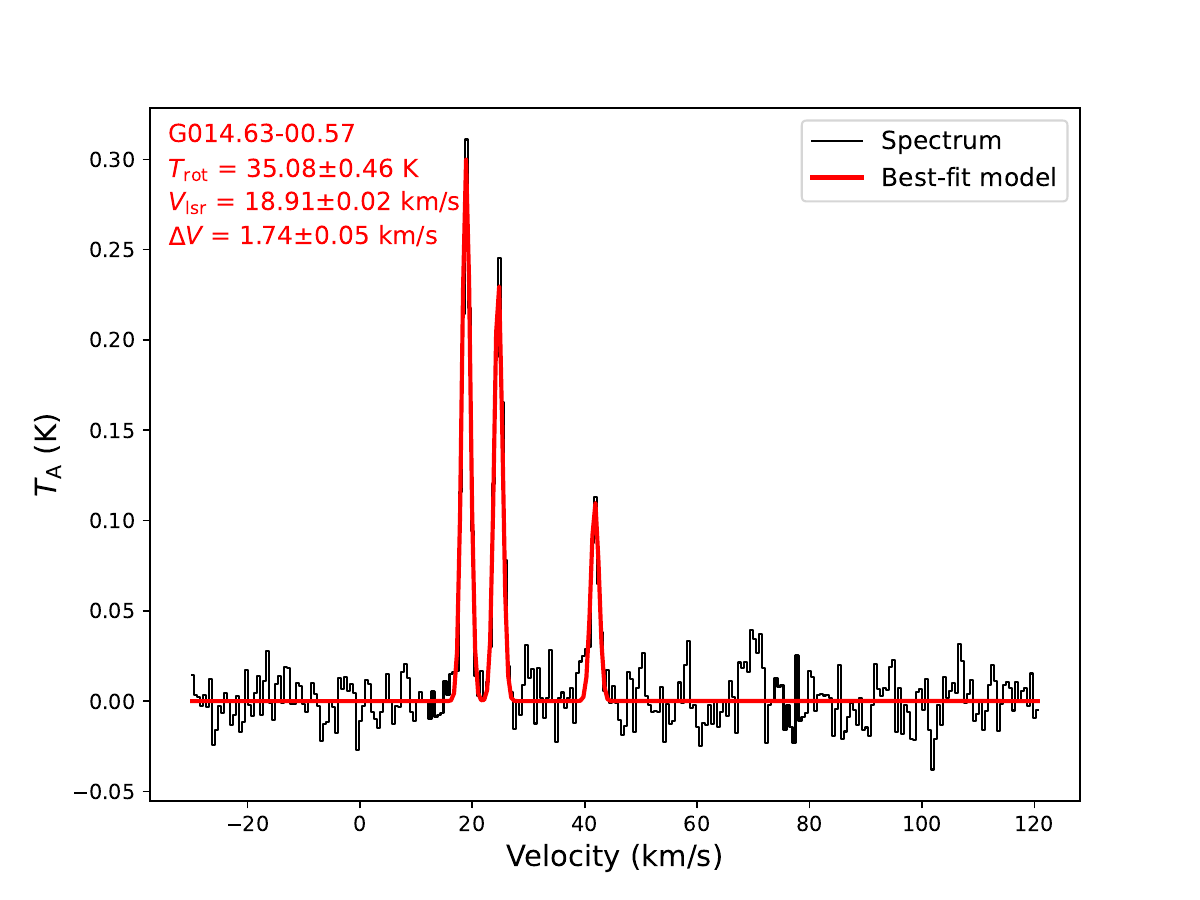}{0.31\textwidth}{}
}
\caption{CH$_3$CCH spectra toward the sample sources (1/6). The observed spectra of CH$_3$CCH 5-4 are in black, while the best fit is in red.}
\label{appendix: CH3CCH_fitting_1}
\end{figure}

\begin{figure}[ht!]
\centering
\gridline{
\fig{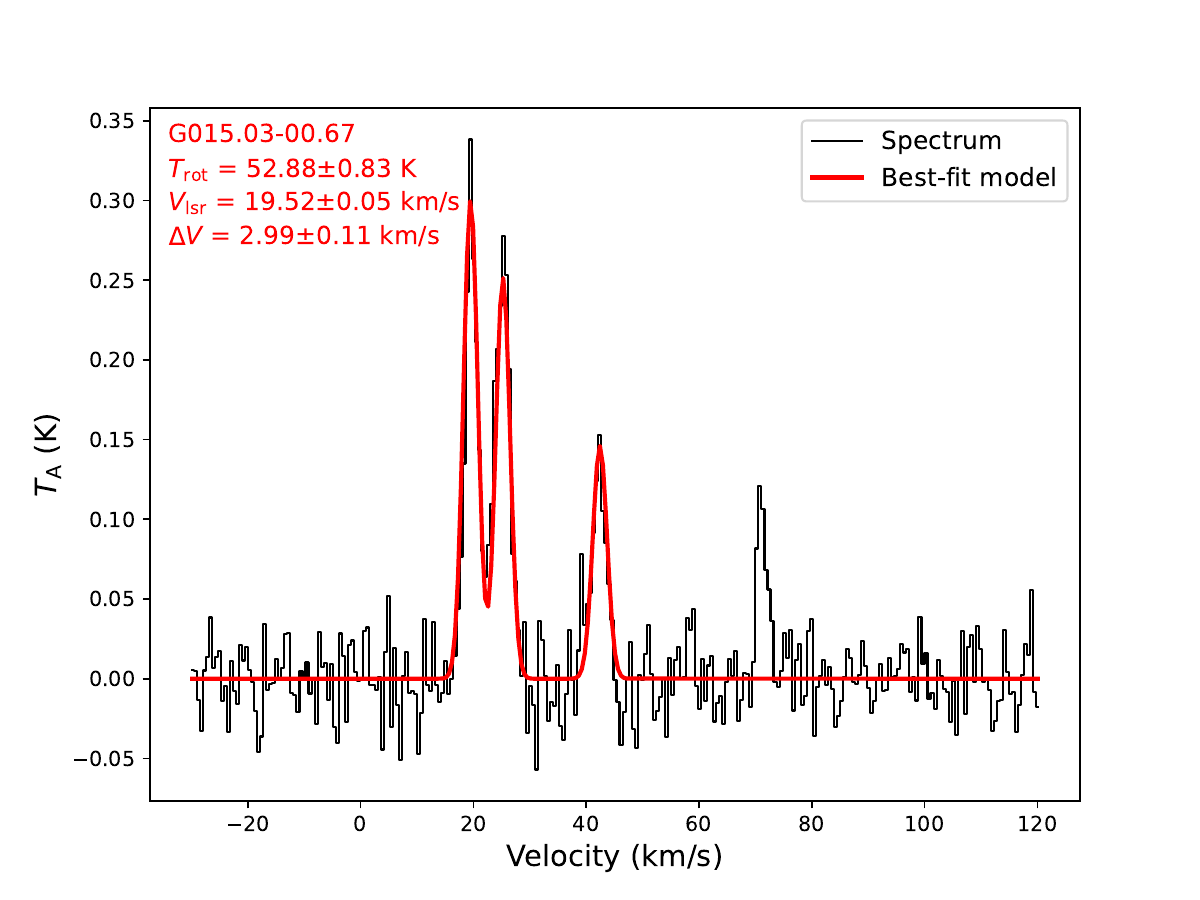}{0.31\textwidth}{}
\fig{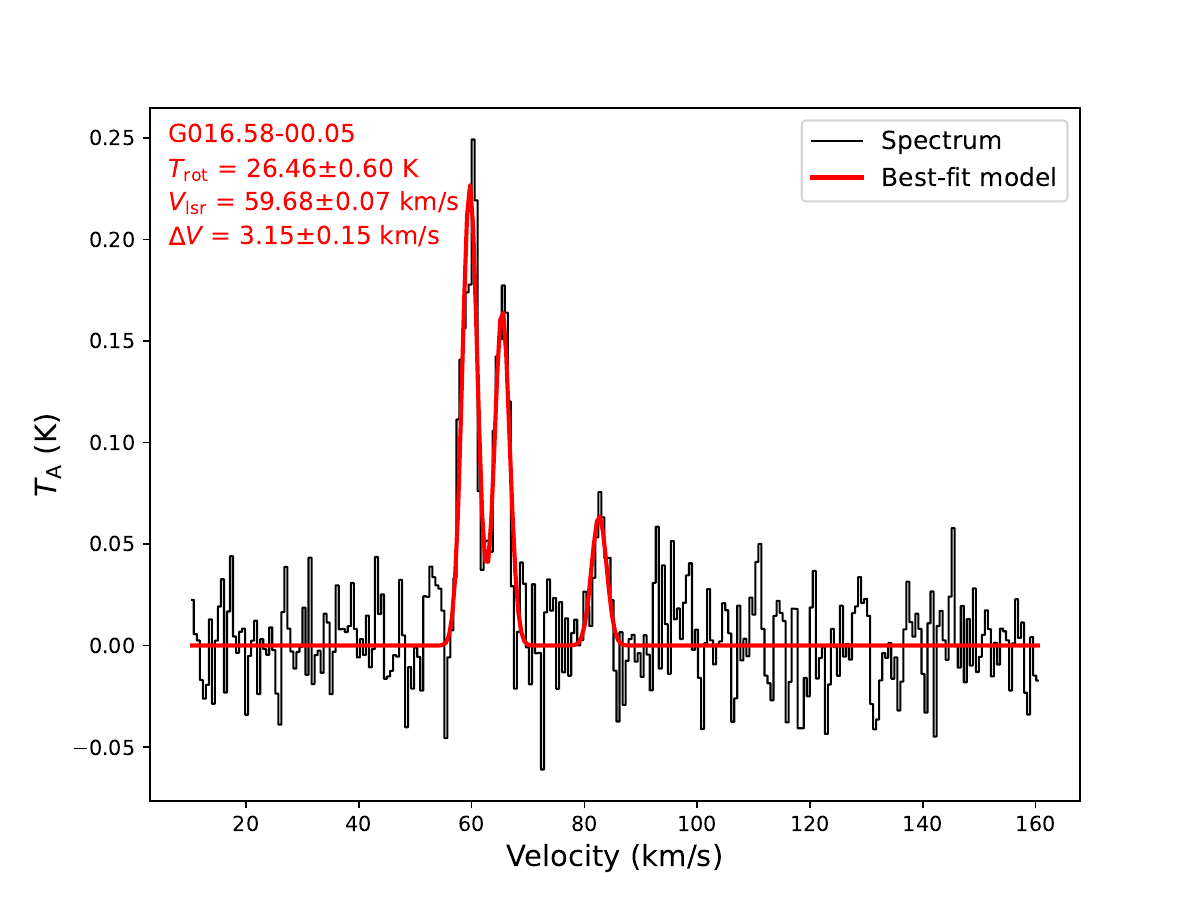}{0.31\textwidth}{}
\fig{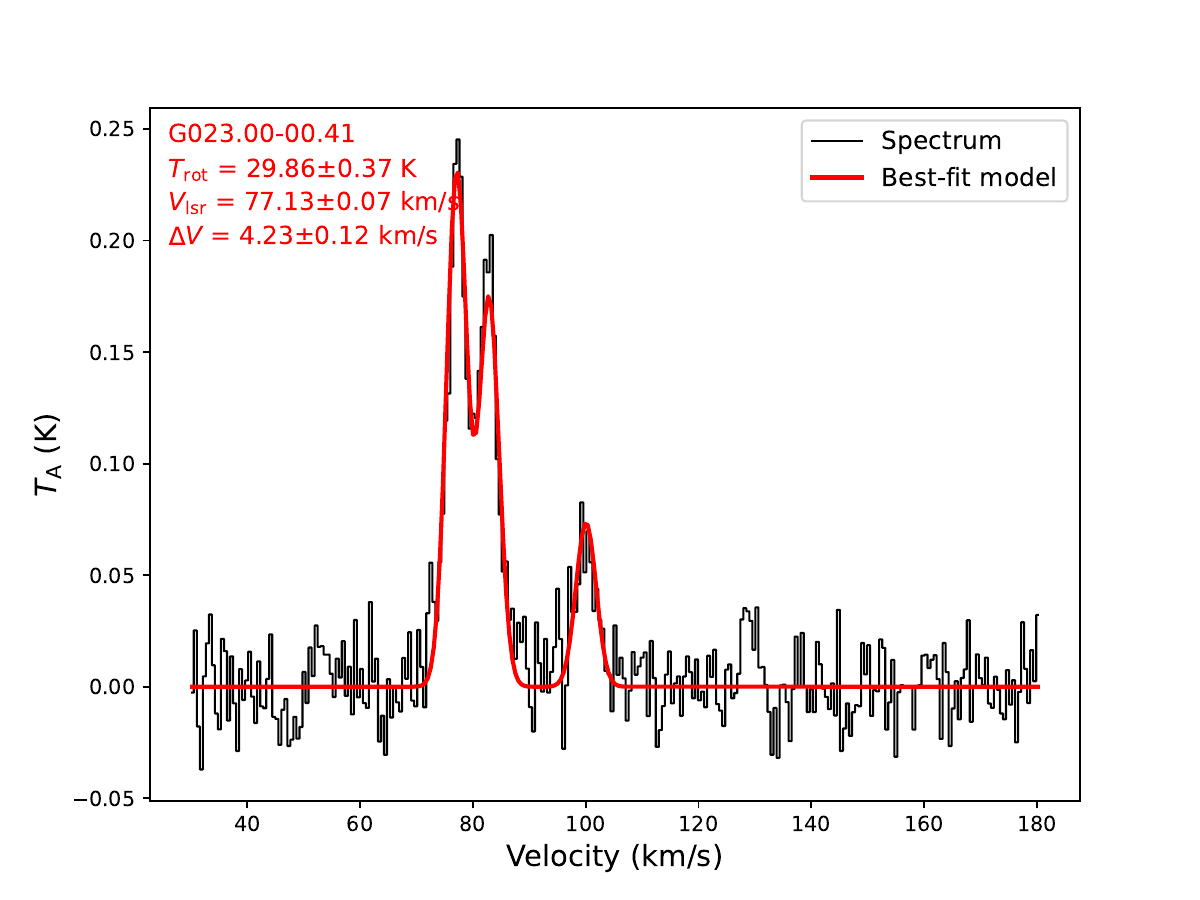}{0.31\textwidth}{}
}
\gridline{
\fig{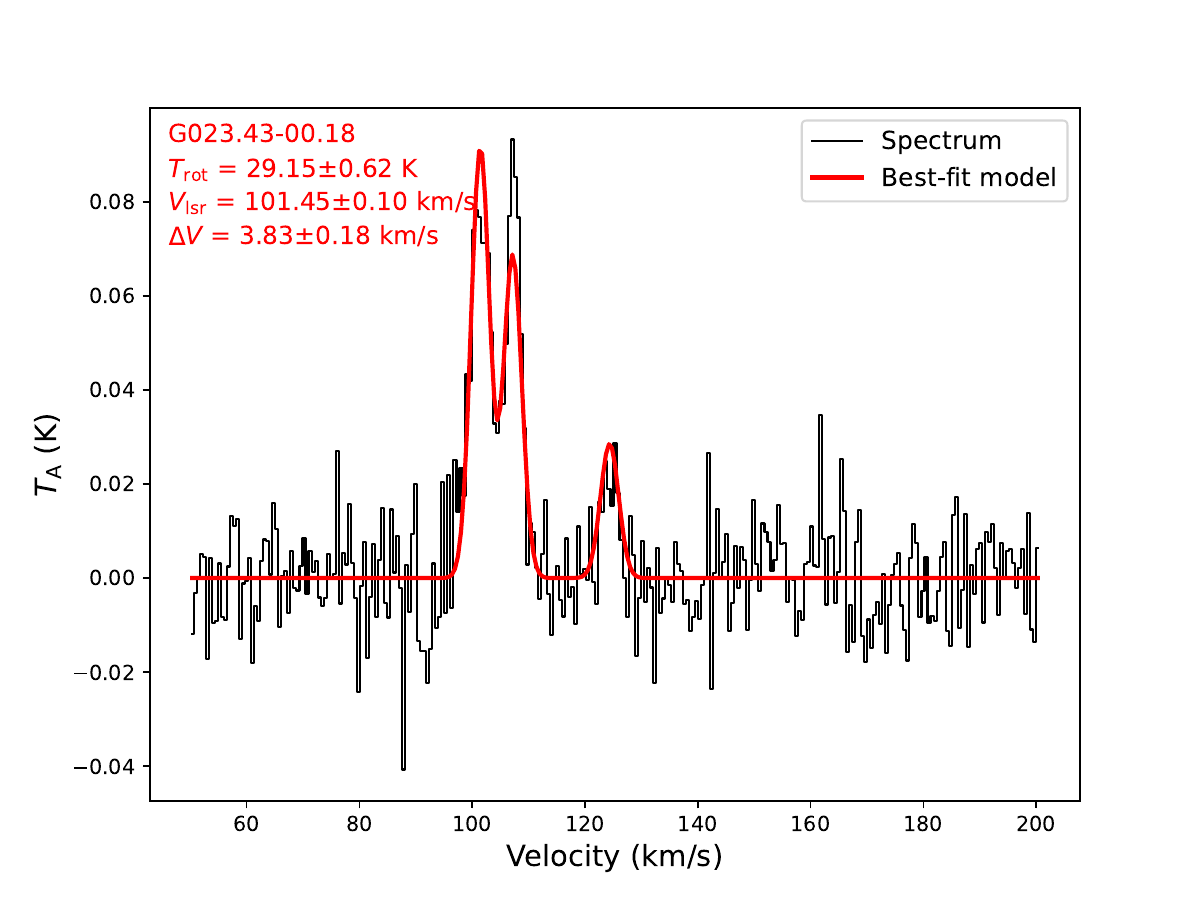}{0.31\textwidth}{}
\fig{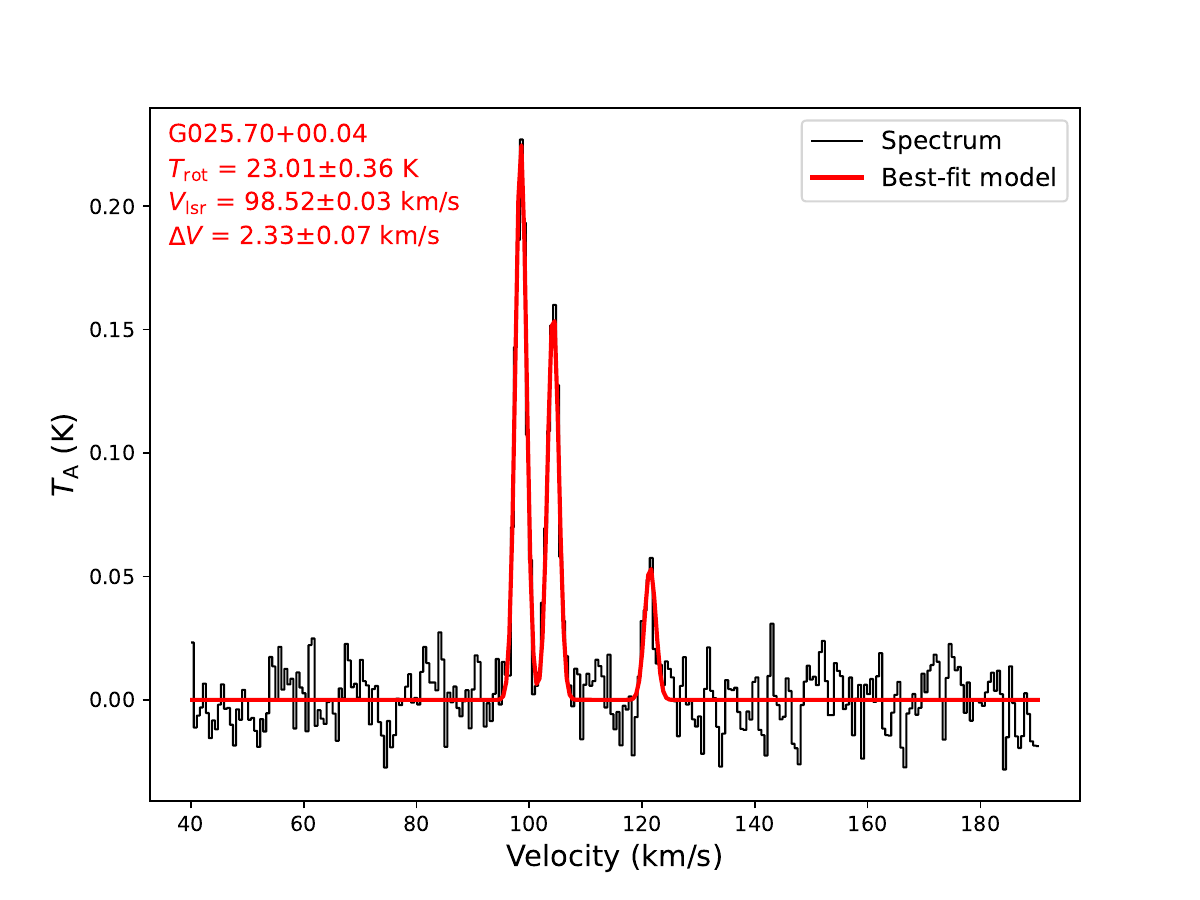}{0.31\textwidth}{}
\fig{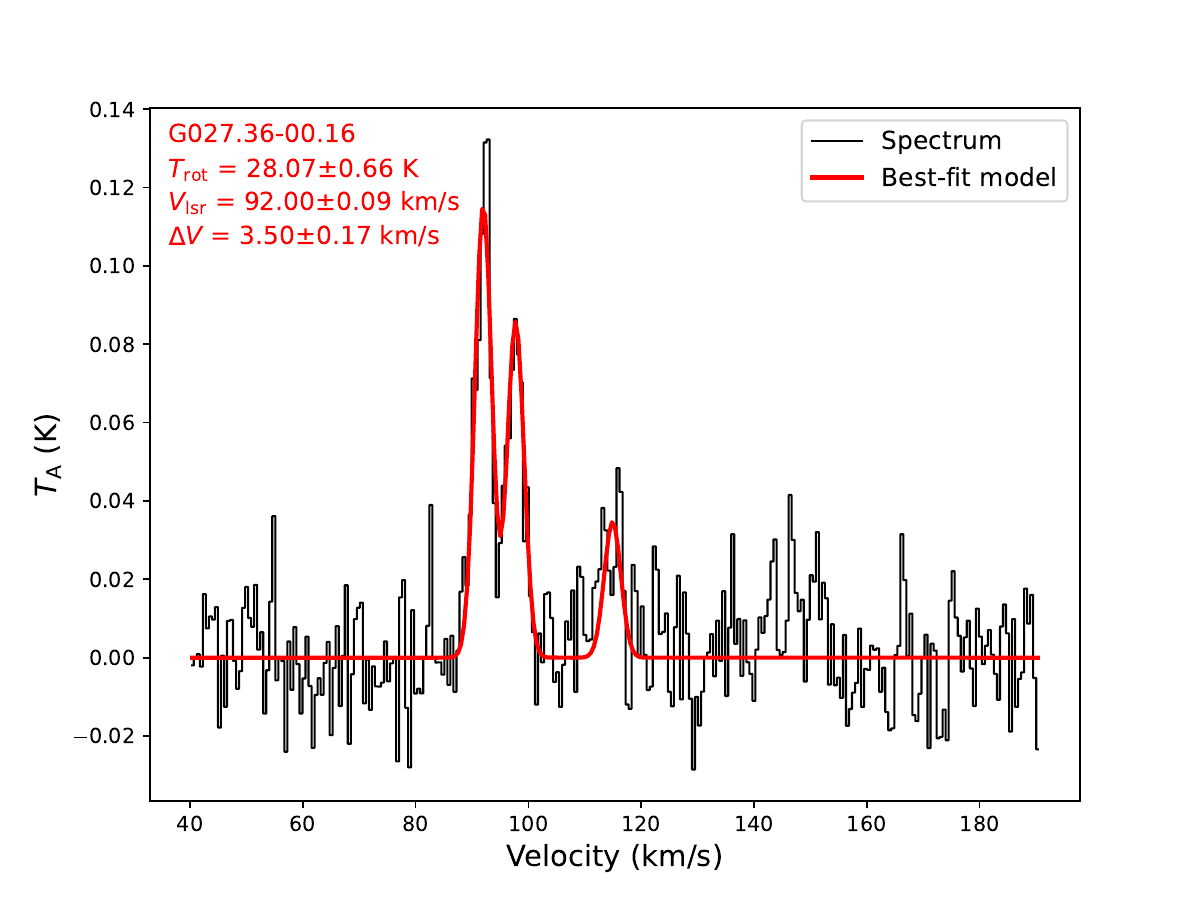}{0.31\textwidth}{}
}
\gridline{
\fig{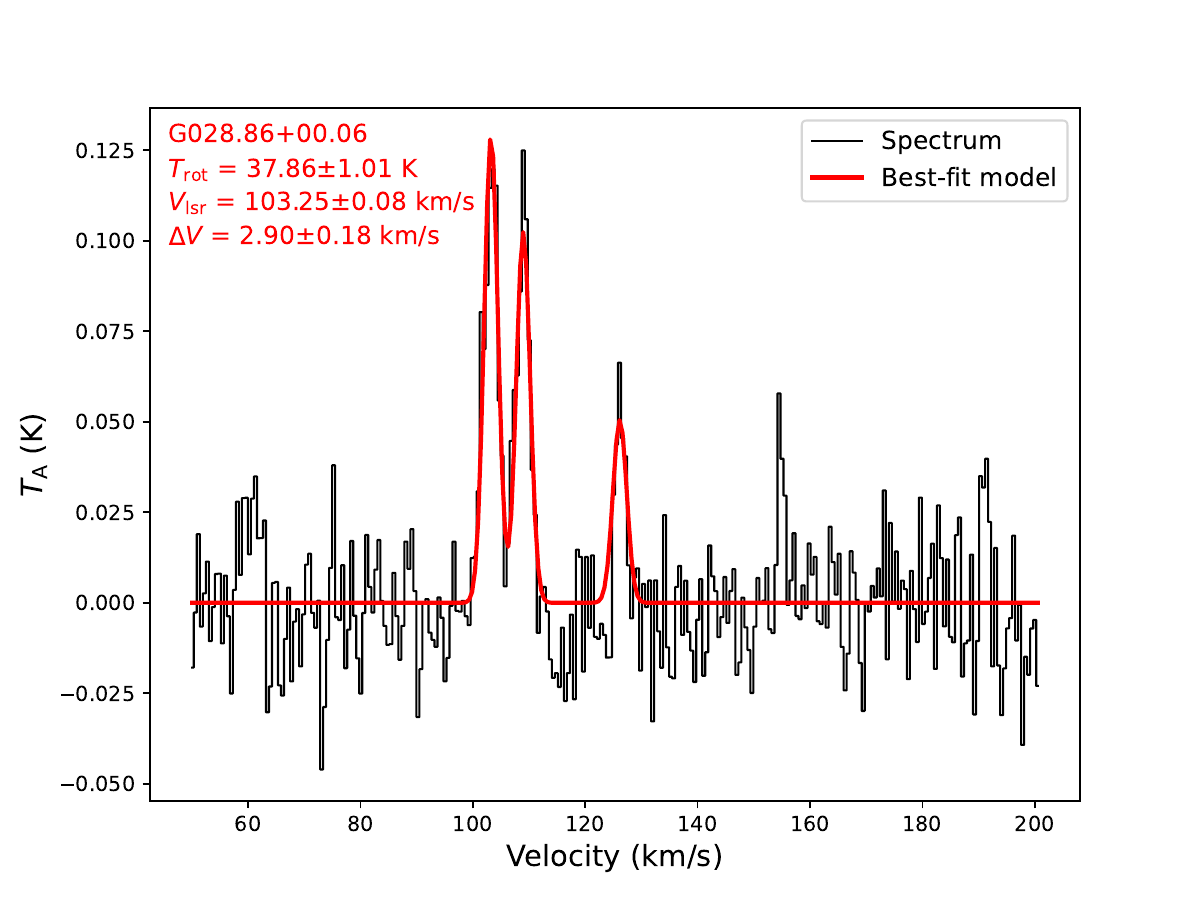}{0.31\textwidth}{}
\fig{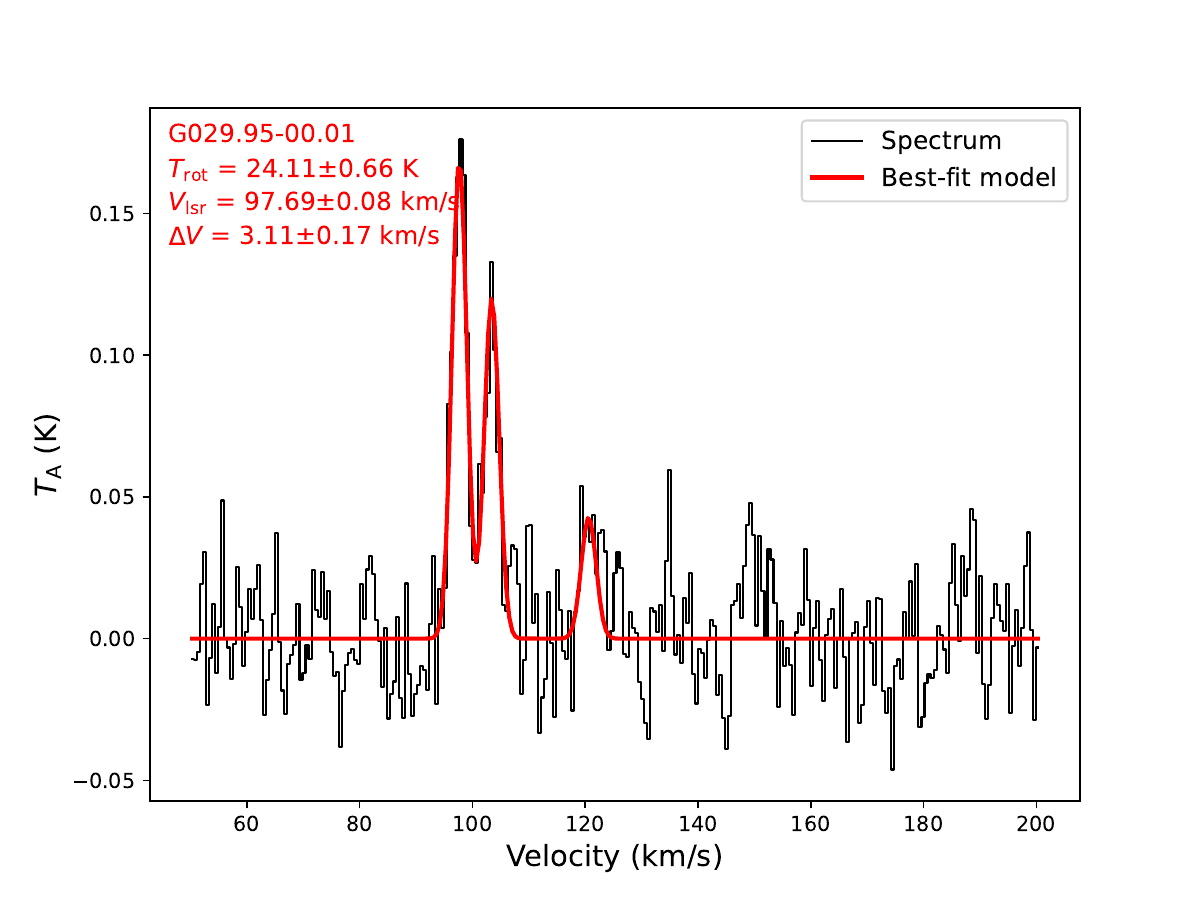}{0.31\textwidth}{}
\fig{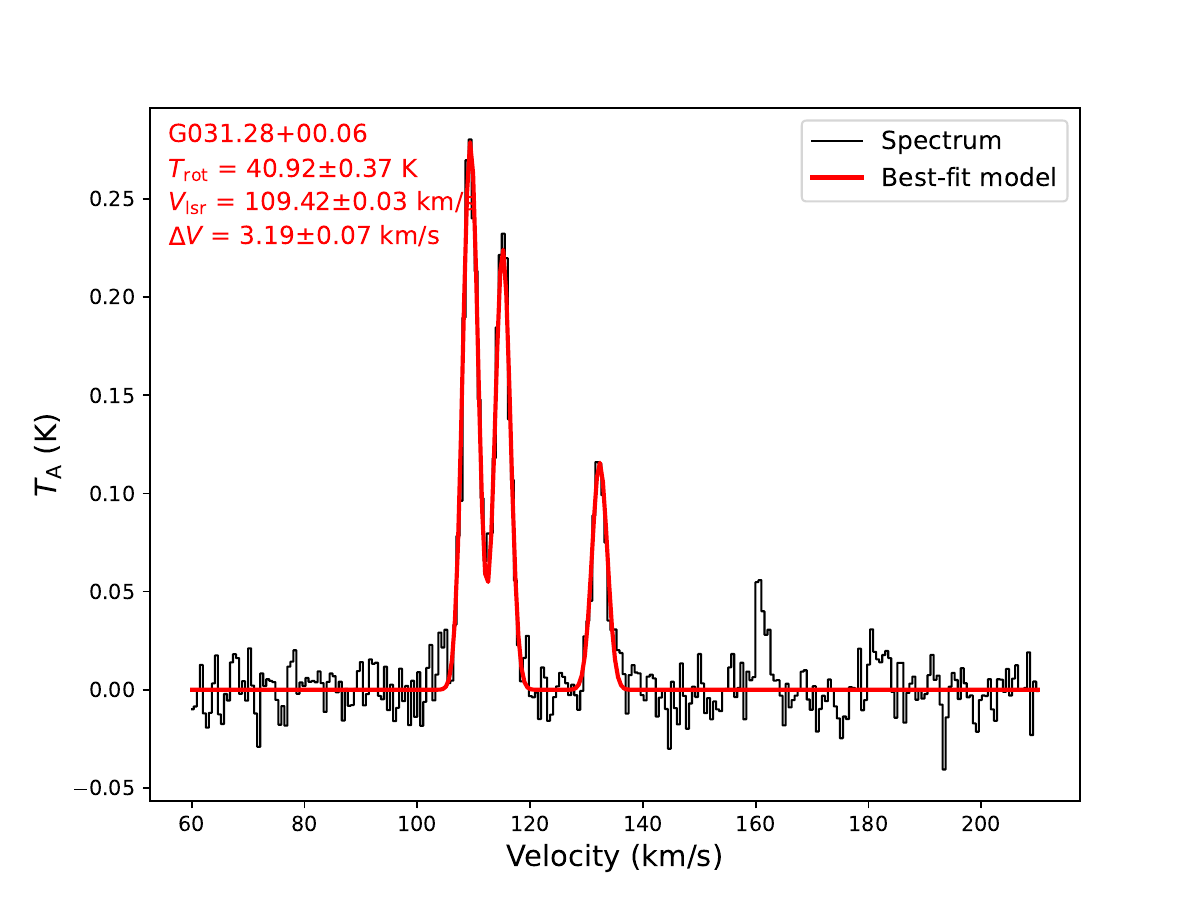}{0.31\textwidth}{}
}
\gridline{
\fig{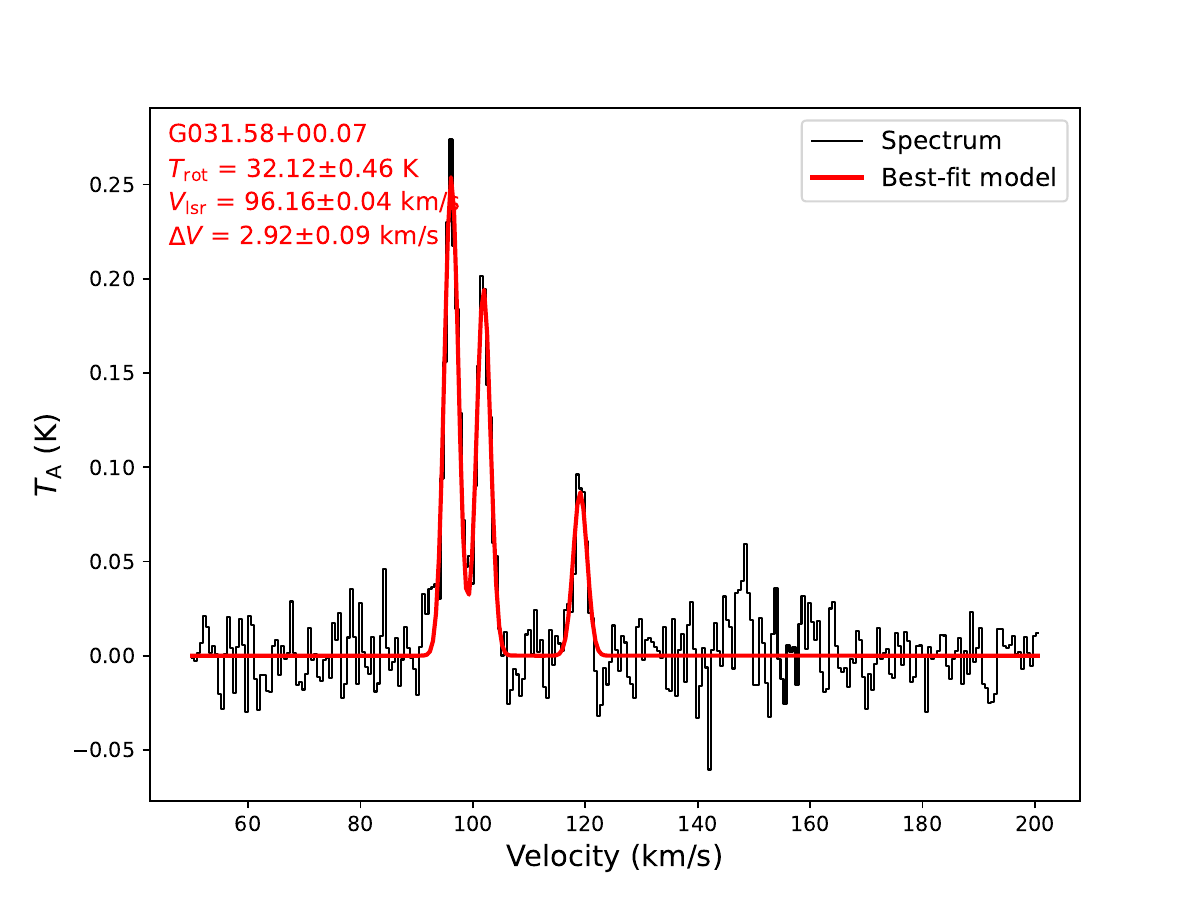}{0.31\textwidth}{}
\fig{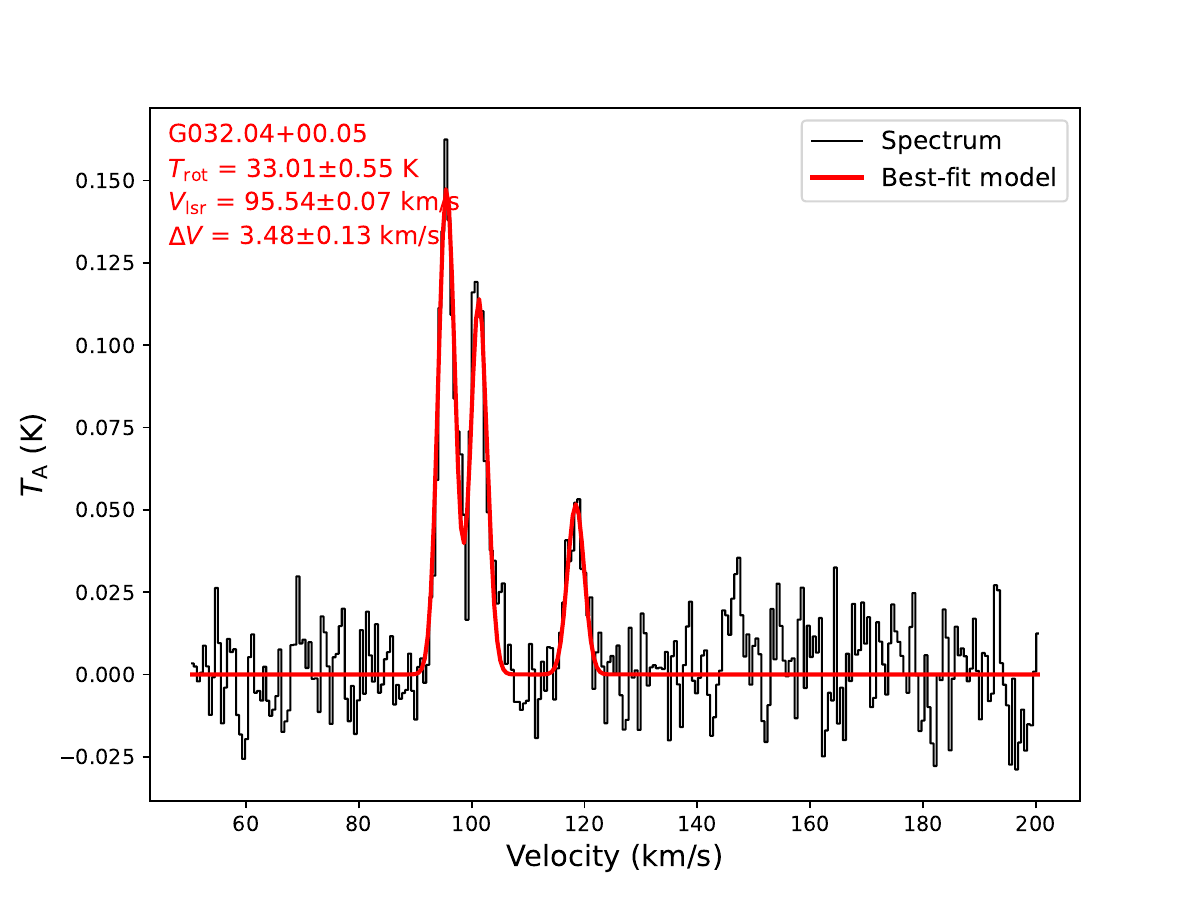}{0.31\textwidth}{}
\fig{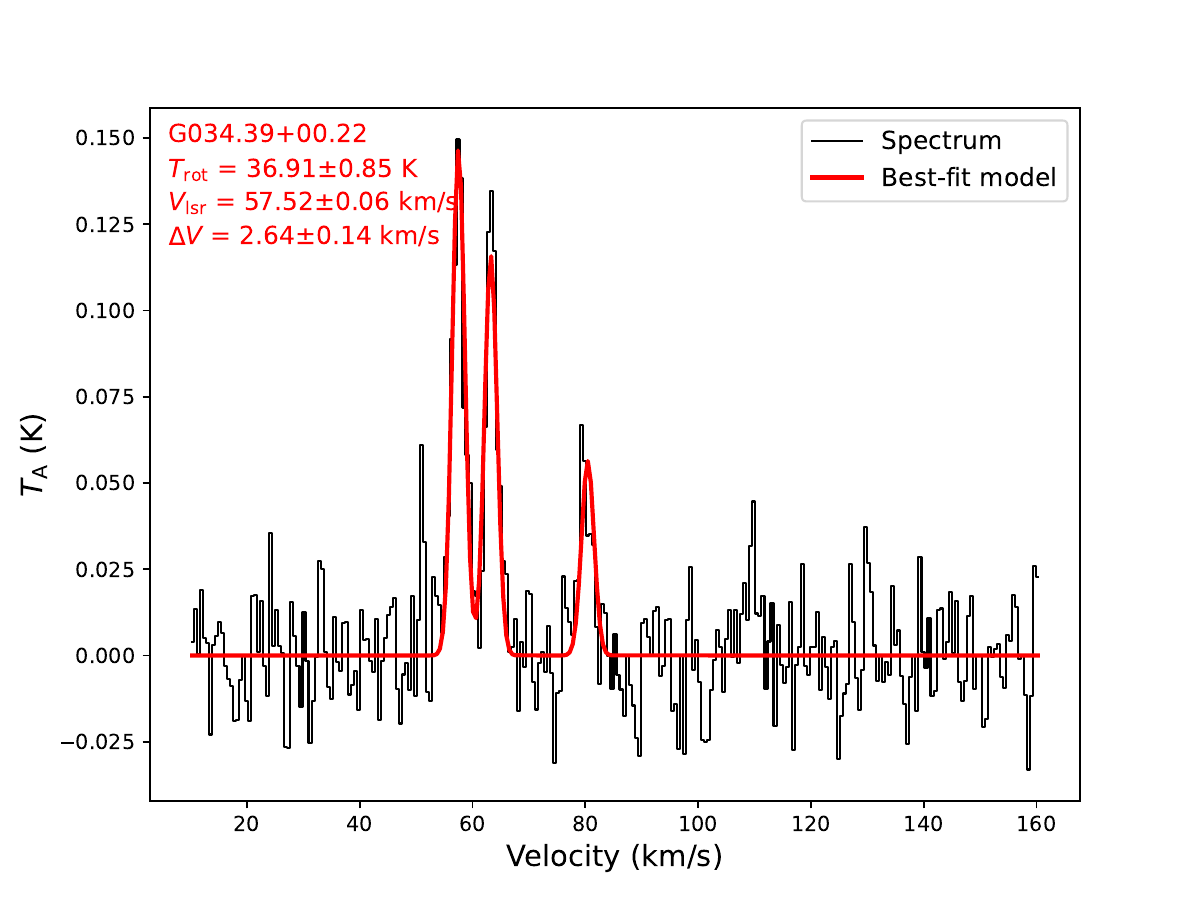}{0.31\textwidth}{}
}
\caption{CH$_3$CCH spectra toward the sample sources (2/6). The observed spectra of CH$_3$CCH 5-4 are in black, while the best fit is in red.}
\label{appendix: CH3CCH_fitting_2}
\end{figure}

\begin{figure}[ht!]
\centering
\gridline{
\fig{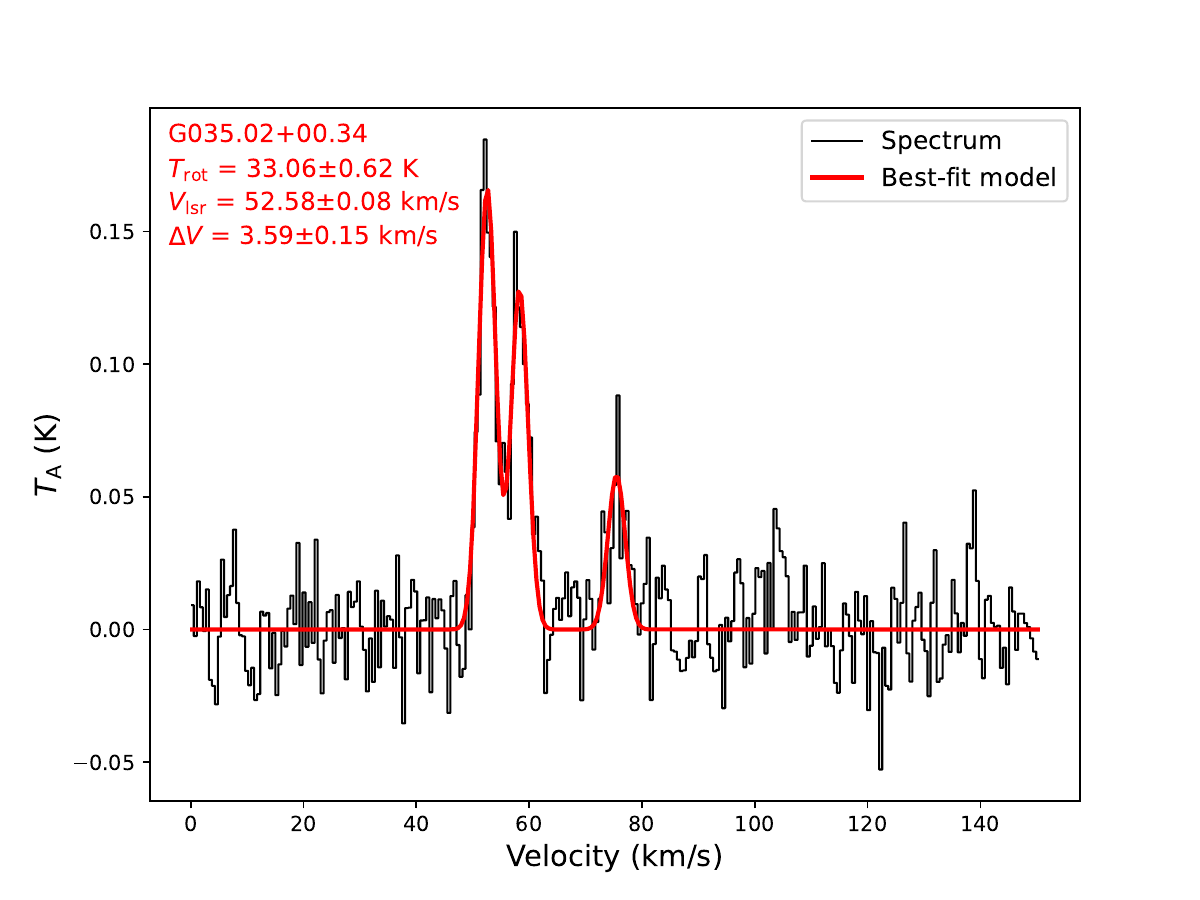}{0.31\textwidth}{}
\fig{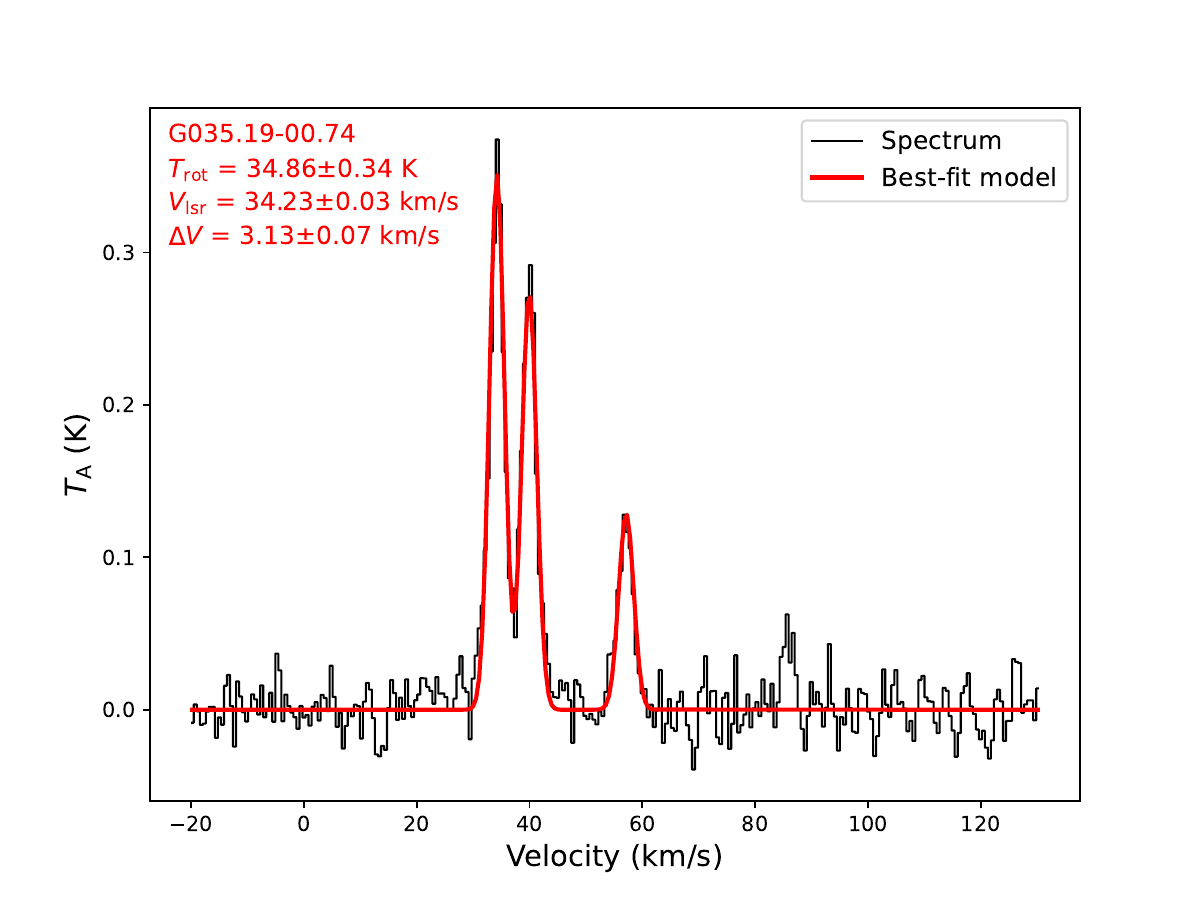}{0.31\textwidth}{}
\fig{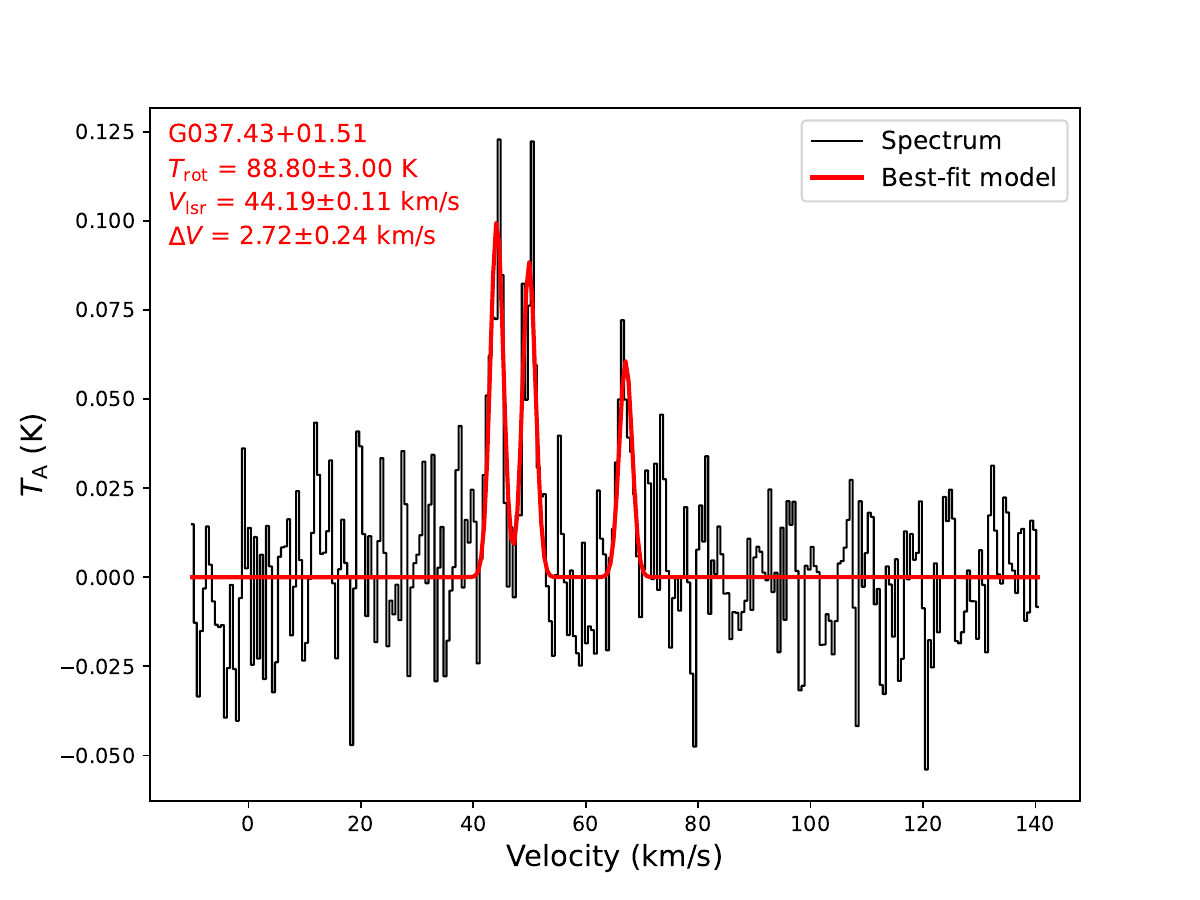}{0.31\textwidth}{}
}
\gridline{
\fig{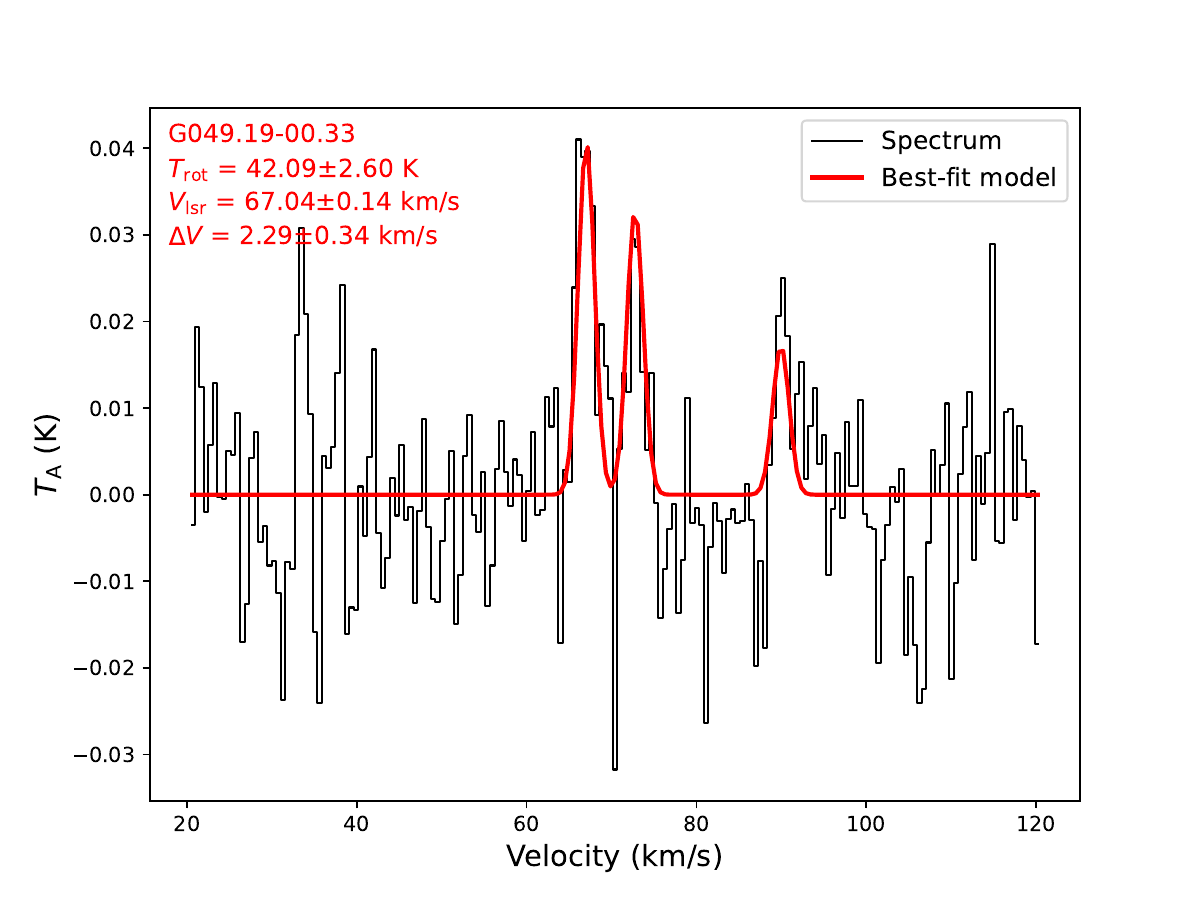}{0.31\textwidth}{}
\fig{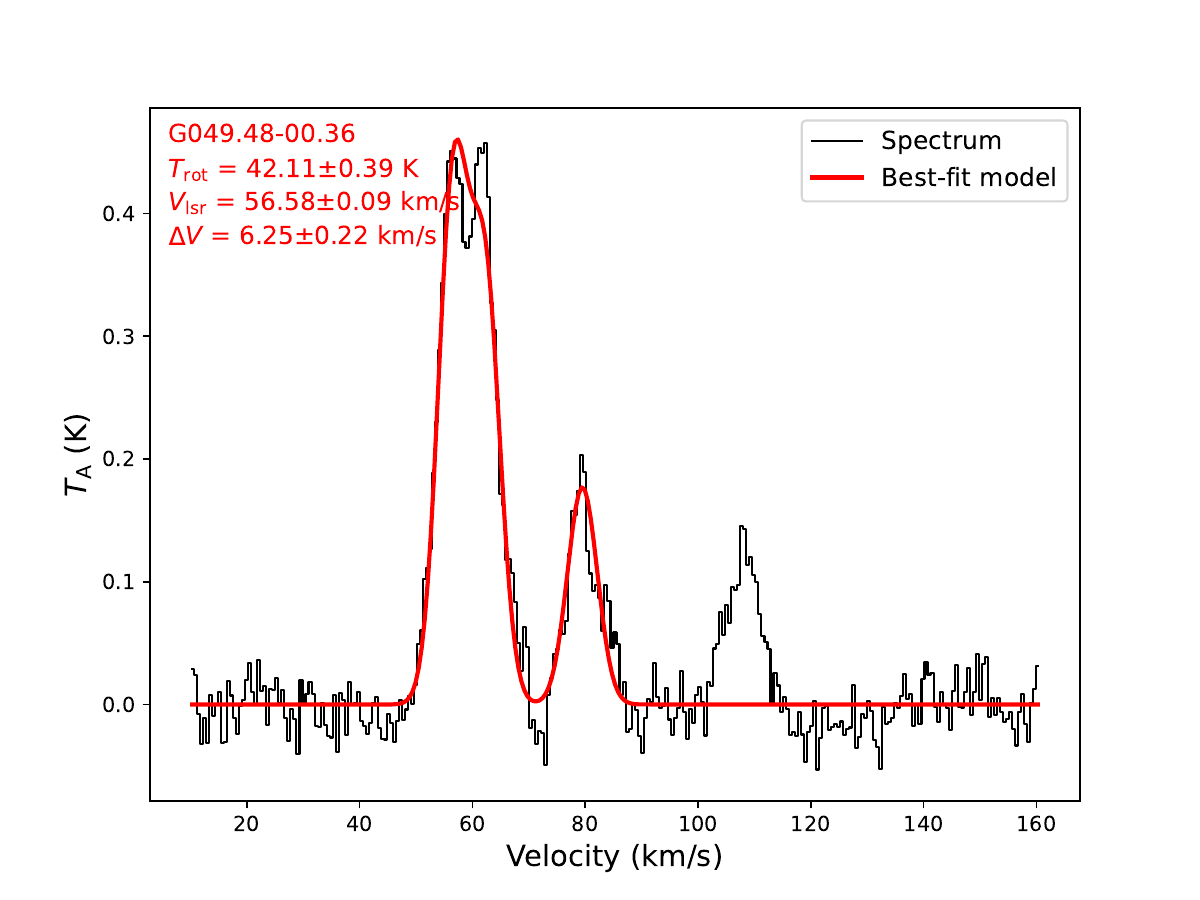}{0.31\textwidth}{}
\fig{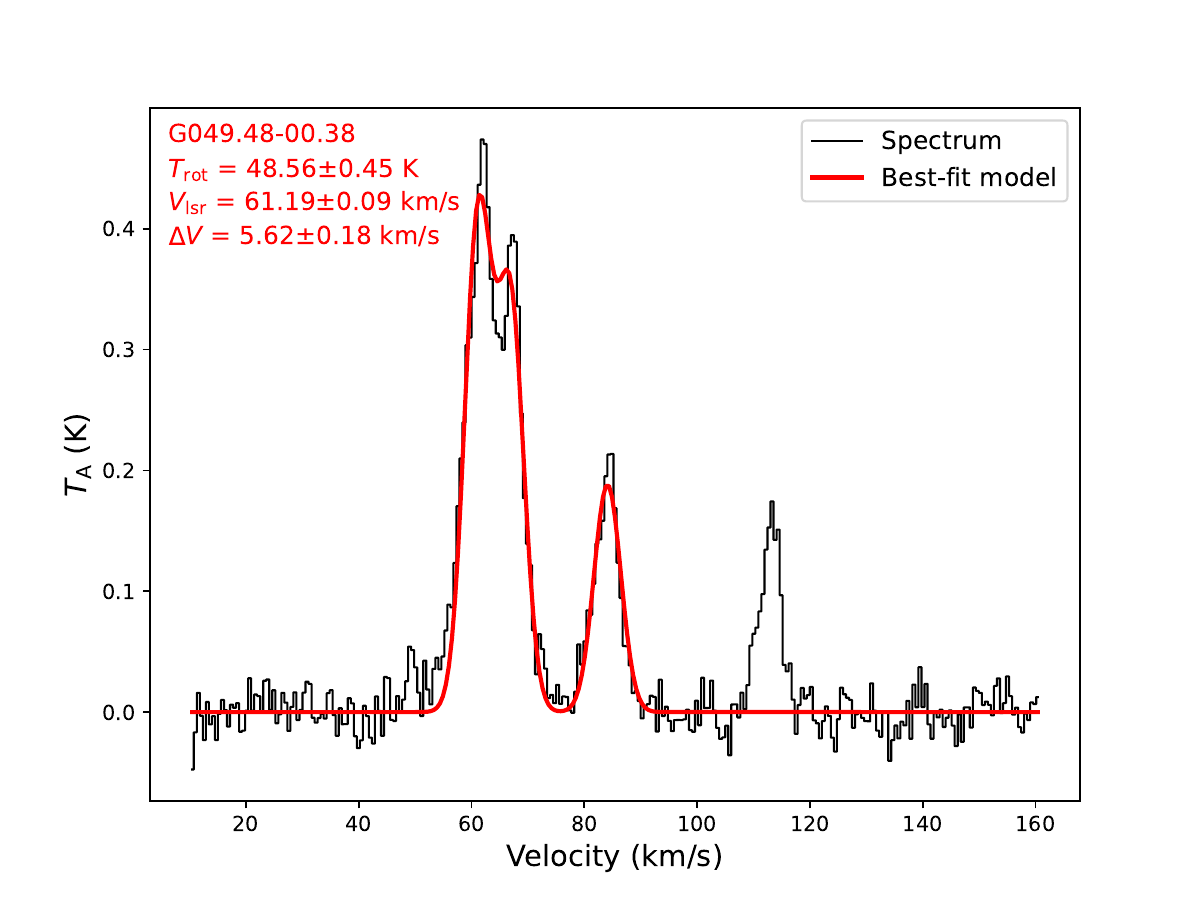}{0.31\textwidth}{}
}
\gridline{
\fig{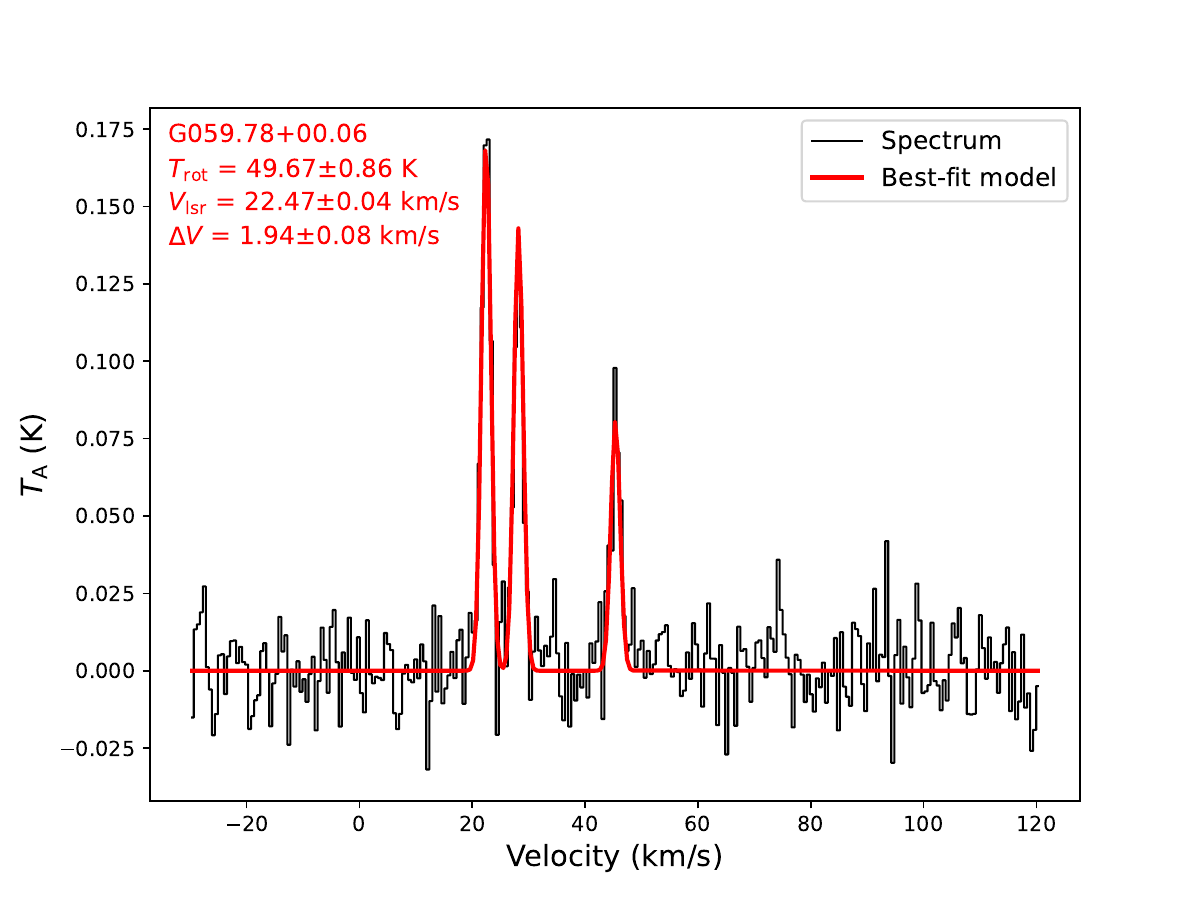}{0.31\textwidth}{}
\fig{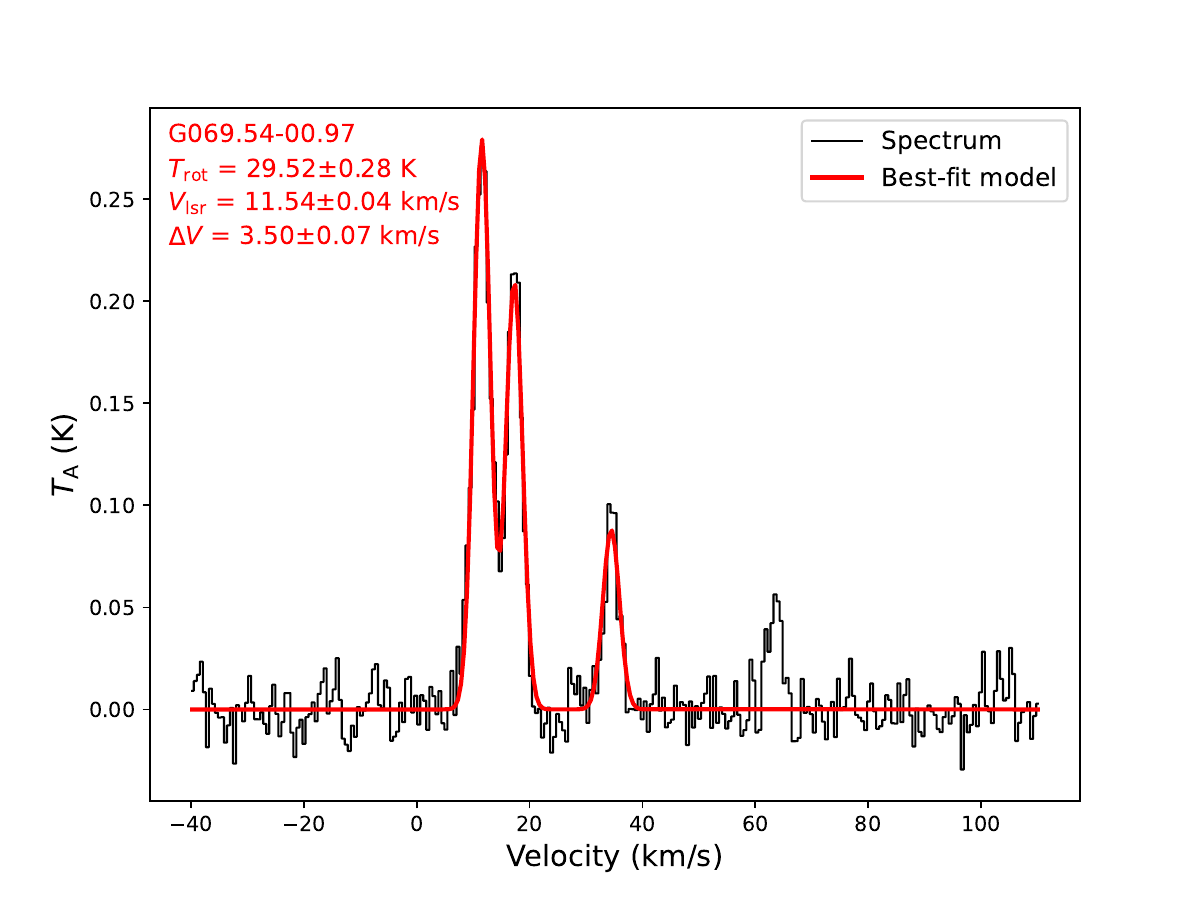}{0.31\textwidth}{}
\fig{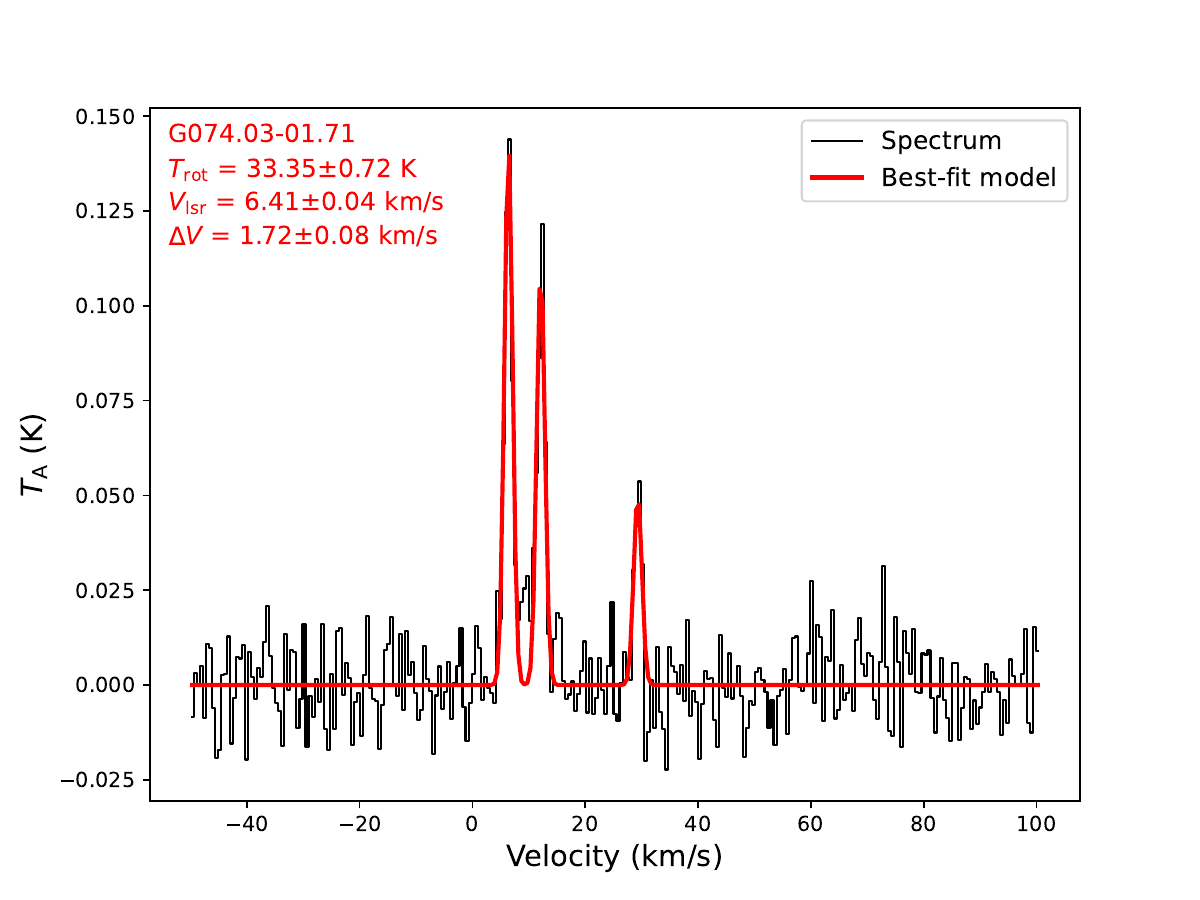}{0.31\textwidth}{}
}
\gridline{
\fig{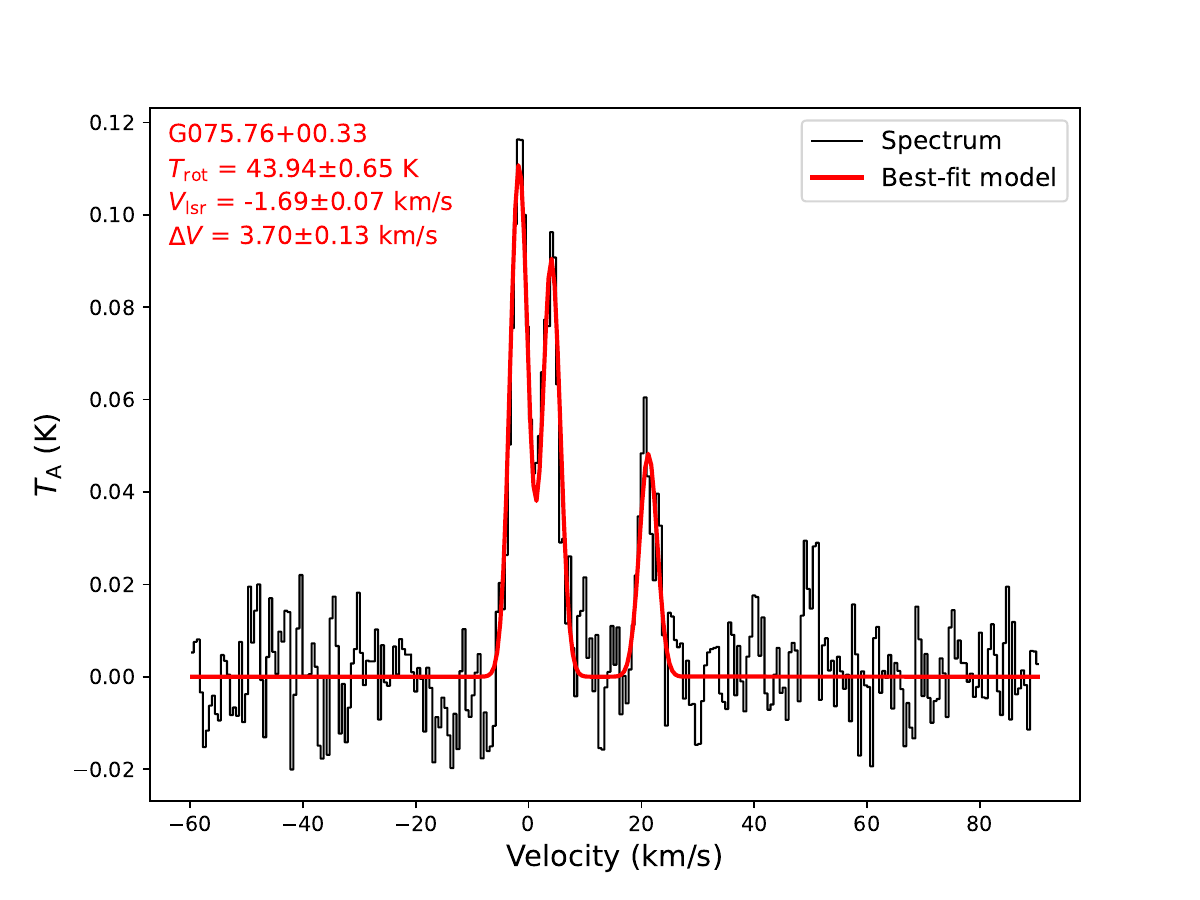}{0.31\textwidth}{}
\fig{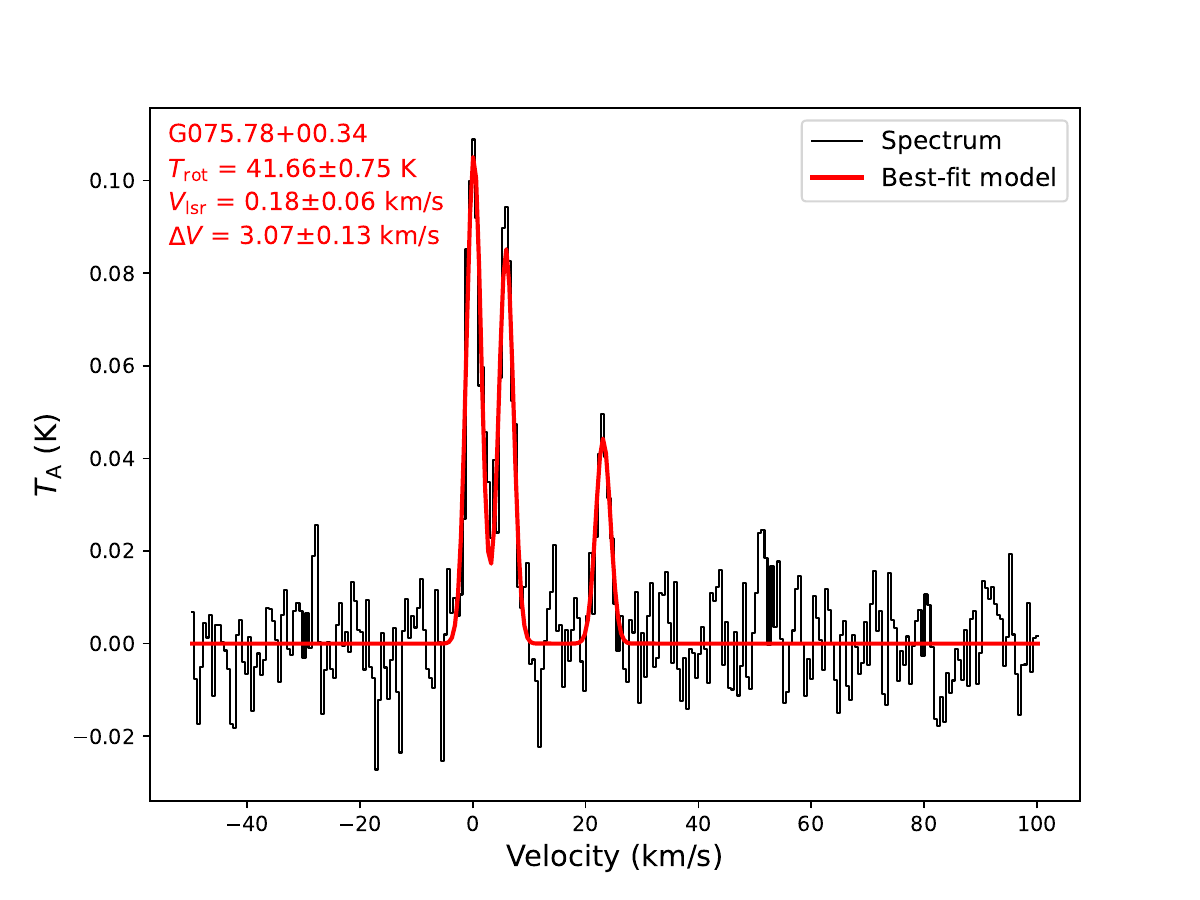}{0.31\textwidth}{}
\fig{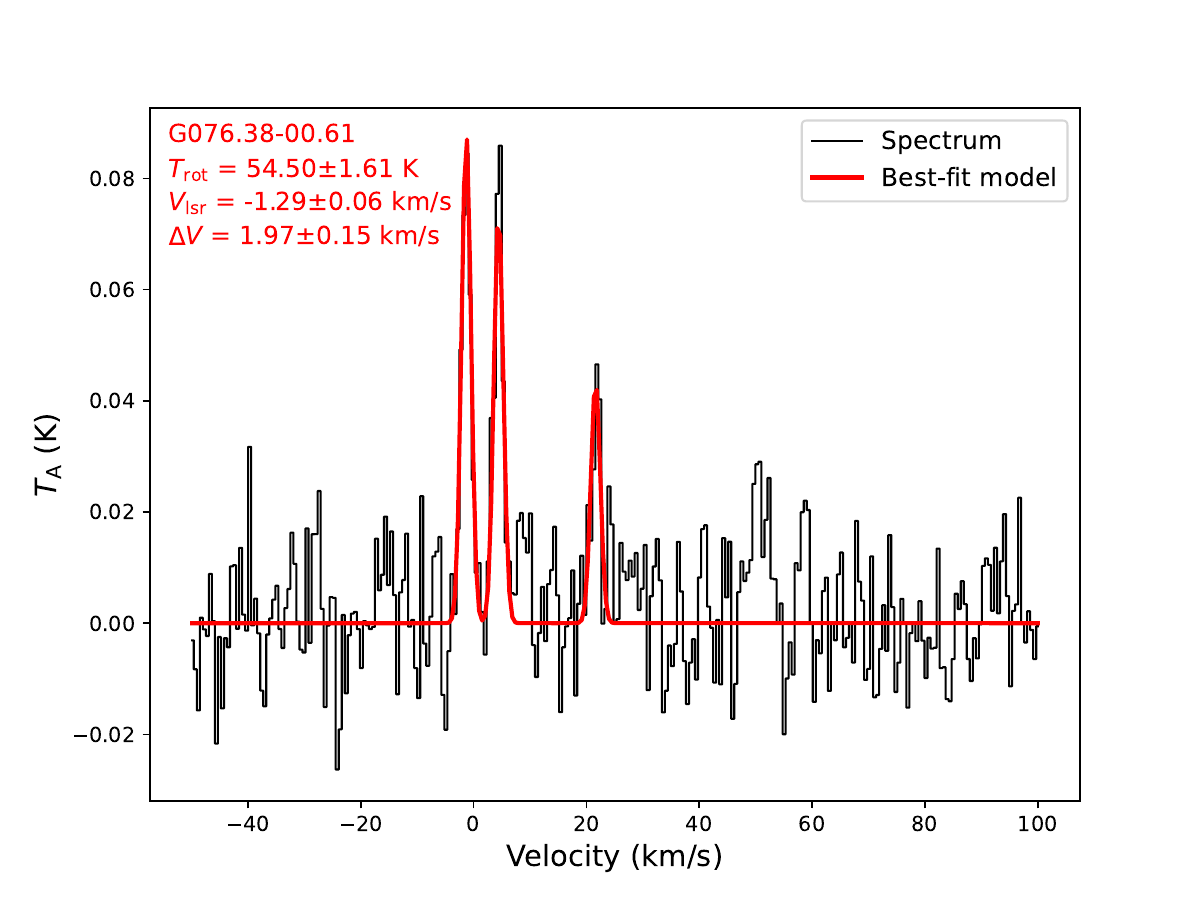}{0.31\textwidth}{}
}
\caption{CH$_3$CCH spectra toward the sample sources (3/6). The observed spectra of CH$_3$CCH 5-4 are in black, while the best fit is in red.}
\label{appendix: CH3CCH_fitting_3}
\end{figure}

\begin{figure}[ht!]
\centering
\gridline{
\fig{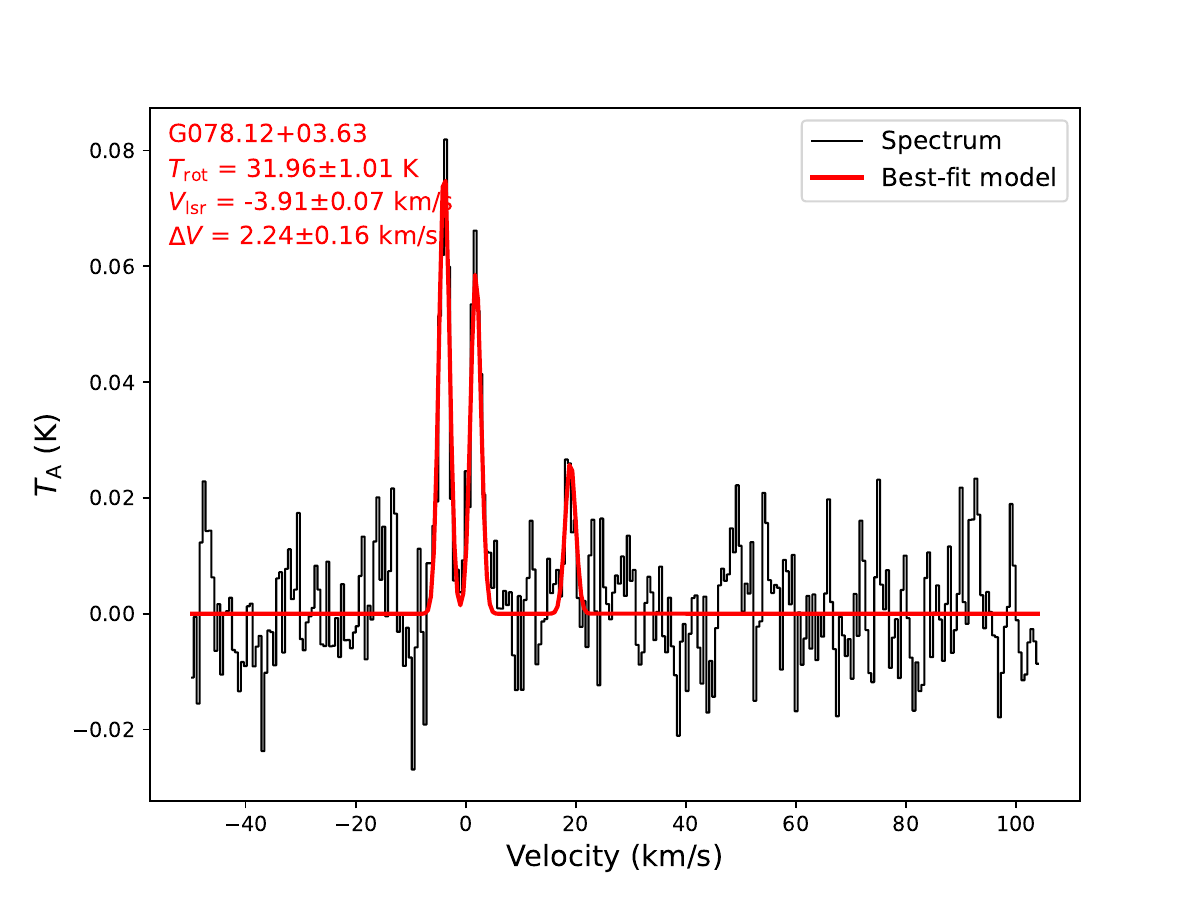}{0.31\textwidth}{}
\fig{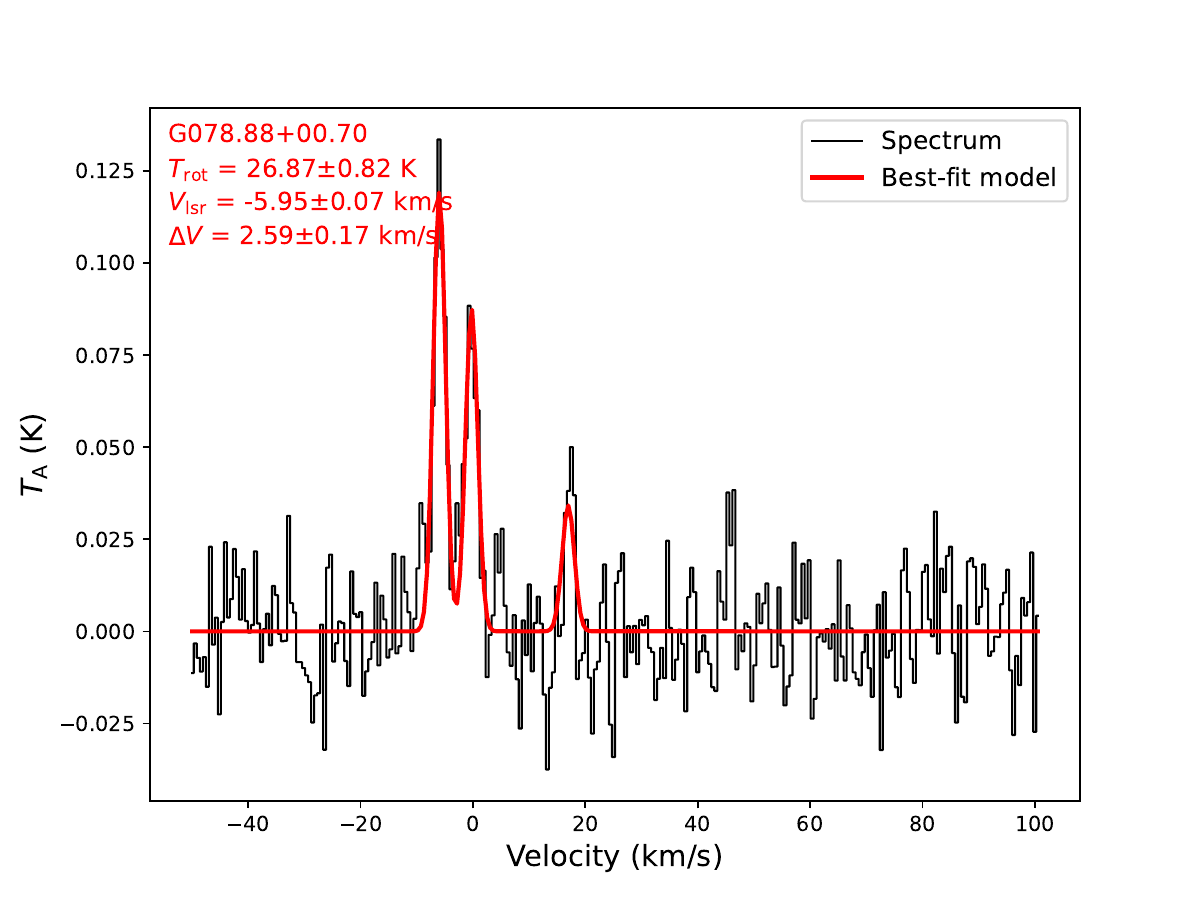}{0.31\textwidth}{}
\fig{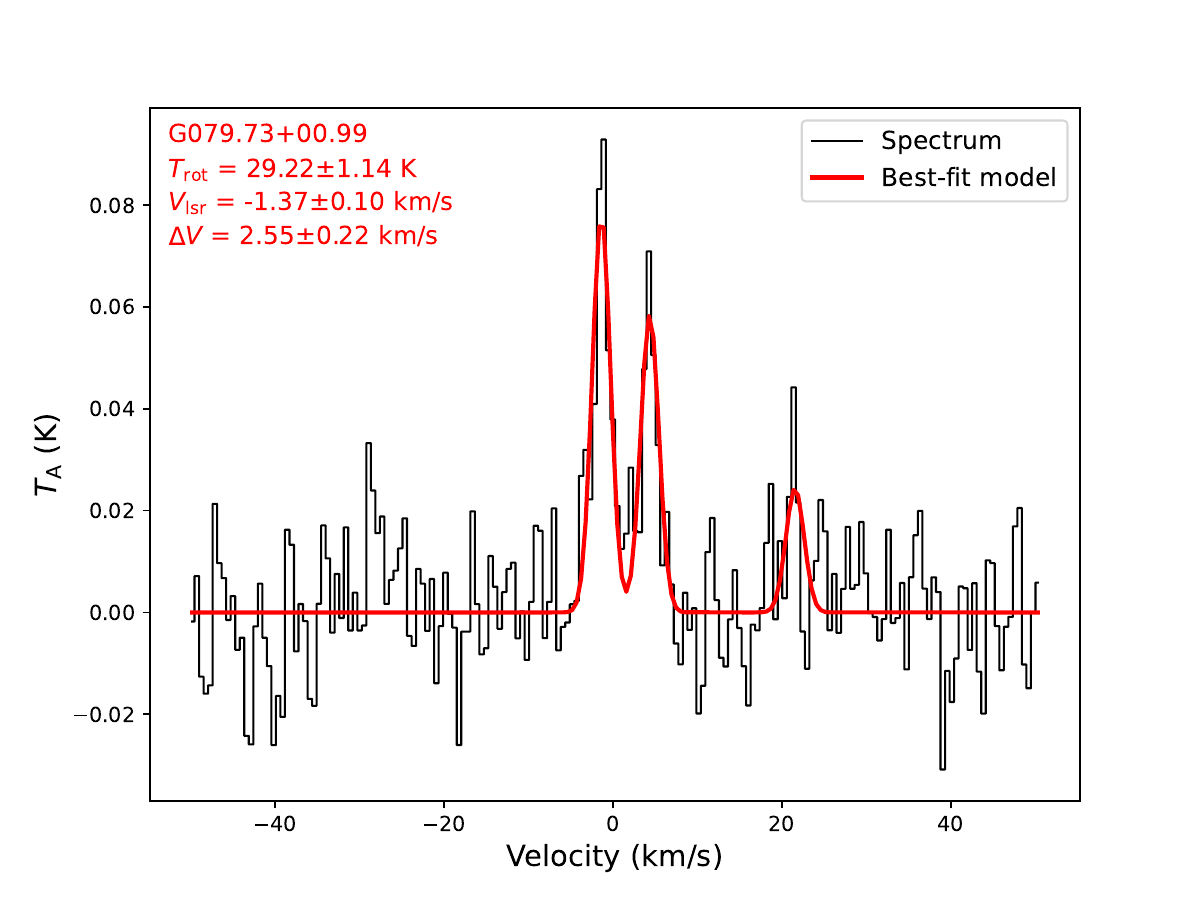}{0.31\textwidth}{}
}
\gridline{
\fig{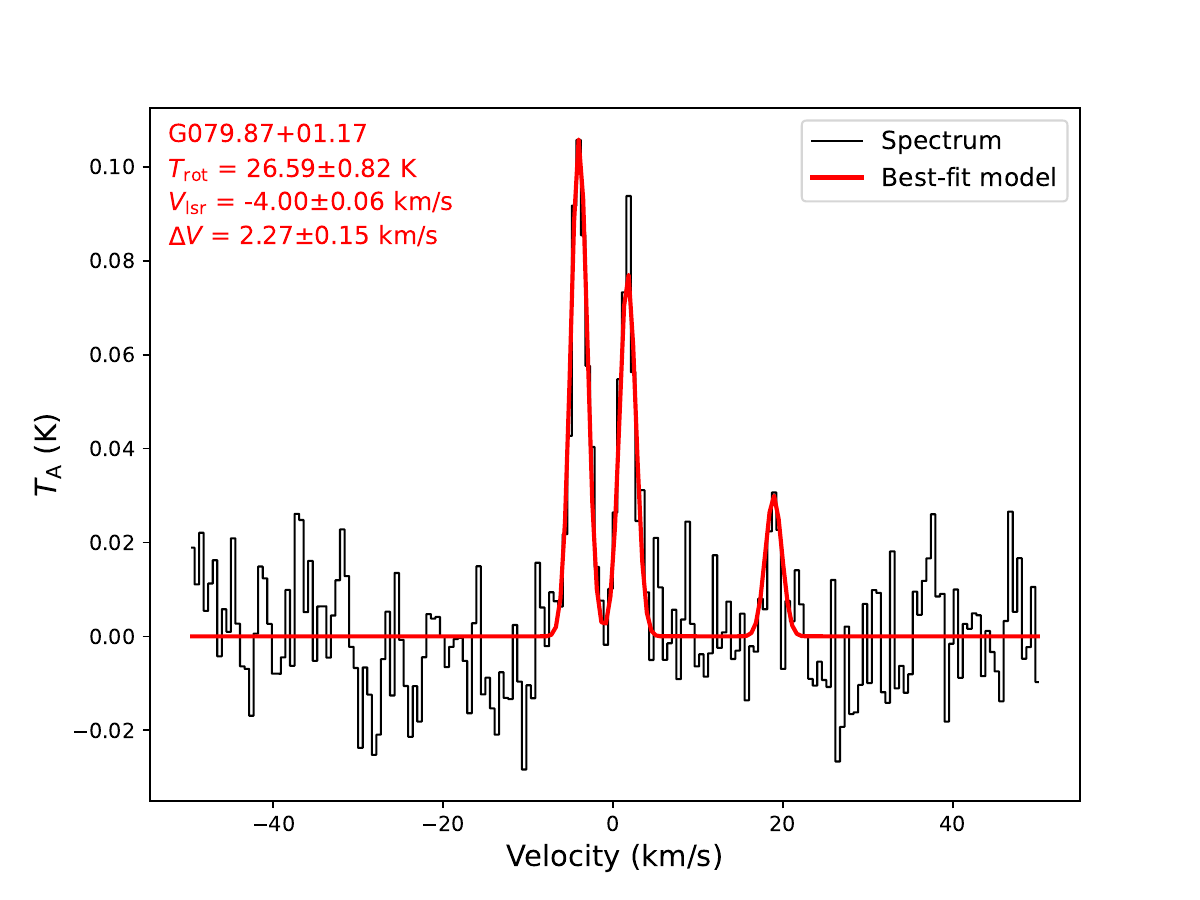}{0.31\textwidth}{}
\fig{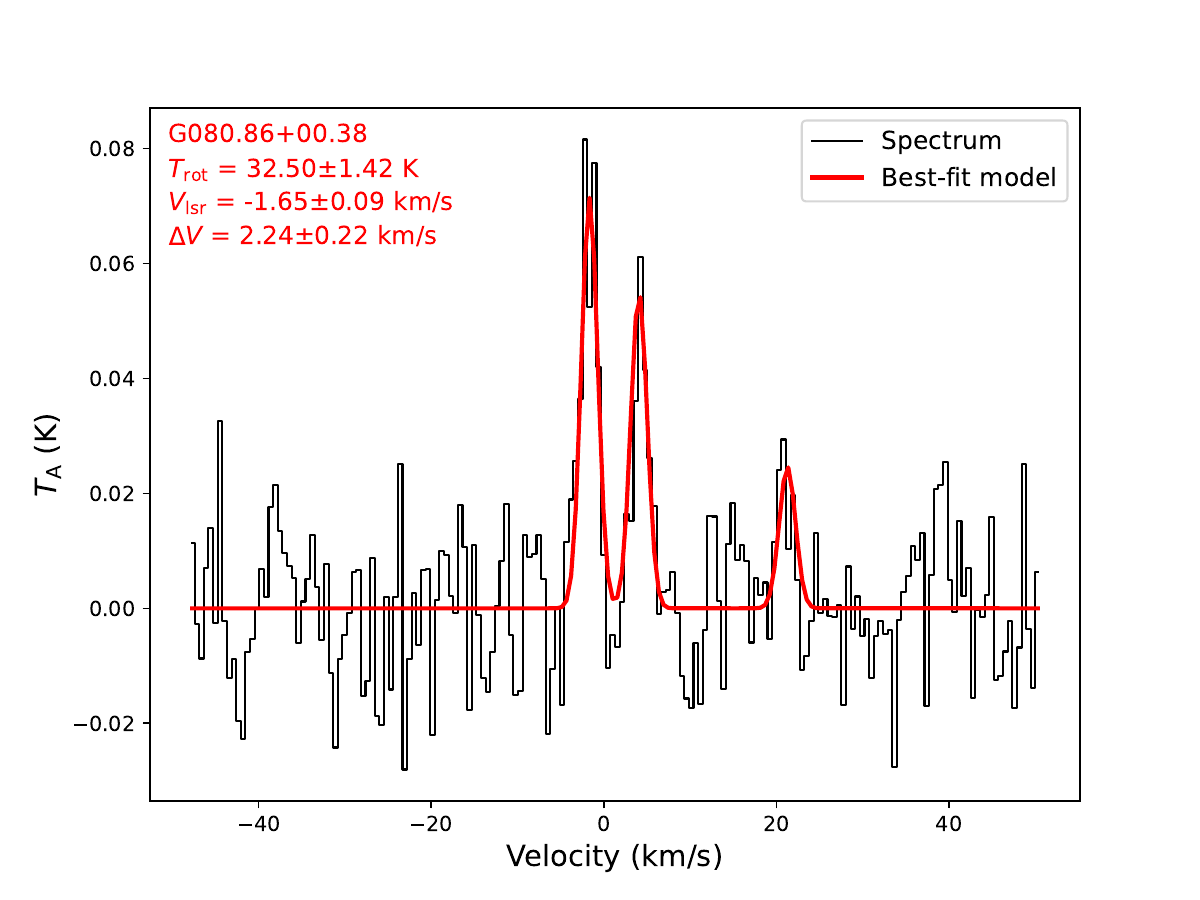}{0.31\textwidth}{}
\fig{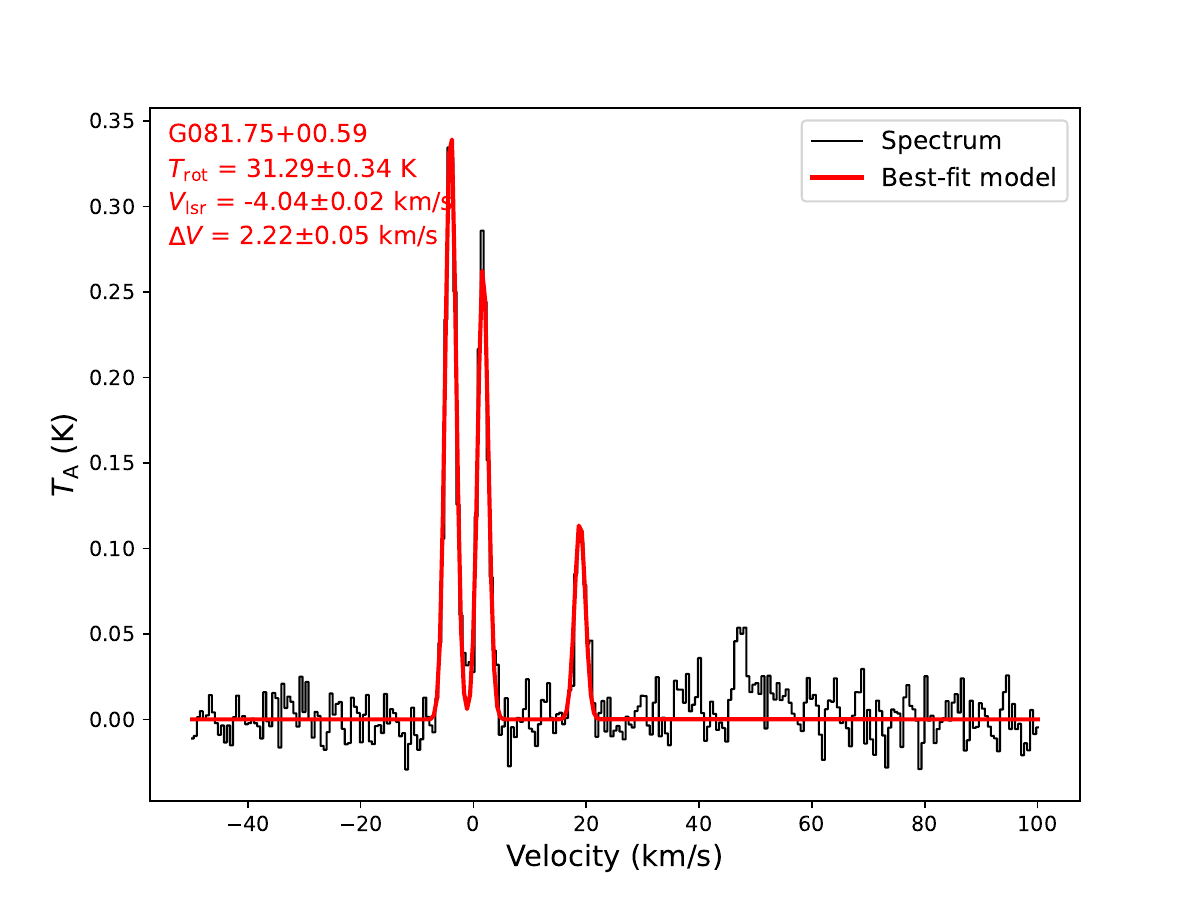}{0.31\textwidth}{}
}
\gridline{
\fig{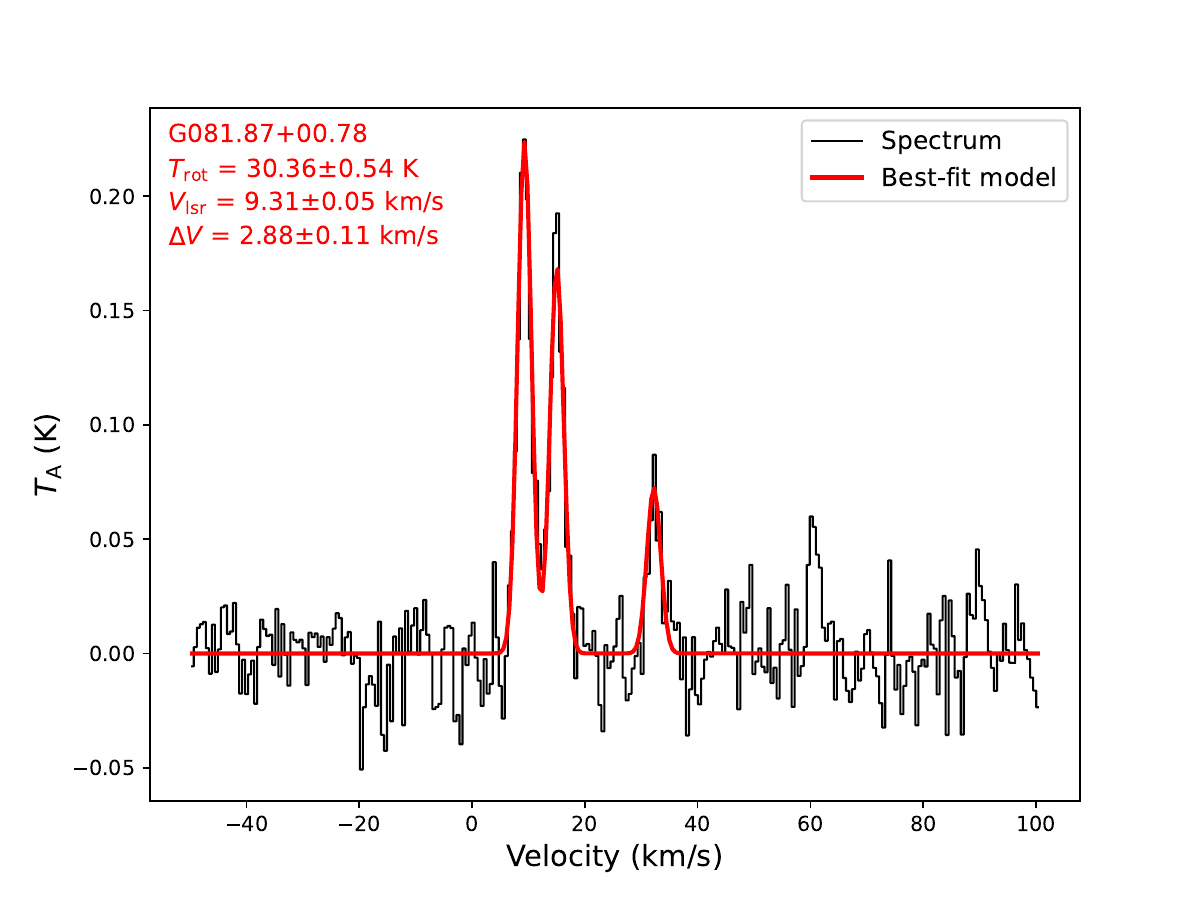}{0.31\textwidth}{}
\fig{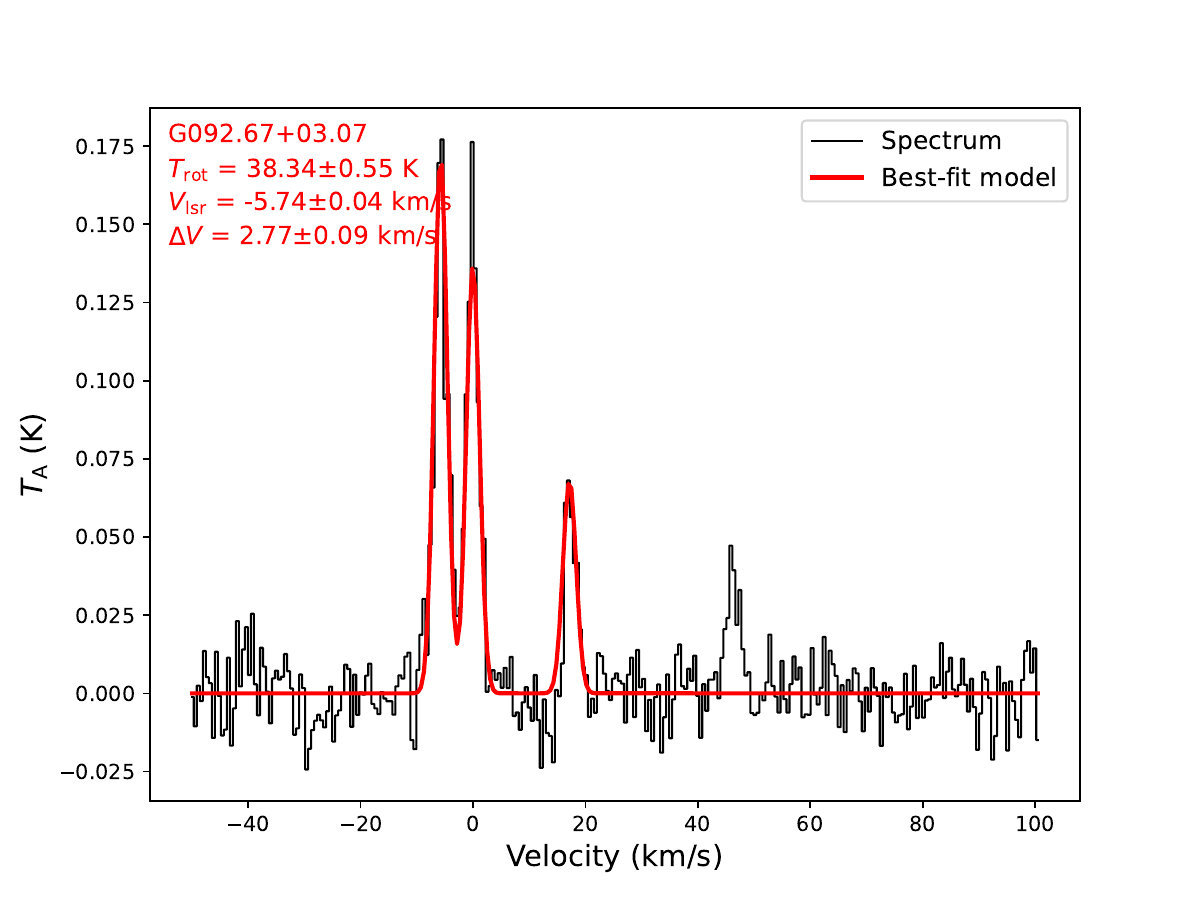}{0.31\textwidth}{}
\fig{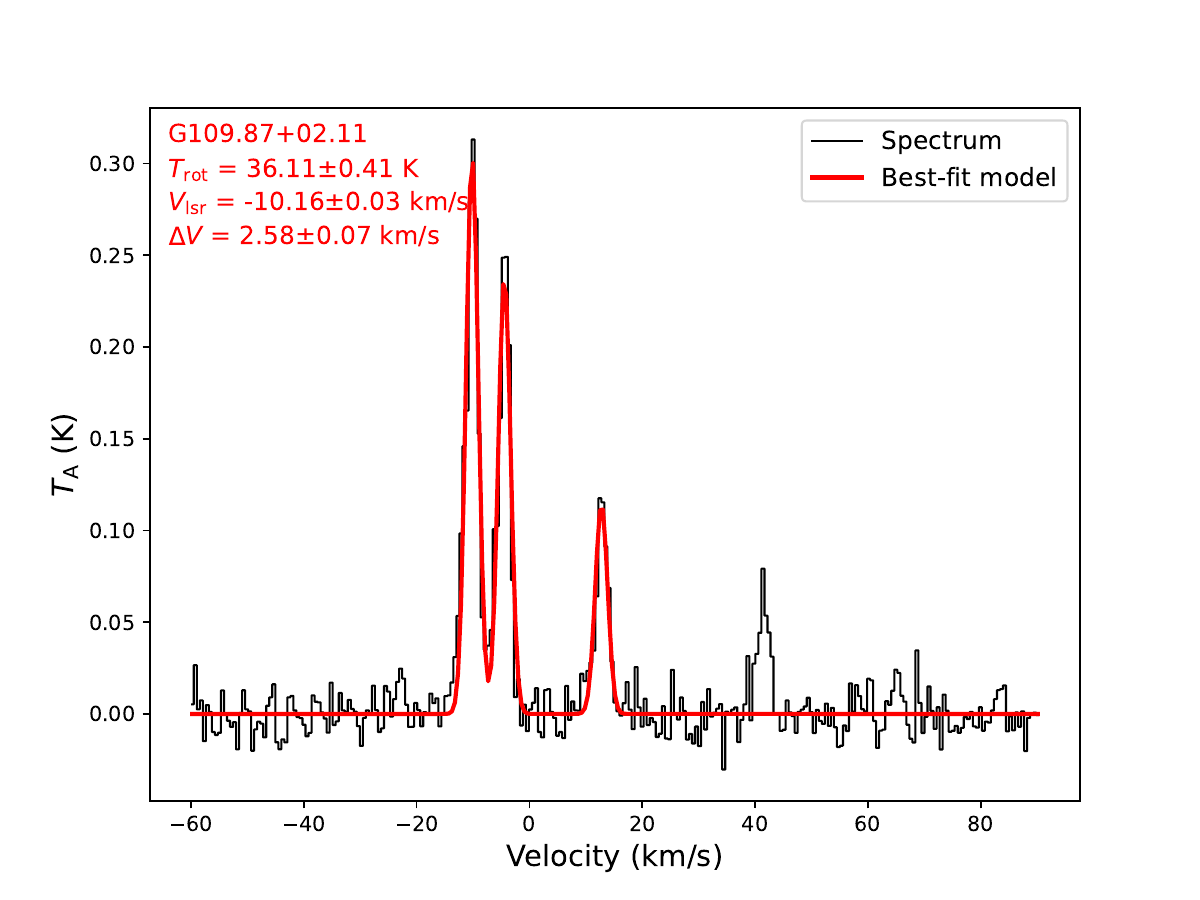}{0.31\textwidth}{}
}
\gridline{
\fig{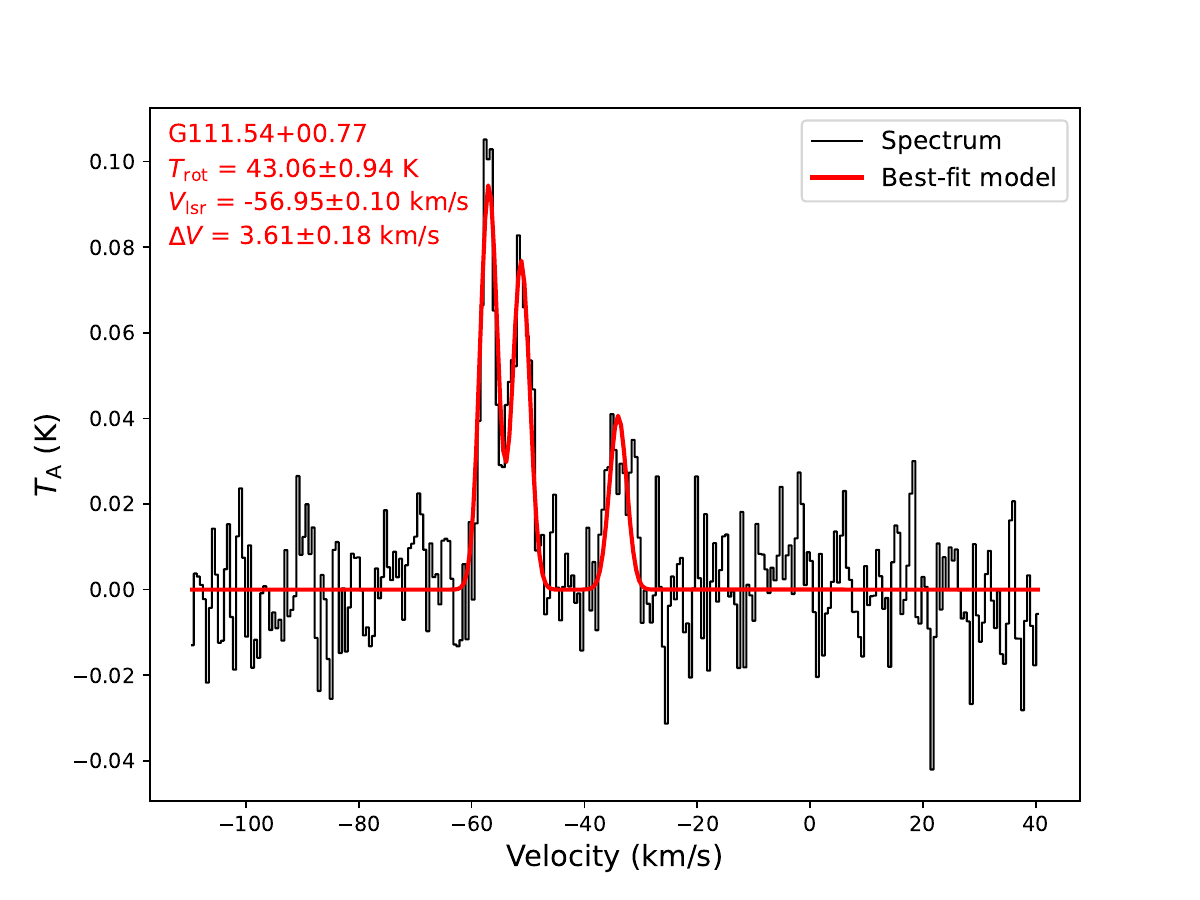}{0.31\textwidth}{}
\fig{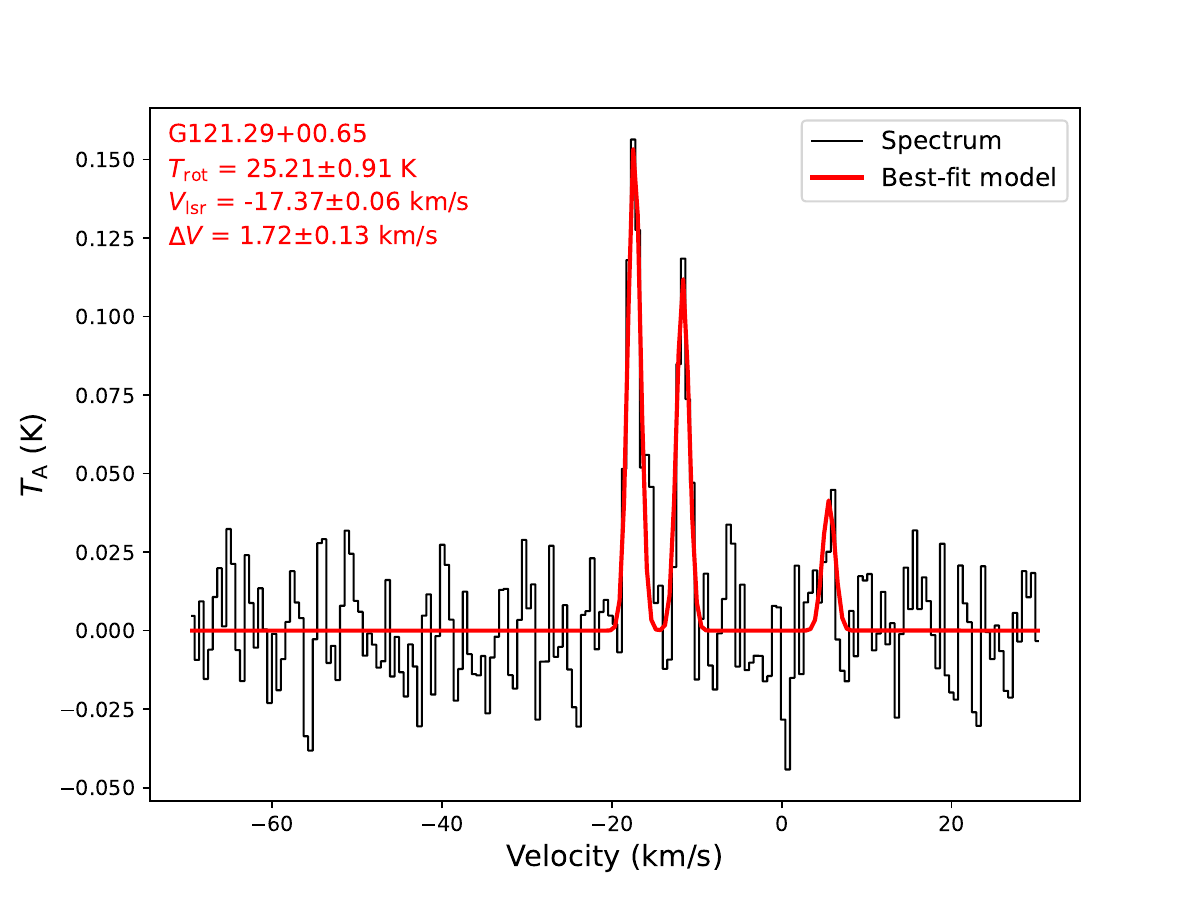}{0.31\textwidth}{}
\fig{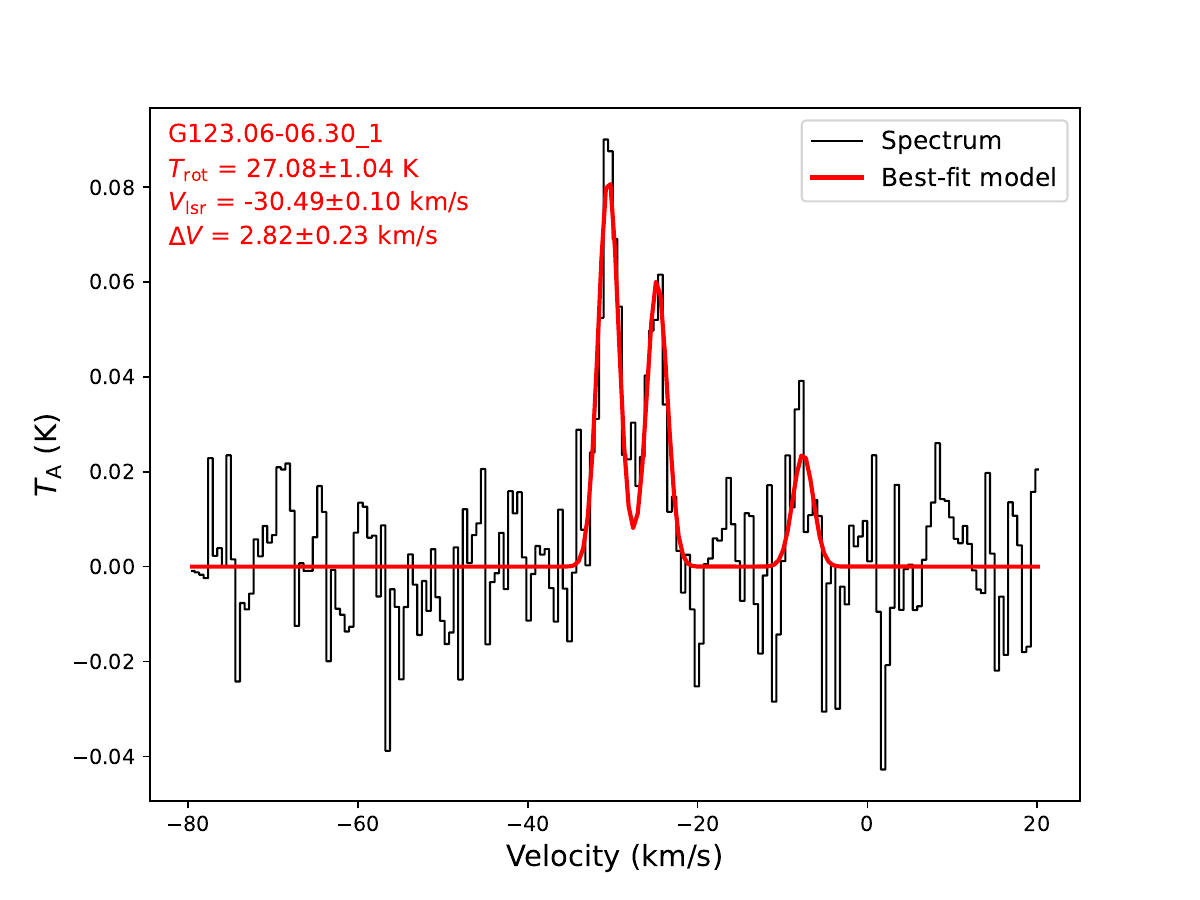}{0.31\textwidth}{}
}
\caption{CH$_3$CCH spectra toward the sample sources (4/6). The observed spectra of CH$_3$CCH 5-4 are in black, while the best fit is in red.}
\label{appendix: CH3CCH_fitting_4}
\end{figure}

\begin{figure}[ht!]
\centering
\gridline{
\fig{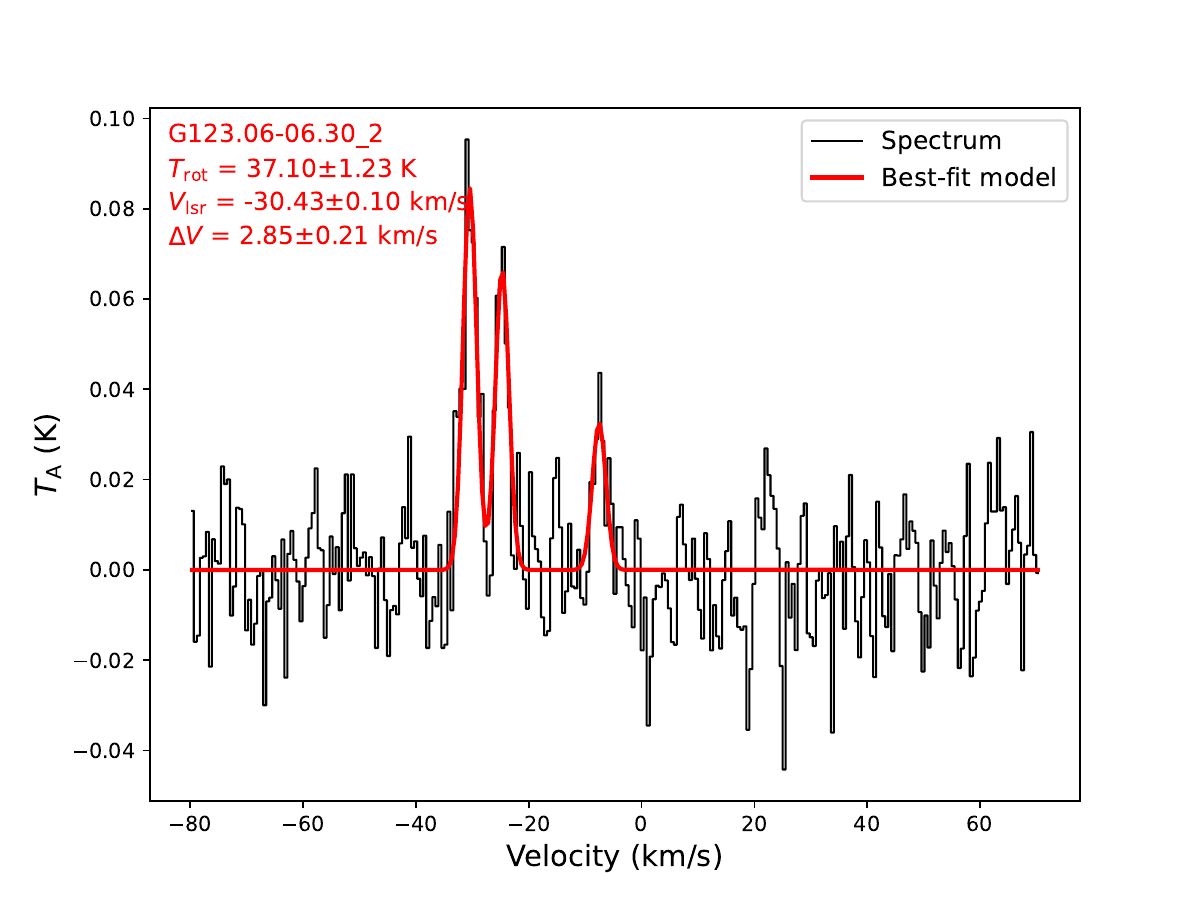}{0.31\textwidth}{}
\fig{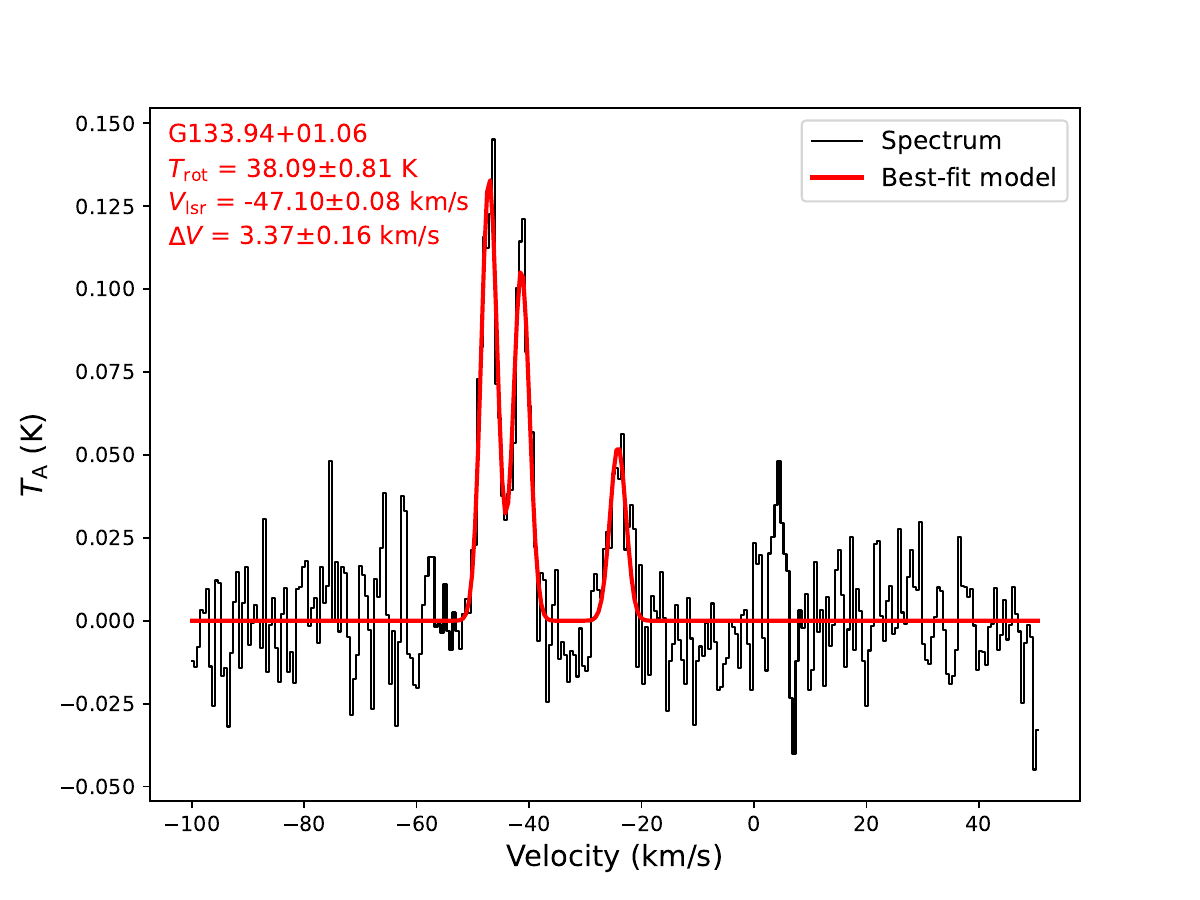}{0.31\textwidth}{}
\fig{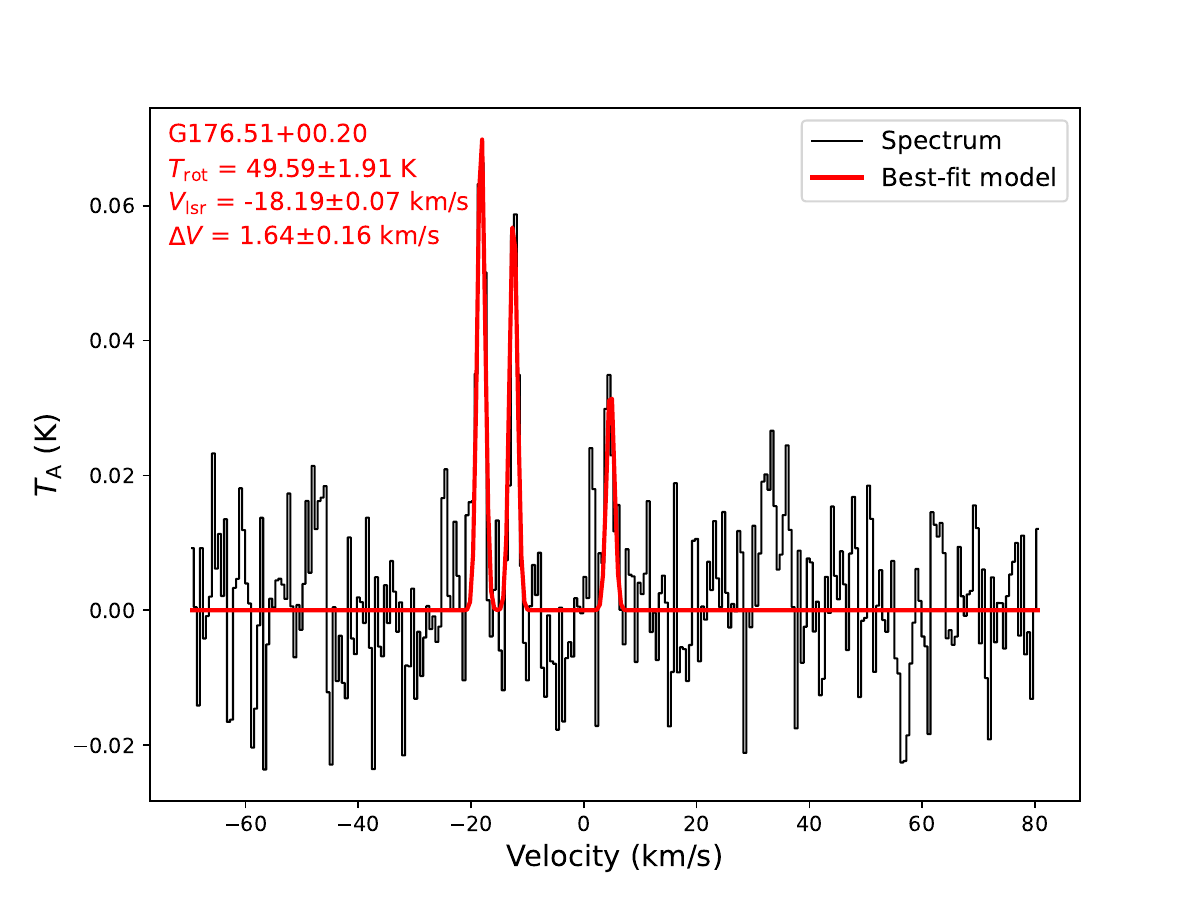}{0.31\textwidth}{}
}
\gridline{
\fig{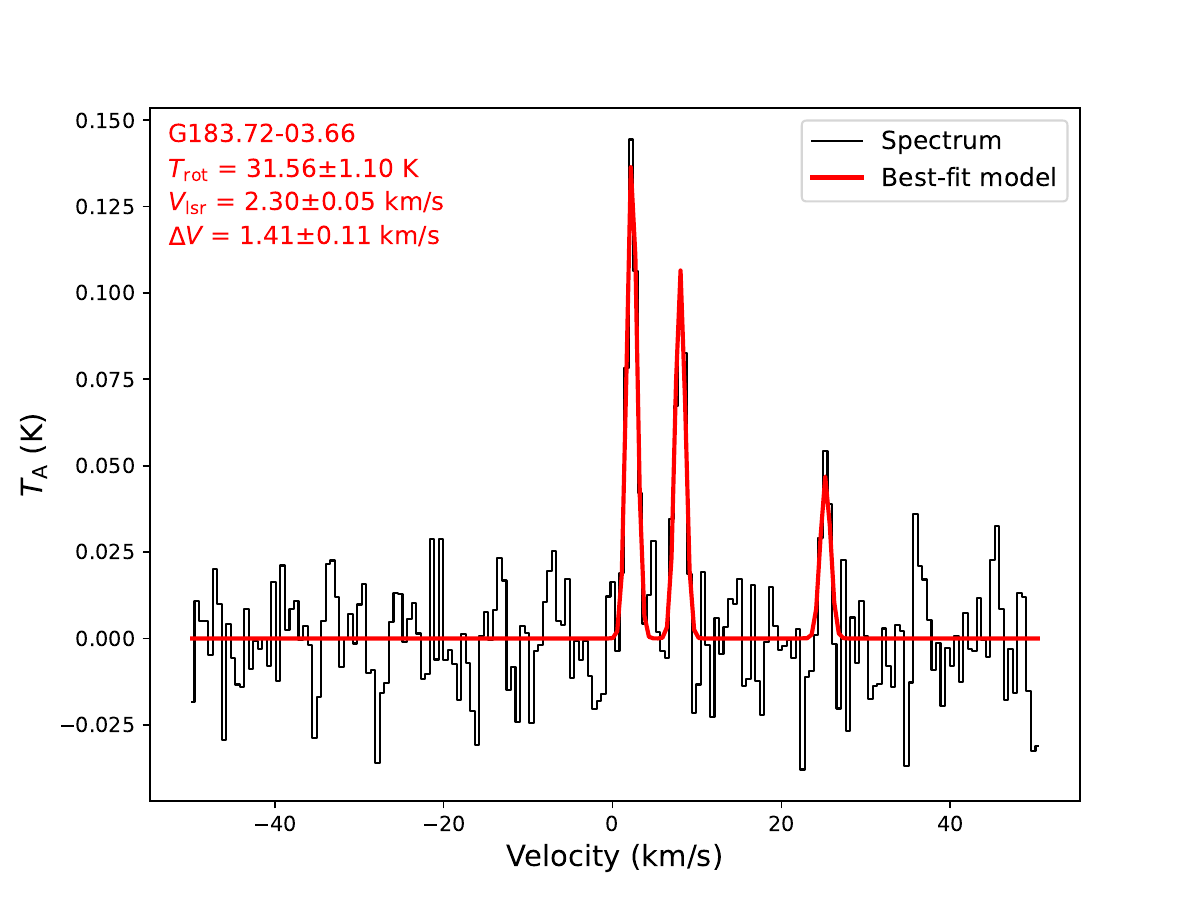}{0.31\textwidth}{}
\fig{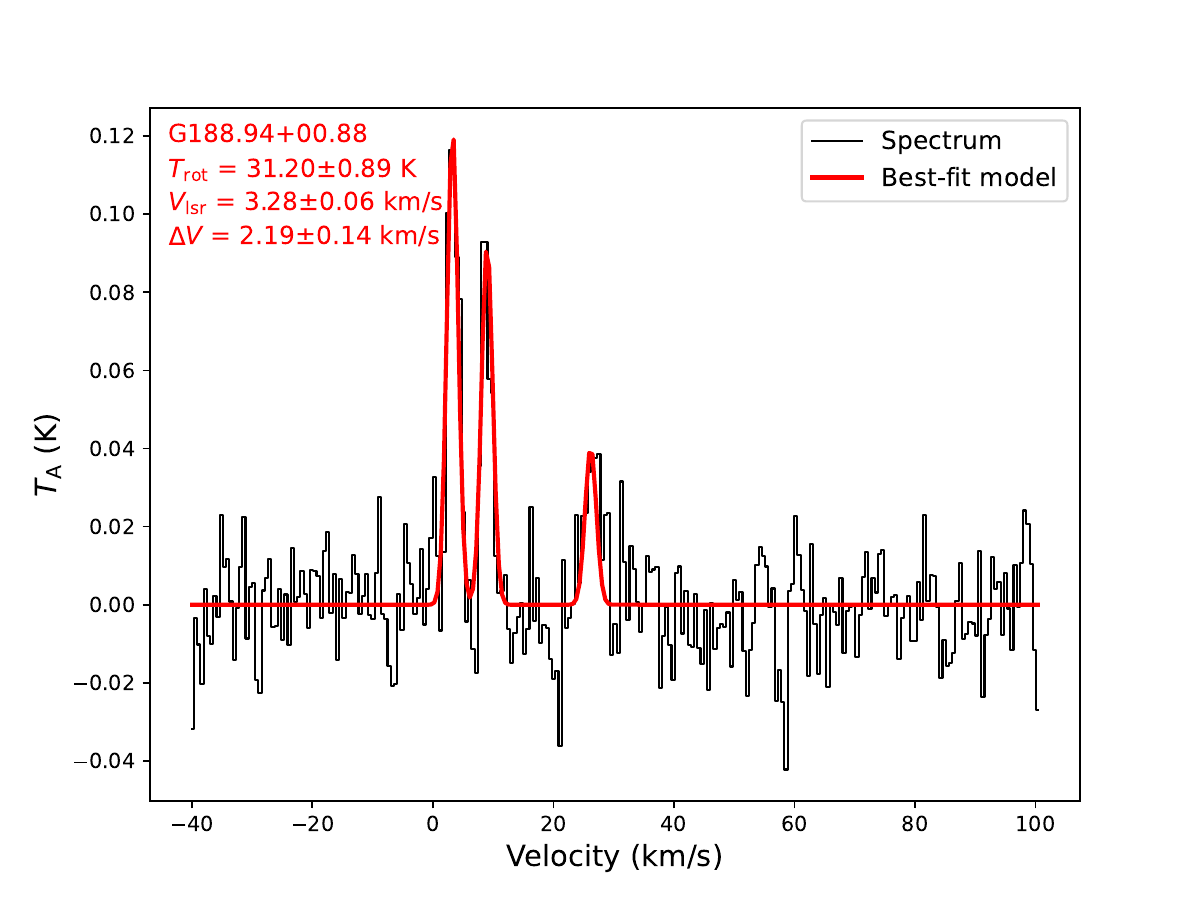}{0.31\textwidth}{}
\fig{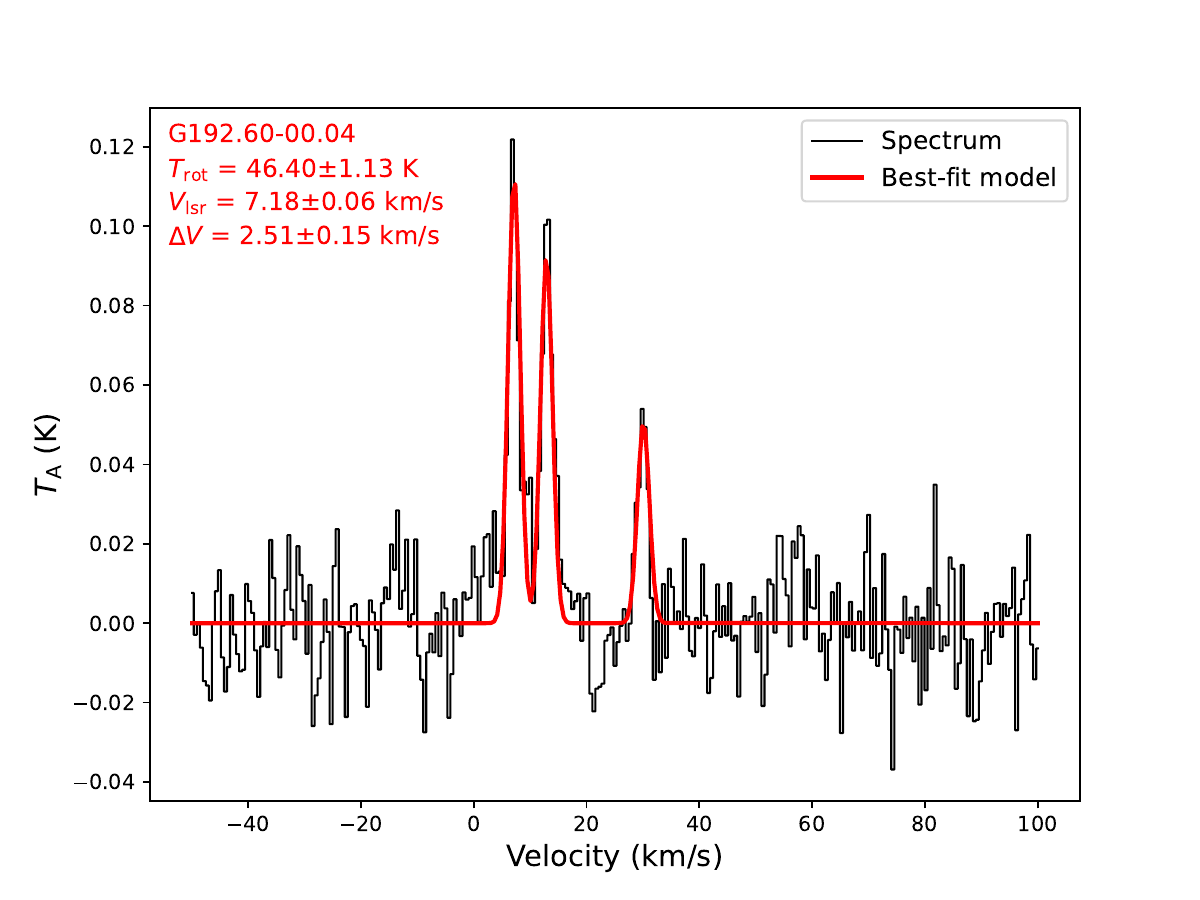}{0.31\textwidth}{}
}
\gridline{
\fig{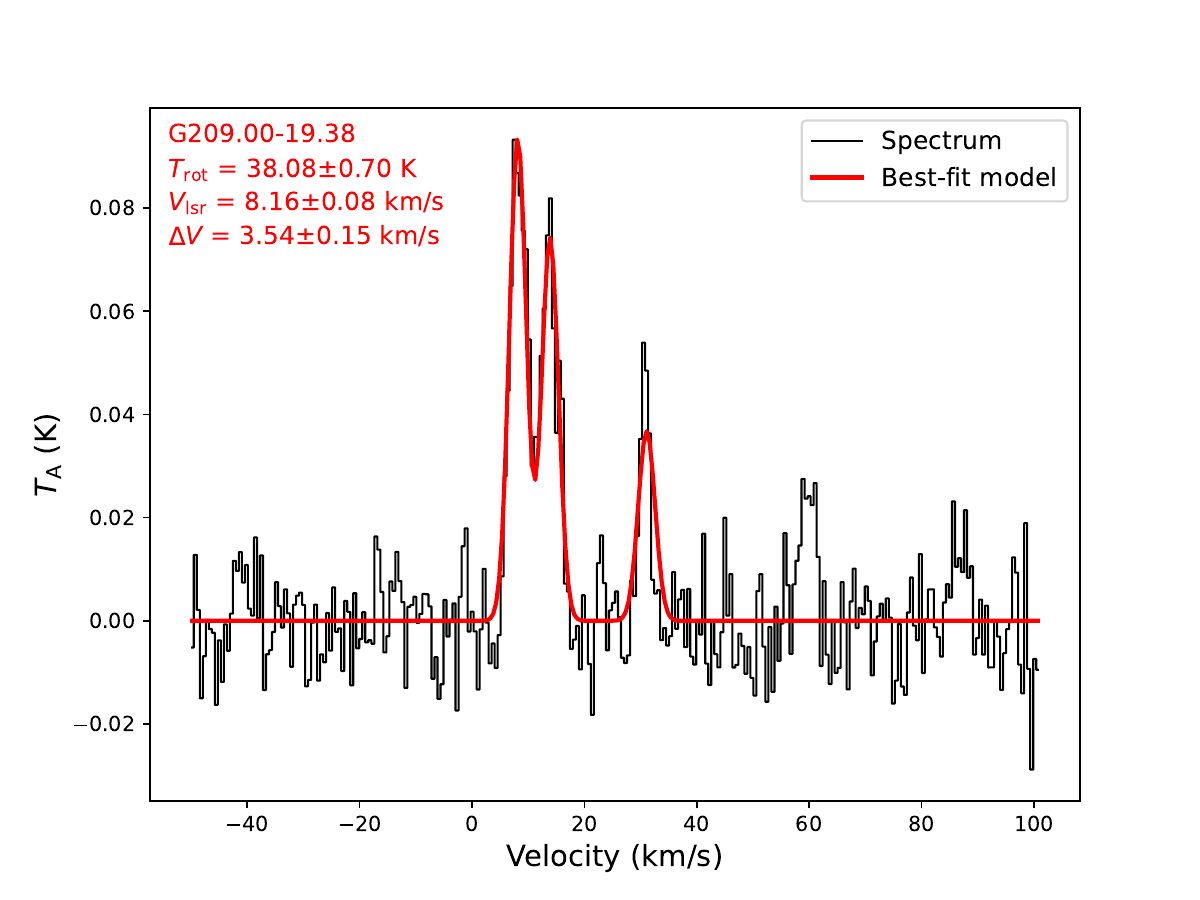}{0.31\textwidth}{}
}
\caption{CH$_3$CCH spectra toward the sample sources (5/6). The observed spectra of CH$_3$CCH 5-4 are in black, while the best fit is in red.}
\label{appendix: CH3CCH_fitting_5}
\end{figure}

\clearpage
\newpage
\section{The fitting result of NH$_3$}
We present the fitting result of NH$_3$(1,1) and NH$_3$(2,2).

\renewcommand{\thefigure}{C\arabic{figure}}  
\setcounter{figure}{0}
\begin{figure*}[ht!]
\centering
\gridline{
\fig{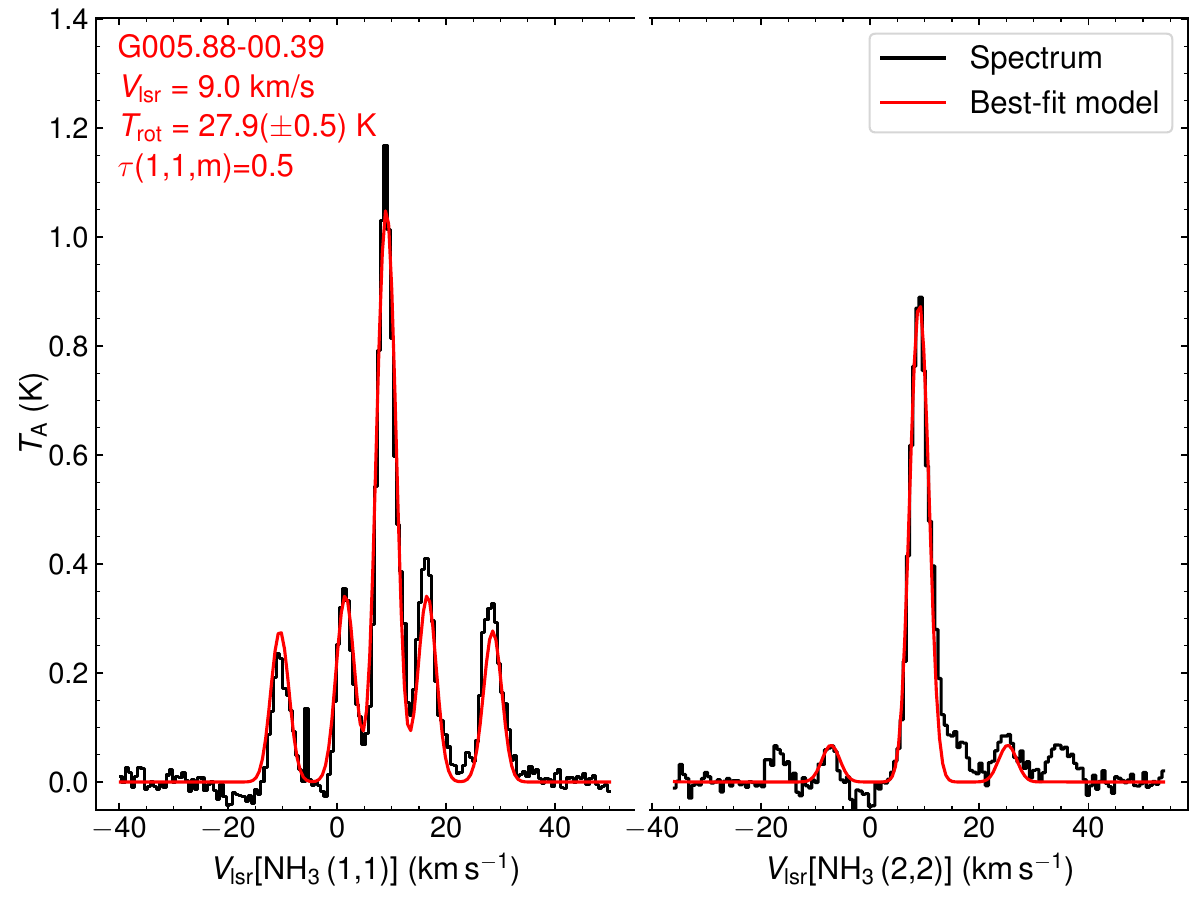}{0.30\textwidth}{}
\fig{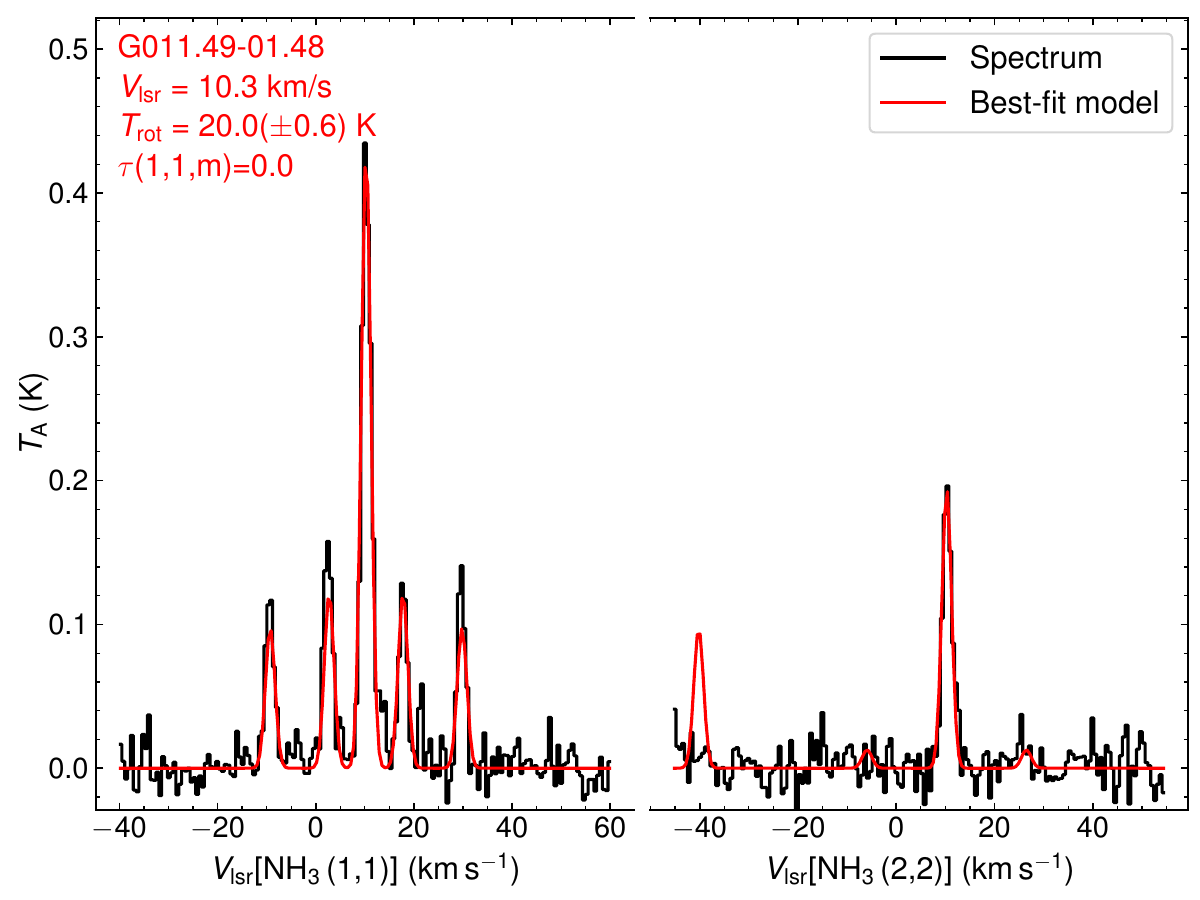}{0.3\textwidth}{}
\fig{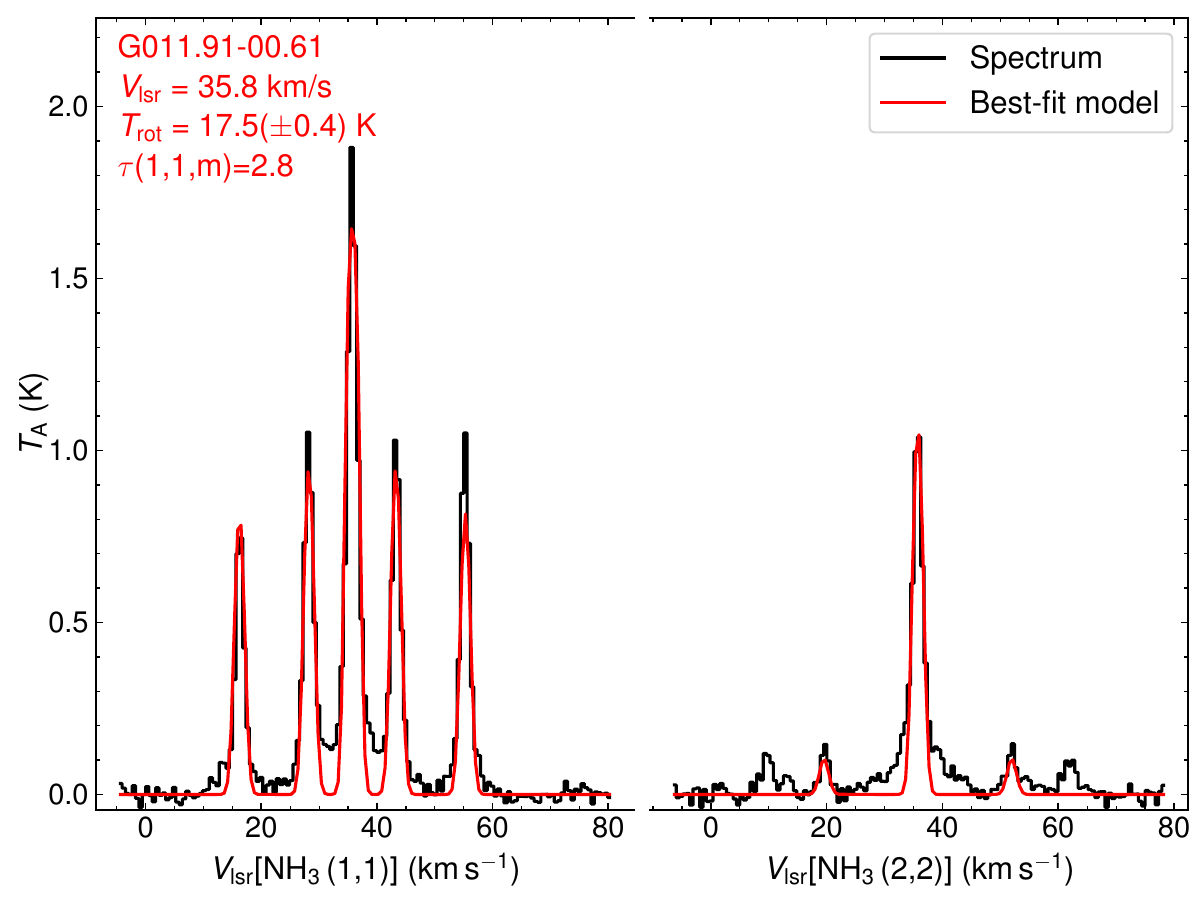}{0.3\textwidth}{}
}
\gridline{
\fig{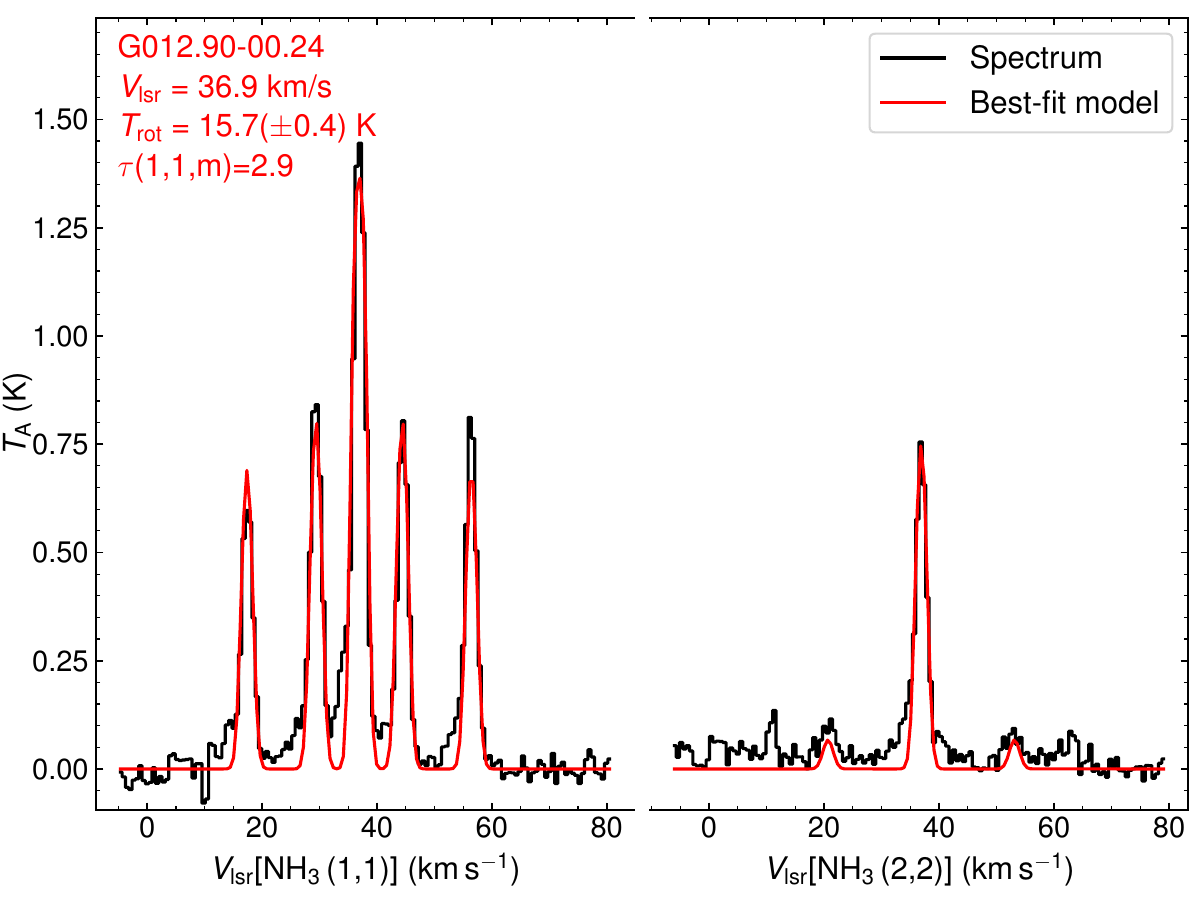}{0.3\textwidth}{}
\fig{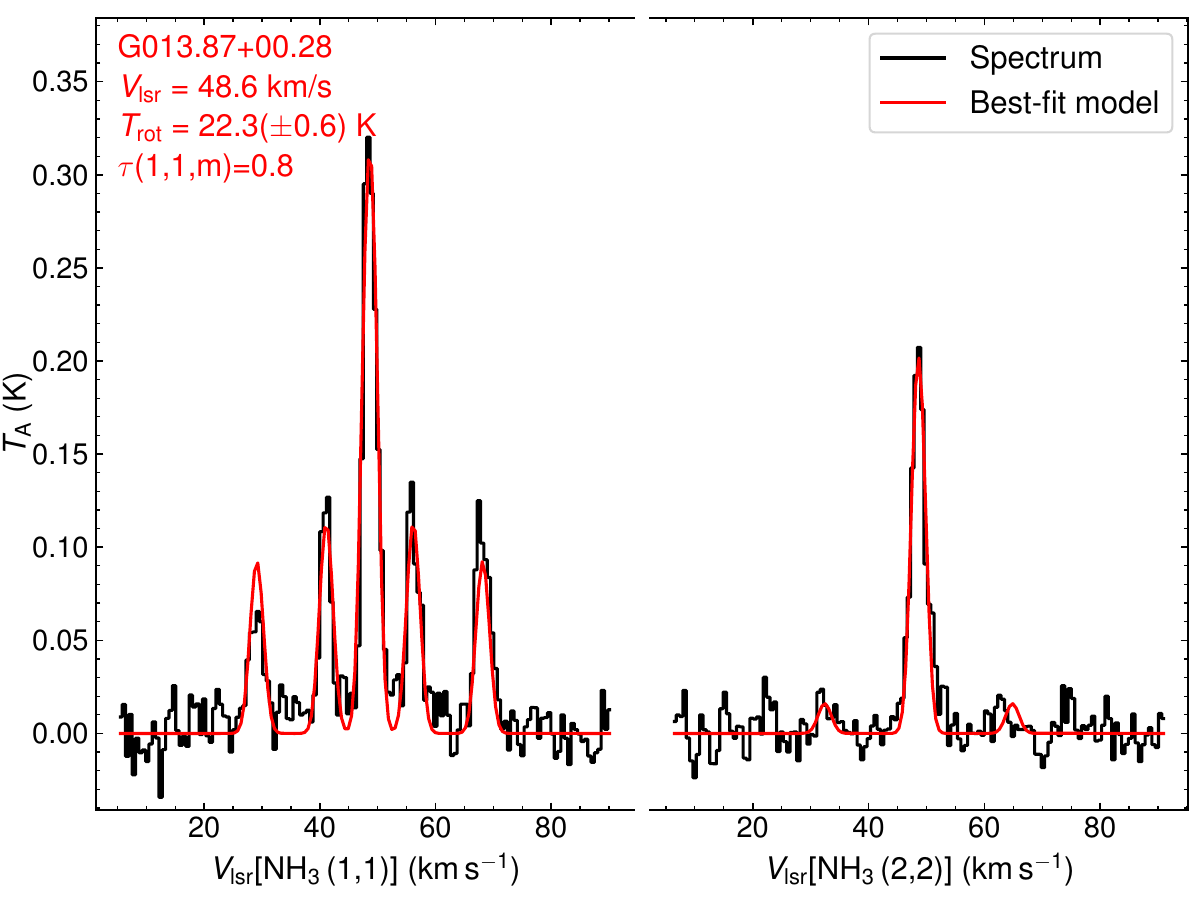}{0.3\textwidth}{}
\fig{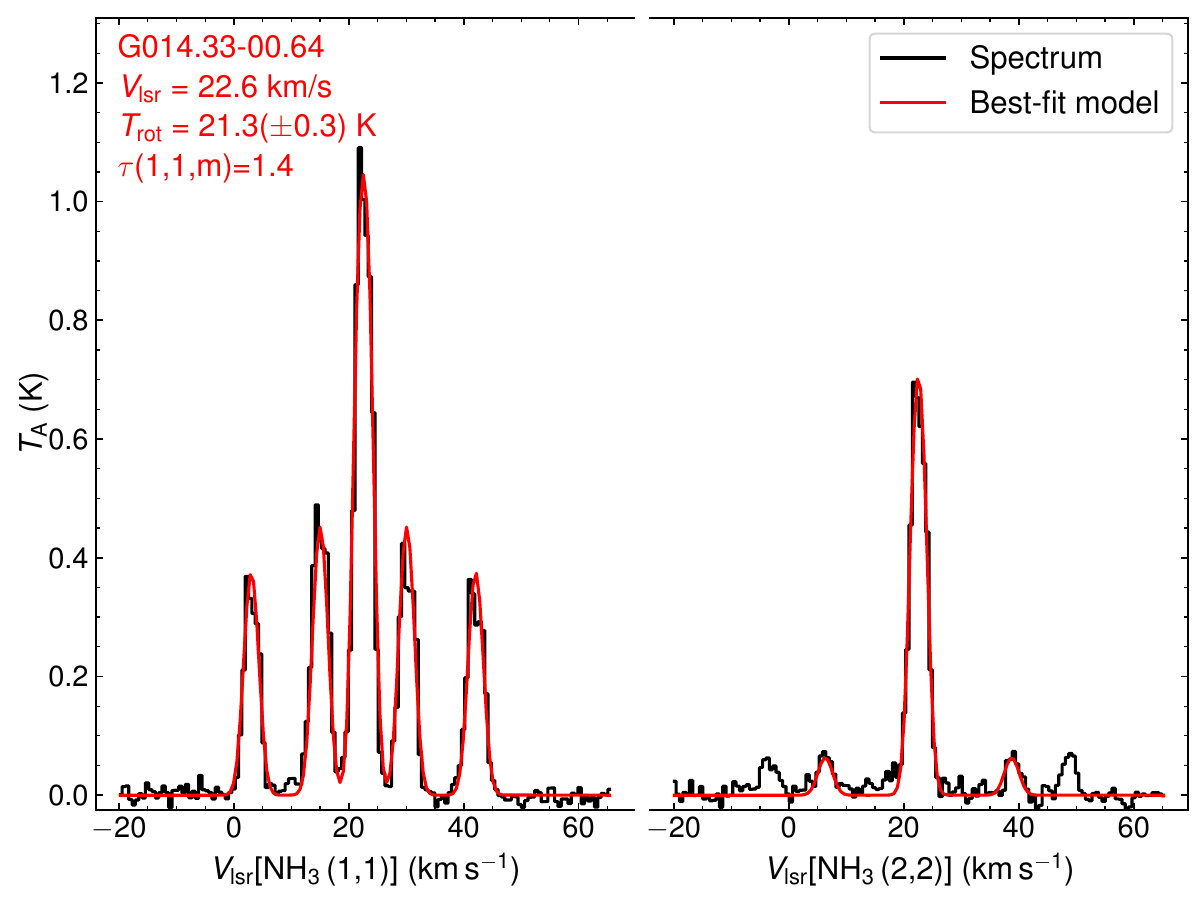}{0.3\textwidth}{}
}
\gridline{
\fig{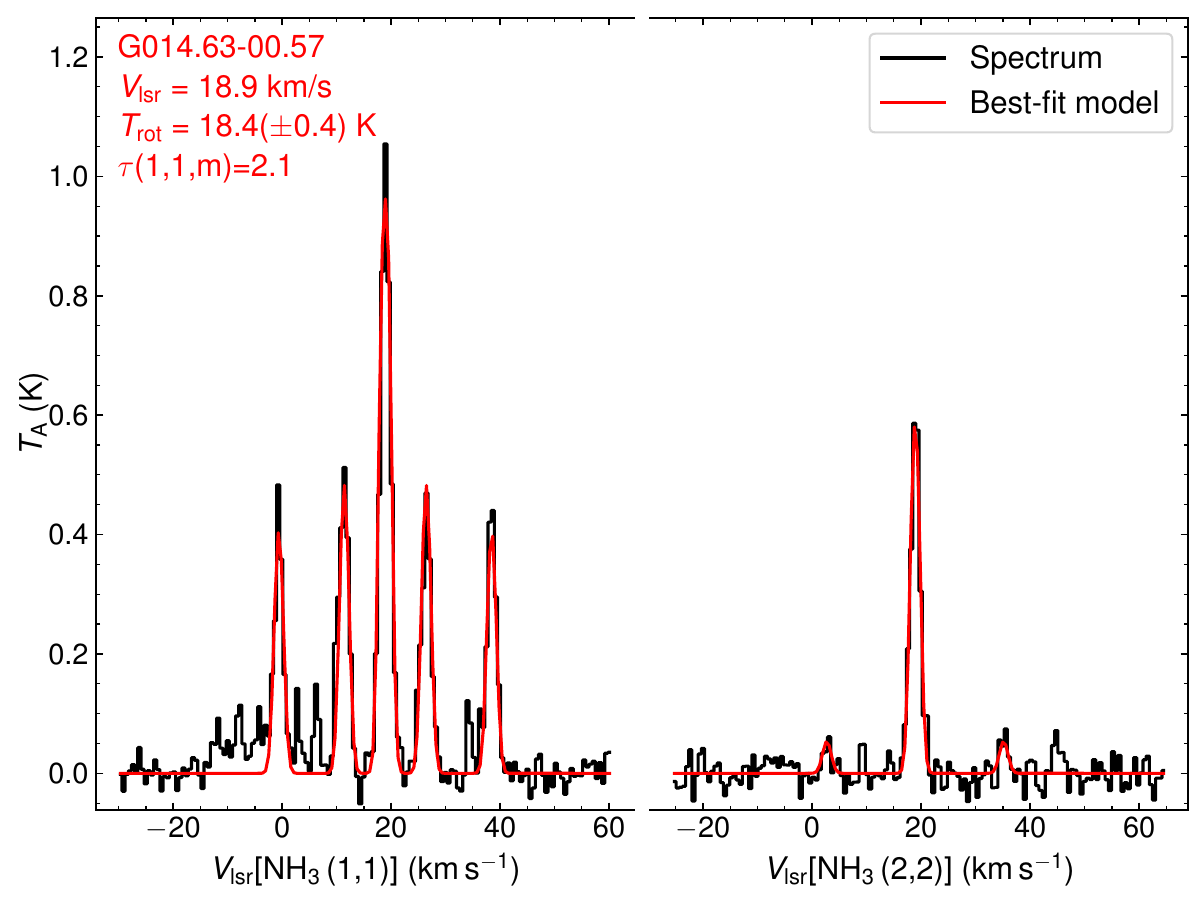}{0.3\textwidth}{}
\fig{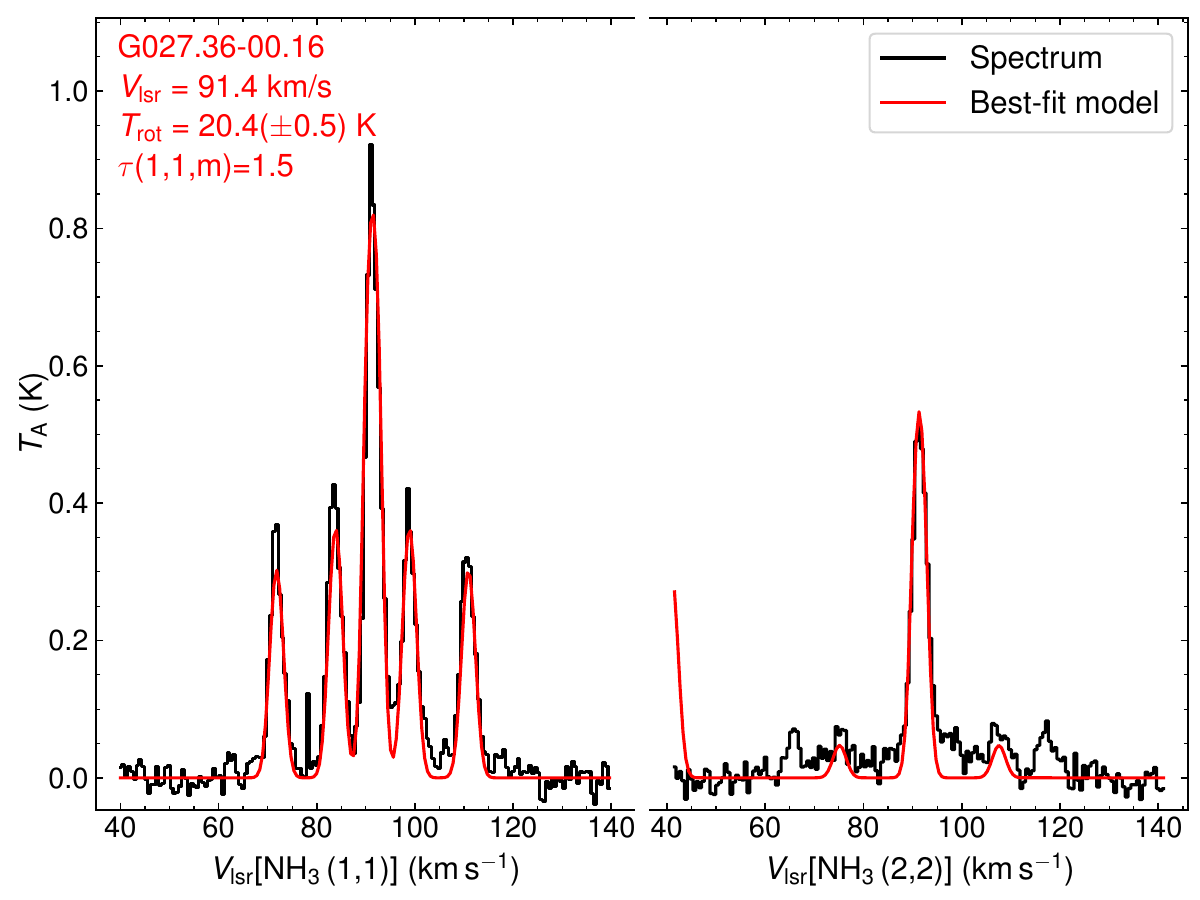}{0.3\textwidth}{}
\fig{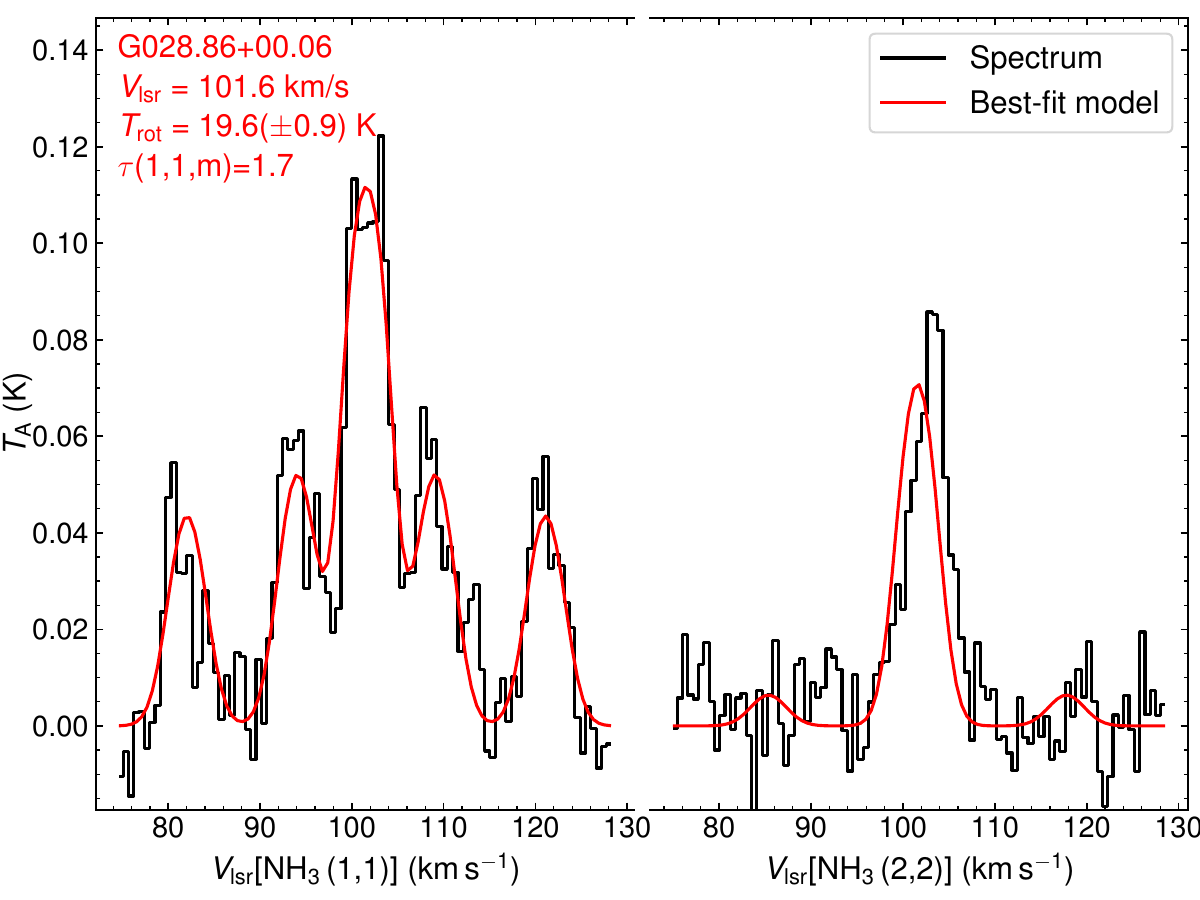}{0.3\textwidth}{}
}
\gridline{
\fig{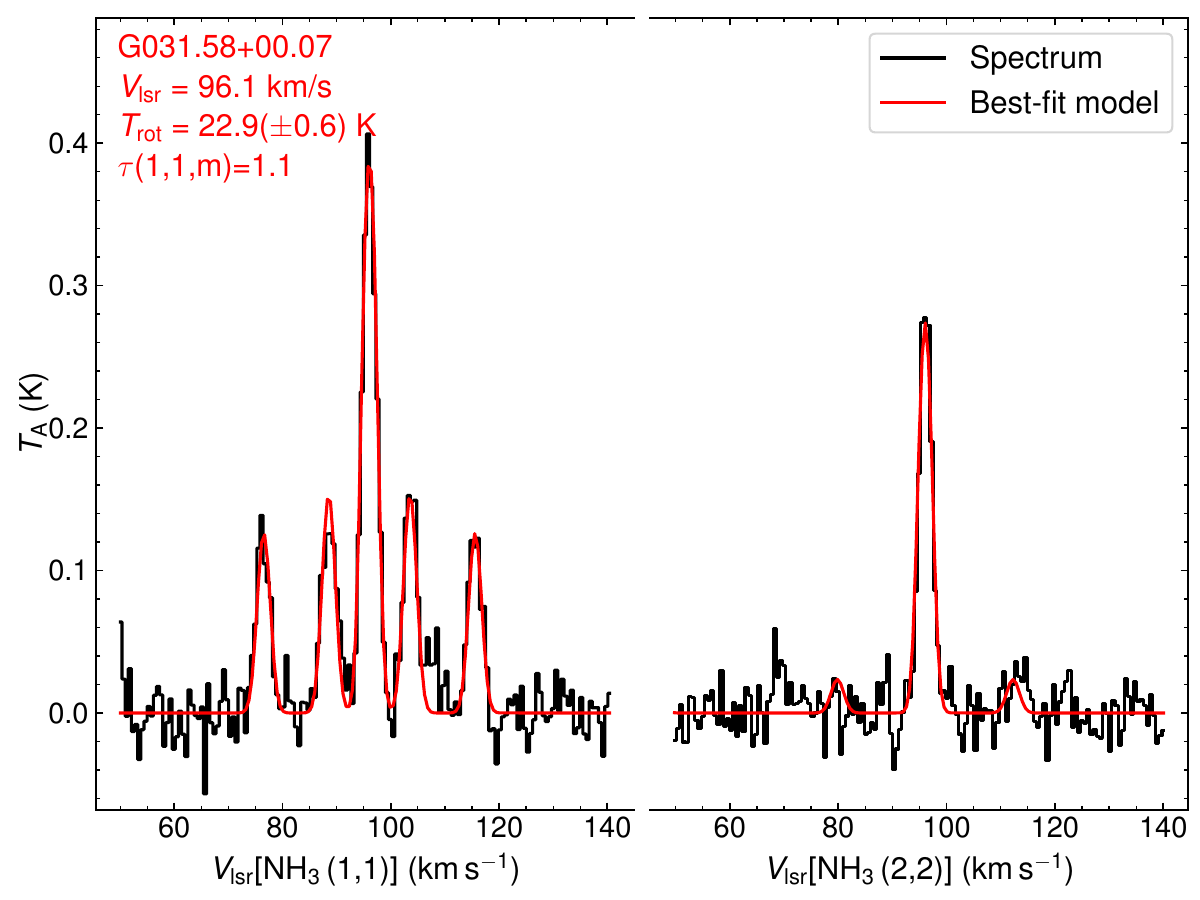}{0.3\textwidth}{}
\fig{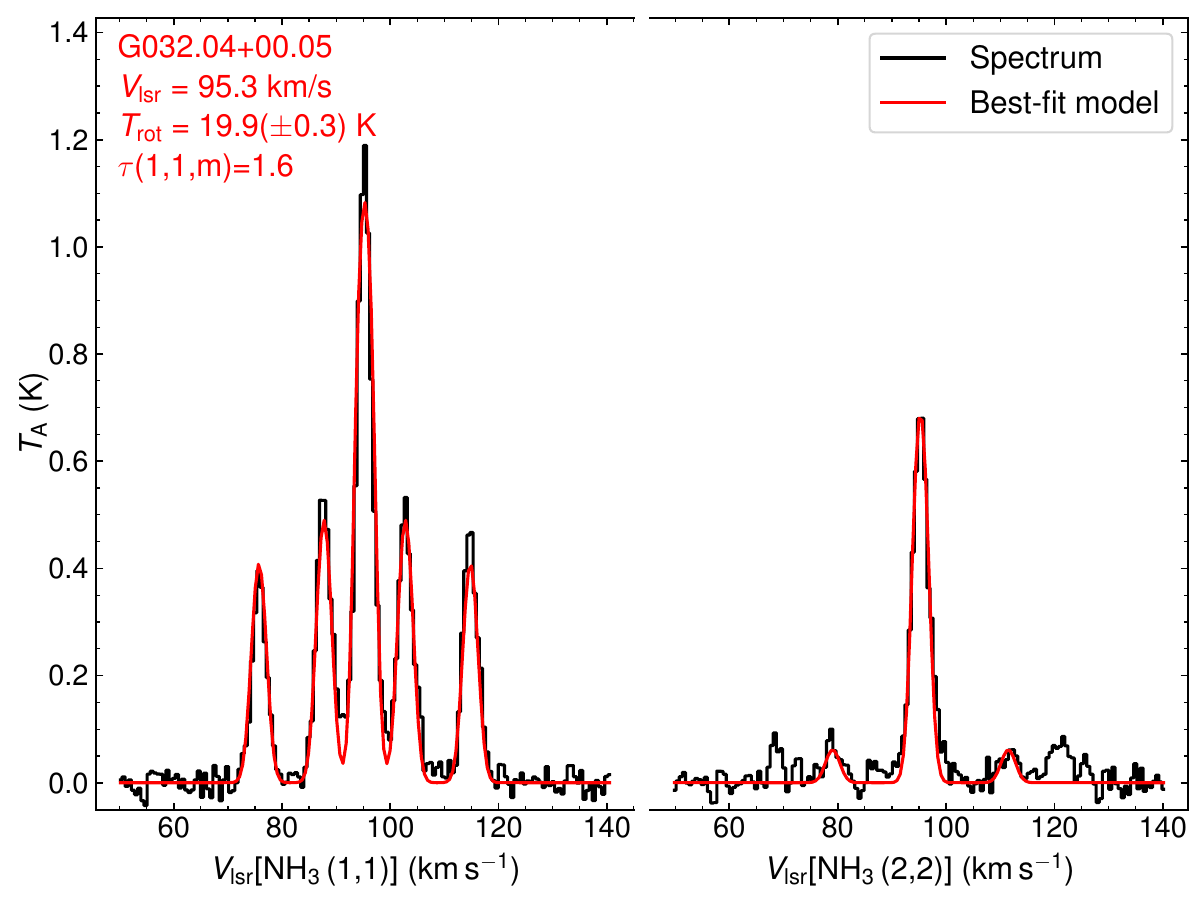}{0.3\textwidth}{}
\fig{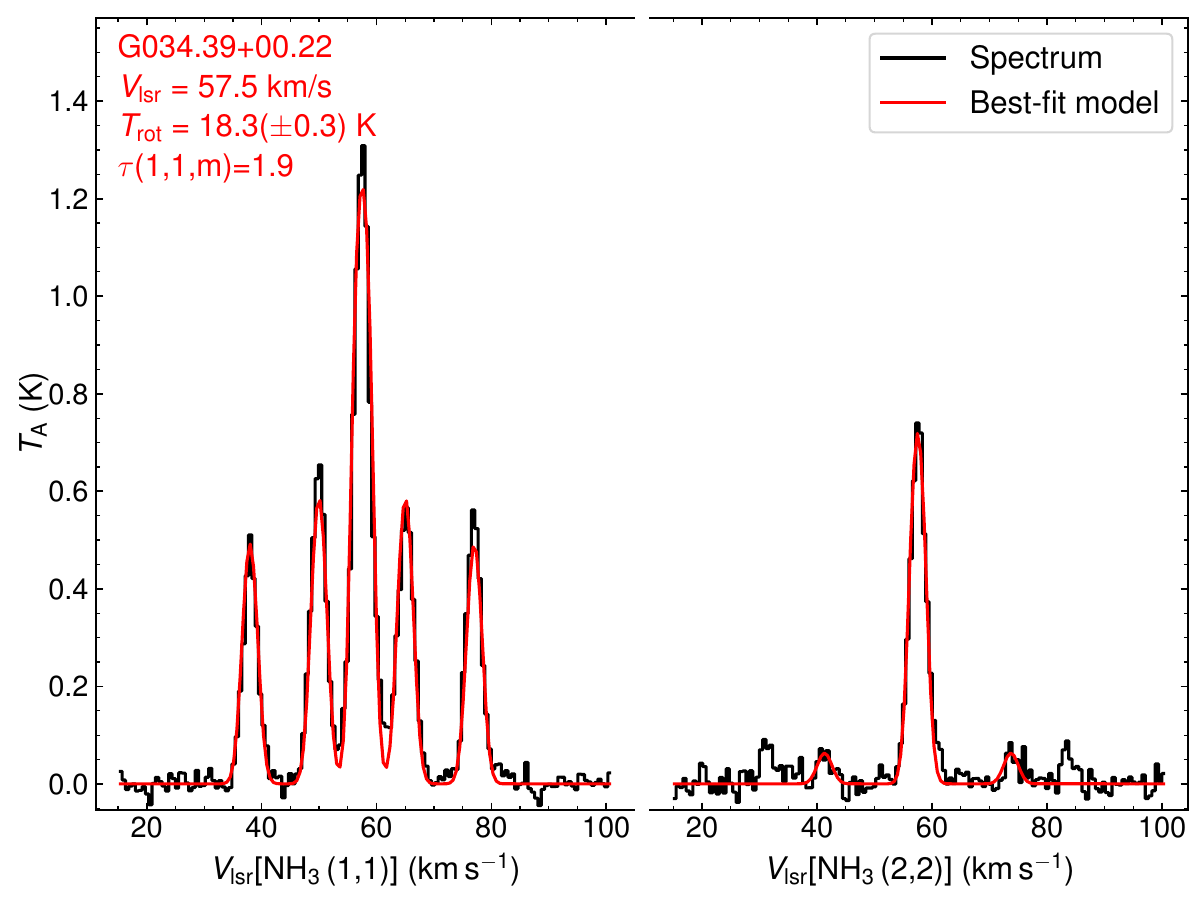}{0.3\textwidth}{}
}
\caption{NH$_3$ spectra toward the sample sources (1/2). The observed spectra of NH$_3$(1,1) and NH$_3$(2,2) are in black, while the best fit from the code of \cite{2015ApJ...805..171L} is in red.}
\label{appendix: NH3_fitting_1}
\end{figure*}

\begin{figure*}[ht!]
\centering
\gridline{
\fig{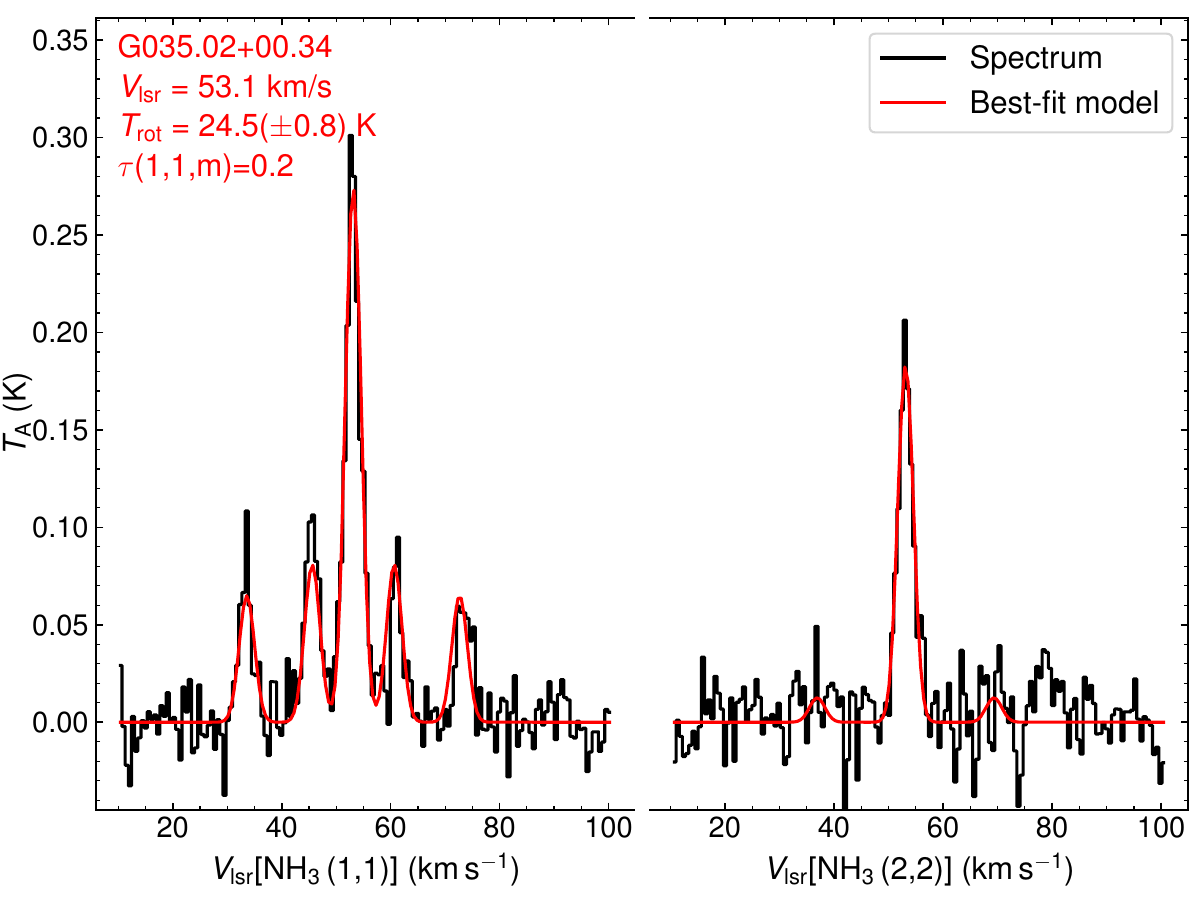}{0.3\textwidth}{}
\fig{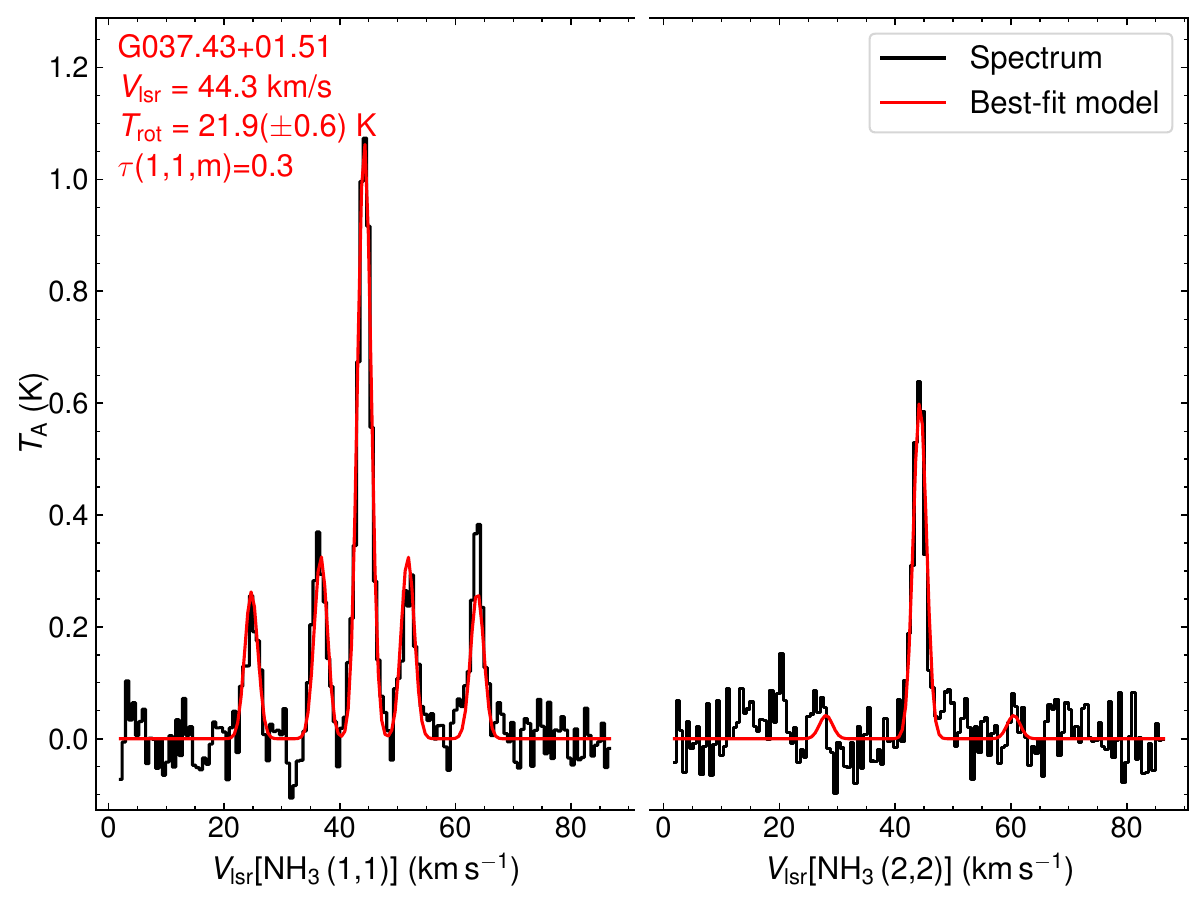}{0.3\textwidth}{}
\fig{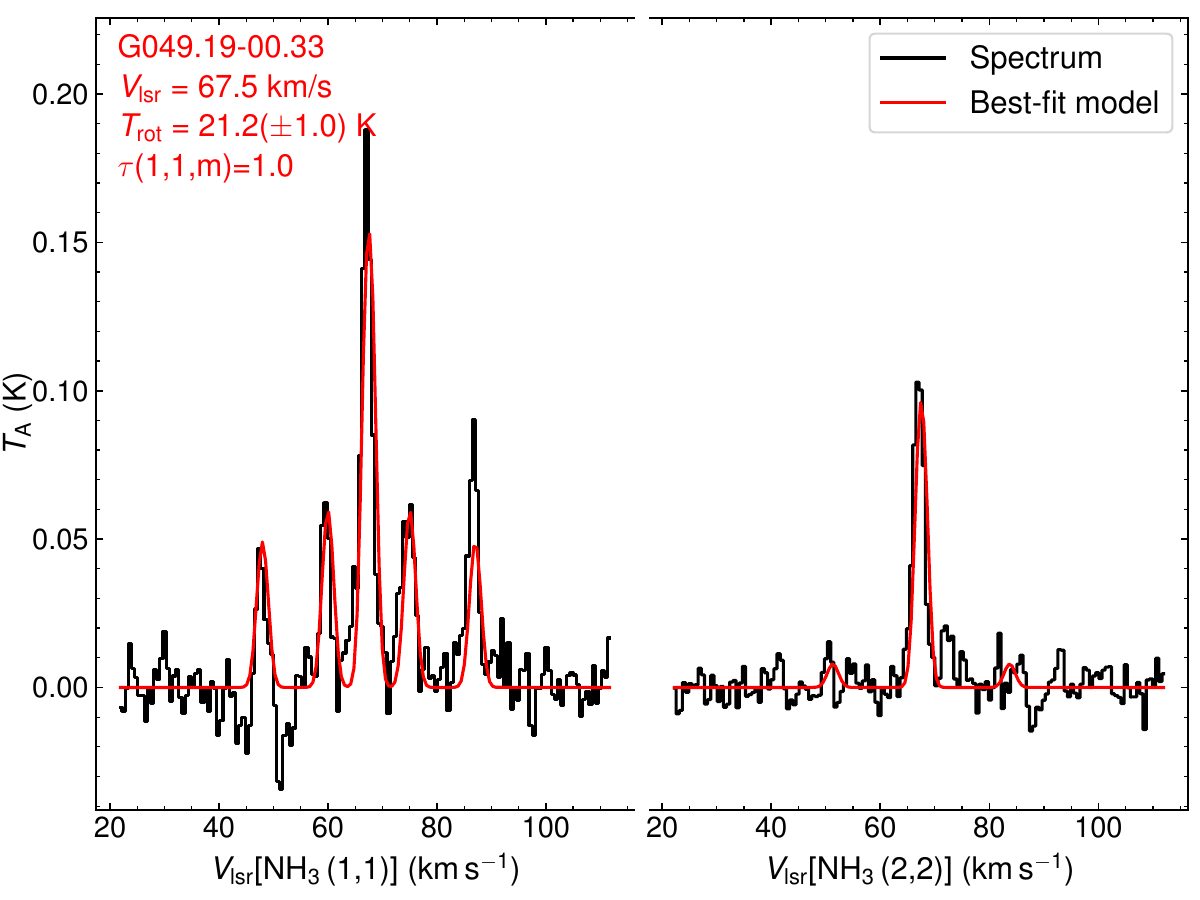}{0.3\textwidth}{}
}
\gridline{
\fig{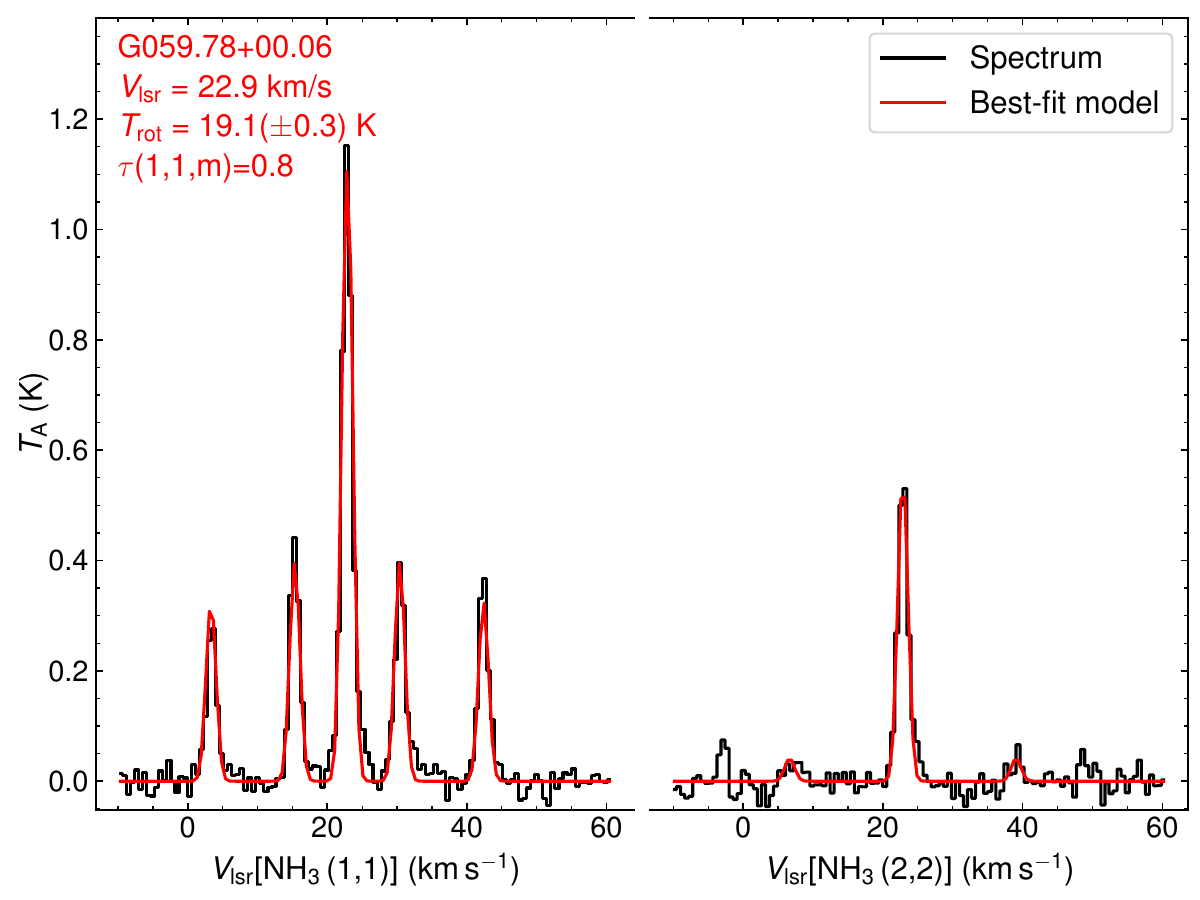}{0.3\textwidth}{}
\fig{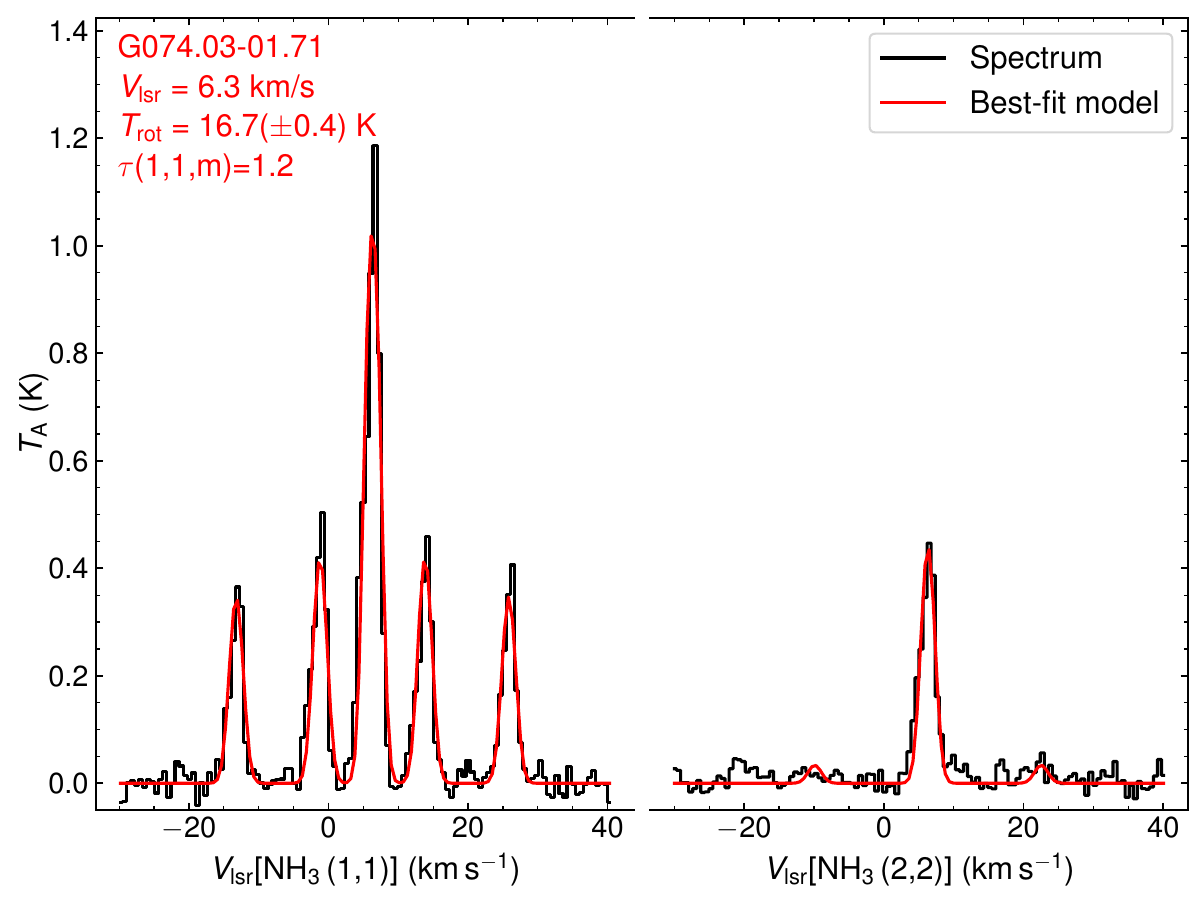}{0.3\textwidth}{}
\fig{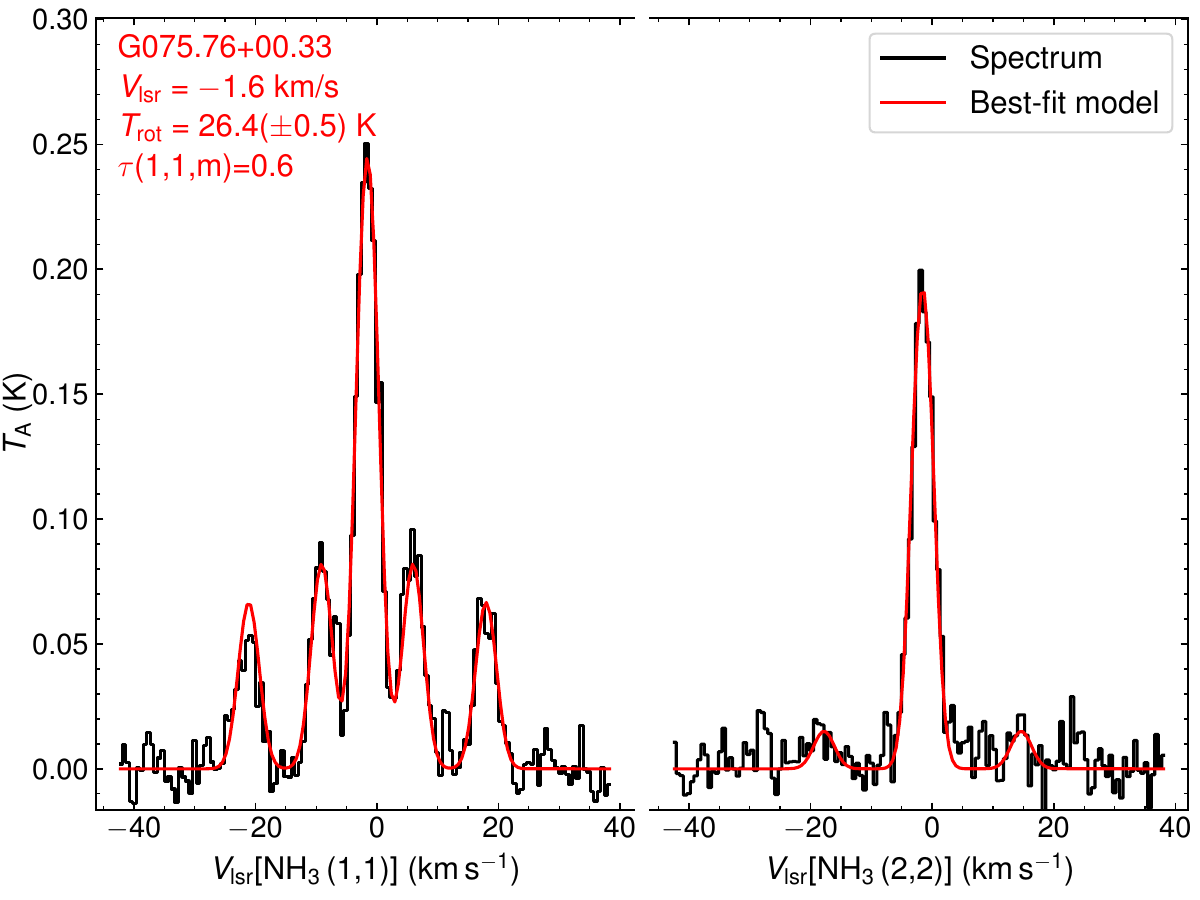}{0.3\textwidth}{}
}
\gridline{
\fig{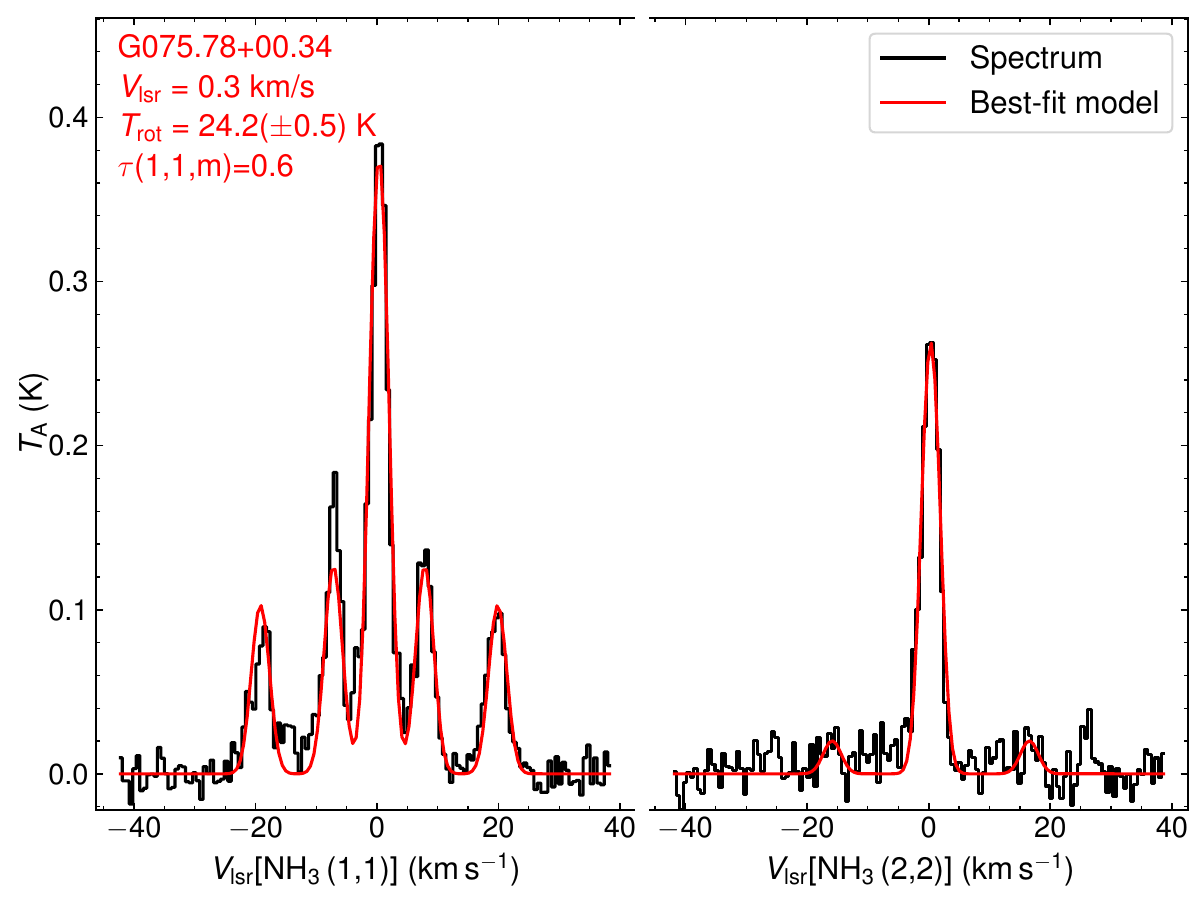}{0.3\textwidth}{}
\fig{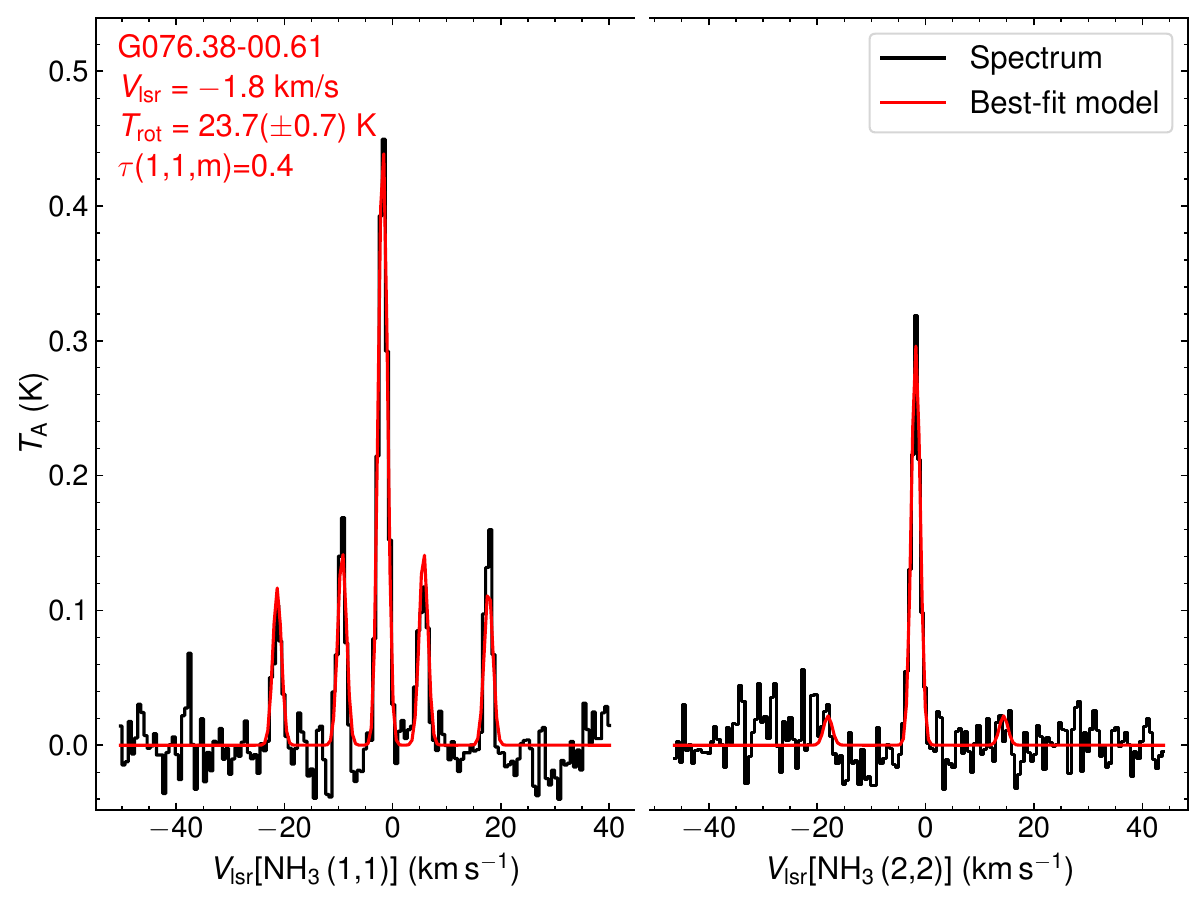}{0.3\textwidth}{}
\fig{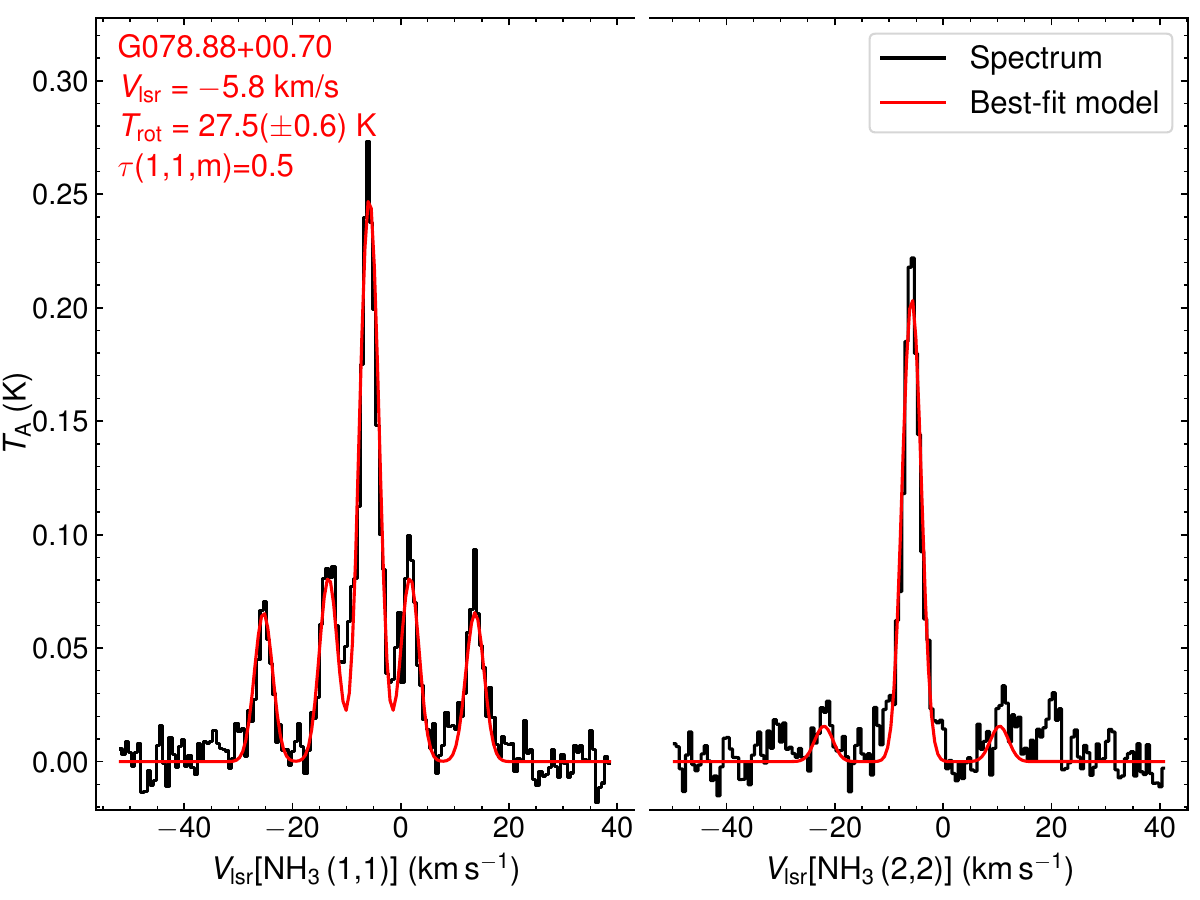}{0.3\textwidth}{}
}
\gridline{
\fig{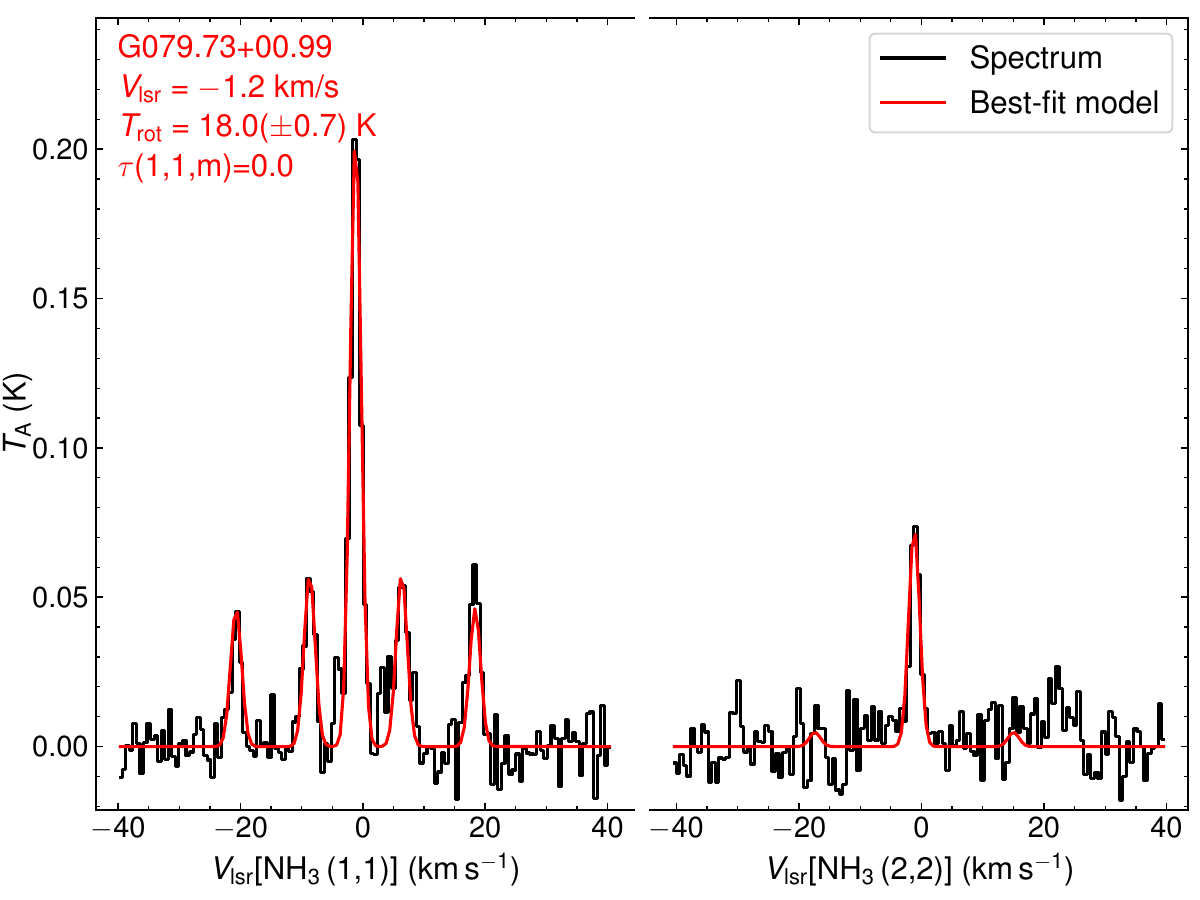}{0.3\textwidth}{}
\fig{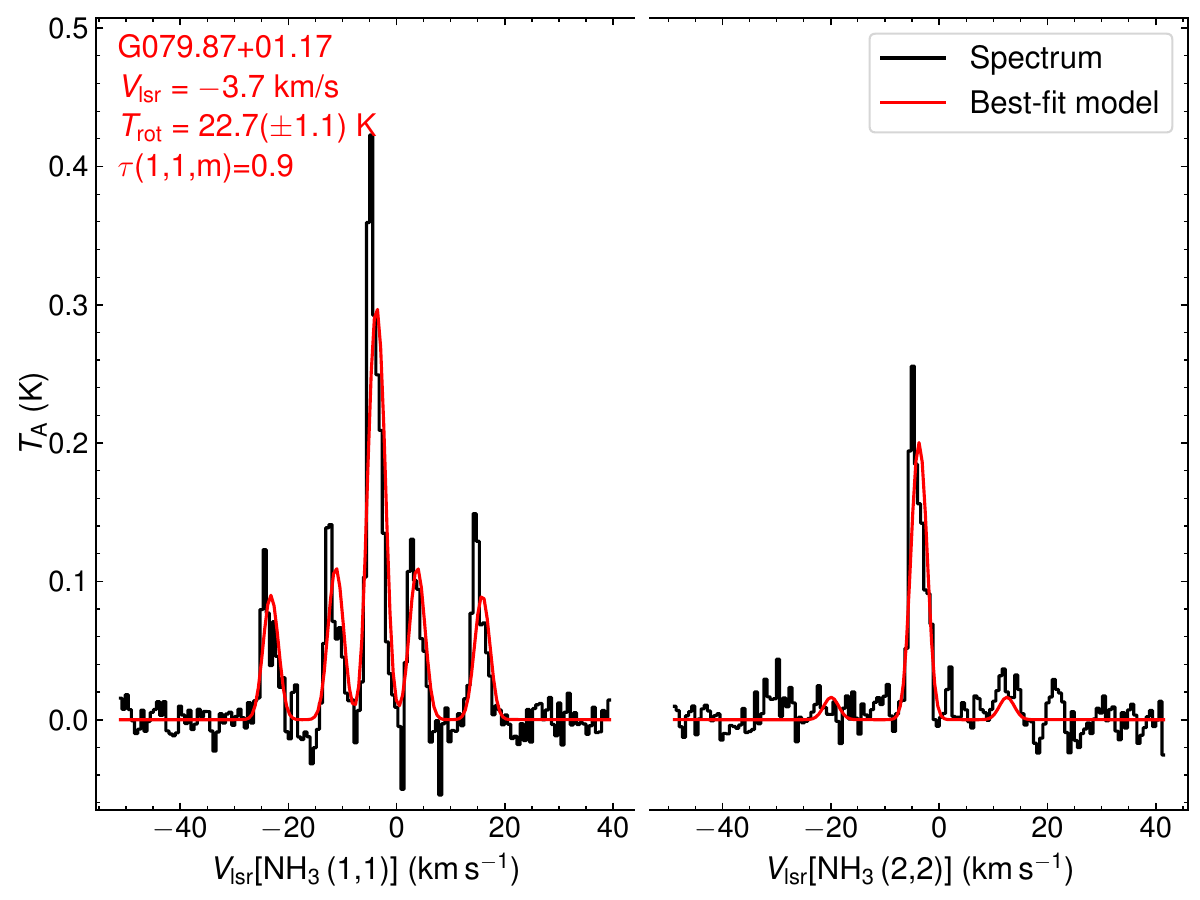}{0.3\textwidth}{}
\fig{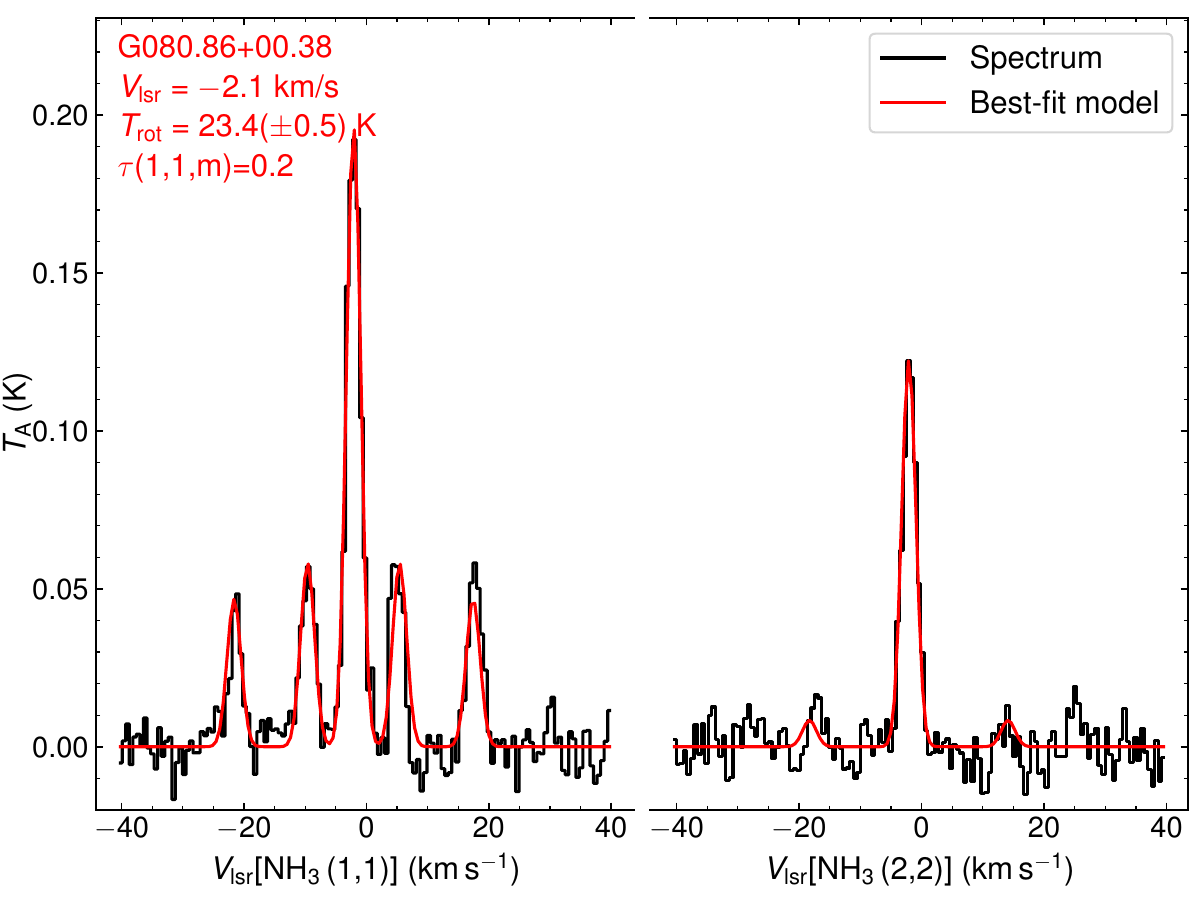}{0.3\textwidth}{}
}
\caption{NH$_3$ spectra toward the sample sources (2/2). The observed spectra of NH$_3$(1,1) and NH$_3$(2,2) are in black, while the best fit from the code of \cite{2015ApJ...805..171L} is in red.}
\label{appendix: NH3_fitting_2}
\end{figure*}

\begin{figure*}[ht!]
\centering
\gridline{
\fig{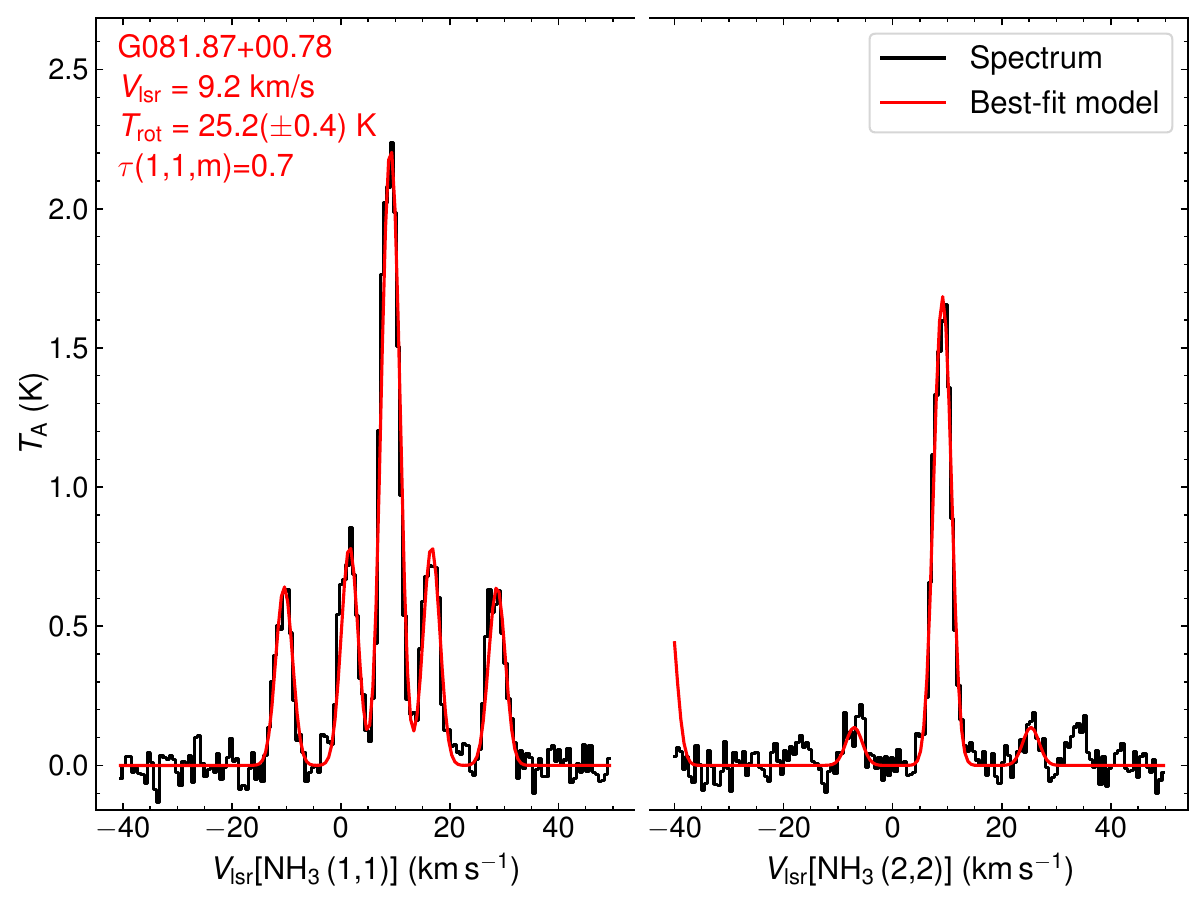}{0.3\textwidth}{}
\fig{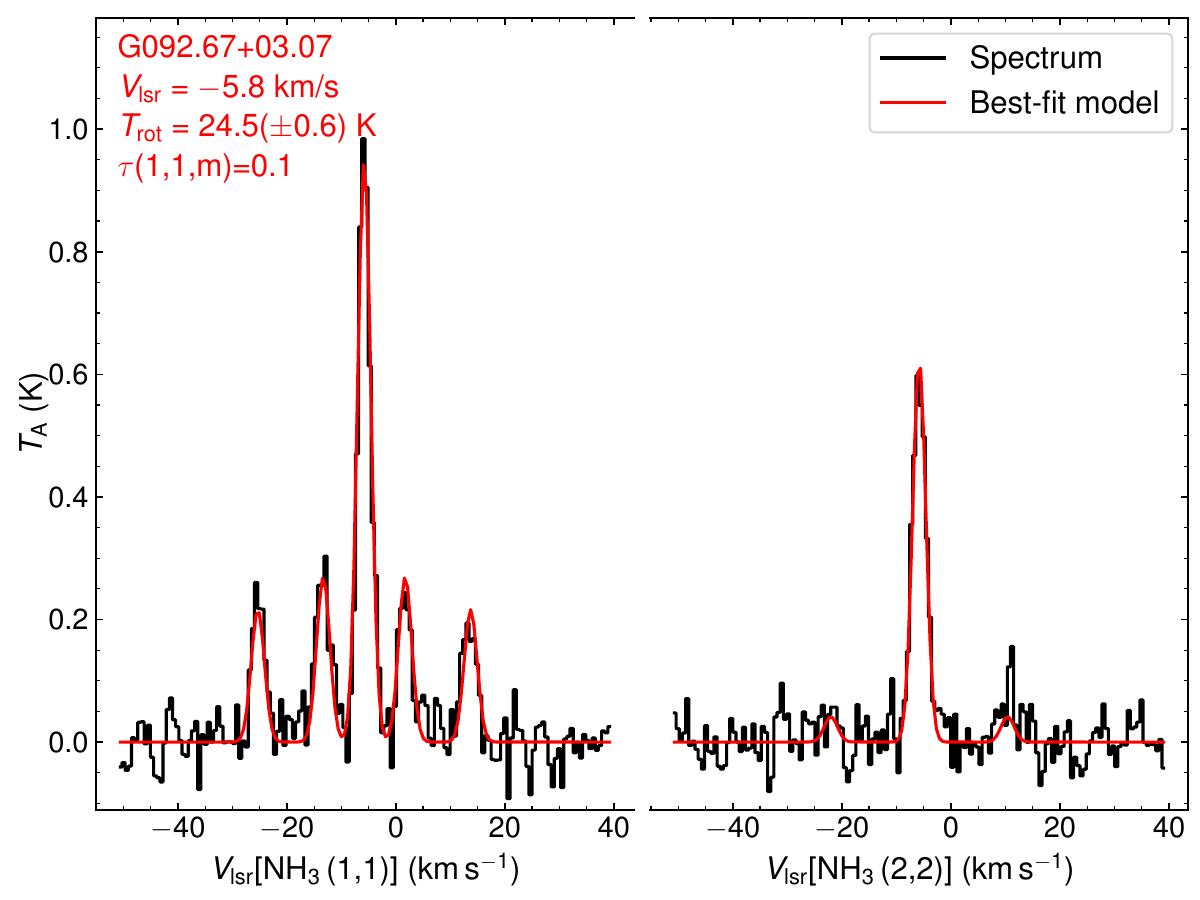}{0.3\textwidth}{}
\fig{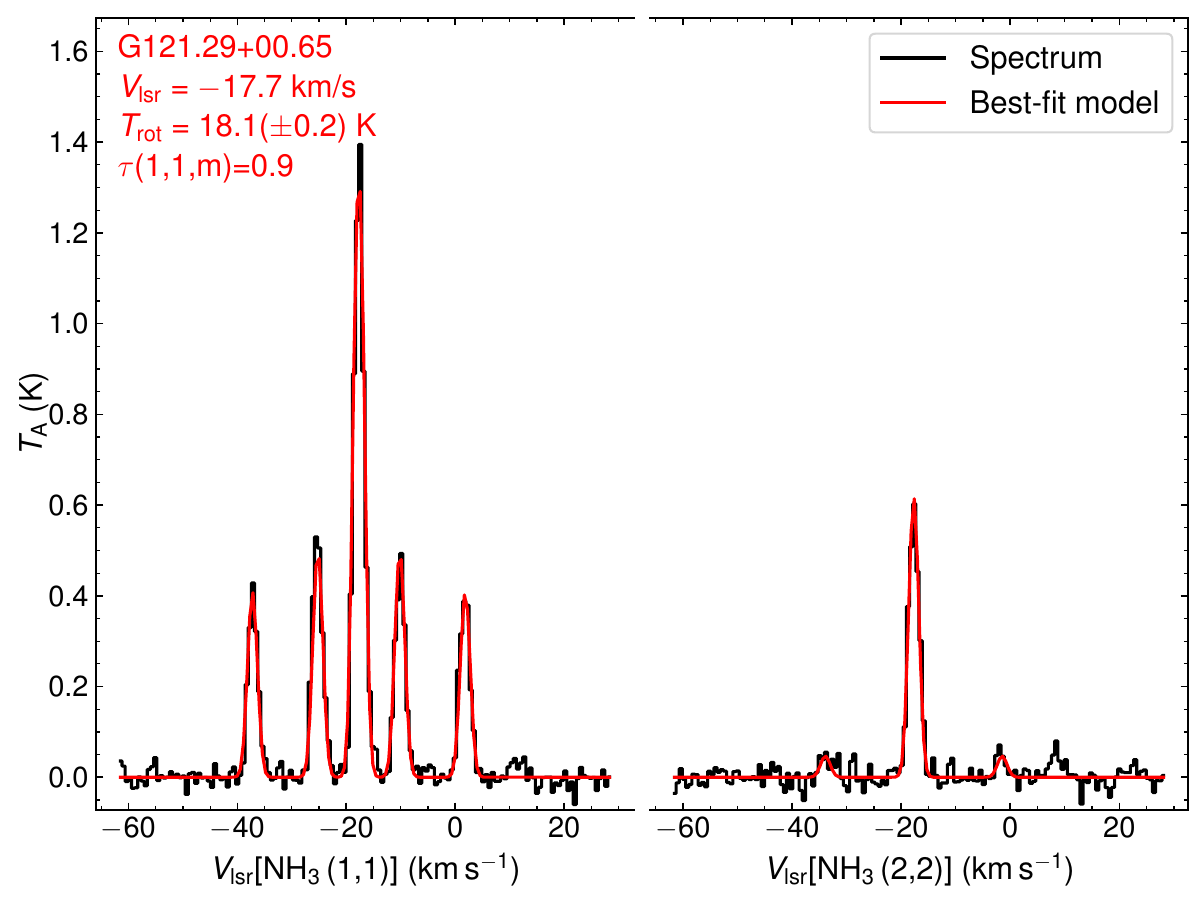}{0.3\textwidth}{}
}
\gridline{
\fig{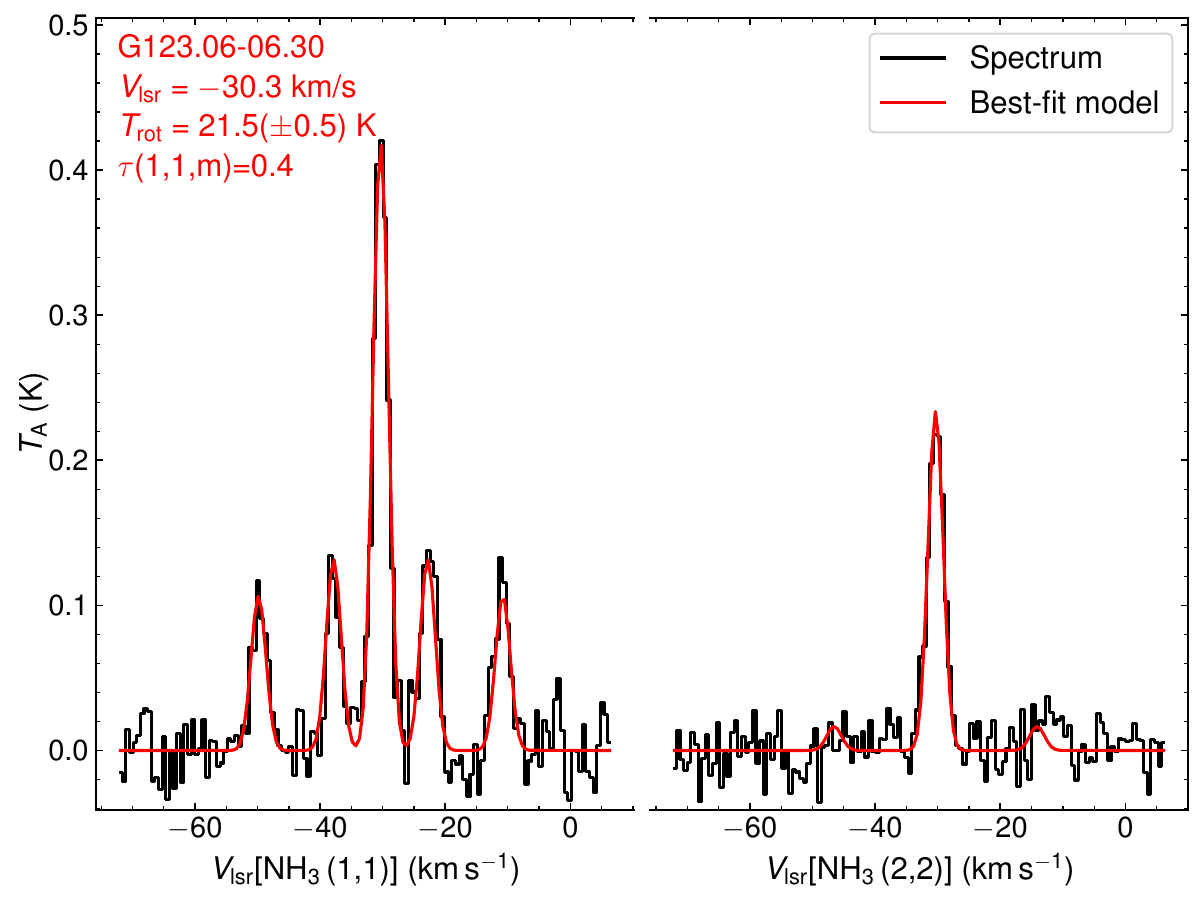}{0.3\textwidth}{}
\fig{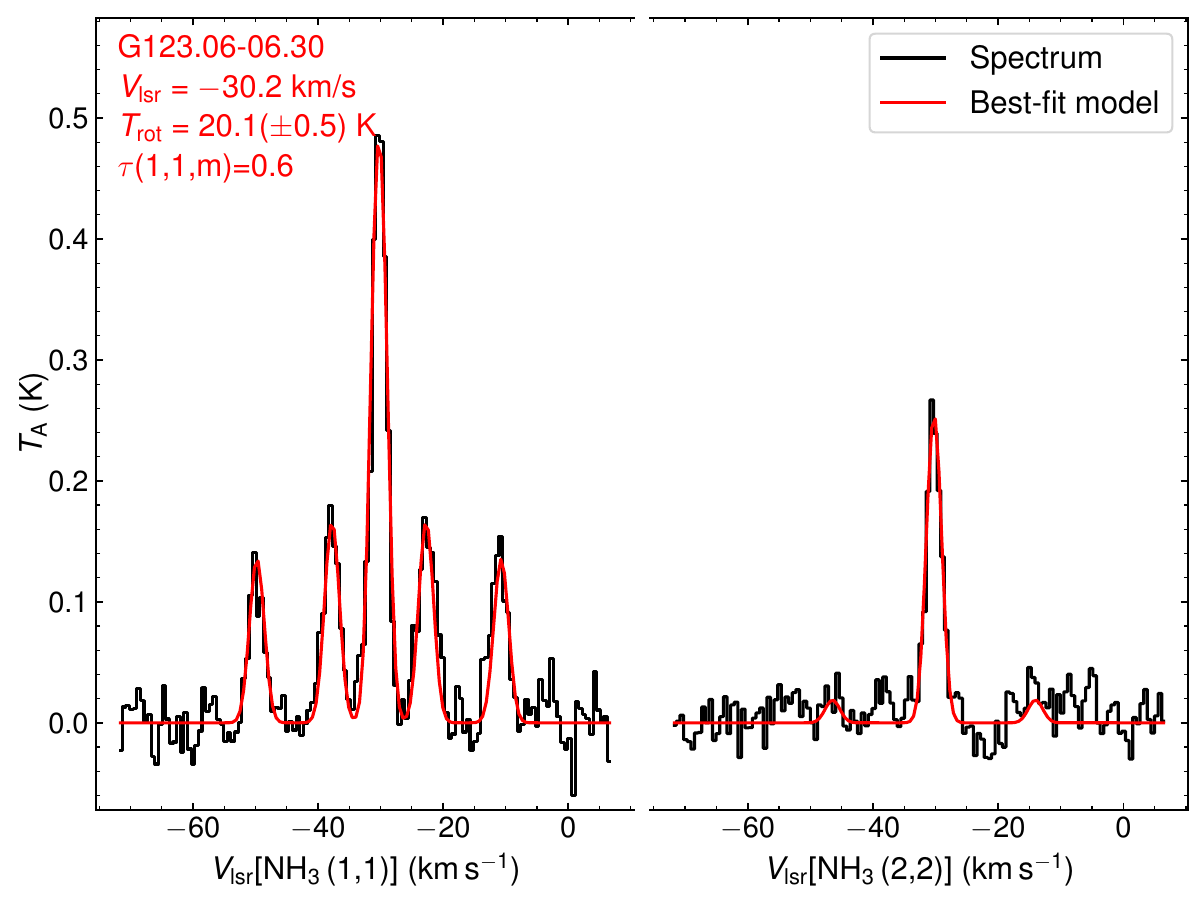}{0.3\textwidth}{}
\fig{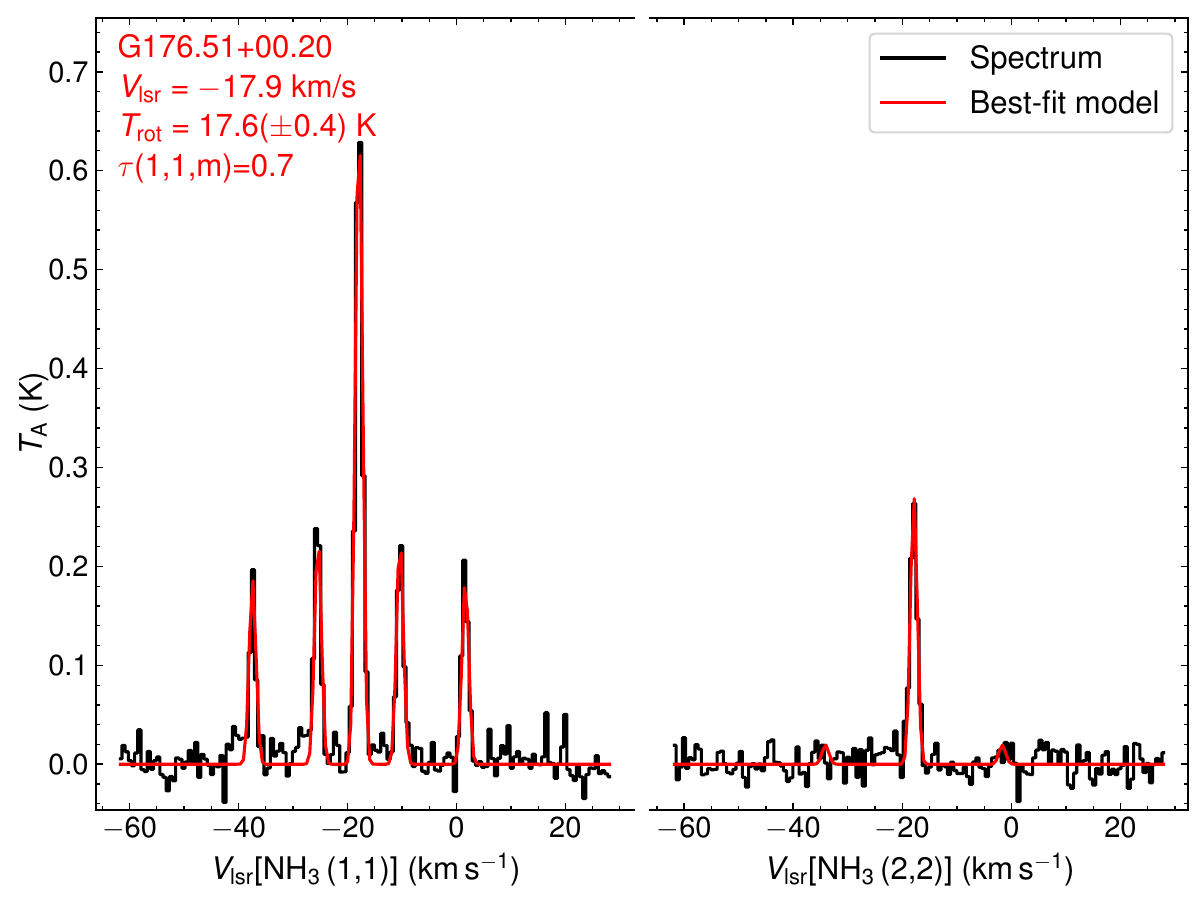}{0.3\textwidth}{}
}
\gridline{
\fig{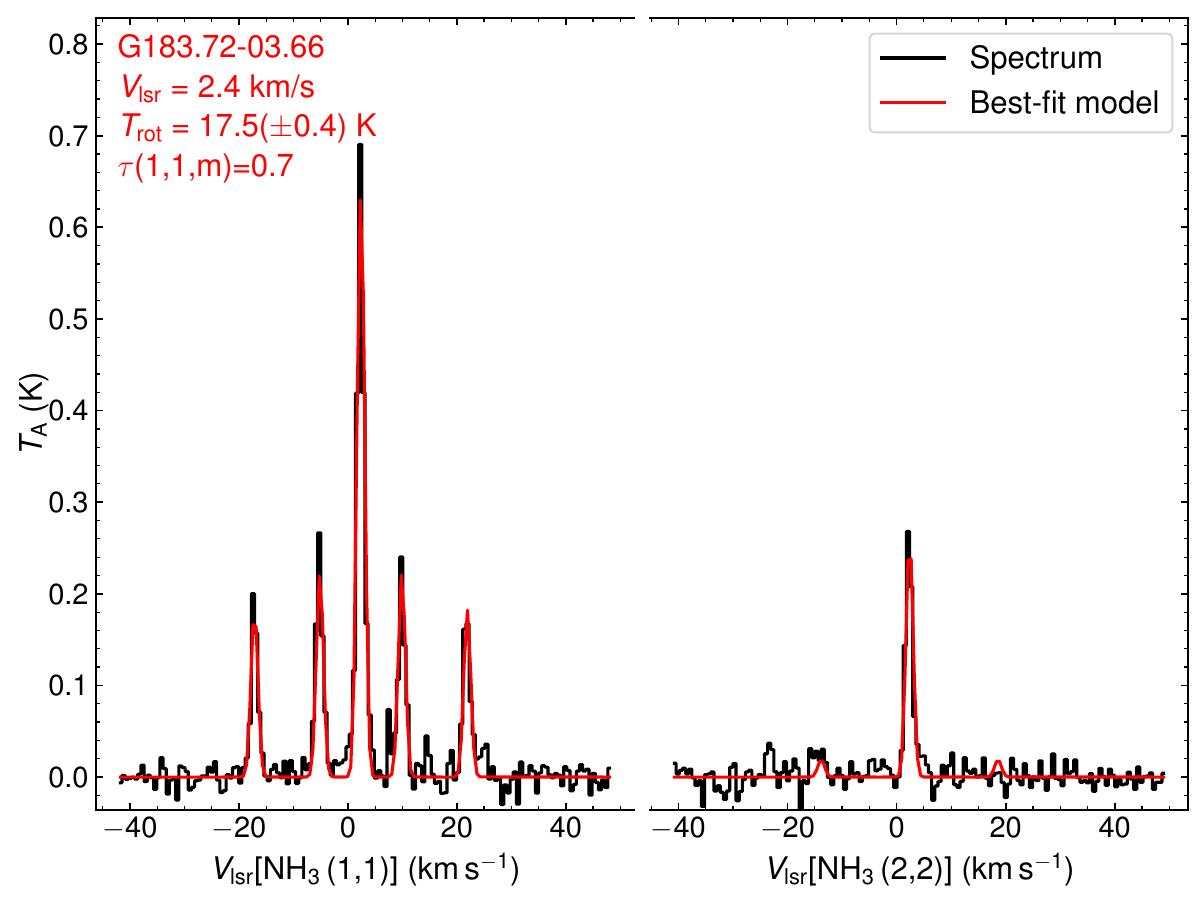}{0.3\textwidth}{}
\fig{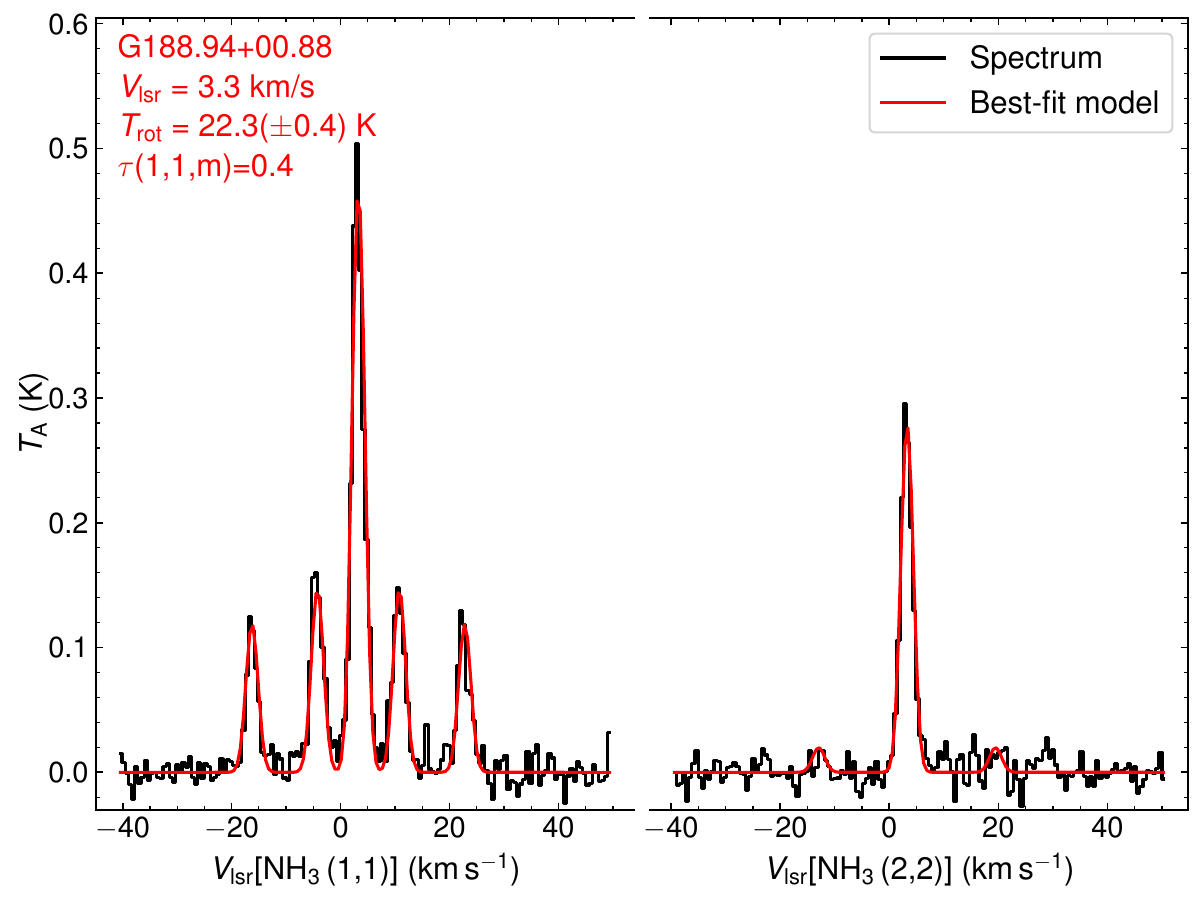}{0.3\textwidth}{}
\fig{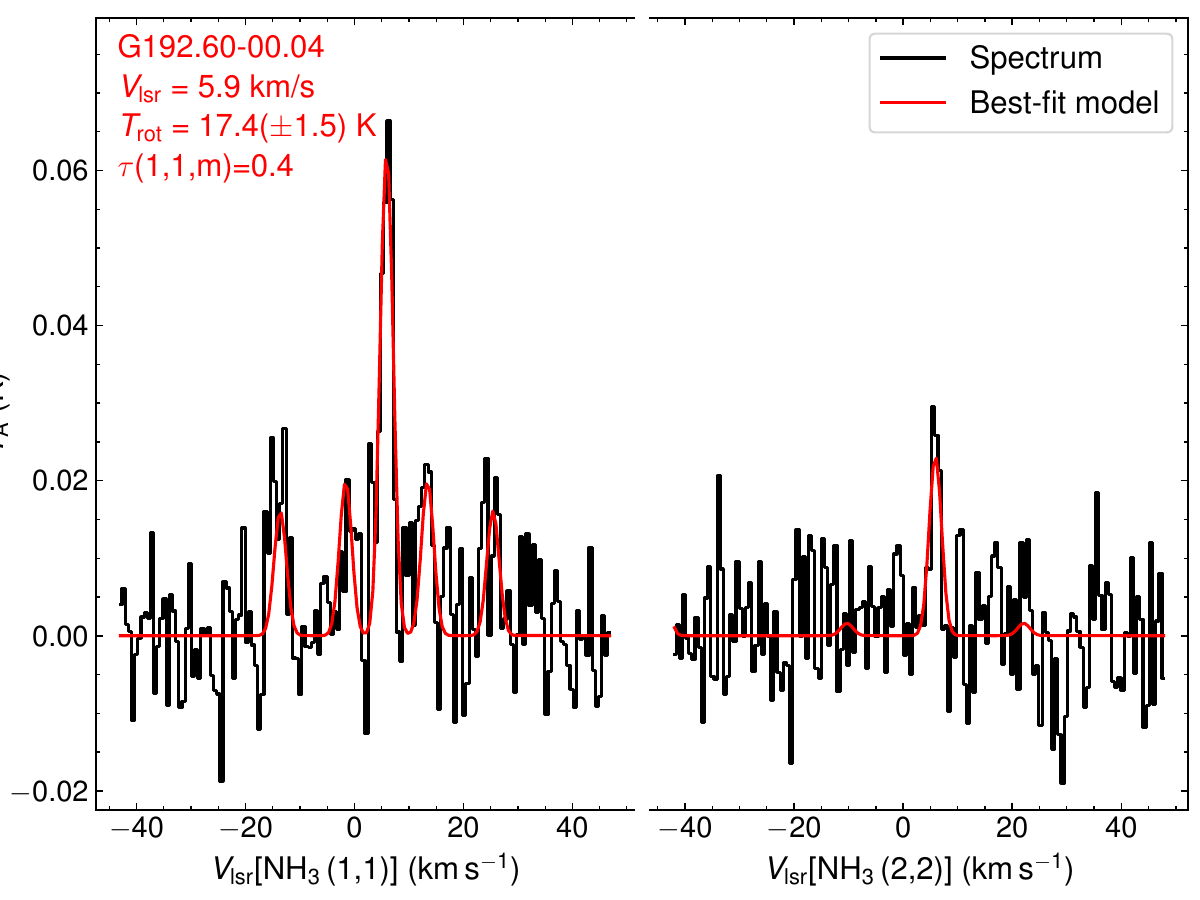}{0.3\textwidth}{}
}
\caption{NH$_3$ spectra toward the sample sources (3/3). The observed spectra of NH$_3$(1,1) and NH$_3$(2,2) are in black, while the best fit from the code of \cite{2015ApJ...805..171L} is in red.}
\label{appendix: NH3_fitting_3}
\end{figure*}

\clearpage
\newpage
\section{Checking the CH$_3$CCH fitting Program}
We used the SciPy Python package to fit the CH$_3$CCH $J$=5-4 spectra and derive the $T_{\rm rot}$(CH$_3$CCH). To ensure the reliability of derived rotation temperatures, we conducted the following tests, which are summarized below.

\textit{Comparison between fits with and without including the $K$=3 transition.} We fit the CH$_3$CCH $J$=5-4 using $K$=0,1,2 transitions in our targets, because the $K$=3 transition was not detected in some sources. To assess whether the inclusion of the $K$=3 transition significantly affects the derived $T_{\rm rot}$, we selected three sources where the $K$=3 line was clearly detected and compared the fitting results with and without this transition. As shown in Figure \ref{transitions}, the $T_{\rm rot}$ derived with and without $K$=3 are generally consistent, except for G005.88-00.39. This target has been reported to show a strong H$^{13}$CN 1-0 emission ($\int T_{\rm mb}$d$\nu\ge$40 K km s$^{-1}$, \citet{2024MNRAS.527.5049L}) with IRAM 30m observations, indicating that the gas density is extremely high. In this target, CH$_3$CCH is more likely to be optically thick, especially in the low $K$ transitions. Therefore, this discrepancy is likely due to the high column density and volume density in G005.88-00.39, which may cause the lower-$K$ transitions ($K$=0 and 1) to become optically thick. 
This optical depth effect flattens the population distribution of the low-$K$ levels and consequently leads to an overestimated $T_{\rm rot}$(CH$_3$CCH) \citep{1999ApJ...517..209G}.

\textit{Comparison between the rotational diagram method and Python fitting results.} The rotational diagram method is often used to derive the $T_{\rm rot}$ from multiple transitions. Our Python program follows a similar theoretical framework with the rotational diagram method to derive $T_{\rm rot}$. To validate the reliability of our code, we compared the $T_{\rm rot}$ obtained from our Python fitting with those derived from the rotational diagram method (see Figure \ref{rotational diagram}). The two methods yield consistent $T_{\rm rot}$, confirming that the theoretical implementation in our code is accurate.

\textit{Is the simulated result for the non-detected line consistent with the noise level?} To test the reliability of our fitting results in cases where the CH$_3$CCH $J$=5-4 $K$=3 was not detected, we selected three sources in which CH$_3$CCH $J$=5–4 $K$=3 was not detected. If the simulated peak intensity of the $K$=3 transition is below the observed noise level, it confirms the correctness of the fitting result. As shown in Figure \ref{noise}, the simulated $K$=3 peaks lie below the corresponding 3$\sigma$ level in the observed spectra, supporting the validity of our fitting approach.

These tests confirm the internal consistency of our CH$_3$CCH fitting results.

\renewcommand{\thefigure}{D\arabic{figure}}  
\setcounter{figure}{0}

\begin{figure}[h]
\centering
\includegraphics[width=0.33\textwidth]{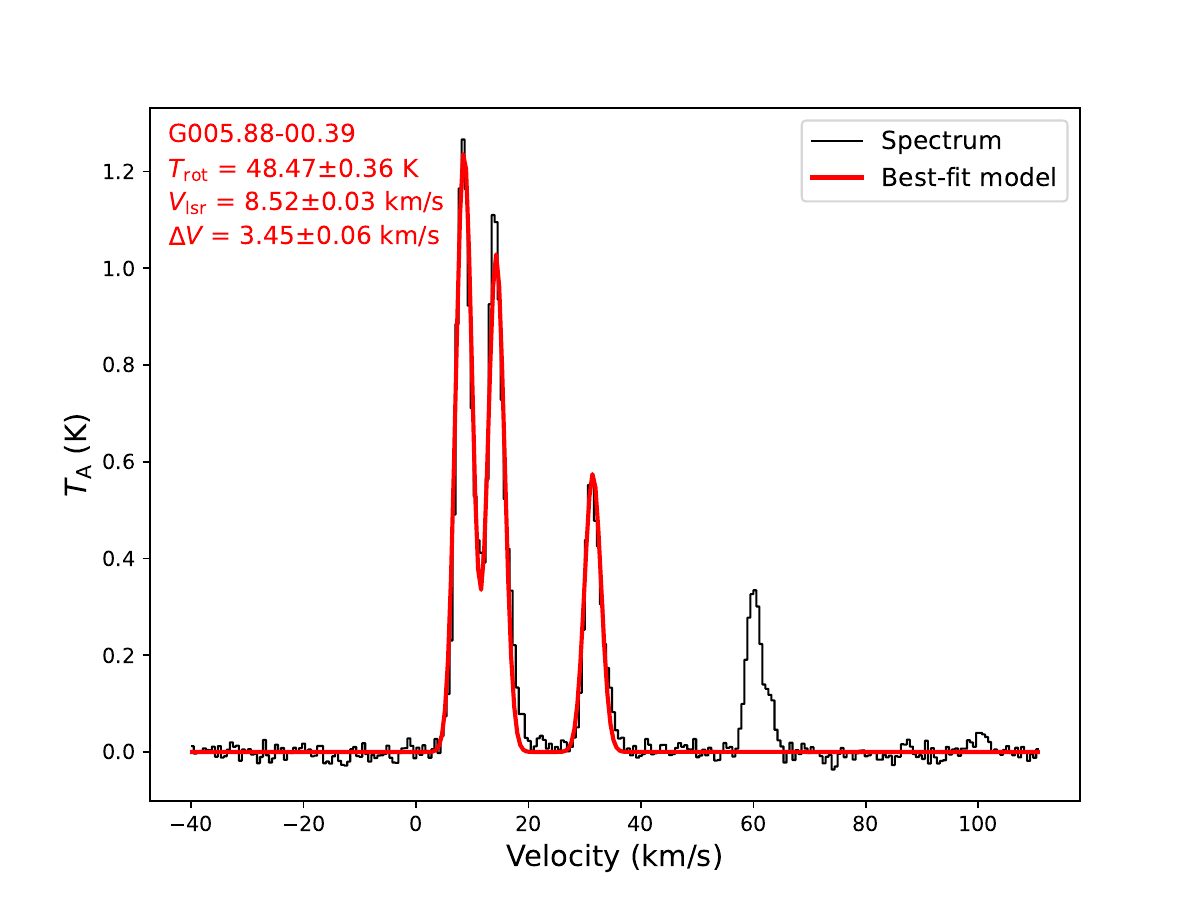}\includegraphics[width=0.33\textwidth]{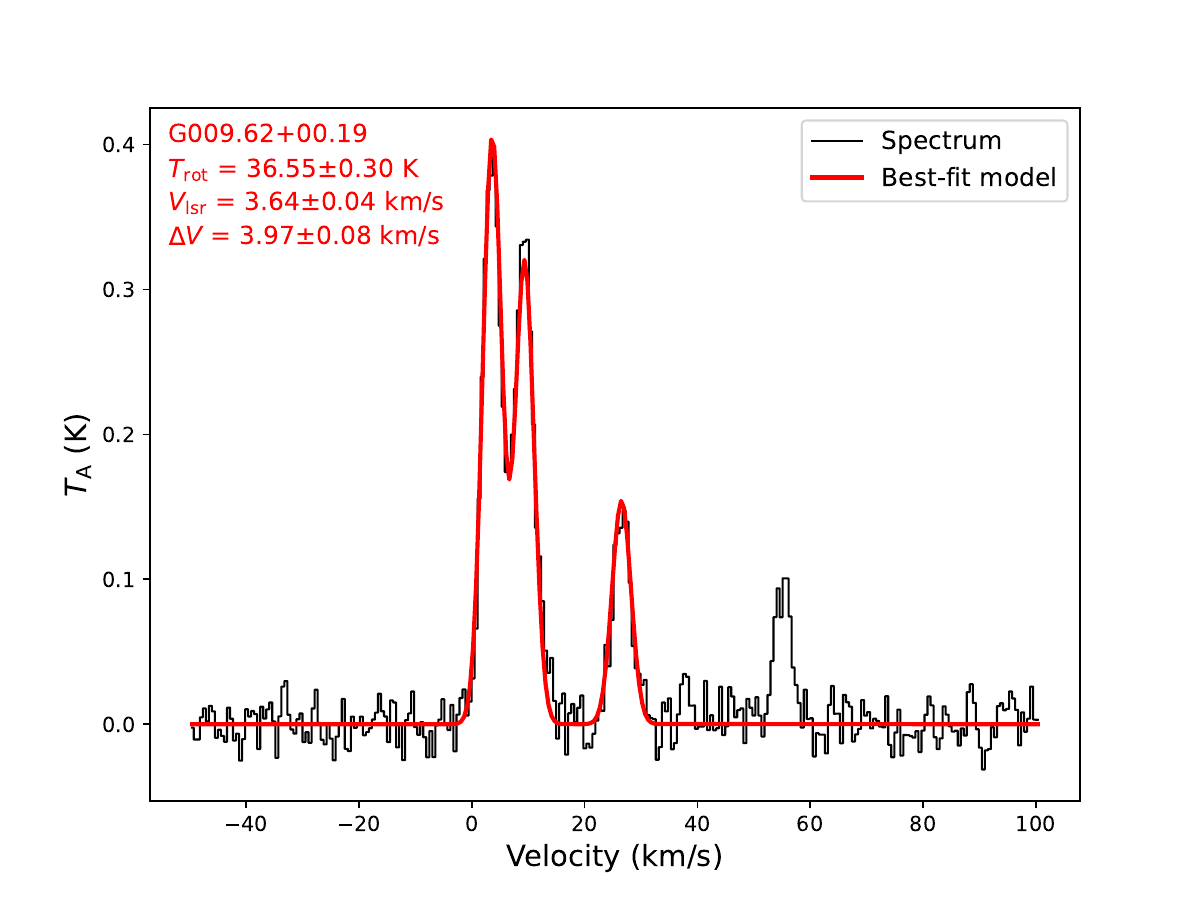}\includegraphics[width=0.33\textwidth]{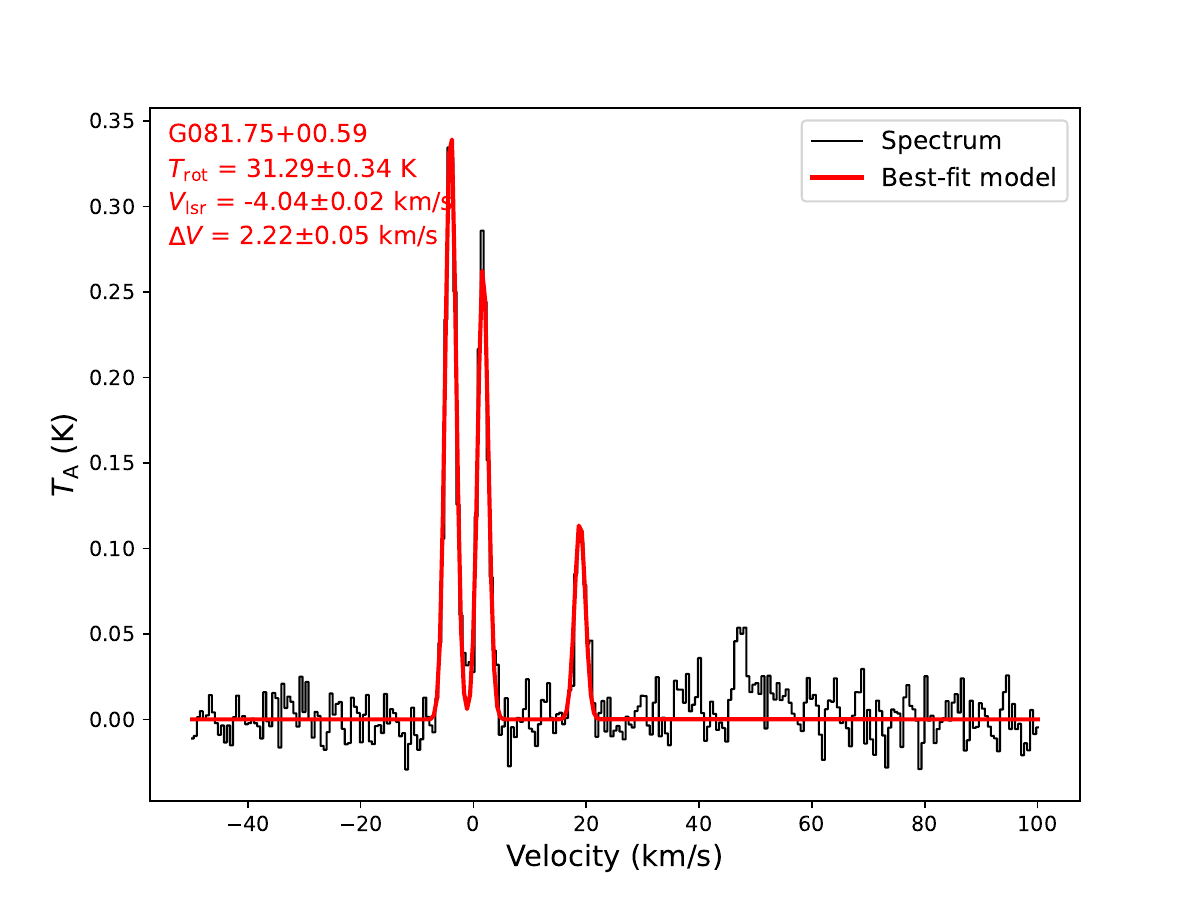}
\includegraphics[width=0.33\textwidth]{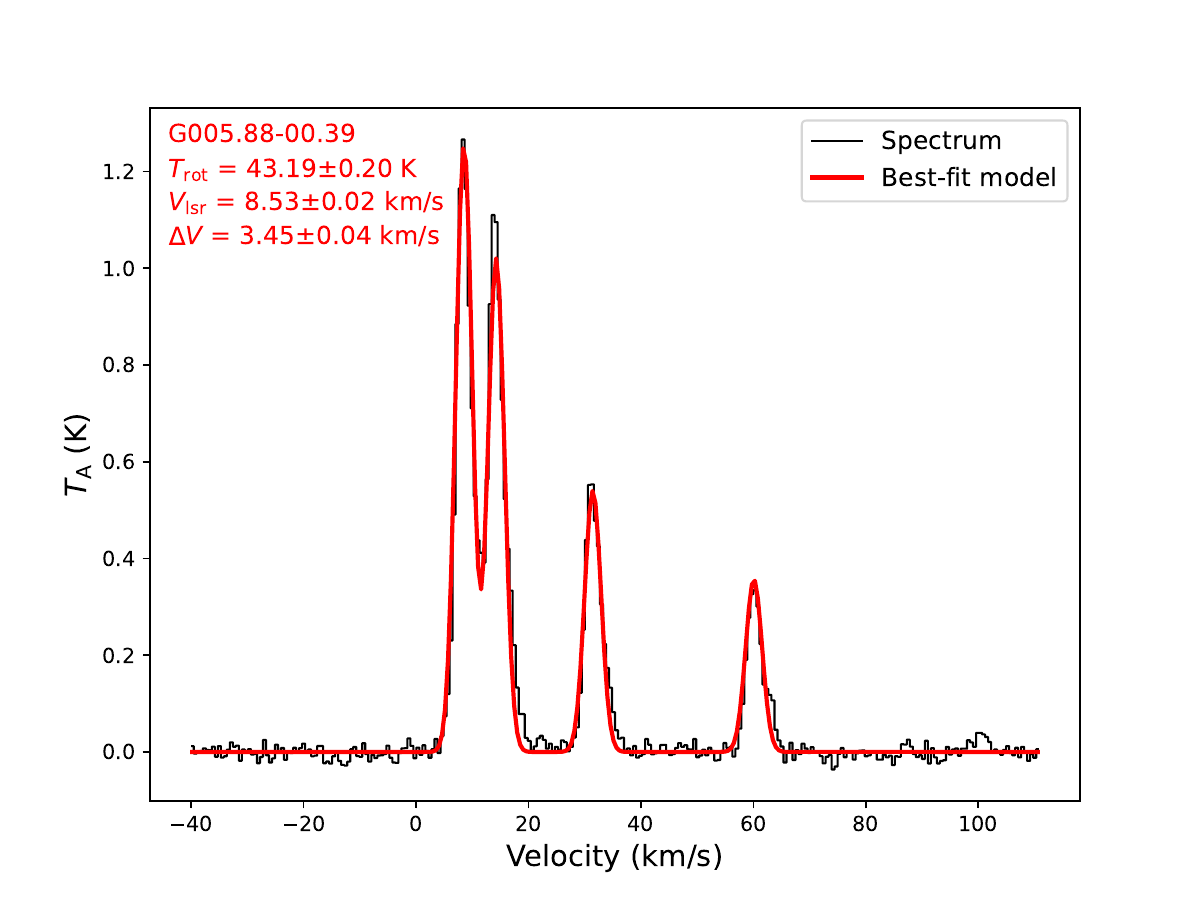}\includegraphics[width=0.33\textwidth]{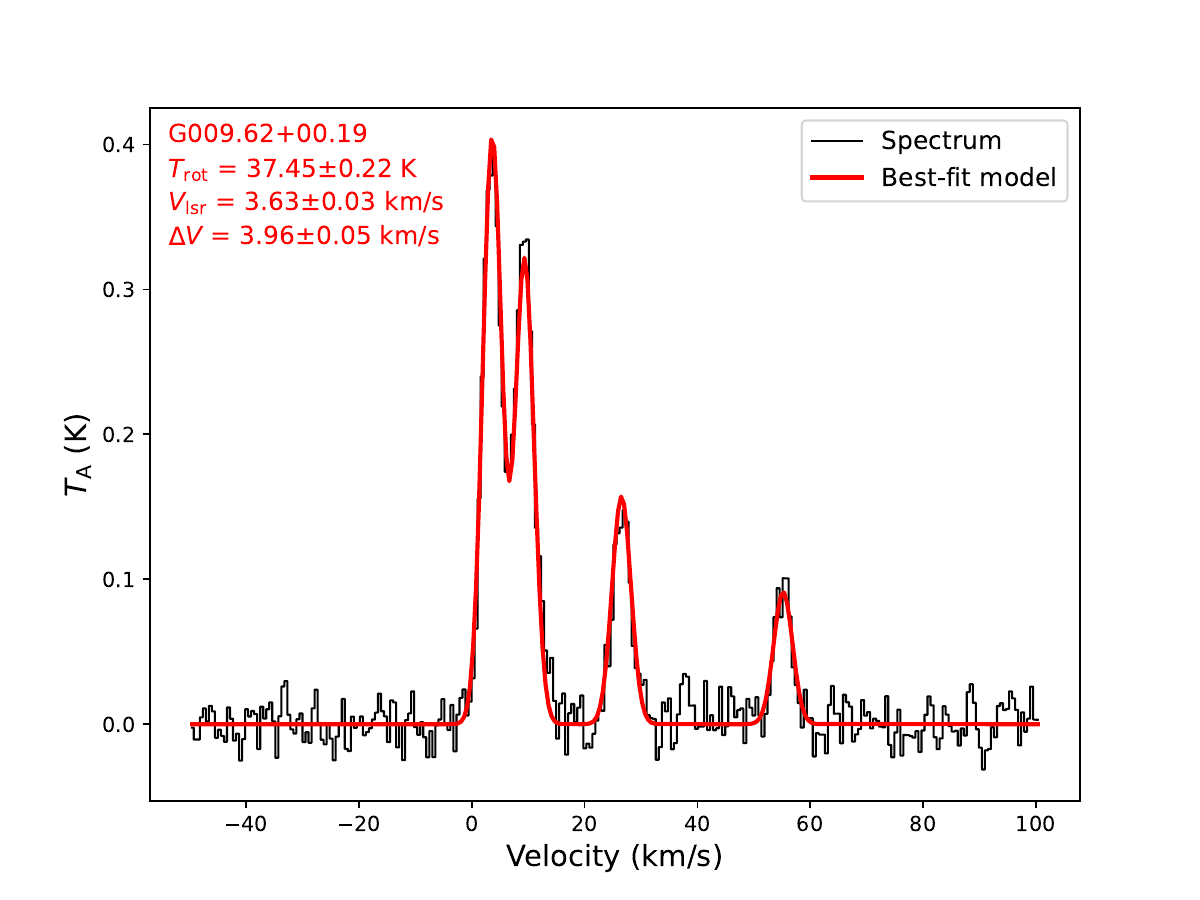}\includegraphics[width=0.33\textwidth]{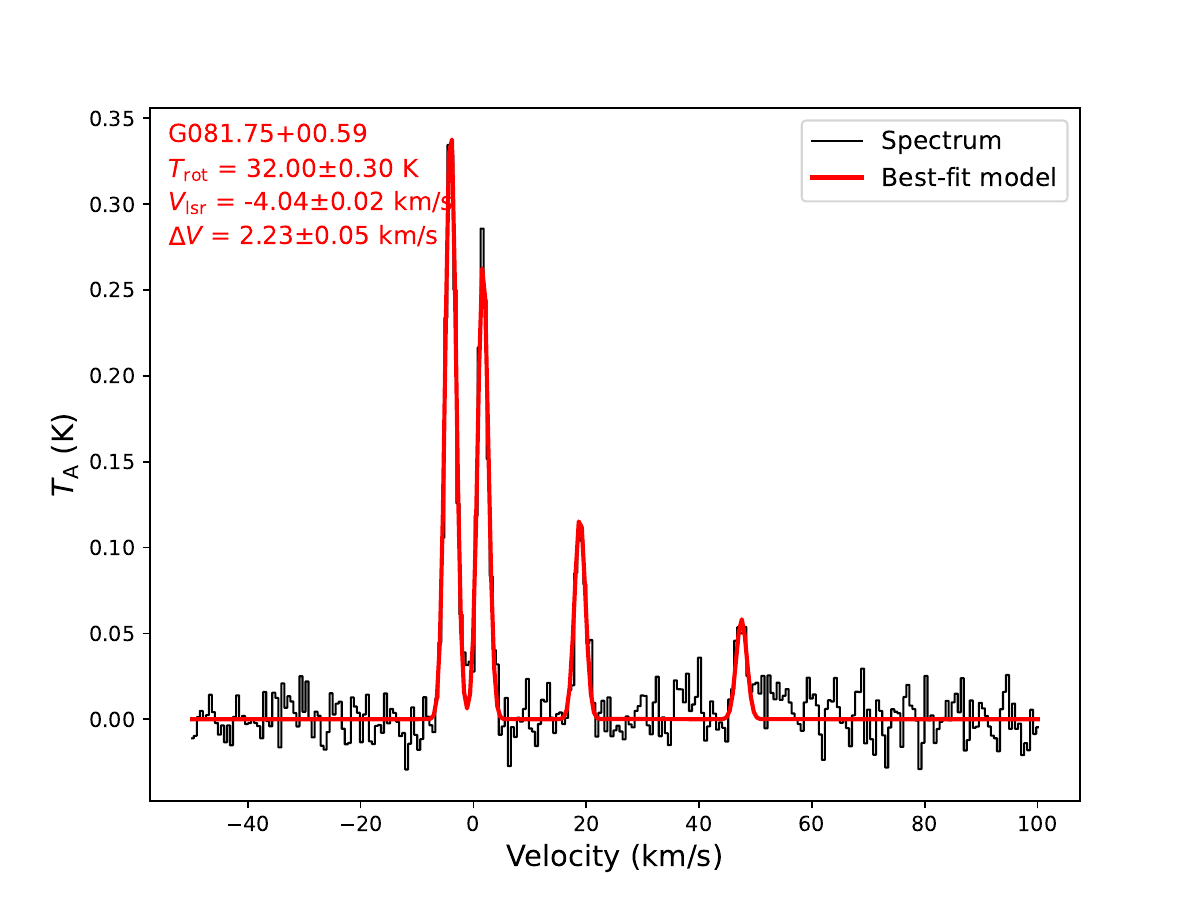}
\caption{The upper panel shows fitting results of CH$_3$CCH $J$=5-4 not considering $K$=3, while the bottom panel considers $K$=3.}
\label{transitions}
\end{figure}

\begin{figure}
\centering
\includegraphics[width=0.33\textwidth]{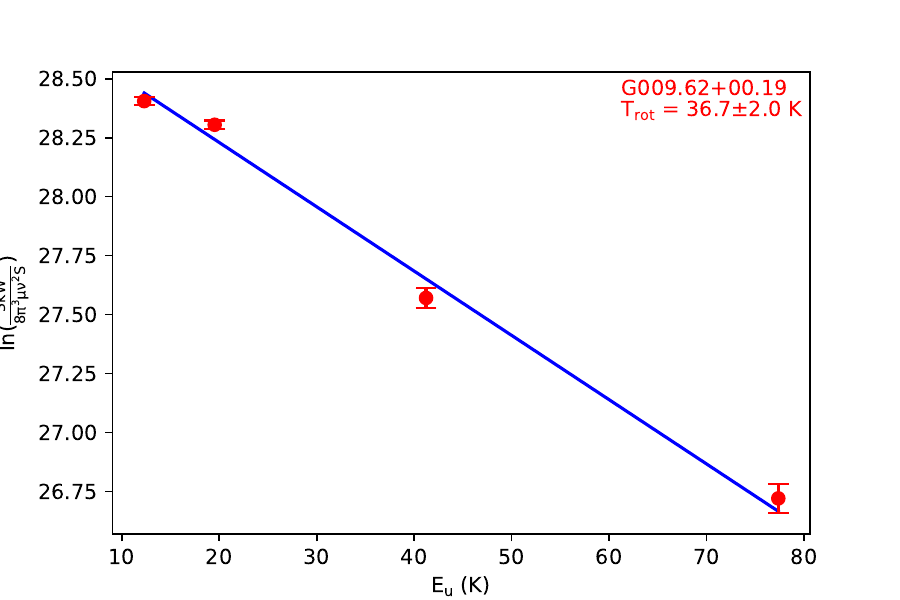}\includegraphics[width=0.33\textwidth]{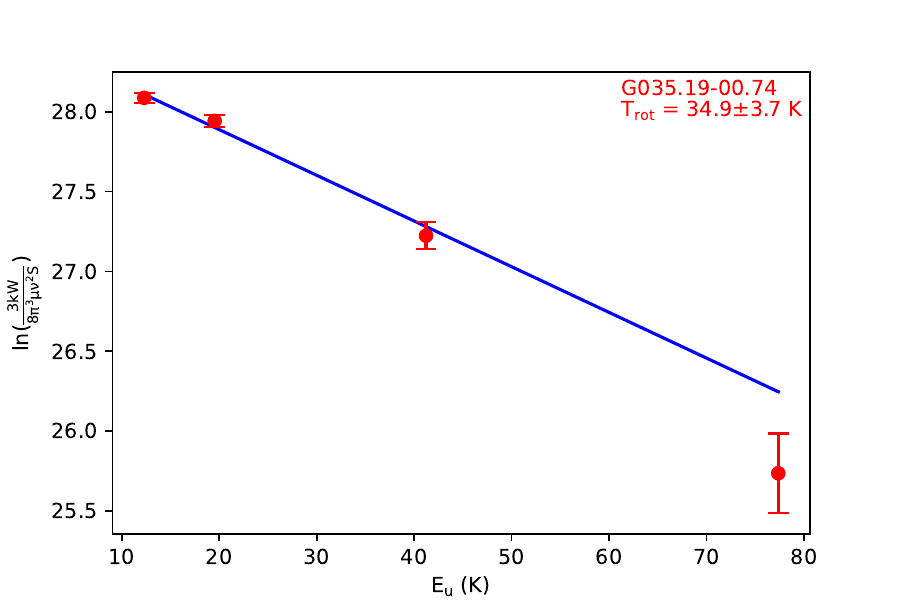}\includegraphics[width=0.33\textwidth]{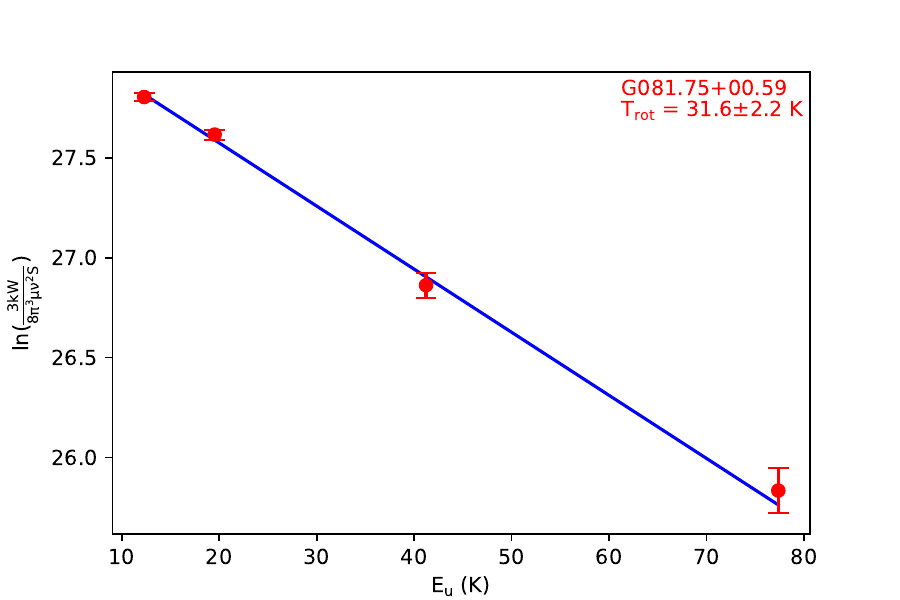}
\includegraphics[width=0.33\textwidth]{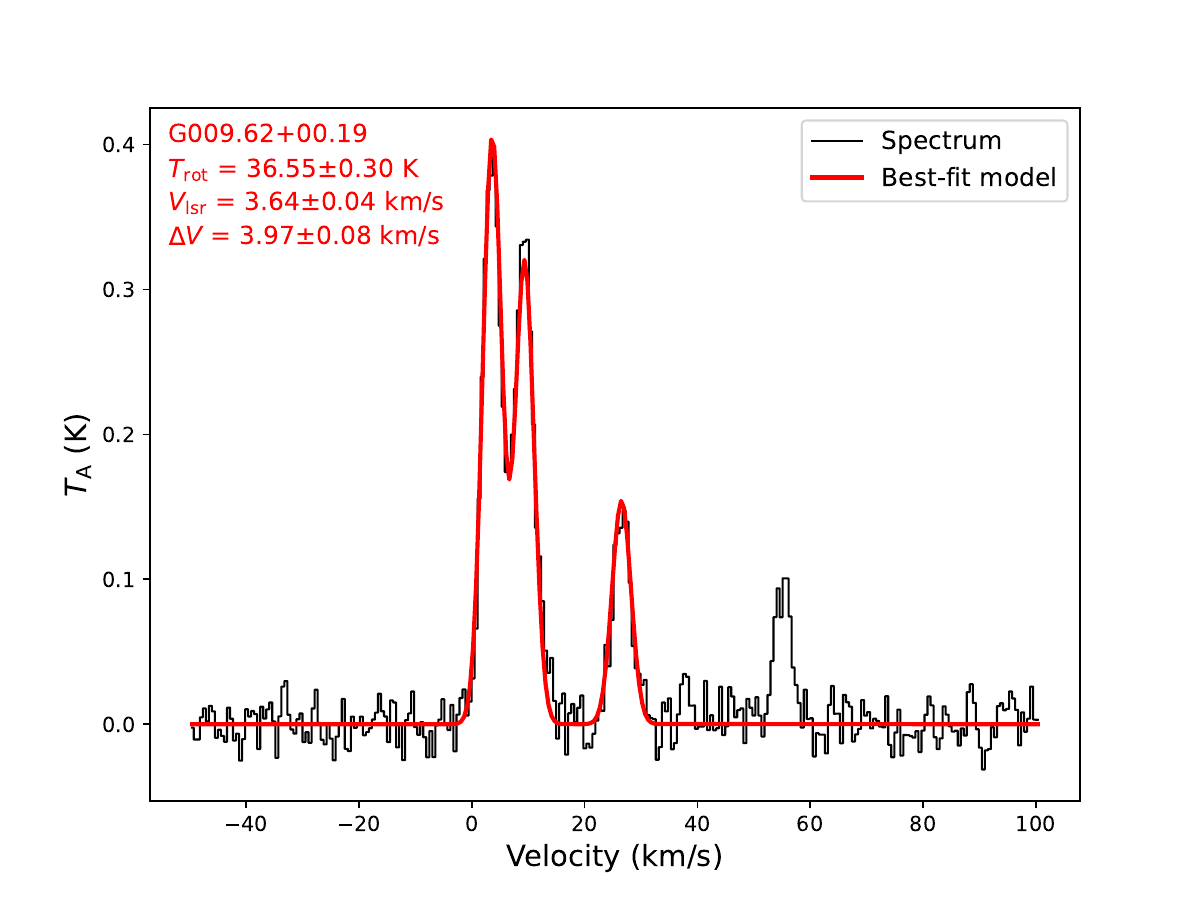}\includegraphics[width=0.33\textwidth]{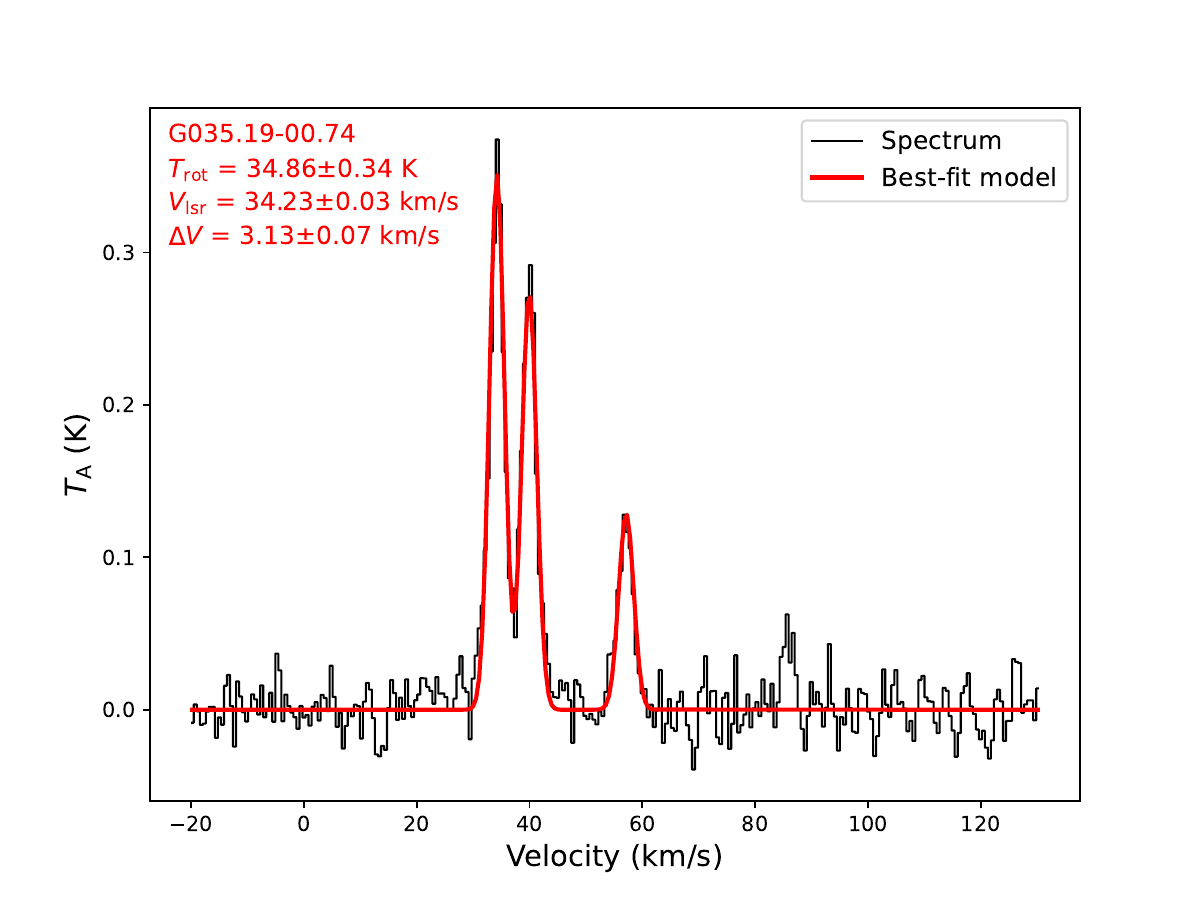}\includegraphics[width=0.33\textwidth]{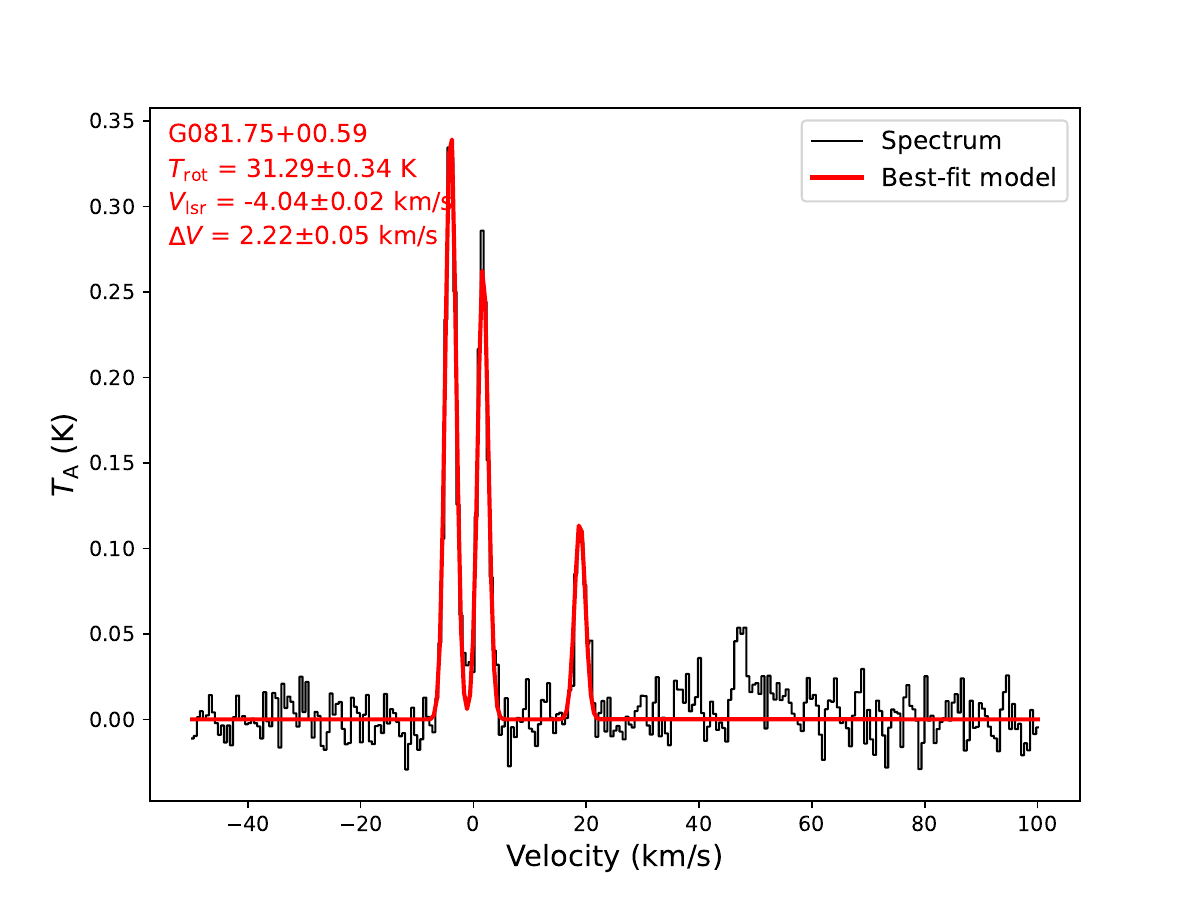}
\caption{The upper panel shows the rotation diagrams of CH$_3$CCH $J$=5-4. Although four $K$-ladder transitions are displayed in the rotation diagrams, only the $K$ = 0, 1, and 2 transitions were used to derive the rotation temperatures. In order to retain all the observational information and comparison, the $K$=3 line is also included. The intensities are given in $T_{\mathrm A}$, so the derived rotation temperatures are unchanged, but the column densities should be corrected by the main beam efficiency to obtain the true values. The bottom panel shows the fitting results of CH$_3$CCH $J$=5-4 corresponding to the upper panel source.}
\label{rotational diagram}
\end{figure}

\begin{figure}
\centering
\includegraphics[width=0.33\textwidth]{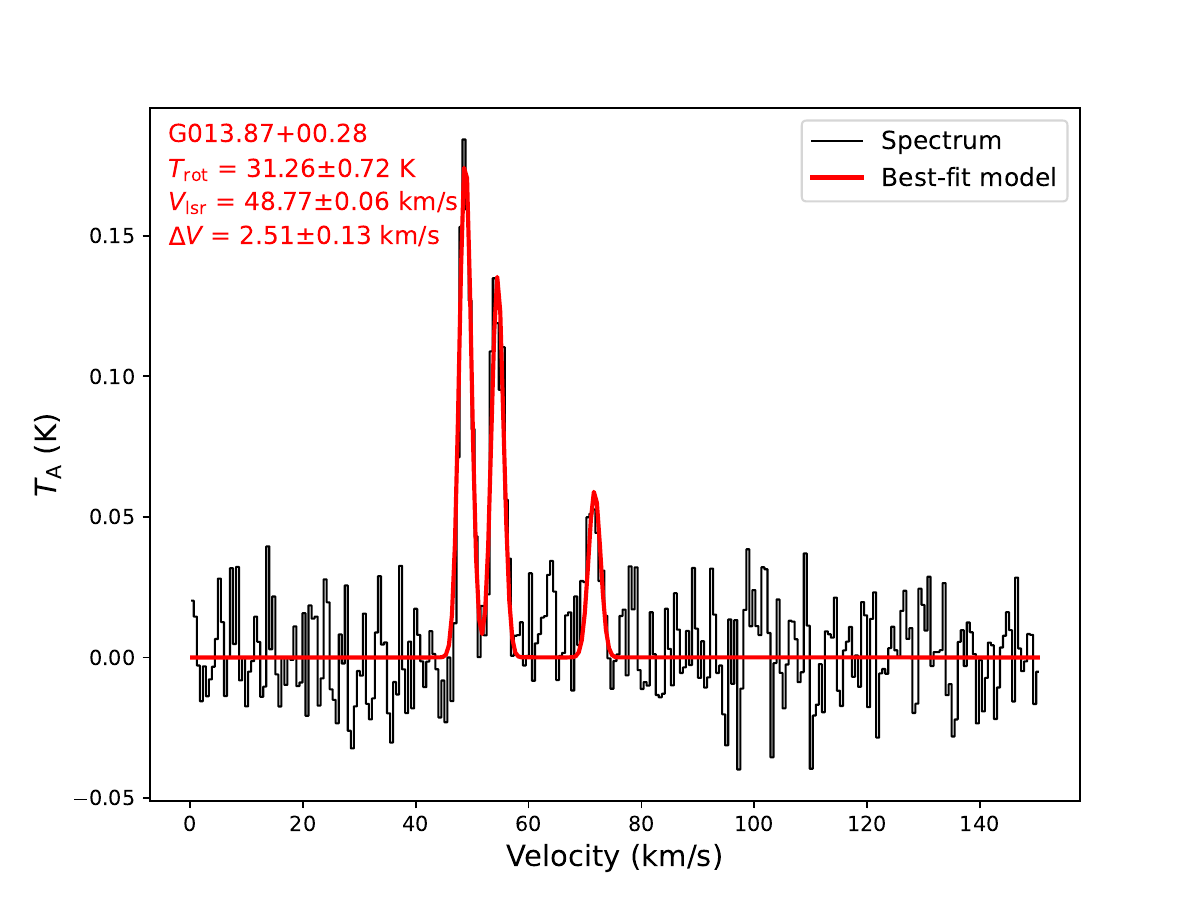}\includegraphics[width=0.33\textwidth]{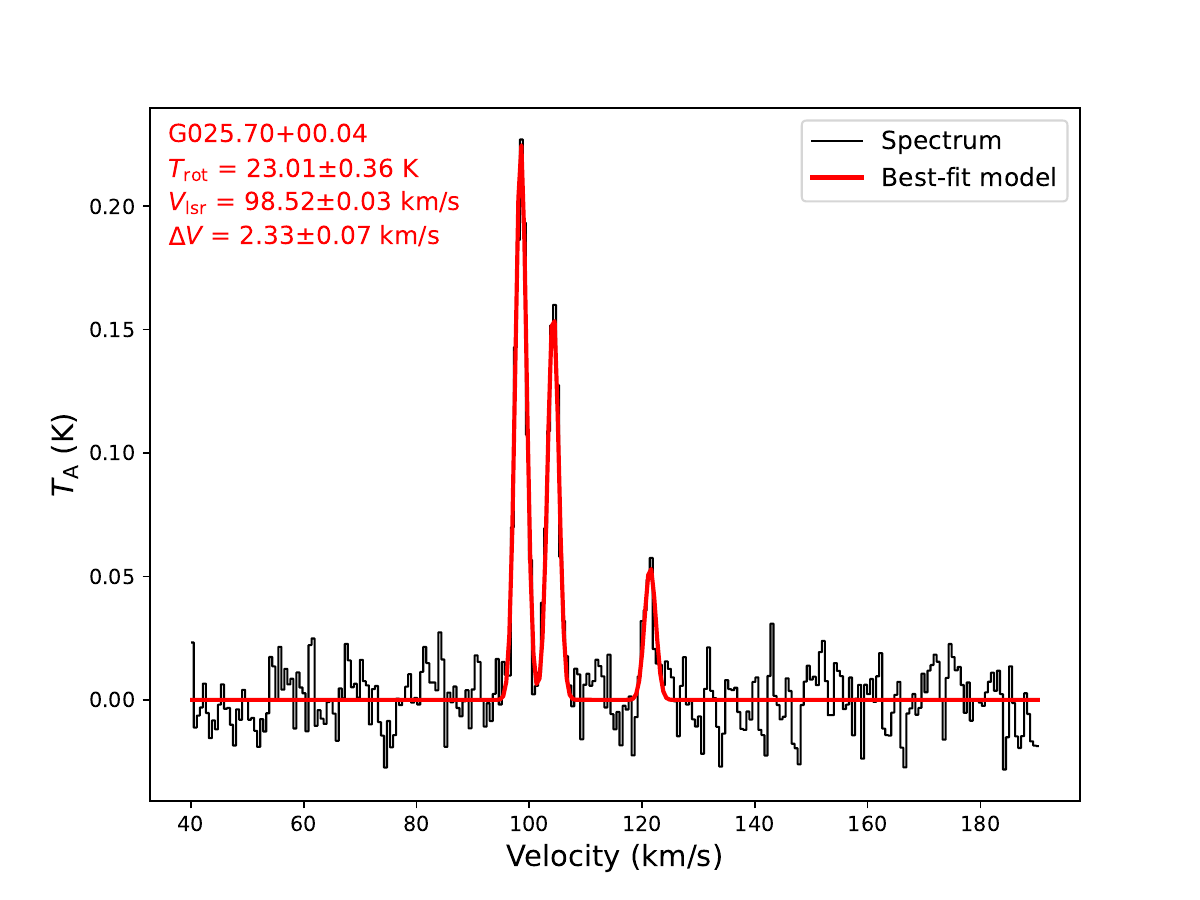}\includegraphics[width=0.33\textwidth]{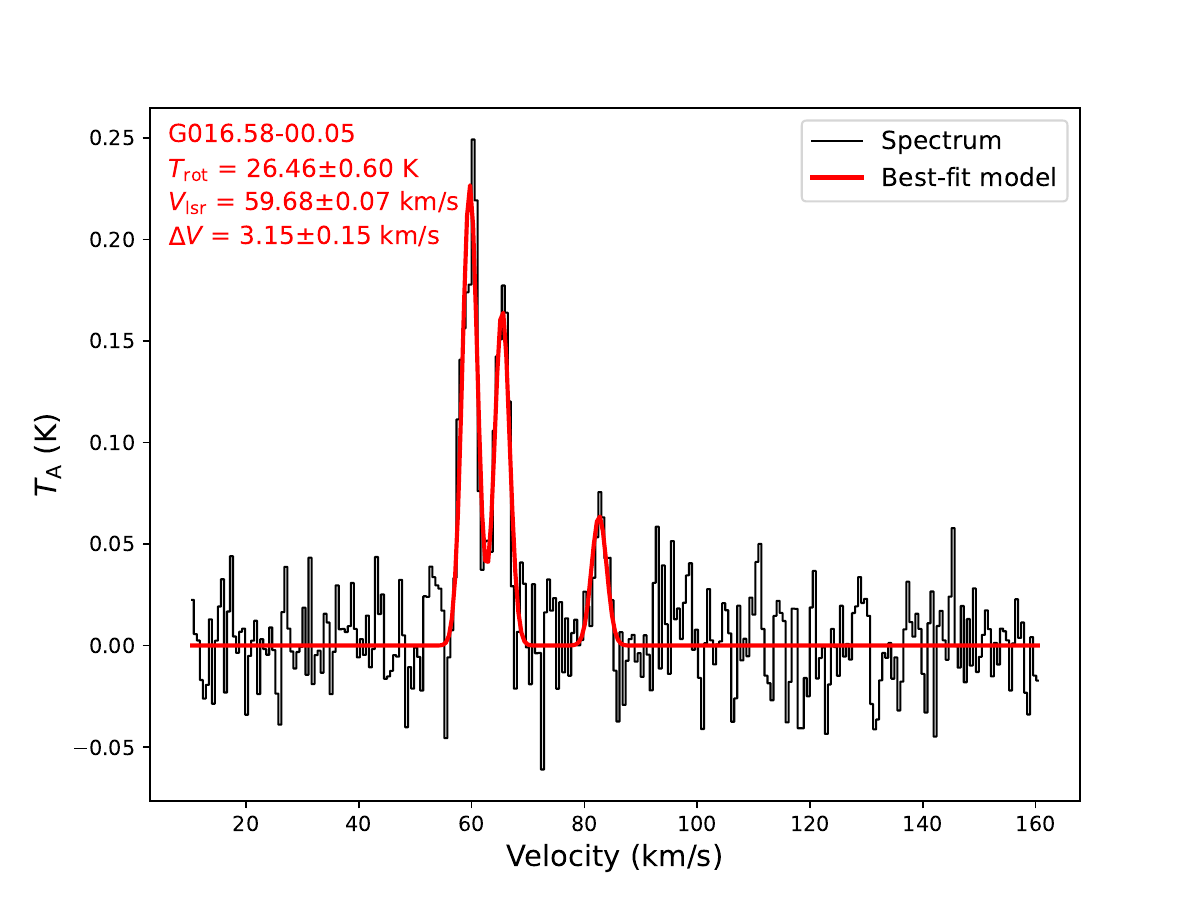}
\includegraphics[width=0.33\textwidth]{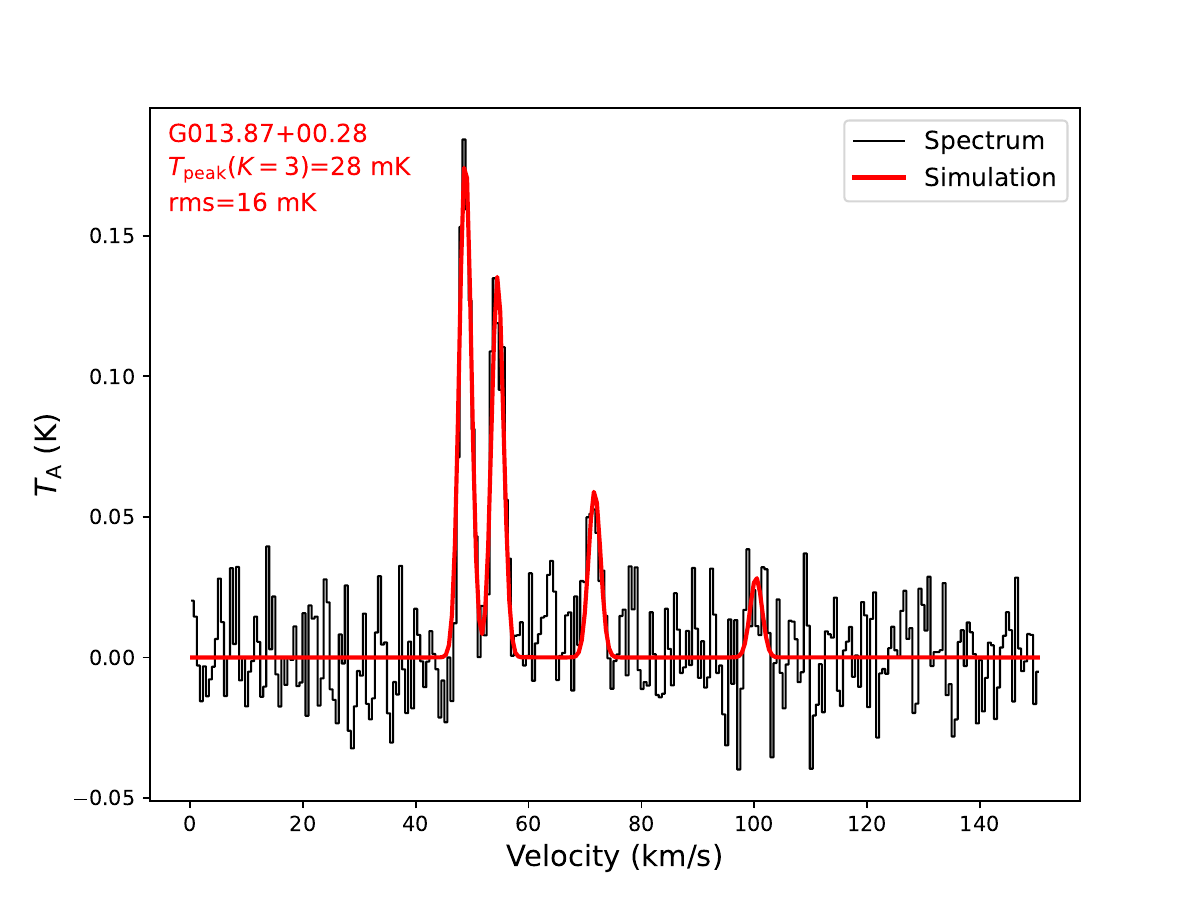}\includegraphics[width=0.33\textwidth]{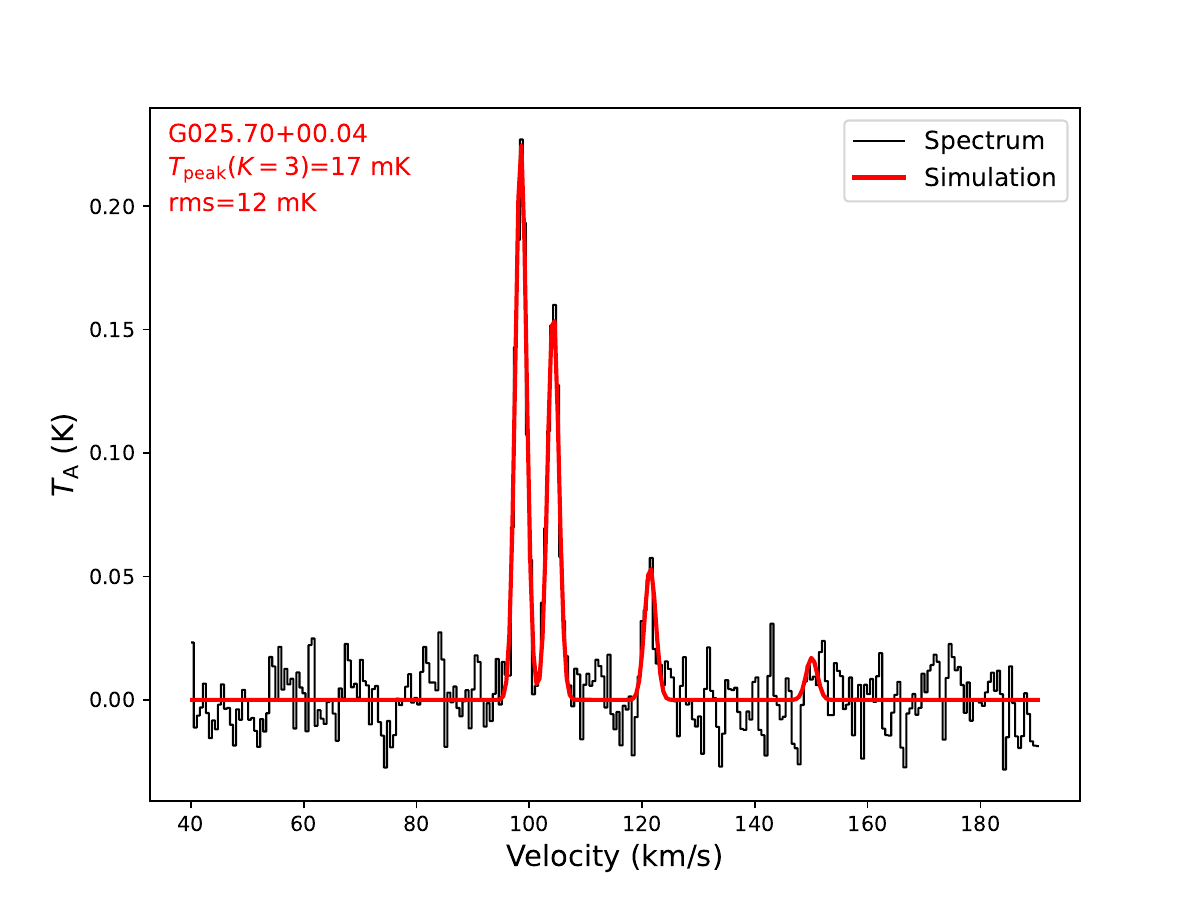}\includegraphics[width=0.33\textwidth]{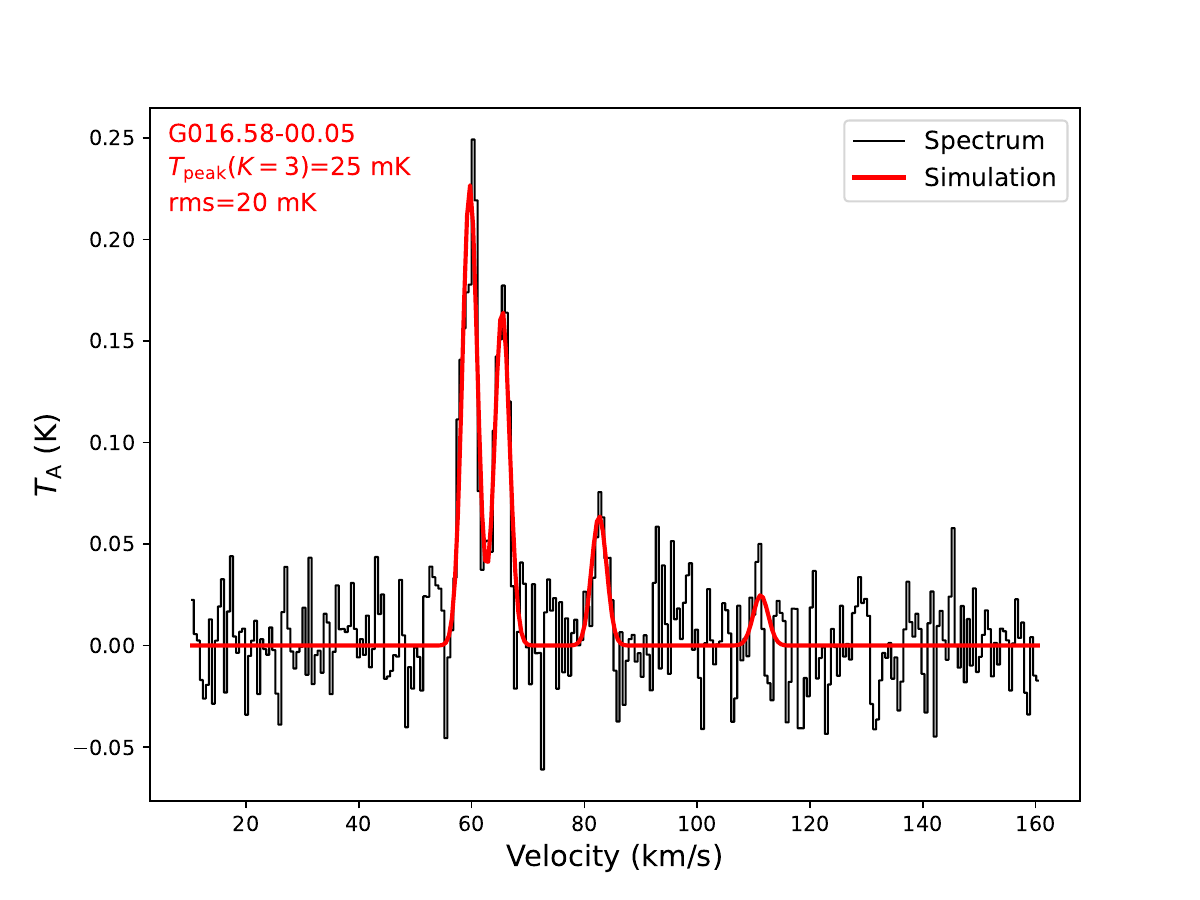}
\caption{The upper panel shows the fitting results of CH$_3$CCH $J$=5-4, while the bottom panel shows the spectra and the simulation results.}
\label{noise}
\end{figure}

\end{document}